# Теорія та методика навчання фундаментальних дисциплін у вищій школі

О. І. Теплицький, І. О. Теплицький,
С. О. Семеріков, В. М. Соловйов

## Професійна підготовка учителів природничо-математичних дисциплін засобами комп'ютерного моделювання: соціально-конструктивістський підхід

Спецвипуск містить монографію О. І. Теплицького, І. О. Теплицького, С. О. Семерікова та В. М. Соловйова, у якій визначені умови професійної підготовки майбутніх учителів природничо-математичних дисциплін засобами комп’ютерного моделювання, розроблено структурно-функціональну модель підготовки, дібрано соціально-конструктивістські форми організації, методи та засоби навчання комп’ютерного моделювання майбутніх учителів природничо-математичних дисциплін.

Для науковців, викладачів та студентів закладів вищої освіти, учителів шкіл та всіх тих, кого цікавлять сучасні теорія та методика навчання.

**Адреса редакції**: а/с 4809, м. Кривий Ріг, 50086, Україна

# ЗМІСТ

## ПЕРЕЛІК УМОВНИХ ПОЗНАЧЕНЬ

| | |
|---|---|
| IDE | Integrated Development Environment (інтегроване середовище програмування) |
| UML | Unified Modeling Language (уніфікована мова моделювання) |
| ЗВО | заклад вищої освіти |
| ГСВО | галузевий стандарт вищої освіти |
| ООА | об’єктно-орієнтований аналіз |
| ООМ | об’єктно-орієнтоване моделювання |
| ООП | об’єктно-орієнтоване програмування |
| ОС | операційна система |
| ПЗ | програмне забезпечення |

# ВСТУП

«Національна стратегія розвитку освіти в Україні на 2012-2021 роки» [174] серед ключових напрямів державної освітньої політики визначає, зокрема, розвиток інноваційної діяльності в освіті та інформатизацію освіти. З метою прискорення реформування освітньої галузі було затверджено Державну цільову соціальну програму підвищення якості шкільної природничо-математичної освіти на 2011-2015 роки [210], спрямовану на стійкий інноваційний розвиток природничо-математичної освіти та його застосування у шкільній практиці. Державний стандарт базової і повної загальної середньої освіти [209] визначає дві комплексні предметні компетентності: 1) природничо-наукову та математичну й 2) проектно-технологічну та інформаційно-комунікаційну, формування яких вимагає оволодіння компетенціями зі створення об’єктів в рамках індивідуальних і колективних проектів; висування та перевірки гіпотез навчально-пізнавального характеру; створення, вивчення та використання об’єктів; використання засобів інформаційно-комунікаційних технологій (ІКТ) для планування, організації індивідуальної і колективної діяльності в інформаційному середовищі. Реалізація вимог стандарту (що впроваджується у частині базової загальної середньої освіти з 1 вересня 2013 р., а у частині повної загальної середньої освіти – з 1 вересня 2018 р.) потребує зміни процесу професійної підготовки майбутніх учителів природничо-математичних дисциплін на основі методу моделювання та дослідницького підходу у навчанні.

Головною спільною рисою природничих наук, основи яких покладено у зміст навчання за природничо-математичними спеціальностями, є використовуваний в них провідний метод дослідження – моделювання, який у процесі навчання стає системотвірною складовою змісту навчання. Враховуючи, що в інформатиці як науці та навчальній дисципліні метод моделювання також є провідним методом дослідження та навчання, у процесі навчання студентів природничих, фізико-математичних та інформатичних спеціальностей педагогічних університетів необхідним є опанування як технології комп’ютерного моделювання, так і технології навчання як дослідження на основі об’єктно-орієнтованого підходу до моделювання та соціально-конструктивістського підходу до навчання.

Вирішення зазначених питань вимагає перебудови системи професійної підготовки майбутніх учителів природничо-математичних дисциплін із урахуванням процесів інтеграції системи освіти України у світовий освітній простір на основі інноваційних педагогічних технологій соціального конструктивізму та конструкціонізму, спрямованих на реалізацію дослідницького підходу в навчанні.

Проблемі навчання моделювання майбутніх учителів природничо-математичних дисциплін присвячені роботи Л. І. Білоусової, О. Г. Колгатіна, К. Є. Рум'янцевої, Г. О. Савченко, А. В. Семенової, Я. Б. Сікори, Ю. Ф. Титової, М. О. Федорової, С. А. Хазіної, О. П. Шестакова та ін.

Філософські та психолого-педагогічні основи застосування дослідницького підходу в навчанні розглядались у роботах Е. Аккерман, Т. І. Бутченка, Е. фон Глазерсфельда, К. Дж. Джерджена, Дж. Дьюї, А. Кея, А. В. Кезіна, О. В. Константинова, П. фон Лоренцена, С. Пейперта, Ж. Піаже, С. А. Ракова, К. В. Рибачука, М. В. Романової, І. Харел, С. А. Цоколова, М. А. Чошанова та інших вітчизняних та зарубіжних дослідників. Соціально-конструктивістські засоби навчання розглядалися в роботах Дж. Адамса, Т. М. Брусенцової, М. Гуздіала, В. Данн, Є. Д. Патаракіна, М. Резника.

Питання навчання об'єктно-орієнтованого моделювання розглядаються в роботах зарубіжних дослідників Й. Бйорстлера, Т. Бринди, З. Шуберт, В. Неллеса, Е. Дж. Корнецькі, Ж.-П. Ріго, С. Хад'єрруїта. Проте методика навчання об'єктно-орієнтованого моделювання майбутніх учителів природничо-математичних дисциплін є нерозробленою.

Вивчення стану професійної підготовки майбутніх учителів природничо-математичних дисциплін дозволило виявити ряд протиріч:

– між потенціалом застосування соціально-конструктивістських ідей у процесі професійної підготовки та відсутністю технології професійної підготовки майбутніх учителів природничо-математичних дисциплін соціально-конструктивістськими засобами комп'ютерного моделювання;

– між спрямованістю природничих наук на міждисциплінарні дослідження об'єктів різної природи та відсутністю уваги до об'єктно-орієнтованого моделювання у вітчизняних підручниках та навчальних і методичних посібниках;

– між потенціалом застосування соціально-конструктивістських засобів інформаційно-комунікаційних технологій навчання об'єктно-орієнтованого моделювання та відсутністю їх систематичного розгляду в курсах моделювання.

Указані суперечності свідчать про необхідність вирішення проблеми ефективної перебудови професійної підготовки майбутніх учителів природничо-математичних дисциплін на засадах комп'ютерного моделювання та ідей соціального конструктивізму і визначають необхідність проведення відповідного дослідження.

Монографія складається з трьох розділів.

У першому розділі проаналізовано розвиток технології комп'ютерного моделювання та сучасний стан її впровадження в професійну підготовку вчителів природничо-математичних дисциплін, досліджено можливості застосування педагогічної технології соціального конструктивізму в процесі професійної підготовки вчителів природничо-математичних дисциплін, обґрунтовано умови підготовки майбутніх учителів природничо-математичних дисциплін засобами комп'ютерного моделювання, описано структуру професійних компетентностей учителя природничо-математичних дисциплін.

У другому розділі розроблено модель підготовки майбутніх учителів природничо-математичних дисциплін засобами комп'ютерного моделювання та систему реалізації умов професійної підготовки в процесі навчання за спецкурсом «Об'єктно-орієнтоване моделювання», визначено критерії, рівні й показники сформованості компетентності з комп'ютерного моделювання майбутніх учителів природничо-математичних дисциплін.

У третьому розділі описано завдання, зміст і результати експериментальної роботи, проведено їх статистичне опрацювання та аналіз.

Автори щиро дякують ініціаторам створення монографії – професорам кафедри інформатики Харківського національного педагогічного університету імені Г. С. Сковороди Людмилі Іванівні Білоусовій, Олександру Геннадійовичу Колгатіну, Сергію Анатолійовичу Ракову та всім співробітникам спільної науково-дослідної лабораторії з питань використання хмарних технологій в освіті ДВНЗ «Криворізький національний університет» та Інституту інформаційних технологій і засобів навчання НАПН України, які вносили пропозиції щодо структури та змісту цієї роботи. Особлива подяка нашим учням Юлії Володимирівни Єчкало та Світлані Вікторівні Шокалюк, які долучались до нашого дослідження із самого його початку. І, нарешті, ми хочемо присвятити цю роботу світлій пам'яті нашого вчителя, співавтора та друга – фундатора криворізької інформатичної школи Олександра Павловича Поліщука.

# РОЗДІЛ 1
# НАУКОВО-ПЕДАГОГІЧНІ ЗАСАДИ ПРОФЕСІЙНОЇ ПІДГОТОВКИ МАЙБУТНІХ УЧИТЕЛІВ ПРИРОДНИЧО-МАТЕМАТИЧНИХ ДИСЦИПЛІН ЗАСОБАМИ КОМП'ЮТЕРНОГО МОДЕЛЮВАННЯ

## 1.1 Компетентнісний підхід як основа професійної підготовки вчителів природничо-математичних дисциплін

У закладах вищої освіти підготовка за напрямами і спеціальностями фахівців всіх освітніх та освітньо-кваліфікаційних рівнів здійснюється за відповідними освітньо-професійними програмами ступенево або неперервно, залежно від вимог до рівня оволодіння певною сукупністю умінь та навичок, необхідних для майбутньої професійної діяльності.

«Великий тлумачний словник сучасної української мови» [84, с. 952] термін «підготовка» визначає як «1. Дія за значенням підготовити. 2. Запас знань, навичок, досвід і т. ін., набутий у процесі навчання, практичної діяльності». Базовий термін «підготовити», у свою чергу, визначається як «1. Забезпечити здійснення, проведення, існування чогось, завчасно роблячи, готуючи для цього все необхідне. 2. Давати необхідний запас знань, передавати навички, досвід і т. ін. в процесі навчання, практичної діяльності». Таким чином, професійну підготовку можна визначити через знання, уміння, навички, досвід і т. ін., набуті у процесі професійного навчання.

Професійна підготовка майбутніх учителів (за В. О. Сластьоніним – «професійно-педагогічна підготовка» [273]) передбачає формування системи компетентностей майбутнього вчителя: як загальнопрофесійних, що відображають особливості його професійної діяльності як учителя, так і спеціальних професійних, що відображають особливості його професійної діяльності як вчителя-предметника. Характеристиками професійних компетентностей учителя є «володіння великим обсягом суспільно-політичних і наукових знань із дисципліни, яка викладається, та з суміжних наук, володіння високим рівнем загальної культури, знання педагогічної теорії, загальної, вікової і педагогічної психології, уміння розв'язувати педагогічну задачу і здійснювати самокритичний аналіз, навички виконання відповідних дій, які є компонентом конкретних видів навчально-виховної діяльності» [273, с. 4-5].

Результат професійної підготовки майбутніх учителів можна виразити також у термінах готовності до професійної (педагогічної) діяльності вчителя, яка, згідно А. Ф. Линенко, є якістю особистості, що охоплює позитивне ставлення до професії, здібності, знання, уміння, навички, стійкі професійно важливі якості [158]. На нашу думку, зміст поняття професійно-педагогічної готовності майбутнього вчителя може бути поданий через його компетентність на основі урахування провідних видів діяльності вчителя, до яких О. Я. Мариновська [162] відносить організаційну, методичну, позашкільну, викладацьку, виховну, управлінську, науково-дослідну, а також самоосвіту і самовиховання вчителя.

В. В. Оніпко результативний компонент професійно-педагогічної підготовки майбутніх учителів визначає через «досягнення випускником визначеного рівня оволодіння системою знань теоретичних основ педагогіки, психології, предметних методик, вікової фізіології, медичних знань, комплексу природничих дисциплін, а також грамотного використання отриманих знань у майбутній практичній педагогічній діяльності» [184, с. 39]. Майбутній учитель має також володіти системою методичних знань, умінь і навичок: знати психолого-педагогічні концепції навчання і використовувати їх як основу у своїй практичній діяльності, знати й розв'язувати проблеми наступності, на високому рівні знати вимоги, які висуваються до обов'язкового мінімуму змісту шкільної освіти; володіти методикою викладання, творчо використовувати методичні інновації й ідеї, уміти самостійно обрати або скласти програму навчання, методично грамотно викласти навчальний матеріал із урахуванням вікових та індивідуальних особливостей учнів.

В. В. Оніпко визначає такі основні види підготовки вчителя [184]:

– *методологічну* – процес формування у студентів системи знань про принципи, методи дослідження й перетворення педагогічної дійсності:

– *культурологічну* та *загальнокультурну*, покликані забезпечити формування й розвиток світогляду вчителя, розширити його загальну освіту і рівень культури;

– *предметну* (*спеціальну*), у ході якої відбувається оволодіння теоретичними і практичними знаннями, притаманними даній професії, формування системи умінь і навичок, розвиток здібностей і надбання професійного досвіду

– *методичну*, що забезпечує знання принципів, змісту, засобів, методів і форм організації навчання;

– *психолого-педагогічну*, спрямовану на професійну освіту вчителя, формування і розвиток педагогічних умінь, розвиток його творчої індивідуальності.

Ураховуючи, що кожен з видів професійної підготовки передбачає формування відповідних складових системи професійних компетентностей, розглянемо принципи застосування компетентнісного підходу до професійної підготовки майбутніх учителів природничо-математичних дисциплін.

Провідна ідея компетентнісного підходу до професійної підготовки полягає в тому, що головним завданням професійної підготовки має стати формування готовності майбутніх учителів до ефективної діяльності в професійній сфері.

Згідно І. С. Мінтій [167], компетентність – це освітній результат (який формується не лише в процесі формальної освіти (навчання у ЗВО), але й неформальної (родина, друзі, робота, політика, релігія, культура й ін.)); сформована особистісна якість, яка містить наступні складові:

– когнітивно-змістову (гносеологічну): знання;

– операційно-технологічну (праксеологічну): навички, уміння, досвід діяльності;

– ціннісно-мотиваційну (аксіологічну): мотивація, ціннісне ставлення;

– соціально-поведінкову: комунікабельність, здатність до адаптації, здатність до інтеграції, вміння спілкуватися, розуміти, поважати та оцінювати різні підходи до розв’язання однієї задачі.

Виокремлені складові не є розрізненими, вони тісно взаємодіють між собою.

Компетентності вчителя А. І. Кузьмінський визначає як комплекс педагогічних здібностей і можливостей, вмотивована спрямованість на освітній процес, система необхідних знань, умінь, навичок і досвіду, які постійно вдосконалюються і реалізуються на практиці [147, с. 146].

Н. Ф. Радіонова та А. П. Тряпіцина [215] професійні компетентності вчителя визначають як інтегральну характеристику, яка визначає здатність вирішувати професійні проблеми та типові професійні задачі, які виникають в реальних ситуаціях професійної педагогічної діяльності, з використанням знань, професійного та життєвого досвіду, цінностей та схильностей. Компетентність завжди «виявляється» в дії, діяльності, поведінці та вчинках, не можна побачити «невиявлену» компетентність. Професійні компетентності вчителя виявляються при вирішенні професійних задач.

Розуміючи професійну підготовку як процес професійного розвитку, оволодіння досвідом майбутньої професійної діяльності, можна визначити, що компетентний спеціаліст завжди орієнтується на майбутнє, передбачає зміни, орієнтований на самоосвіту. Особливістю професійних компетентностей є те, що вони реалізуються в сьогоденні, а орієнтовані на майбутнє.

*Професійні компетентності* – це сукупність ключових, базових та спеціальних компетентностей [215].

*Ключові компетентності* – це компетентності, необхідні для будь-якої професійної діяльності, сприяють успіху людини у сучасному мінливому світі.

*Базові компетентності* відображають специфіку визначеної професійної діяльності (педагогічної, медичної, технічної та ін.).

Для професійної педагогічної діяльності базовими є компетентності, необхідні для організації професійної діяльності в контексті вимог до системи освіти на визначеному етапі розвитку суспільства.

*Спеціальні компетентності* відображають специфіку конкретної предметної чи надпредметної сфери професійної діяльності. Спеціальні компетентності можна розглядати як реалізацію ключових та базових компетентностей в галузі навчального предмету, конкретної галузі професійної діяльності.

Всі три види компетентностей взаємопов'язані та розвиваються одночасно, що забезпечує становлення професійних компетентностей як визначеної цілісної, інтегративної особистісної характеристики фахівця.

На нашу думку, для означення компетентностей, які відображають загальну специфіку професійної діяльності, краще підходить характеристика «загальнопрофесійні». Адже, зазвичай, характеризуючи певні властивості, як базові, ми маємо на увазі не лише найзагальніші, але й найпростіші.

На думку В. О. Адольфа, «професійні компетентності (педагога)– складне утворення, що містить комплекс знань, вмінь, властивостей і якостей особистості, які забезпечують варіативність, оптимальність та ефективність побудови навчально-виховного процесу» [65, с. 118].

Як зазначає Н. Г. Ничкало, професійні компетентності вчителя – це «гармонійне поєднання знань навчальної дисципліни, методики і дидактики викладання, а також умінь і навичок культури педагогічного спілкування» [178, с. 8]. У структурі професійних компетентностей вчителя дослідник вирізняє дві підструктури: діяльнісну (знання, уміння, навички і здібності особистості, необхідні для здійснення педагогічної діяльності) і комунікативну (знання, уміння, навички, необхідні особистості для здійснення педагогічного спілкування). Таким чином, у структурі загальнопрофесійної компетентності вчителя Н. Г. Ничкало виокремлює такі компетентності: предметні, методичні, дидактичні, комунікативні.

В. М. Гриньова вважає, що професійні компетентності педагога є чинником підвищення якості освіти і включають професійно-змістовий, технологічний і професійно-особистісний компоненти [96, с. 25].

Професійно-змістовий компонент передбачає наявність у викладача цінностей – знань з предмету, який він викладає, суміжних дисциплін, з дисциплін, що виражають квінтесенцію спеціальності, якими має оволодіти студент, теоретичних знань з основ наук, які вивчають особистість людини, що забезпечує усвідомленість при визначенні педагогом змісту його професійної діяльності з виховання, навчання та освіти студентів. Технологічний компонент включає професійні цінності – знання, апробовані в дії, тобто цінності-вміння. Забезпечують цей компонент інформаційно-інноваційні технології, які ґрунтуються на комплексному діагностико-дослідному осмисленні педагогічної ситуації і перспективному її прогнозуванні. Професійно-особистісний компонент включає особистісні здібності-цінності. Отже, структура загальнопрофесійних компетентностей вчителя за В. М. Гриньовою наступна: предметні; психологічні; методичні; ІКТ-компетентності.

Ґрунтовно розглядаються професійні компетентності педагога і в роботі С. О. Дружилова: «професійні компетентності педагога – це багатофакторне явище, що містить в собі систему теоретичних знань учителя та способів їх застосування в конкретних педагогічних ситуаціях, ціннісні орієнтації педагога, а також інтегративні показники його культури (мова, стиль спілкування, відношення до себе та своєї діяльності, до суміжних галузей знань та ін.)» [105, с. 28]. У професійних компетентностях педагога науковець виокремлює такі компоненти: мотиваційно-вольовий, функціональний, комунікативний та рефлексивний.

Мотиваційно-вольовий компонент включає в себе: мотиви, цілі, потреби, ціннісні установки, стимулює творчий прояв особистості в професії; передбачає наявність інтересу до професійної діяльності.

Функціональний компонент в загальному випадку проявляється у вигляді знань про способи педагогічної діяльності, необхідних вчителю для проектування та реалізації тієї чи іншої педагогічної технології.

Комунікативний компонент компетентностей включає вміння ясно і чітко викладати думки, переконувати, аргументувати, будувати докази, аналізувати, висловлювати судження, передавати раціональну і емоційну інформацію, встановлювати міжособистісні зв'язки, погоджувати свої дії з діями колег, вибирати оптимальний стиль спілкування в різних ділових ситуаціях, організовувати і підтримувати діалог.

Рефлексивний компонент виявляється в умінні свідомо контролювати результати своєї діяльності і рівень власного розвитку, особистісних досягнень; сформованість таких якостей і властивостей, як креативність, ініціативність, націленість на співробітництво, співтворчість, схильність до самоаналізу. Рефлексивний компонент є регулятором особистісних досягнень, пошуку особистого сенсу у спілкуванні з людьми, самоврядування, а також збудником самопізнання, професійного зростання, вдосконалення майстерності та формування індивідуального стилю роботи.

Л. Г. Карпова професійні компетентності учителя визначає як «інтегративне особистісне утворення на засадах теоретичних знань, практичних умінь, значущих особистісних якостей та досвіду, що зумовлює готовність учителя до виконання педагогічної діяльності та забезпечує високий рівень її самоорганізації». Професійні компетентності вчителя не мають вузько професійних меж, оскільки від нього вимагається постійне осмислення розмаїття соціальних, психологічних, педагогічних та інших проблем, які пов’язані з освітою [128, с. 28].

В. О. Сластьонін стосовно визначення професійних компетентностей педагога зауважує: «поняття професійних компетентностей педагога виражає єдність його теоретичної і практичної готовності до здійснення педагогічної діяльності і характеризує його професіоналізм» [273, с. 30].

Структура професійних компетентностей учителя, згідно зі В. О. Сластьоніним, може бути розкрита через педагогічні вміння. Педагогічні вміння він об’єднує в чотири групи.

1. Уміння «переводити» зміст об’єктивного процесу виховання в конкретні педагогічні завдання: вивчення особистості і колективу для визначення рівня їх підготовленості до активного оволодіння новими знаннями і проектування на цій основі розвитку колективу й окремих учнів; виділення комплексу освітніх, виховних і розвиваючих завдань, їх конкретизація і визначення домінуючого завдання (психолого-педагогічні компетентності).

2. Уміння побудувати і привести в рух логічно завершену педагогічну систему: комплексне планування освітньо-виховних завдань; обґрунтований відбір змісту освітнього процесу; оптимальний вибір форм, методів і засобів його організації (дидактичні компетентності).

3. Уміння виокремлювати і встановлювати взаємозв’язки між компонентами і факторами виховання, приводити їх у дію: створення необхідних умов (матеріальних, морально-психологічних, організаційних, гігієнічних та ін.); активізація особистості школяра, розвиток його діяльності, що перетворює його із об’єкта в суб’єкт виховання, організація і розвиток спільної діяльності, забезпечення зв’язку школи із середовищем, регулювання зовнішніх, не програмованих впливів (психолого-педагогічні, організаційні компетентності).

4. Уміння обліку та оцінки результатів педагогічної діяльності: самоаналіз і аналіз освітнього процесу і результатів діяльності вчителя; визначення нового комплексу домінуючих і підпорядкованих педагогічних завдань (психолого-педагогічні компетентності).

Теоретична діяльність, у свою чергу, виявляється в узагальненому умінні педагогічно мислити, що передбачає наявність у вчителя аналітичних, прогностичних, проективних, а також рефлексивних умінь.

Зміст практичної готовності виражається у зовнішніх (предметних) уміннях, тобто в діях, які можна спостерігати. До них належать організаторські і комунікативні уміння.

Таким чином, В. О. Сластьонін виокремлює у загальнопрофесійних компетентностях вчителя психолого-педагогічні, дидактичні, організаційні та комунікативні компетентності.

Розглядаючи питання професійних компетентностей, І. О. Зімняя зазначає, що в результаті навчання в людини мають бути сформовані деякі соціально-професійні якості, які надають можливість їй успішно виконувати виробничі задачі та взаємодіяти з іншими людьми [121, с. 16]. Ці якості можуть бути визначені як соціально-професійні компетентності людини.

У такому розумінні соціально-професійні компетентності людини є його особистісні, інтегративні, сформовані якості, які проявляються в адекватності розв’язання (стандартних та особливо нестандартних, творчих) задач у всьому розмаїтті соціальних та професійних ситуацій.

О. Г. Ларіонова у структурі професійних компетентностей вчителя виокремлює інформаційно-методологічні, теоретичні, методичні, соціально-комунікативні та особистісно-валеологічні [153, с. 32].

О. В. Лебедєва визначає таку структуру професійних компетентностей вчителя [154]:

1) науково-теоретичні компетентності:

– спеціальні компетентності (фундаментально-наукова підготовка) – знання й уміння у галузі фахового предмета, наукові основи шкільного курсу відповідної шкільної дисципліни;

– методологічні компетентності – знання філософії науки як методологічної основи пізнавальної діяльності (методи наукового пізнання у фаховій галузі);

– інформаційні компетентності – навички й уміння орієнтуватися в інформаційному просторі, використовувати комп’ютерні технології на різних етапах освітнього процесу;

2) методичні компетентності:

– загальнопедагогічні;

– дидактичні;

– конкретно-методична підготовка, яка представлена через специфічні методи і прийоми навчання;

3) психолого-педагогічні компетентності:

– комунікативні компетентності – знання й уміння щодо побудови сприятливого психологічного клімату, відносин співробітництва в системах «учитель-учень», «учень-учень», продуктивних стосунків з колегами;

– диференціально-психологічні компетентності, які пов'язані з виявленням особистісних якостей і спрямованості учнів, з формуванням і розвитком їхньої пізнавальної активності.

У даному випадку вважаємо, що загальнопрофесійні компетентності вище описаної структури складаються з таких компонентів: інформаційні, загальнопедагогічні, дидактичні, методичні, комунікативні та психологічні компетентності.

Згідно з О. М. Спіріним, система компетентностей учителя наступна [292, с. 212]:

1) загальні компетентності: компетентності щодо індивідуальної ідентифікації й саморозвитку; міжособистісні компетентності; суспільно-системні компетентності;

2) професійні компетентності: загальнопрофесійні компетентності; предметно-орієнтовані (профільно-орієнтовані) компетентності (науково-предметні, предметно-педагогічні); технологічні компетентності (компетентності в галузі педагогічних технологій та інформаційно-технологічні компетентності); професійно-практичні компетентності.

Згідно визначеної нами структури професійних компетентностей вчителя, до загальнопрофесійних компетентностей за О. М. Спіріним віднесемо загальнопрофесійні, предметно-педагогічні та технологічні, а до спеціальних професійних – науково-предметні та професійно-практичні компетентності.

Н. Ф. Радіонова та А. П. Тряпіцина [215] виокремлюють 5 основних груп задач, досвід вирішення яких характеризує базові (в нашому трактуванні – загальнопрофесійні) компетентності вчителя: 1) бачити учня в навчальному процесі (психолого-педагогічні компетентності); 2) будувати навчальний процес, орієнтований на досягнення мети конкретного ступеню навчання (дидактичні компетентності); 3) встановлювати взаємодію з іншими суб'єктами навчального процесу, партнерами школи (організаційно-управлінські компетентності); 4) створювати та використовувати з педагогічною метою навчальне середовище (загальнопедагогічні компетентності); 5) проектувати та здійснювати професійну самоосвіту (навчальні компетентності).

Дещо по-іншому підходить до структурування загальнопрофесійних компетентностей вчителя А. К. Маркова [163, с. 82]. Згідно з нею, до складу загальнопропрофесійних компетентностей входять процесуальні та результативні. У процесуальних вона виокремлює педагогічну діяльність, педагогічне спілкування та особистість вчителя; в результативних – навченість та вихованість учнів.

Для оцінки рівня сформованості компетентностей А. К. Маркова для кожної з складових пропонує розглядати сукупність необхідних знань, вмінь та психологічні вимоги до їх виконання.

Н. В. Кузьміна [146] структуру професійно-педагогічних (загальнопрофесійних – у нашому трактуванні) компетентностей вчителя визначає наступним чином:

– методичні компетентності щодо способів формування знань, вмінь і навичок учнів;

– соціально-педагогічні компетентності у сфері процесів спілкування;

– диференціально-психологічні компетентності у сфері мотивів, здатностей і спрямувань учнів;

– аутопсихологічні компетентності у галузі переваг і недоліків власної діяльності й особливостей власних особистісних якостей;

– спеціальні і професійні компетентності у сфері тієї дисципліни, що викладається.

Експерти Міжнародної організації ЮНЕСКО також активно працюють над розробкою професійних компетентностей учителів. Особливу увагу вони приділяють ІКТ-компетентностям [14]. Вони зазначають, що для більш успішного життя, навчання та роботи в інформаційному суспільстві студенти та викладачі повинні використовувати існуючі технології.

У рамках освітнього процесу використання ІКТ надає студентам можливість:

– здійснювати пошук даних, їх аналіз; виконувати обчислення;

– вирішувати проблеми та приймати рішення;

– творчо та ефективно використовувати всі можливі інструменти для підвищення власної продуктивності;

– стати інформованими, відповідальними, свідомими громадянами.

Завдяки ефективному використанню ІКТ в процесі навчання студенти можуть значно підвищити свої можливості. Ключовою фігурою в допомозі студентам з розвитку власних можливостей має стати викладач.

Тому сучасним викладачам необхідно бути готовими, щоб забезпечити підтримку студентам щодо вивчення ІКТ, ознайомити їх з перевагами, які вони можуть надати та бути готовими використовувати ІКТ. Традиційні освітні методи вже більше не забезпечують в повній мірі майбутніх викладачів всіма необхідними навичками.

Проект ЮНЕСКО «ICT Competency Standards for Teachers (ICT-CST)» («Стандартні ІКТ-компетентності для викладачів») [14] створений з метою розробки повної структури компетентностей з ІКТ, проектування освітніх стандартів існуючого навчання і освітніх програм на ICT-CST, здійснення спроби прискорення глобальних змін в цій області. Запропонований навчальний план з курсу, задачею якого є формування ІКТ-компетентностей у вчителів розвиває три підходи: освіченість, поглиблення і створення знань; з шістьма компонентами освітньої системи: спрямованість, навчальний план, педагогіка, ІКТ, організація й управління, розвиток професіоналізму вчителя.

Отже, підсумовуючи, в системі професійних компетентностей учителя (у тому числі – природничо-математичних дисциплін) можна виокремити такі складові:

– ключові компетентності: навчальні; соціальні; загальнокультурні; здоров'язберігаючі; ІКТ-компетентності; громадянські; підприємницькі;

– загальнопрофесійні: методичні; науково-дослідницькі; психолого-педагогічні; організаційно-управлінські; комунікативні; ІКТ-компетентності вчителя;

– спеціальні професійні (предметні).

Професійна підготовка вчителів природничо-математичних дисциплін у ЗВО України здійснюється в межах галузей знань 0401 «Природничі науки» (напрями підготовки 6.040101 «Хімія», 6.040102 «Біологія», 6.040104 «Географія», 6.040106 «Екологія, охорона навколишнього середовища та збалансоване природокористування»), 0402 «Фізико-математичні науки» (напрями підготовки 6.040201 «Математика», 6.040203 «Фізика») та 0403 «Системні науки та кібернетика» (напрям підготовки 6.040302 «Інформатика»). Найбільш поширені ці спеціальності зі спеціалізацією «Інформатика», уведення якої відображає необхідність професійно орієнтованої ІКТ-підготовки майбутнього вчителя природничо-математичних дисциплін. У додатку Д представлено статистичні відомості щодо професійної підготовки майбутніх учителів природничо-математичних дисциплін у ЗВО України.

Із табл. Д.1 видно, що природничо-математичні напрями підготовки користуються значною популярністю: так, у 2012 р. кількість абітурієнтів, зарахованих на перший курс, була на 24,3 % вище обсягу державного замовлення. Аналіз назв спеціальностей [124] показує поширеність так званих «подвійних спеціальностей» – напрямів підготовки («перша спеціальність») з певною спеціалізацією («друга спеціальність»), найбільш популярною з яких є спеціалізація «Інформатика», основою якої виступає напрям підготовки 6.040302 «Інформатика». Підготовка бакалаврів за цим напрямом у ЗВО України виконується в межах галузі знань «Системні науки та кібернетика»; відповідні складові галузевого стандарту вищої освіти України [88; 89] затверджено Наказом МОН України № 880 від 16 вересня 2010 року.

Рис. 1.1 відображає суспільний попит на різні напрями природничо-математичної підготовки. Так, у 2012 році найпопулярнішим серед вступників був напрям 6.040201 «Математика», на другому місці – 6.040102 «Біологія», а на третьому – 6.040104 «Географія». Слід зазначити, що суспільний попит та державне замовлення на підготовку вчителів природничо-математичних дисциплін за різними напрямами підготовки суттєво різниться.

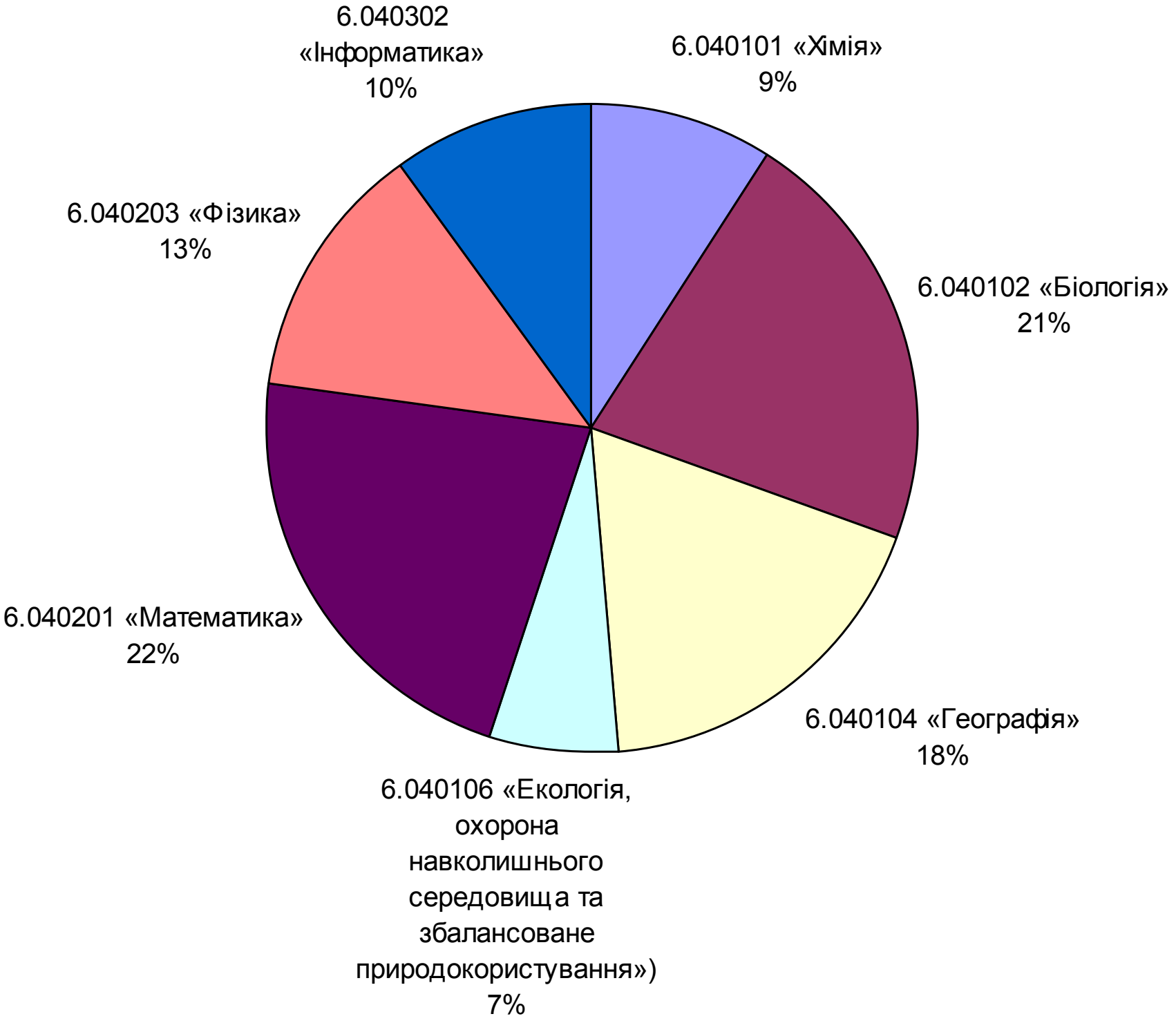

Рис. 1.1. Розподіл абітурієнтів, зарахованих у 2012 р. на перший курс, за різними природничо-математичними напрямами підготовки

Із рис. 1.2 видно, що за напрямами підготовки 6.040302 «Інформатика», 6.040203 «Фізика», 6.040201 «Математика», 6.040101 «Хімія» перевищення кількості зарахованих абітурієнтів над державним замовленням знаходиться у межах 1,63 %-11,41 %, а за напрямами підготовки 6.040102 «Біологія», 6.040104 «Географія», 6.040106 «Екологія, охорона навколишнього середовища та збалансоване природокористування») – у межах 36,30 %-102,91 %.

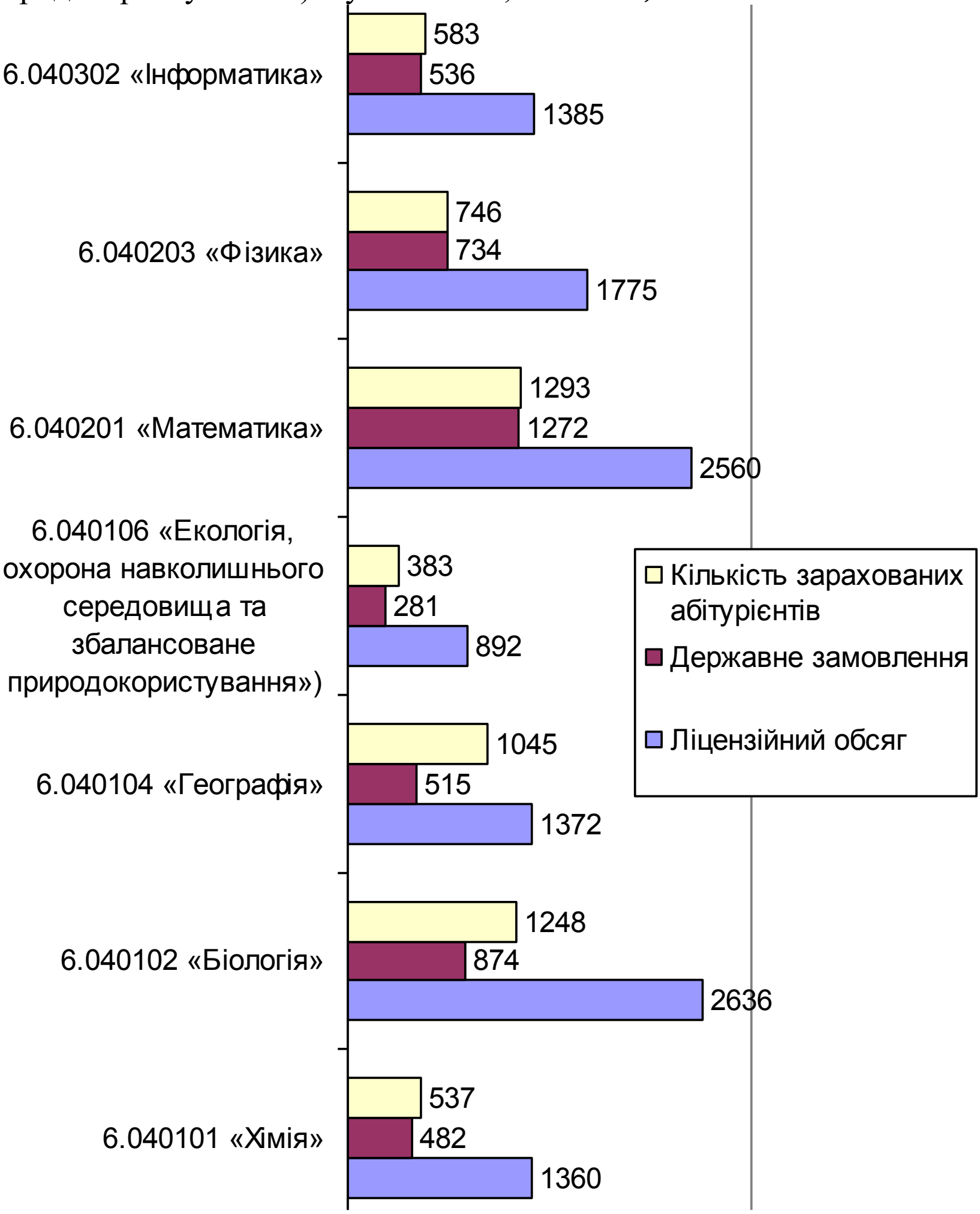

Рис. 1.2. Співвідношення кількості зарахованих абітурієнтів, державного замовлення та ліцензійного обсягу на природничо-математичні напрями підготовки

Саме ці показники і змушують педагогічні ЗВО уводити «подвійні спеціальності», додаючи до суспільно затребуваних напрямів державно важливі спеціалізації, провідною з яких є спеціалізація «Інформатика». Крім того, уведення спеціалізацій обумовлюється вимогою повноти природничо-математичної підготовки: напрям 6.040201 «Математика» є єдиним, за яким відбувається підготовка у всіх 35 ЗВО, які готують майбутніх учителів природничо-математичних дисциплін (рис. 1.3).

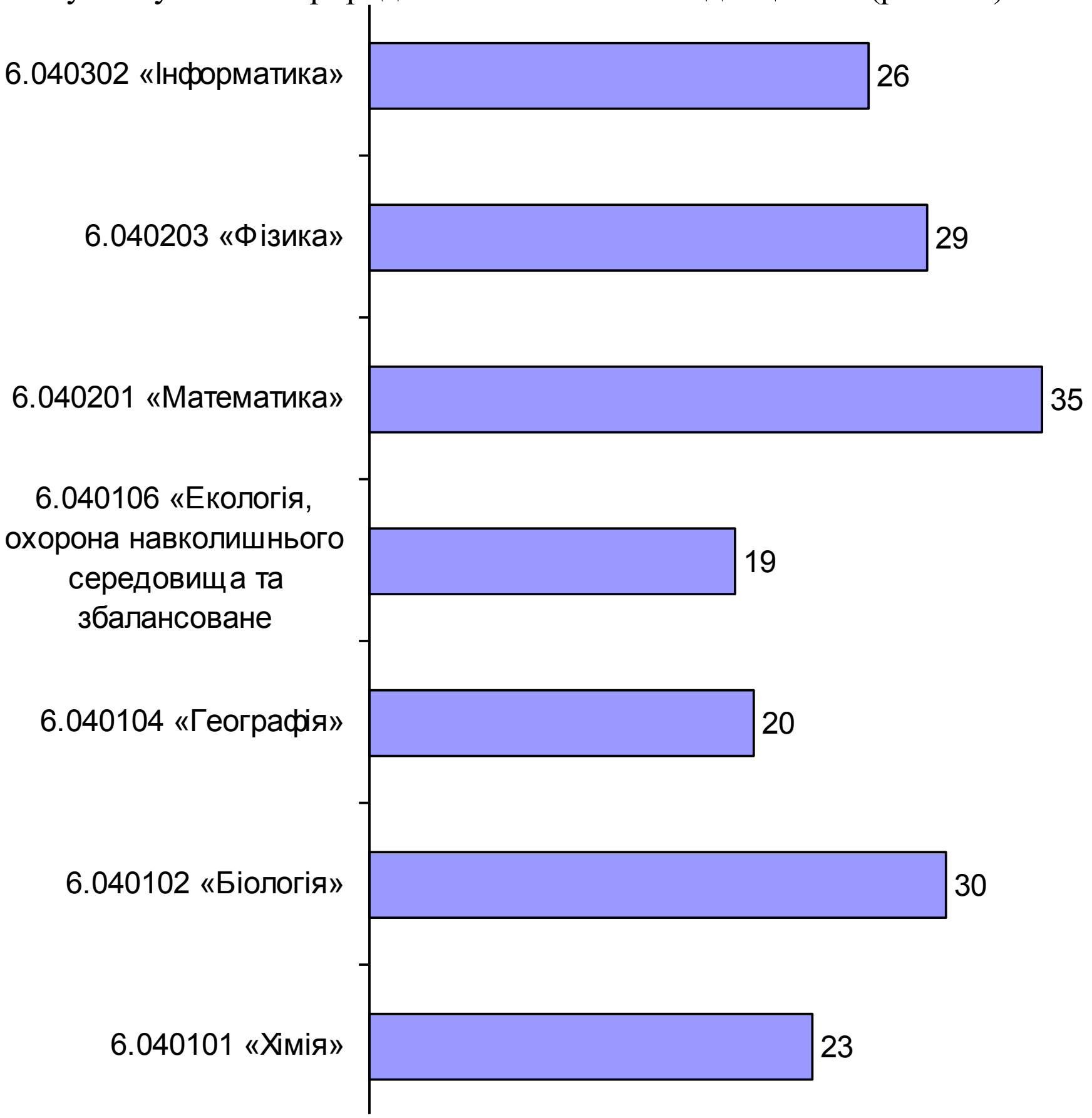

Рис. 1.3. Кількість ЗВО, що забезпечують підготовку вчителів природничо-математичних дисциплін за різними напрямами підготовки

На другому місці за кількістю ЗВО – напрям 6.040102 «Біологія», на третьому – 6.040203 «Фізика» (у той час як третій за популярністю серед абітурієнтів напрям підготовки 6.040104 «Географія» за кількістю ЗВО займає передостаннє місце).

Із метою виявлення спільних рис у професійній підготовці майбутніх учителів природничо-математичних дисциплін проаналізуємо цілі та зміст шкільної природничо-математичної освіти.

Державний стандарт базової і повної загальної середньої освіти [209] визначає дві комплексні предметні компетентності: 1) природничо-наукова і математична; 2) проектно-технологічна та інформаційно-комунікаційна.

Метою освітньої галузі «Природознавство» є формування в учнів природничо-наукової компетентності як базової складової загальної культури особистості і розвитку її творчого потенціалу через розв'язання таких завдань:

1) забезпечення оволодіння учнями апаратом природничих наук з метою розуміння перебігу природних явищ і процесів;

2) забезпечення усвідомлення учнями фундаментальних ідей і принципів природничих наук;

3) набуття досвіду практичної та експериментальної діяльності, здатності застосовувати знання в процесі пізнання світу.

Математична компетентність формується в освітній галузі «Математика», до завдань якої відносяться:

1) розкриття можливостей математики в описанні реальних процесів і явищ дійсності, забезпечення усвідомлення математики як універсальної мови природничих наук;

2) забезпечення розуміння учнями математичних моделей як таких, що дають змогу описувати загальні властивості об'єктів, процесів та явищ;

3) розвиток умінь опрацьовувати, шукати і використовувати інформацію, критично оцінювати її, виокремлювати головне, аналізувати, робити висновки;

4) формування здатності приймати рішення в умовах неповної, надлишкової, точної та ймовірнісної інформації.

Формування і розвиток проектно-технологічної та інформаційно-комунікаційної компетентностей для реалізації творчого потенціалу учнів і їх соціалізації в суспільстві відбувається через формування в учнів навичок і вмінь проводити основні операції з інформаційними об'єктами, зокрема:

1) створювати об'єкти, фіксувати, записувати, спостерігати за ними і вимірювати їх, зокрема, в рамках реалізації індивідуальних і колективних проектів;

2) висувати і перевіряти нескладні гіпотези навчально-пізнавального характеру, створювати, вивчати та використовувати об'єкти;

3) використовувати засоби ІКТ для планування, організації індивідуальної і колективної діяльності в інформаційному середовищі.

Зокрема, основними завданнями навчання інформатики у старшій школі є формування в учнів здатності виявляти й аналізувати інформаційні процеси в технічних, біологічних і соціальних системах та будувати і використовувати інформаційні моделі, а також засоби опису та моделювання явищ і процесів.

Таким чином, формування в учнів природничо-наукової і математичної, проектно-технологічної та інформаційно-комунікаційної компетентностей вимагає комплексного використання математичного та комп'ютерного моделювання як складової їхньої фундаментальної підготовки. Враховуючи, що зазначений Державний стандарт впроваджується в частині базової загальної середньої освіти з 1 вересня 2013 р., а в частині повної загальної середньої освіти – з 1 вересня 2018 року, вкрай необхідною є перебудова процесу професійної підготовки майбутніх учителів природничо-математичних дисциплін на основі посилення ролі методу моделювання.

Отже, професійно-педагогічна підготовка вчителя природничо-математичних дисциплін характеризується багатоаспектністю й варіативністю змісту і носить інтегративно-міждисциплінарний характер. Тому її сутнісною ознакою є інтегративність, яку розглядаємо як один з пріоритетних напрямків удосконалення професійно-педагогічної освіти, оскільки вона забезпечує комплексність, повноту, цілісність знань, формування системного мислення майбутнього педагога, сприяє глибокому осмисленню педагогічної діяльності, забезпеченню мобільності й варіативності змісту педагогічної освіти. Оскільки професійна підготовка майбутніх учителів природничо-математичних дисциплін повинна, з одного боку, давати знання про найсучасніші технології в даній області, а з іншого – формувати узагальнені навички опанування нових засобів ІКТ, то для успішної професійної діяльності у майбутніх учителів природничо-математичних дисциплін має бути сформована спеціальна професійна компетентність з комп'ютерного моделювання.

## 1.2 Комп'ютерне моделювання в системі професійної підготовки учителів природничо-математичних дисциплін

Бурхливий розвиток методів моделювання, що відбувається особливо інтенсивно в останні десятиліття, спричинив формування ряду специфічних понять, уявлень і прийомів, пов'язаних з побудовою, аналізом і використанням моделей різних класів. Сьогодні є всі підстави говорити про моделювання і модельні методи як про самостійну галузь знань, сфера додатків яких простягається від питань теорії пізнання до вирішення суто практичних виробничих проблем.

Підвищений інтерес до модельної проблематики обумовлений тією роллю, яку методи моделювання, особливо математичного, набули в сучасних дослідженнях. Крім того, цей інтерес стимулюється, з одного боку, прогресуючою складністю проблем що їх доводиться вирішувати людині в своїй діяльності, а з іншого – великими успіхами в розвитку прикладної математики, обчислювальної техніки і програмування, дослідження операцій, наукових дисциплін кібернетичного циклу. В сукупності з традиційними галузями науки вони дають можливість для вирішення значної частини прикладних задач науковими методами з метою отримання оптимальних результатів.

Першим в історії прикладом науково обґрунтованого застосування методу моделювання, можна вважати роботи з дослідження гідродинамічних характеристик суден у випробувальних каналах, які розгорнулися в другій половині XIX століття [308, с. 20-21]. Проте навіть у перші десятиліття двадцятого століття найбільш поширене тлумачення поняття «модель» як і раніше було пов’язане не з наукою, а з виробництвом.

Переломними в розвитку моделювання виявилися 40-50-ті роки XX століття, роки становлення наукових дисциплін кібернетичного спрямування, методів дослідження операцій, бурхливого розвитку прикладної математики, обчислювальної техніки і комп’ютерного програмування. У цей період значно розширився перелік дослідницьких і прикладних завдань, які вирішуються науковими методами, і це спричинило необхідність виявлення і опису істотних для кожної задачі властивостей і рис безлічі найрізноманітніших об’єктів, тобто необхідність усвідомленої побудови їх моделей. Безперервно зростаючі можливості чисельних математичних методів і автоматизованих обчислювальних систем, що реалізовують ці методи, не лише стимулювали математичну формалізацію різноманітних моделей, але й сприяли розвитку концептуального апарату, пов’язаного з моделюванням. Саме в ці роки склалися сучасні поняття аналогової і алгоритмічної математичної моделі, отримали значний розвиток методи чисельного, у тому числі статистичного імітаційного моделювання на комп’ютерах, були розроблені спеціальні прийоми для модельного дослідження систем складної структури, методи ідентифікації об’єктів-оригіналів і таке інше. По суті, тільки починаючи з 40-х років двадцятого століття стало можливим говорити про моделювання як про свідомо використовуваний науковий метод, уживаний для вирішення різних проблем, пов’язаних з об’єктами довільної природи [176].

Перші роботи з комп'ютерного моделювання були пов'язані з фізикою, де за допомогою моделювання розв'язувався ряд задач гідравліки, фільтрації, теплопереносу й теплообміну, механіки твердого тіла тощо. Моделювання у той час являло собою переважно розв'язання складних нелінійних задач математичної фізики за допомогою ітераційних схем або методу Монте-Карло.

Розвиток методу моделювання пов'язується насамперед із чотирма найвидатнішими досягненнями наукового пізнання XX ст. – дослідним вивченням мікросвіту, тенденцією до математизації усього сучасного природознавства, розробкою кібернетичного підходу до вивчення складних систем і розвитком загальної теорії систем, основні положення якої вперше були сформульовані у 1950-ті рр. Людвіг фон Берталанфі (Ludwig Von Bertalanffy) [5].

Розвиток кібернетичних засобів та прийомів моделювання і особливо удосконалення засобів інформаційно-комунікаційних технологій дозволяють комплексно моделювати взаємопов'язані об'єкти. Певні труднощі, що зустрічаються при створенні подібних моделей взаємодії об'єктів різної природи, долаються завдяки системному підходу. Принципи системного підходу (декомпозиції, цілісності, принцип багатоаспектності розгляду об'єкта, єдності цілей та ін.) і швидкодіючі комп'ютери дозволяють моделювати взаємодію окремих «зрізів» (страт, елементів об'єкту) між собою і усього об'єкту – із середовищем. Моделі, побудовані відповідно до цих принципів, називають імітаційними. Саме тому останнім часом комп'ютерне моделювання пов'язують не тільки з фундаментальними дисциплінами, а, в першу чергу, із системним аналізом.

Центральною процедурою в системному аналізі є побудова узагальненої моделі, яка відображає усі фактори і взаємозв'язки реальної ситуації, що можуть проявитися в процесі дослідження. Іншими словами, основою всього системного аналізу або головним етапом дослідження при проектуванні будь-якої системи є побудова математичних моделей.

Метод моделювання є одним із найдавніших методів пізнання, що має ґрунтовні традиції у розвитку природничо-математичних наук. Для того, щоб пов'язати ці традиції із сучасними видами моделювання і при цьому більш повно розкрити своєрідність сучасних прийомів моделювання, необхідно мати широке означення даного методу.

Перше, що звертає на себе увагу при методологічному розгляді моделювання, – це виключне розмаїття його форм та видів. Проте незалежно від форм побудови і специфічних особливостей тих галузей об'єктивного світу, що відтворюються в моделях, процедура моделювання у своїй основі залишається однаковою. Цей факт дозволяє виявити принципову можливість узагальненого філософського підходу до моделювання, яка охоплює його різноманітні форми.

Як зазначає В. О. Штофф [391], у всіх сферах людської діяльності моделювання виступає як деякий вид опосередкування, тобто практичне або теоретичне освоєння об’єкта дослідження здійснюється за допомогою проміжної ланки – моделі. Процес моделювання є особливою формою опосередкування, коли дослідник ставить між собою та об’єктом, що його цікавить, деяку проміжну ланку – модель. Модель у такому випадку виступає як представник (замінник) об’єкта. Об’єктивною основою модельного опосередкування є деяка схожість моделі й об’єкта дослідження.

Моделювання, як і будь-яка інша пізнавальна процедура, не є чисто суб’єктивним вольовим актом, що здійснюється суб’єктом пізнання (дослідником) довільно. Вирішальною об’єктивною основою операції моделювання є наявність незалежної від суб’єкта відповідності між моделлю й об’єктом, що моделюється. У найбільш абстрактному вигляді ця відповідність може трактуватися як вираження їх деякої структурної відповідності. В самому процесі встановлення відповідності виявляється єдність об’єктивного й суб’єктивного, причому об’єктивний зміст цього процесу пов’язується з деякою спільністю структур моделі й оригіналу в певному відношенні, тоді як суб’єктивний елемент пов’язаний не з довільністю суб’єкта, а з реальною практичною потребою. Ця потреба і визначає необхідність у кожному конкретному випадку зіставляти об’єкт-модель з об’єктом-оригіналом.

На основі таких міркувань формулюється основна умова теоретичного обґрунтування і практичної можливості моделювання. А саме: при моделюванні має місце деяка спільність у певному відношенні між моделлю й об’єктом, що моделюється (об’єктивний фактор обґрунтування моделювання), при цьому міра і форма даної спільності задаються тією практичною потребою, для задоволення якої здійснюється операція моделювання (практично-суб’єктивний фактор обґрунтування моделювання). Така умова повинна знайти відображення в узагальненому означенні моделювання як його найважливіша якість.

В узагальненому означенні моделювання слід також урахувати і гносеологічну умову моделювання, яка полягає в тому, що модель як самостійний предмет дослідження є одночасно і певним відображенням (за виразом В. О. Штоффа – «гносеологічним образом» [391, с. 149]) об’єкта-оригінала. При дослідженні побудована модель виступає як відносно самостійний від оригіналу «об’єкт-замінювач», що є немов би «другою реальністю», і через посередництво якого ми осягаємо предмет дослідження. При цьому на передньому плані виявляється не «образність», а «об'єктність» моделі, її здатність слугувати предметом дослідження, замінюючи в певних межах самий об'єкт дослідження.

Отже, в узагальненому означенні моделювання слід враховувати і основну онтологічну умову моделювання, і єдність «об'єктної» та «образної» сторін моделі – гносеологічну умову моделювання.

З урахуванням обох умов І. Б. Новик пропонує таке означення: «Моделювання – це метод практичного або теоретичного опосередкованого оперування об'єктом, в ході якого безпосередньо досліджується не сам об'єкт, а деяка проміжна система – модель, яка: а) знаходиться в деякій об'єктивній відповідності із самим об'єктом пізнання; б) може в ході пізнання на його окремих етапах заміщувати в певних відношеннях сам виучуваний об'єкт; в) здатна давати в процесі її дослідження нову інформацію про самий об'єкт. Ця допоміжна система – модель може виступати як у вигляді речовинного об'єкта, так і у вигляді деякого сполучення знаків» [179, с. 15-16].

Дещо доступнішим для розуміння студентів, на нашу думку, є означення моделі, яке пропонує В. О. Штофф: «Модель – це мислено уявлювана або матеріально реалізована система, яка, відображаючи або відтворюючи об'єкт дослідження, здатна замінювати його так, що її вивчення дає нам нову інформацію про цей об'єкт» [391].

Таким чином, модель містить у собі найважливіші, найістотніші для даної конкретної задачі характеристики (параметри) досліджуваного об'єкта. Вона абстрагується від неістотного, другорядного. Модель – це деяка ідеалізація дійсності, деяке спрощення її. Проте ступінь спрощення та ідеалізації при побудові моделей може з часом змінюватись у відповідності до мети дослідження. З означення моделі випливає, що вона має чимось відрізнятись від об'єкта дослідження, бо інакше перестане бути моделлю й ототожниться з ним. Усяке абстрагування відбувається шляхом нехтування якимись індивідуальними особливостями досліджуваного явища, проте в процесі абстрагування можуть виявлятися й певні закономірності. Таким чином, при створенні моделей звичайно прагнуть, щоб модель була ізоморфним або хоч би гомоморфним образом системи, що моделюється.

Процес абстрагування при створенні моделі, тобто включення до її опису суттєвих і виключення несуттєвих ознак (властивостей) є найбільш принциповим і складним у методології моделювання. Дійсно, для будь-якого опису важлива не лише його інформативність, а й узагальненість. А усяке узагальнення потребує відкидання того, що не є суттєвим з огляду на мету дослідження. Чим менше ознак виявляються суттєвими і згадуються в описі, тим ширше коло задач, тим частіше такий опис може бути використаний. Однак вимоги інформативності й узагальненості опису антагоністичні: чим більший обсяг опису (узагальненість), тим меншим виявляється його зміст (інформативність). Це за суттю відома теза формальної логіки стосовно обсягу та змісту понять: у міру збільшення обсягу поняття (узагальненості висловлювання) відбувається зменшення його змісту (кількість відомостей, якими воно визначається). Тому при моделюванні й вимагається, щоб відповідно до поставленої мети дослідження в опис моделі включалося або свідомо вводилося все суттєве і виключалося все, що для цієї мети не є суттєвим.

Включення до опису моделі якомога більшої кількості ознак об'єкта-оригінала пов'язане з намаганням дослідника забезпечити достатність системи ознак, навіть не звертаючи уваги на те, чи всі вони необхідні. Цим задовольняється бажання передбачити або забезпечити певний хід експерименту. Однак для розкриття причин явищ особливо важливою є перевірка необхідності введених параметрів. Це одна з важливих вимог до наукового знання. Історія науки доводить, що невключення навіть деяких суттєвих ознак мало шкодить розвиткові науки, у той час як включення ознак несуттєвих найчастіше виявляється неприпустимим. Так, наприклад, механіка Ньютона після доповнення її новими суттєвими параметрами квантової механіки не перестала існувати як наука. Офіційна ж медицина, перевівши заклинання до умов несуттєвих, рішуче відкинула знахарство, хоч і використовує лікування травами, введене в ужиток саме знахарями.

Коли проведено опис об'єкта або створено його модель, то деякі параметри (властивості) об'єкта в описі моделі вже обговорені або свідомо відтворені, а решта не згадуються або не вводяться і цієї «решти» завжди дуже багато. Проблема адекватності моделі об'єктові дослідження, тобто відповідності між моделлю й об'єктом – найчастіше це питання про те, чи не виявилося щось суттєве серед того, що не включено до опису моделі, і чи не введено до опису щось несуттєве. Питання залишається нерозв'язаним доти, доки не сформульовано мету. Перевірка відповідності між описом і спостереженням – це одна з найважливіших функцій моделі, яка полягає у з'ясуванні відповідності поведінки моделі й об'єкта за певних умов або для певної групи задач. Якщо при багаторазовому повторенні натурного експерименту відбувається варіювання у широких межах несуттєвих ознак (саме для цього експеримент і здійснюють багаторазово) і при цьому результат відтворюється у межах обговореної точності, то робиться висновок, що несуттєві ознаки дійсно не повинні враховуватись. Так само шляхом варіацій кожного із суттєвих (за припущенням) параметрів перевіряється опис «на надлишковість».

Домогтися невідтворюваності в моделі того, що не передбачено, – задача нелегка, оскільки властивості, що не повинні відтворюватися, в більшості випадків невідомі. Особливо важко це з’ясувати в матеріальних моделях, які мають безліч своїх власних характеристик. Ось чому найбільш ефективними є моделі, в яких замість копіювання суттєвих властивостей об’єкта встановлюється лише система взаємно однозначних відповідностей між параметрами об’єкта, з одного боку, і для моделі – з іншого.

Останнє твердження відображає факт існування широкого класу моделей, до яких, зокрема, належать інформаційні та математичні моделі, які взагалі не містять жодних непередбачених властивостей, оскільки все, що не обговорене, втрачає у таких моделях сенс. Адже несхожість елементів, із яких складаються модель і об’єкт, добре гарантує від випадкового відтворення в моделі властивостей об’єкта. Так, відповідні елементи об’єкта й, наприклад, математичної моделі аж ніяк не схожі між собою. Наприклад, функціональна залежність «не схожа» на графічне зображення цієї залежності або на ті процеси, що нею описуються. «Схожість» полягає лише в наявності системи однозначних відповідностей.

Таким чином, у сучасному науковому пізнанні процедура моделювання узагальнюється, в ній постійно з’являються нові риси, невпинно зростає розмаїття видів моделей. Це узагальнення стосується об’єктивної схожості моделі й системи, що моделюється: розвиток методу моделей іде від геометричної до фізичної схожості, що розкриває певну спільність різних процесів і сил, а далі і до ще більш складних форм встановлення схожості об’єктів. Як зазначає І. Б. Новик, «процес узагальнення моделювання, що проходить через усю історію науки, має принципове значення для розуміння тенденцій розвитку всього сучасного наукового пізнання» [179, с. 12].

І хоча навести достатньо повну класифікацію можливих видів моделювання досить важко хоча б через багатозначність поняття моделі, тим не менш стосовно природничих та технічних наук прийнято розрізняти наступні види моделювання:

– концептуальне моделювання, за якого сукупність вже відомих фактів або уявлень про досліджуваний об’єкт чи систему трактується за допомогою певних спеціальних знаків, символів, операцій над ними або за допомогою природної чи штучної мови;

– фізичне моделювання, за якого модель та об’єкт, що моделюється, являють собою реальні об’єкти або процеси спільної або різної фізичної природи, причому між процесами в об’єкті-оригіналі й у моделі виконуються деякі співвідношення подібності, що випливають із схожості фізичних явищ;

– структурно-функціональне моделювання, за якого моделями є схеми (блок-схеми), графіки, креслення, діаграми, таблиці, малюнки, доповнені спеціальними правилами їхнього об’єднання й перетворення;

– математичне (логіко-математичне) моделювання, за якого моделювання, включаючи побудову моделі, здійснюється засобами математики й логіки;

– імітаційне (програмне) моделювання, за якого логіко-математична модель досліджуваного об’єкта являє собою алгоритм функціонування об’єкта, реалізований у вигляді програмного комплексу для комп’ютера.

Перераховані види моделювання не є взаємовиключними й можуть застосовуватися при дослідженні складних об’єктів або одночасно, або в деякій комбінації. Крім того, в певному сенсі концептуальне й, скажімо, структурно-функціональне моделювання нерозрізнені між собою, тому що ті ж блок-схеми, звичайно ж, є спеціальними знаками із установленими операціями над ними.

Фахівці із системного аналізу [150; 170] стверджують, що імітаційні моделі є найбільш універсальними. Вони використовуються для кількісного передбачення властивостей будь-яких складних об’єктів, оскільки засновані на відображенні апріорних уявлень дослідника про їхню структуру та динаміку. Найбільш загальний підхід до побудови таких моделей ґрунтується на принципі декомпозиції, що веде до ієрархічної рівневої структури моделі, компонентами якої є окремі підсистеми об’єкта. Таку модель реалізують на комп’ютері у вигляді пакета прикладних програм за модульним принципом. Кожен модуль являє собою окрему модель – компоненту ієрархічної структури.

Традиційно під комп’ютерним моделюванням розумілося лише імітаційне моделювання. Проте в останні десятиліття, завдяки розвитку графічного інтерфейсу та зростанню потужності комп’ютерів, широкого розвитку набуло комп’ютерне, структурно-функціональне та об’єктно-орієнтоване моделювання.

Під *комп’ютерною моделлю* найчастіше розуміють умовний образ об’єкта або деякої системи об’єктів (чи процесів), описаний за допомогою взаємозалежних комп’ютерних таблиць, блок-схем, діаграм, графіків, малюнків, анімаційних фрагментів, гіпертекстів, що відображає структуру і взаємозв’язки між елементами об’єкта. Комп’ютерні моделі такого виду прийнято називати структурно-функціональними.

*Комп’ютерне моделювання* – метод розв’язування задач аналізу або синтезу складної системи на основі використання її комп’ютерної моделі. Сутність комп'ютерного моделювання полягає в отриманні кількісних і якісних результатів за моделлю. Якісні висновки, отримані за результатами аналізу, дозволяють виявити невідомі раніше властивості складної системи: її структуру, динаміку розвитку, стійкість, цілісність тощо. Кількісні висновки в основному носять характер прогнозу деяких майбутніх або пояснення минулих значень змінних, що характеризують систему.

Предметом комп'ютерного моделювання можуть бути: економічна діяльність певного підприємства чи закладу, інформаційно-обчислювальна мережа, технологічний процес, будь-який реальний об'єкт або процес, і взагалі – будь-яка складна система. Цілі комп'ютерного моделювання можуть бути різними, однак найчастіше моделювання є, як уже відзначалось вище, центральною процедурою системного аналізу, причому під системним аналізом будемо далі розуміти сукупність методологічних засобів, що використовуються для підготовки й прийняття рішень економічного, організаційного, соціального або технічного характеру.

Сучасна методологія системного аналізу складних об'єктів передбачає такі етапи побудови та використання математичної моделі: 1) змістовий опис об'єкта моделювання; 2) формулювання цілей дослідження; 3) аналіз попередньої інформації про об'єкт і побудова його концептуальної моделі; 4) побудова математичної (формалізованої, абстрактної) моделі об'єкта; 5) розробка моделюючого алгоритму; 6) побудова у відповідності до алгоритму програмної системи моделювання об'єкта; 7) ідентифікація (оцінювання параметрів) моделі; 8) перевірка адекватності моделі; 9) розробка сценаріїв для машинних експериментів із моделлю, які відбивають систему цілей дослідження моделі та її практичного використання; 10) застосування моделі для розв'язування практичних задач.

Пункти 7-9 у цьому переліку реалізуються шляхом виконання специфічного виду дослідницької роботи за допомогою комп'ютера. Провідний фахівець з комп'ютерного моделювання – академік А. О. Самарський [230] назвав цей вид досліджень обчислювальним експериментом. На відміну від натурного, в обчислювальному експерименті шукані характеристики моделі обчислюються за її параметрами.

Підсумовуючи в найбільш загальних рисах характеристику структури процесу моделювання, зазначимо такі її складові:

– актуалізація вже відомих знань про об'єкт-оригінал, зафіксованих в описі об'єкту моделювання;

– вибір інформаційної моделі з числа існуючих або створення такої

моделі (матеріальної або ідеальної);

– дослідження моделі (теоретичне та / або експериментальне);

– для того, щоб модель змогла виконати роль засобу вивчення об'єкту, дані, що їх було одержано при її дослідженні, в подальшому мають бути відповідним способом перенесені на оригінал;

– перевірка істинності даних, отриманих за допомогою моделі (адекватності моделі відносно оригіналу) і включення їх до системи знань про оригінал.

Використання моделювання в навчанні допомагає у вирішенні багатьох педагогічних завдань, серед яких активізація розумової діяльності, формування науково-теоретичного мислення, підвищення ефективності засвоєння знань, дотримання принципу єдності теорії і практики [335]. У процесі навчання у свідомості студента створюється картина, що відповідає рівню засвоєних знань, по суті, деяка модель дійсності. Тому володіння технологією математичного і комп'ютерного моделювання є важливим компонентом вищої природничо-наукової освіти, оскільки саме математичне моделювання дозволяє розкрити зв'язки абстрактних математичних понять з реальністю. Здійснюючи в структурі математичного моделювання поетапний перехід від формалізації завдання на початку розв'язування задачі до інтерпретації результатів наприкінці, ми здійснюємо деяке унаочнення математичних засобів. Саме тому роль математичного моделювання як засобу наочності є загальновизнаною.

Уявлення про структуру математичного моделювання (про його компоненти), специфіку окремих його етапів створюють базу для розвитку загальних навичок застосування математики до вирішення практичних завдань. Використання методу комп'ютерного математичного моделювання дозволяє наблизитись до вирішення ще однієї проблеми – посилення міжпредметних зв'язків. Головне, що може бути досягнуто в цьому напрямі – це ілюстрація прикладів застосування математики, розширення арсеналу математичних моделей і засобів моделювання. Цілеспрямоване вивчення математичного моделювання сприяє формуванню професійних компетенцій майбутніх учителів природничо-математичних дисциплін (сьогоднішніх студентів хіміко-біологічних та фізико-математичних факультетів педагогічних ЗВО), ефективно впливає на мотиваційні, орієнтаційні, операційні та оцінні компоненти навчання.

Аналіз теоретичних досліджень [47; 67; 73 та ін.] переконливо показує важливий вплив математичного моделювання на розвиток розумових, творчих і математичних здібностей студентів, на прискорення процесу розв'язування дослідницьких задач, формування науково-теоретичного мислення, підвищення ефективності засвоєння знань, забезпечення високого рівня підготовки фахівців.

Проблема моделювання – одна з найважливіших методологічних проблем, висунутих на передній план розвитком природничих наук у XX ст., особливо фізики, хімії, кібернетики. Минуле століття принесло методу моделювання нові успіхи, але одночасно поставило його перед серйозними випробуваннями. З одного боку, кібернетика виявила нові можливості і перспективи цього методу в розкритті спільних закономірностей і структурних особливостей систем різної фізичної природи, що належать до різних рівнів організації матерії, тобто до різних форм руху. З іншого ж боку, такі сучасні теорії, як теорія відносності і особливо квантова механіка, вказали на неабсолютний, відносний характер механічних моделей, на труднощі, пов’язані з моделюванням. Але як би там не було, інтерес до побудови моделей, до їх дослідження і до аналізу результатів моделювання став загальним, і тепер немає жодної науки, жодної галузі знання, де не намагалися б говорити про моделі, займатися моделюванням. Природно, що моделі і модельний (обчислювальний) експеримент не обійшли увагою і освіту. В багатьох випадках ми говоримо про моделювання, розглядаючи в якості його об’єктів широке коло задач не тільки навчального змісту, але й організаційних проблем, починаючи від моделювання вчителем уроку і всього навчального процесу зі свого предмету і аж до моделювання системи управління освітою.

Застосування методу моделювання дозволяє показати універсальність математичного апарату, дає можливість уніфікувати опис різноманітних за своєю природою процесів. Використання понять, пов’язаних з моделюванням, безпосередньо в процесі навчання дозволяє удосконалювати методику викладання, уникати формального підходу до навчання, реалізувати інтеграційні зв’язки. У студентів формуються уявлення про роль математичних методів у перетворюючій діяльності, про характер віддзеркалення математикою явищ навколишнього світу.

Психологічні проблеми моделювання розглядалися, зокрема в роботах М. М. Амосова, де наголошується, що використання моделювання можна розглядати як засіб пізнання й осмислення нового знання. Модель розглядається як продукт психічної діяльності, а моделювання – як процес відтворення певних сторін чи властивостей прототипу (об’єкта-оригінала) [70, с. 46].

Як зазначає П. В. Кійко [130], моделювання в навчанні має два аспекти: перший включає до змісту освіти поняття моделі, яке має бути засвоєне студентами в процесі навчання, і другий – такий, що відображає учбову дію, якою студенти мають оволодіти в навчально-пізнавальній діяльності.

Перший аспект обґрунтовує необхідність включення до змісту освіти понять моделі й моделювання. Побудова і вивчення моделей реальних об'єктів є основним методом наукового пізнання. Модельний характер сучасної науки показує, що завдання навчання студентів моделювання може бути вирішене в тому випадку, якщо наукові моделі виучуваних явищ займуть у змісті навчання належне їм місце і вивчатимуться з використанням відповідної термінології, з роз'ясненням студентам суті таких понять, як модель і моделювання.

Другий аспект полягає в застосуванні методу моделювання для виявлення структури і суттєвих зв'язків у виучуваних явищах, а також у формуванні вмінь використовувати цей метод для побудови узагальнених схем дій в процесі вивчення складних абстрактних понять (зокрема, понять квантової механіки, генетики, теорії штучного інтелекту тощо). Цей аспект можна реалізувати в процесі навчання студентів побудові, дослідженню і застосуванню моделей відповідних процесів.

У процесі моделювання об'єктів, що належать різним формам руху матерії, якісно нового характеру набувають інтегративні (міжпредметні) зв'язки, що об'єднують різні галузі знання за допомогою спільних законів, понять, методів дослідження.

Питанню навчання моделювання майбутніх учителів природничо-математичних дисциплін присвячені роботи М. В. Алексеєва [67], Л. І. Білоусової [76], О. Г. Колгатіна [134], І. І. Кондратенко [137], І. А. Левіної [155], К. Є. Рум'янцевої [226], Г. О. Савченко [229], Е. І. Сарафанюка [231], А. В. Семенової [232], Я. Б. Сікори [272], М. О. Федорової [369], С. А. Хазіної [372], О. П. Шестакова [383] та ін.

Зокрема, в роботах О. Г. Гейна [92], О. В. Князєвої [133], Н. В. Макарової [122], О. П. Шестакова [383], Ю. П. Штепи [390] розкриваються методичні аспекти навчання інформаційного моделювання, в роботах Л. В. Кавурко [125], І. В. Каменської [127], Н. А. Тарасової [298] – математичного моделювання; в роботах І. В. Бабичевої [73], Д. Д. Бичкової [82], П. В. Кійка [130], Л. М. Шенгелії [381] показана інтегративна роль комп'ютерного моделювання.

Проблемі міжпредметних зв'язків присвячені роботи Д. Д. Бичкової [82], О. Г. Гейна [92], П. В. Кійка [130], Л. М. Шенгелії [381] та ін. У цих дослідженнях в групах навчальних предметів вивчаються складні взаємозв'язки, направлені на формування наукового стилю мислення:

– проводиться дидактичний аналіз відповідних галузей знань, що обумовлюють їх інтеграцію в рамках одного предмету – інформатики;

– інформатика надає засоби створення комп’ютерних математичних та інформаційно-логічних моделей.

Сьогодні персональний комп’ютер є необхідним і невід’ємним елементом процесу підготовки майбутніх вчителів, проте комп’ютерно-орієнтоване навчання може і повинне бути взаємозв’язаним з класичними методами навчання. Зокрема, при проведенні технічного (фізичного) лабораторного експерименту перевагу слід надавати натурному, а не віртуальному варіанту.

Таким чином, виникає нагальна потреба у розробці умов професійної підготовки майбутніх учителів природничо-математичних дисциплін засобами комп’ютерного моделювання.

### 1.3 Об’єктно-орієнтоване моделювання як технологія дослідження складних систем та як навчальна дисципліна

Відомо, що разом із зростанням можливостей середовищ розробки зростає їхня складність, і як наслідок, – зростає складність їх вивчення, що може призводити до появи технологічного ухилу в навчанні інформатики студентів природничих, фізико-математичних та інформатичних спеціальностей.

Прискорення розвитку технологій (особливо інформаційно-комунікаційних) створює утруднення для системи освіти: так, упровадження засобів комп’ютерного моделювання у процес професійної підготовки майбутніх учителів природничо-математичних дисциплін змінює всі компоненти системи професійної підготовки (мету, зміст, засоби, організаційні форми і методи професійної підготовки). При цьому змінюється і зміст навчання, що вимагає розробки нових методик навчання майбутніх учителів на основі інноваційних технологій організації спільного дослідницького навчання.

Зокрема, сьогодні традиційні підходи до професійно орієнтованої ІКТ-підготовки майбутніх учителів природничо-математичних дисциплін не дають змоги відстежити швидкоплинну дійсність, обумовлену бурхливим розвитком засобів обчислювальної техніки, системного та прикладного програмного забезпечення, парадигм програмування, інформаційних систем та технологій організації, аналізу та подання навчальних відомостей. Розвиток технологій програмування зумовив необхідність практичного вивчення не лише сучасних програмних засобів, але й технологій їх розробки.

В умовах ІКТ-орієнтованого навчання математики, фізики, хімії, інформатики, біології та географії вплив зміни засобів навчання на зміну змісту навчання є більшим, ніж для інших дисциплін, і це породжує проблему неусталеності змісту навчання (особливо це є актуальним для інформатики). Розв'язання цієї проблеми можливе за двома основними напрямами:

1) усталення змісту навчання через ігнорування впливу зміни технологій;

2) усталення змісту навчання через усталення технологій.

При цьому жоден з них у повній мірі не розв'язує проблему.

Перспективним напрямом вбачається фундаменталізація навчання природничо-математичних дисциплін через посилення ролі основного методу дослідження в природничих науках – методу моделювання, що одночасно виступає в якості провідного універсального методу навчання. При цьому урахування психологічних особливостей відображення свідомістю людини об'єктів оточуючої дійсності вимагає відповідної їх інтерпретації у комп'ютерних моделях. Таким чином, вирішення проблеми фундаменталізації змісту навчання природничо-математичних дисциплін в умовах швидкої зміни засобів ІКТ потребує об'єднання методу моделювання з об'єктно-орієнтованою технологією, що разом утворюють якісно нову концепцію – *об'єктно-орієнтоване моделювання*.

Об'єктно-орієнтоване програмування (ООП), що отримало широке розповсюдження як потужна технологія програмування, містить у собі традиційні процедурні методи структурного програмування, а об'єктний підхід включає підхід структурний.

Популярність ООП у чималій мірі визначається більш сильною формою структуризації програмного забезпечення (ПЗ), що створюється на його основі. Використання ООП прискорює процес розробки програм, надаючи можливість швидкої модифікації створюваного ПЗ. Підґрунтям ООП є більш загальна технологія пізнання, що має назву об'єктного підходу до відображення дійсності. За об'єктного підходу реальні об'єкти відображаються у свідомості в об'єктно-орієнтовані моделі.

Основне утруднення, що постає перед розробником складного програмного продукту, полягає в неспроможності одночасно утримувати в пам'яті усі необхідні деталі. Дж. Міллер (George A. Miller) та його послідовники стверджують (див., наприклад, [79; 43]), що максимальна кількість об'єктів, з якою здатен одночасно оперувати людський мозок, не перевищує 7±2 (так званий «гаманець Міллера»). Це «магічне число», скоріше за все, пов'язане з обсягом короткострокової пам'яті у людини. Ще одним обмежуючим фактором тут виступає швидкість опрацювання мозком нової інформації: йому потрібно приблизно 5 секунд на сприймання кожного нового об'єкту. Як бачимо, природна здатність людського мозку до роботи зі складними системами є низькою.

Проте, як услід за Е. Дейкстрою (Edsger Wybe Dijkstra) зазначає Б. Страуструп (Bjarne Stroustrup) [60], ще з давніх часів людству відомий простий та ефективний спосіб управління складними системами: «Розділяй та володарюй». Тому при проектуванні складної системи слід складати її з окремих невеликих підсистем – у цьому випадку ми не виходимо за межі можливостей людини: при розробці системи будь-якого рівня складності необхідно одночасно утримувати в пам'яті інформацію лише про деякі її частини.

Такий підхід називається алгоритмічною декомпозицією і забезпечує психологічне підґрунтя для процедурного програмування, визначаючи головну вимогу до написання підпрограми: «усі дії, що виконуються в підпрограмі, повинні усвідомлюватися одночасно», і якщо ця вимога не виконується, підпрограму слід поділити на дрібніші блоки.

Проте число подій, що їх одночасно може опрацювати людина, не залежить від обсягу інформації, яка міститься у кожній події, і це дає людині надзвичайно ефективний механізм опрацювання складних повідомлень – абстрагування. Не маючи можливості відтворити у всіх деталях складний об'єкт, ми ігноруємо несуттєві для нас деталі і, таким чином, маємо справу з узагальненою, ідеалізованою моделлю об'єкта. І, хоча при цьому ми, як і раніше, змушені охоплювати одночасно значну кількість властивостей об'єкту, та завдяки абстрагуванню ми використовуємо узагальнені властивості суттєво більшого семантичного обсягу. Це особливо доцільно, коли ми розглядаємо світ з позицій об'єктно-орієнтованої взаємодії, оскільки об'єкти як абстракції реального світу являють собою насичені зв'язні інформаційні одиниці. При цьому ми також обмежені кількістю об'єктів, яку можемо сприйняти у кожний окремий момент. Все одно, використовуючи абстрактні поняття, ми отримуємо можливість працювати із складними системами, а, отже, і створювати складні програмні продукти. У процесі моделювання абстрагування (спрощення моделі через накладання певних припущень стосовно природи модельованого об'єкту) відбувається на етапі формалізації моделі. При цьому виділяються суттєві властивості об'єктів та відношення між ними.

Складні системи можна досліджувати, концентруючи основну увагу або на об'єктах, що фігурують у системі, або на процесах, що протікають в ній. Проте доцільніше розглядати систему як впорядковану сукупність об'єктів, які в процесі взаємодії один з одним забезпечують функціонування системи як єдиного цілого. Об'єкти, що складають систему, можуть утворювати ієрархії. За такого підходу основним способом дослідження складної системи є об’єктна декомпозиція, яка в процесі моделювання реалізується як через ієрархії наслідування об’єктів, так і через ієрархії моделей (за принципом спрощення / ускладнення).

Таким чином, з’являється змога розширити межі пізнавальних можливостей людини, використовуючи методи декомпозиції, виділення абстракцій та створення ієрархій. Саме ці методи покладені в основу об’єктного підходу.

Методичні засади застосування об’єктно-орієнтованого підходу в процесі професійної підготовки майбутніх вчителів розглянуті в роботах М. А. Алтухової [69], О. Г. Степанова [293], Л. Ю. Ханіпової [374].

Застосуванню об’єктно-орієнтованого підходу до моделювання присвячені роботи С. В. Беневольського [74], В. І. Болдонова [77], Г. Буча (Grady Booch) [183], В. В. Васілакіна [83], Я. М. Гаращенка [91], С. І. Гоменюка [95], О. Т. Гур’єва [98], І. А. Жирякової [116], Е. Йордона (Edward Nash Yourdon) [123], І. М. Килимник [131], Ю. Б. Колесова [74; 135], Ю. В. Кошарної [141], З. А. Крєпкої [143], О. І. Лісовиченка [161], С. Меллора (Stephen J. Mellor) [385], В. В. Мухортова [173], І. Б. Поліщук [198], Дж. Рамбо (James E. Rumbaugh) [217], М. П. Сіліч [271], І. Й. Труба [368], Т. Л. Фомічевої [370], Т. Фрауннштейна (Thomas Frauenstein) [26], І. О. Чмиря [115], А. Ю. Чудінова [379], В. М. Шека [380], Є. О. Шергіна [382], С. Шлеєр (Sally Shlaer) [385], С. М. Щедріна [392].

Реалізацією об’єктного підходу до дослідження складних систем є *об’єктно-орієнтоване моделювання* – це вид комп’ютерного моделювання, за якого середовищем моделювання є деяке середовище програмування, що надає можливість конструювання об’єктів, їх використання та обміну повідомленнями між ними.

Вид програмування, що використовується в такому середовищі, називають *об’єктно-орієнтованим програмуванням* – це технологія програмування, заснована на поданні програми у вигляді сукупності об’єктів, кожен з яких є реалізацією деякого класу, а класи утворюють ієрархію за принципами наслідуваності.

Таким чином, об’єктно-орієнтоване моделювання є видом комп’ютерного моделювання, яке, враховуючи особливості перебігу психічних процесів та відображення дійсності, надає можливість конструювати чітко структуровані та компактні комп’ютерні моделі.

А. П. Єршов у роботі «Про об’єктно-орієнтовану взаємодію з ЕОМ» [107] надає об’єктно-орієнтованому програмуванню більш широкого змісту, ніж програмуванню лише з використанням об’єктно-орієнтованих мов. У якості одного з прикладів об’єктно-орієнтованої взаємодії програмуючого користувача з ЕОМ А. П. Єршов посилається на Е-практикум як реальну систему автоматизованого конструювання програм, особливо підкреслюючи при цьому тезу про перспективність та універсальність об'єктно-орієнтованої взаємодії.

Об'єктний підхід дозволяє приховувати деталі реалізації компоненту і робити видимими ззовні тільки ті атрибути й операції, які потрібні іншим компонентам. С. Шлеєр та С. Меллор [385] визначають об'єктний підхід як послідовний ітеративний процес, що дозволяє у простий спосіб вносити зміни до вже налагодженого програмного продукту і в якому результати одного з етапів можуть впливати на рішення, прийняті на попередніх.

Об'єктний підхід відомий ще з давніх часів. Так, давнім грекам належить ідея про те, що світ можна розглядати як у термінах об'єктів, так і подій. А в XVII ст. Р. Декарт (René Descartes) підкреслював, що люди зазвичай мають погляд на світ, який сьогодні міг би називатися об'єктно-орієнтованим. У «Роздумах про метод...» він наголошує на чотирьох правилах об'єктно-орієнтованого підходу: «Перше – ніколи не приймати за істинне нічого, що не є очевидним, тобто ретельно уникати поспішності та упередження і включати у свої судження тільки те, що представляється розумові настільки ясно і чітко, що жодним чином не зможе дати привід до сумніву. Друге – ділити кожну із розглядуваних мною труднощів на стільки частин, скільки буде потрібно, щоб краще їх розв'язати. Третє – розташовувати свої думки у певному порядку, починаючи з предметів найпростіших і легко пізнаваних, і сходити мало-помалу, як по східцях, до пізнання найбільш складних, допускаючи існування порядку навіть серед тих, які в природному ході речей не передують один одному. І останнє – робити скрізь переліки настільки повні і огляди настільки всеохопні, щоб бути впевненим, що нічого не пропущено» [102, с. 260].

У 60-70-х рр. XX ст. ця думка була розвинена в одній з течій когнітивної філософії – об'єктивістській епістемології – розділу філософії, який досліджує процеси мислення та пізнання – (А. Ренд (Ayn Rand) [227]), а у 1980-х р.р. М. Л. Мінскі (Marvin Minsky) разом із С. Пейпертом (Seymour Papert) запропонував модель людського мислення, в якій розум людини розглядається як спільнота агентів, що по-різному мислять [44]. На його думку, лише спільні дії таких агентів приводять людину до осмисленої поведінки. Реалізація таких моделей можлива в мультиагентних середовищах об'єктно-орієнтованого моделювання, таких як NetLogo, з використанням ще однієї концепції ООП – паралелізму, що знайшла своє втілення в архітектурі сучасних комп'ютерних систем.

Основні ідеї об'єктного підходу (абстрагування, типізація, ієрархія тощо) в тому чи іншому вигляді були присутні в практиці програмування, починаючи з перших мов високого рівня.

*Об'єктно-орієнтоване проектування* – це вид проектування, що поєднує в собі процес об'єктної декомпозиції та прийоми подання логічної та фізичної, а також статичної та динамічної моделей проектованої системи [42]. Об'єктно-орієнтоване проектування відображає процес конструювання об'єктно-орієнтованої моделі.

Саме об'єктно-орієнтована декомпозиція відрізняє об'єктно-орієнтоване проектування від структурного; у першому випадку логічна структура моделі відображається абстракціями у вигляді класів і об'єктів, у другому – алгоритмами.

*Об'єктно-орієнтований аналіз* (ООА) – це вид аналізу, за якого вимоги до моделі системи сприймаються з точки зору класів та об'єктів, виявлених у предметній галузі [366]. Об'єктно-орієнтований аналіз спрямований на створення об'єктно-орієнтованих моделей реальної дійсності. Основними етапами об'єктно-орієнтованого аналізу є:

1) розробка інформаційної моделі. На цьому етапі центральним є абстрагування концептуальної суті в завданні в термінах об'єктів і атрибутів. Стосунки між сутностями формалізуються у зв'язках, які ґрунтуються на лініях поведінки, правилах і фізичних законах, що превалюють у реальному світі;

2) розробка моделі станів. Цей етап методу пов'язаний з поведінкою об'єктів і зв'язків у часі. В об'єктно-орієнтованому аналізі кожен об'єкт і зв'язок має життєвий цикл – регулярну складову частину динамічної поведінки. Часто використовують моделі станів для формалізації життєвих циклів об'єктів і зв'язків. Моделі станів, що виражаються в діаграмах і таблицях переходу, взаємодіють між собою за допомогою подій; їх організовують в ієрархію, щоб зробити систему взаємодії впорядкованою і зрозумілою;

3) розробка моделі процесів. На цьому етапі об'єктно-орієнтованого аналізу визначаються дії, що їх необхідно виконати для переходу об'єкту з одного стану у інший. Отримувані в такий спосіб процеси надалі можуть бути перетворені безпосередньо на оператори (методи) об'єктно-орієнтованого проектування;

Між всіма цими визначеннями існує тісний взаємозв'язок: на результатах об'єктно-орієнтованого аналізу формуються моделі, на яких ґрунтується об'єктно-орієнтоване проектування; у свою чергу, об'єктно-орієнтоване проектування створює фундамент для реалізації об'єктно-орієнтованої моделі.

Великого значення набуває розгляд сфер застосування ООП і предметних областей, де воно може виявитися найбільш ефективним, тобто мати явні переваги в порівнянні, наприклад, з процедурним підходом. Принципова відмінність двох підходів визначається тим, що є первинним – об'єкт чи процес.

Авторами [368] ці критерії уточнюються: «Якщо ідейним стрижнем програми є процес, навряд чи вживання ООП може принести якусь користь і підвищити наочність коду, ефективність розробки і супроводу. Проте існують такі розділи прикладного програмного забезпечення, де висока ефективність і наочність вживання засобів ООП не викликає сумнівів, наприклад, імітаційне моделювання складних систем. Застосування ООП дозволяє побудувати ефективну і гранично узагальнену схему імітаційного моделювання складних систем. У цій прикладній області об'єктний підхід повною мірою виявляє ті свої переваги, заради досягнення яких він, власне, і був розроблений: природність і наочність процесу програмування, первинність даних по відношенню до процесів і головне – легкість супроводу і подальшої модифікації моделюючої програми, якщо система зазнає зміни» [368, с. 15-16].

У одній з перших робіт [18], де було дано формальне визначення поняття «об'єктно-орієнтоване моделювання», зазначено: «Коли говорять про об'єктно-орієнтовану концепцію опису систем, то мають на увазі програмне забезпечення у вигляді сукупності дискретних об'єктів, які містять і структуру даних, і реалізацію поведінки. З іншого боку, імітаційну модель можна розглядати як деяку множину сутностей, що взаємодіють одна з одною. З цієї точки зору можна навіть сказати, що імітаційна модель завжди об'єктно-орієнтована, оскільки об'єктна парадигма – це цілком природний спосіб моделювання, якщо відображувати реальні сутності на об'єкти».

Саме це стало однією з причин включення блоку імітаційних моделей до змісту спецкурсу «Об'єктно-орієнтоване моделювання» для майбутніх учителів природничо-математичних дисциплін.

Природний зв'язок ООП та моделювання був відображений вже в першій мові ООП – Simula-67, сама назва якої походить від слова *simulation* – моделювання. Автори Simula-67 запропонували використовувати спеціальні моделі – класи, які описують множину близьких за своїми властивостями об'єктів, що мають внутрішню структуру і поведінку, та надають можливість вибирати конкретний елемент цієї множини, створюючи конкретний екземпляр класу – об'єкт, і наділяючи його конкретними значеннями параметрів [26].

Практично всі сучасні професійні мови програмування (Object Pascal, C++, Java та ін.) засновані на об'єктно-орієнтованому підході.

Значного поширення в процесі інформатичної підготовки вчителів природничо-математичних дисциплін набули системи візуального програмування, також засновані на об’єктно-орієнтованій технології (Delphi, C++Builder, JBuilder та ін.). Вже розроблені й розробляються у великій кількості засоби об’єктно-орієнтованого аналізу та проектування, що базуються переважно на уніфікованій мові моделювання UML (Unified Modeling Language). При цьому останні версії багатьох сучасних систем програмування почали включати підтримку методів об’єктно-орієнтованого проектування (паттернів проектування).

Сучасний стан розвитку об’єктного підходу як технології опису і проектування складних багатокомпонентних систем визначає уніфікована мова моделювання UML [217]. Модифікації об’єктно-орієнтованої технології стосовно моделювання складних багатокомпонентних динамічних систем в UML називаються об’єктно-орієнтованим моделюванням (ООМ [45]) і представлені мовою моделювання Modelica, а також вхідними мовами пакетів моделювання AnyLogic, MvStudium та ін. За допомогою ООМ зручно розв’язувати багато типових завдань моделювання, а саме:

– створювати бібліотеки типових компонентів як бібліотеки класів;

– повторно використовувати компоненти, здійснюючи за необхідності їх спеціалізацію за допомогою наслідування класів;

– природним чином будувати моделі з безліччю однотипних об’єктів;

– здійснювати параметризацію моделей за допомогою поліморфізму;

– при моделюванні систем із змінним складом створювати і знищувати екземпляри об’єктів у ході обчислювального експерименту.

Не дивлячись на те, що автори UML вважають її мовою дискретного моделювання, яка не призначена для розробки неперервних (суцільних) систем, що зустрічаються у фізиці й механіці, UML заслуговує на увагу в силу наступних причин.

По-перше, автори UML є провідними фахівцями з об’єктного підходу, і ця мова зафіксувала всі основні досягнення об’єктного підходу до проектування програмних систем, стала фактичним стандартом об’єктного підходу. Тому розробникам спеціальних мов ООМ є сенс використовувати основні поняття UML як свого роду метамову.

По-друге, частина конструкцій UML може бути використана безпосередньо як для попереднього опису структури проекту, так і в якості основи для розширення.

Спрямування UML на моделювання складних багатокомпонентних динамічних систем визначило включення блоку динамічних моделей до змісту навчального курсу об'єктно-орієнтованого моделювання.

Поняття класу і екземпляра є ключовими в ООП. Об'єкт – це деяка сутність, що має атрибути і поведінку. Клас – це формалізований опис множини об'єктів, що мають однакове функціональне призначення, атрибути і поведінку. З поняттям класу нерозривно пов'язане поняття екземпляра класу, тобто конкретного об'єкта з множини всіх об'єктів, що описуються даним класом. Конкретний об'єкт відрізняється від решти всіх об'єктів того ж класу унікальними значеннями атрибутів.

Серед атрибутів виділяються параметри – такі атрибути, значення яким надається тільки один раз при створенні екземпляра класу і далі не змінюються протягом всього часу існування об'єкту.

В UML виділяються «активні» і «пасивні» об'єкти і відповідно «активні» і «пасивні» класи, екземплярами яких вони є. Активним об'єктом у UML називається об'єкт (процес), що виконується (розвивається в часі) і може ініціювати управляючі дії, надсилаючи повідомлення іншим об'єктам (впливаючи на них у такий спосіб). Пасивним називається об'єкт, що сам не виконується, а всі зміни його стану є реакціями на повідомлення активного об'єкту. Стосовно дискретних систем ці визначення цілком достатні, проте стосовно ООМ складних динамічних систем вони уявляються дещо вузькими. Для компонентів складних динамічних систем уводять спеціальне поняття «активний динамічний об'єкт» – активний об'єкт, що ініціюється через певні наперед визначені проміжки часу з достатньою для адекватного моделювання частотою [36].

Проведений аналіз принципів застосування об'єктного підходу до моделювання надав можливість визначити новий вид комп'ютерного моделювання – об'єктно-орієнтоване моделювання, а дослідження змісту поняття об'єктно-орієнтованого моделювання надає можливість визначити його як навчальну дисципліну [7; 10; 11] у такий спосіб: **об'єктно-орієнтоване моделювання – це навчальна дисципліна, предметом вивчення якої є способи конструювання та дослідження об'єктно-орієнтованих моделей.**

Зауважимо, що, хоча у дослідженні Т. Бринди (Torsten Brinda) [11] і представлена дидактична система навчання об'єктно-орієнтованого моделювання в середній школі, застосувати її у професійній підготовці майбутніх учителів без суттєвої адаптації до вітчизняної системи вищої педагогічної освіти неможливо.

Проведене дослідження надає можливість обґрунтованого вибору *упровадження об'єктно-орієнтованого моделювання у процес навчання*

*інформатичних дисциплін* у якості умови професійної підготовки майбутніх учителів природничо-математичних дисциплін засобами комп'ютерного моделювання.

## 1.4 Педагогічна технологія соціального конструктивізму

### *1.4.1 Теоретичні засади конструктивістського підходу до процесу навчання*

Нагальна потреба в інтеграції системи освіти України у світовий освітній простір, у пошуку варіантів зближення позицій освіти України і розвинених країн Європи та США вимагає суттєвої модернізації концептуальних і методологічних засад та змісту вітчизняної освіти. Адже сучасна освіта має відображати рух до інформаційного суспільства, зазнавати постійного вдосконалення за рахунок використання вітчизняними дослідниками власних передових освітніх ідей та запозичення плідних філософських та педагогічних ідей західних психологів і педагогів, а також за рахунок інтенсивного використання сучасних засобів ІКТ.

Як зазначає С. А. Раков, «... діалектика розвитку методології навчання є рухом від передавання системи знань від викладача до студента до самостійного конструювання студентом особистої системи знань у навчальному процесі на основі дослідницьких підходів у навчанні. При цьому функції викладача перетворюються з функції демонстратора готових теорій у менеджера процесу пошуку та конструювання нових знань, а функції студента – з реципієнта готових теорій до активного конструктора власної системи знань. Це стосується зовсім нової парадигми навчального процесу, у якому активними співтворцями стають і студенти, і викладачі» [216, с. 5].

Опрацювання теоретичних аспектів педагогіки Заходу показує, що чи не найбільша увага там приділяється філософським засадам тих або інших педагогічних теорій. Найбільш прогресивні з цих теорій спираються на основні положення соціально-конструктивістського навчання, які С. А. Раков формулює за Л. С. Виготським [86] у такий спосіб: 1) визнання різноманіття талантів; 2) колективний резонанс; 3) колективна рефлексія.

Дослідницький підхід у навчанні С. А. Раков визначає як підхід, при якому ідеями досліджень просякнуті всі форми навчальної роботи: лекції, практичні заняття, лабораторні заняття, індивідуальна та самостійна робота, курсові та дипломні проекти. Дослідницький підхід у навчанні реалізується через дослідницьку діяльність та навчальні дослідження, через рефлексування яких набувається індивідуальна, особистісна методологія проведення досліджень [216, с. 41]. Поняття дослідження пов’язано із тим напрямом у розвитку освіти, який спирається на загальнофілософське поняття пізнання, зокрема, на конструктивістську епістемологію [216, с. 46].

Психолого-педагогічні основи застосування дослідницького підходу в навчанні (узагальнена вітчизняна назва ряду зарубіжних течій у філософії психології й педагогіки, відомих під назвами конструктивізм і конструкціонізм) розглядалися в роботах Е. Аккерман (Edith Ackermann) [2], Л. Бабак [71], Т. І. Бутченка [81], Е. фон Глазерсфельда (Ernst von Glasersfeld) [93], С. Далена (Siv Dahlén) [101], К. Дж. Джерджена (Kenneth J. Gergen) [103], Дж. Дьюї (John Dewey) [99], Я. Кафай (Yasmin Kafai) [16], А. В. Кезіна [129], А. Кея (Alan C. Kay) [33], О. В. Константинова [139], П. фон Лоренцена (Paul von Lorenzen) [37], Дж. В. Максвелла (John W. Maxwell) [41], А. В. Пашкової [190], С. Пейперта [15; 192], Ж. Піаже (Jean Piaget) [52], М. Резника (Mitchel Resnick) [16; 54], К. В. Рибачука [222; 223; 224], М. В. Романової [225], М. В. Смагіної [278], І. Харел (Idit Harel) [15], С. А. Цоколова [376; 377], М. А. Чошанова [378], К. В. Якімової [393], І. М. Януш [394].

Державний стандарт [209] визначає вимоги щодо формування в учнів навичок і вмінь створювати об’єкти в рамках реалізації індивідуальних і колективних проектів, висувати і перевіряти гіпотези навчально-пізнавального характеру, створювати, вивчати та використовувати об’єкти, використовувати засоби ІКТ для планування, організації індивідуальної і колективної діяльності в інформаційному середовищі. Реалізація цих вимог потребує зміни професійної підготовки майбутніх учителів природничо-математичних дисциплін не лише на основі комп’ютерного (зокрема, об’єктно-орієнтованого) моделювання, а й на основі соціально-конструктивістського підходу до навчання.

Розглянемо більш докладно психолого-педагогічні засади використання соціально-конструктивістського підходу до навчання у професійній підготовці майбутніх учителів природничо-математичних дисциплін.

Починаючи з кінця сімдесятих – початку вісімдесятих років XX ст. в Західній Європі почали поширюватись думки стосовно природи знання, які утворили нову науково-філософську проблематику.

С. А. Цоколов відзначає: «І кібернетичні ідеї Фьорстера, і конструктивістська психологія Піаже, і біо-когнітивні погляди Юекскюля, розвинені згодом У. Матураною і Ф. Варелою, і які відомі ... здебільшого як робочі гіпотези або теорії, були покликані пояснювати емпіричний матеріал, здобутий в рамках тієї або іншої науки. У єдину (хай і неоднорідну) проблематику їх уперше об’єднали в США і в

Німеччині, внаслідок чого в теорії пізнання утворився новий напрям – радикальний конструктивізм», основоположником якого вважають німецького філософа Ернста фон Глазерсфельда. Проте С. А. Цоколов вважає термін «конструктивізм» невдалим «... тому що, по-перше, він уже вживався в області традиційної філософії з дещо іншим смисловим відтінком; по-друге, на початку двадцятих років XX ст. він позначав рух в області образотворчого мистецтва і архітектури, що протягом нетривалого часу існував у Радянському Союзі і, по-третє, – внаслідок свого чужого німецькій мові звучання. І якби дитя, що народилося, до цього часу вже не носило даного імені, то позначення «наука про дійсність» (Wirklichkeitsforschung) було б значно придатнішим» [377].

С. А. Раков, відзначаючи роль Е. фон Глазерсфельда, вказує, що саме на основі його робіт «педагоги та дослідники, які працюють разом з учителями, розробили конструктивістську модель для того, щоб прояснити та сприяти розвиткові мислення вчителів» [216, с. 47].

Центральна парадигма радикального конструктивізму може бути передана наступною цитатою з роботи Е. фон Глазерсфельда: «...(a) знання не знаходиться пасивним способом, воно активно конструюється суб'єктом, що пізнає; (b) функція пізнання носить адаптивний характер і служить для організації дослідного світу, а не для відкриття онтологічної реальності» [93]. С. А. Раков підкреслює, що «навчальні дослідження, або дослідницький метод навчання математики узгоджується із конструктивістськими поглядами на математичні знання та вивчення математики, оскільки вони відповідають умовам стимулювання математичного мислення та створюють умови для критичної рефлексії у процесах усвідомлення математики» [216, с. 47].

Поняття «конструктивізм» і «радикальний конструктивізм» об'єднує твердження про те, що будь-яке знання конструюється суб'єктом (когнітивною системою [9], спостерігачем, живим організмом і т. п.). Те, як це сприймається в кожному конкретному випадку і які з цього випливають висновки, визначає вид конструктивізму, що його дотримується та чи інша група філософів. «Радикальність» радикального конструктивізму полягає, за словами Е. фон Глазерсфельда, в його радикальному відмежуванні від усіх форм традиційної епістемології, яка допускає в тій чи іншій мірі відповідність знання об'єктивній реальності. Філософська позиція, на якій жорстко наполягає Е. фон Глазерсфельд, свідчить, що знання принципово не може відображати або відповідати ніякому реальному світу з огляду на те, що єдиний доступний суб'єктові «реальний світ» – це і є той світ, який суб'єкт сам конструює в процесі пізнання. В рамках радикального конструктивізму два твердження – «конструювання знань» і

«конструювання реальності» [393] – мають однаковий сенс.

Філософія радикального конструктивізму базується на певних наукових концепціях. Як зазначалося вище, за найзагальнішими ознаками виділяють три основні концептуальні підходи: психологічний, кібернетичний і біологічний. Так, найістотніше, що було запозичене Е. фон Глазерсфельдом з робіт Ж. Піаже [52] – це твердження про те, що будь-яке знання конструюється суб'єктом в процесі формування власного досвіду. Великий вплив на формування конструктивістської позиції мав розвиток кібернетичного способу мислення (саме так він визначає кібернетику в «Декларації американського кібернетичного товариства» 1985 року: «Кібернетика – це образ думки, а не зібрання фактів» [29]). Особливе значення в рамках даного підходу мають роботи Хайнца фон Фьорстера (Heinz von Foerster), що приділив у своїх дослідженнях спеціальну увагу кібернетичному аспекту конструктивістської теорії пізнання [25].

Біологічний аспект конструктивізму практично невіддільний від кібернетичного, оскільки все, що говориться в конструктивізмі про організацію нервової системи і мозку, про когнітивні властивості живих систем, концептуально було сформульовано, хоча і на біологічному матеріалі, але виключно в рамках кібернетичного підходу. Теза У. Матурани (Humbert R. Maturana) та Ф. Варели (Francisco J. Varela) «життя є пізнання» [40] могла стати значущою лише в контексті узагальненого вчення про біосистеми. Ще один біологічний аспект радикального конструктивізму змикається, з іншого боку, з психологічним підходом.

Ж. Піаже конструктивізм розглядає як такий підхід у психології навчання, що описує когнітивну діяльність, завдяки якій індивідууми розвивають та модифікують свої індивідуальні когнітивні моделі представлення знань через процеси акомодації та рефлексивного абстрагування [216, с. 46]. Як у генетичній епістемології Ж. Піаже, так і в сучасних трактуваннях еволюційного учення, поняття адаптації розуміють не як відповідність пристосованого організму (його біологічних і когнітивних якостей) своєму навколишньому середовищу, а виключно як придатність (viability) для продовження власного існування і виживання.

Таким чином, всі три позначені гілки конструктивізму – психологічна, кібернетична і біологічна – можуть бути розділені лише умовно, оскільки зрештою служать одним і тим самим концепціям радикального конструктивізму.

Конструктивізм – це загальне позначення для різних спрямувань у науці, мистецтві й філософії, які ставлять в центр поняття конструкції,

для позначення вироблюваного в цих областях продукту. Тому у філософії конструктивізмом називають теоретико-пізнавальну концепцію, яка трактує пізнання як, перш за все, конструювання, і тим самим формулює точку зору, протилежну метафізичним і реалістичним теоріям пізнання. Згідно з [376] у даний час існує величезна кількість напрямів у різних областях знання, які відносять себе до конструктивізму, і ці напрями можуть істотно відрізнятися один від одного.

В. Деніелс (Victor Daniels) показав, що історію ідей конструктивізму можна прослідкувати аж до часів античності [19]. Ідея про те, що людина сама створює (конструює) свої філософські системи і моделі світу зустрічається в багатьох системах філософської думки – цим підкреслюється активна роль суб'єкта пізнання в протилежність його пасивній ролі в теоріях емпіричного типу (сенсуалізмі, теорії віддзеркалення тощо і так далі).

Конструктивізм протистоїть філософському уявленню про пізнання як про віддзеркалення об'єктивної реальності. Конструктивізм виходить із того, що інформація (знання) не міститься в об'єкті і в ході пізнання не витягується з нього, а є продуктом деякого суб'єкт-об'єктного відношення, що включає позицію спостерігача, його практичну діяльність і засоби пізнання. В результаті суб'єкт, котрий пізнає, активно вибудовує знання у вигляді різного роду конструктів, що моделюють і передумовлюють його (суб'єкта) досвід.

Термін «конструктивізм» у цьому значенні почав використовувати Ж. Піаже в кінці 1960-х рр. XX ст., і цей термін набув поширення в 1980-ті рр. для позначення широкого спектру теоретичних і методологічних побудов, що акцентують роль минулого досвіду в побудові картини світу, роль соціальних, історичних і культурних чинників у продукуванні наукових знань.

Конструктивізм (зокрема, конструкціонізм) знайшов застосування в ряді прикладних областей, перш за все, в теорії навчання, розглядуваного як процес активного і заснованого на раніше засвоєних знаннях з конструювання учнями ментальних моделей світу та з їх практичної діяльності. На цій теорії заснована система конструктивістської педагогіки, яка прагне створити для учнів розвивальне середовище. Воно має забезпечити доступ до різних описів реальності, здатне навчити способам конструювання знань, виходячи з індивідуальності і неповторного досвіду кожного учня.

П. Вацлавік (Paul Watzlawick) [61] ставить завдання навчити людей створювати альтернативні, більш адаптивні конструкції – описи світу, в якому вони живуть, і свого місця в ньому. У нарративній психотерапії, що працює з описом життєвого шляху, конструюються біографічні версії минулого, виходячи із завдань сьогодення. Методи дискурс-аналізу використовуються в критичному аналізі продуктів масової комунікації.

Конструктивізм виходить із того, що навчання – це активний процес, в ході якого люди активно конструюють знання на основі власного досвіду: не можна передати знання суб'єкту навчання у готовому вигляді, а можна тільки створити умови для успішного самоконструювання знань у процесі навчання. Е. фон Глазерсфельд основне положення конструктивістської дидактики формулює у такий спосіб: «Знання, як таке, ніколи не може бути передане від однієї людини до іншої. Єдиний спосіб, яким індивід може набувати знання, це створювати його самому або конструювати його для себе. Діяльність викладача має розглядатися як спроба так змінити навколишнє відносно того, хто навчається, середовище, щоб той зміг побудувати саме такі когнітивні структури, які хоче передати йому вчитель» [93]. Класик математичного конструктивізму П. фон Лоренцен радикалізує це твердження: «Ми тільки тоді що-небудь розуміємо, коли самі можемо це створити» [37].

Конструктивістська епістемологія розуміє учіння як процес самоорганізації знання, що виникає на основі побудови сенсів і дійсності кожним окремим учнем і тим самим це знання є індивідуальним і непередбачуваним. Викладач повинен створити багатообразне оточення (інформаційне середовище), за можливості багате і мультимодальне (звернене до багатьох чуттєвих якостей), цікаве і орієнтоване на комунікацію (взаємодію). Це середовище, з одного боку, звернене до досвіду, що вже є у студента, а з іншого – повинне бути загадковим і містити в собі потенційні відкриття, повинне приваблювати дітей, спонукати їх до пошуку, дослідження, самоорієнтації, до виявлення проблем і пошуку їх рішень. Таким чином, у конструктивістській дидактиці має місце принципова відмова від навчання, спрямованого на учня: вчитель відмовляється від прямого повідомлення знання, але надає учневі можливість самому сконструювати своє знання на основі наданого матеріалу. Вчитель, застосовуючи навчальний матеріал, повинен допомогти учневі самостійно будувати своє знання.

Конструктивістська дидактика є напрямом сучасної дидактики, що в останні десятиліття завоювала провідні позиції в розвинених країнах Заходу. Своїм корінням цей напрям іде в епоху реформ педагогіки початку XX століття. Завдяки дослідженням когнітивного розвитку дитини в науковій школі Ж. Піаже багато елементів у практиці його послідовників знайшли своє обґрунтування з боку конструктивізму. Окрім Ж. Піаже, до «батьків-засновників» сучасної конструктивістської дидактики (прагматичного конструктивізму) відносять Дж. Дьюї [99] і Л. С. Виготського [86].

Конструктивістська педагогіка ґрунтується на теорії Ж. Піаже і об'єднує ряд підходів до навчання і виховання дітей. У них, як було відмічено вище, виходять із того, що знання, інтелект, мислення, самостійність та інші характеристики особистості не можуть бути привнесені ззовні, але мають бути активно сконструйовані самою дитиною. Останнє відбувається у взаємодії з матеріальним і соціокультурним навколишнім світом (середовищем). При цьому Ж. Піаже розрізняє два основні процеси [194]:

1) асиміляція: дитина сприймає інформацію зі свого оточення і інтерпретує її відповідно до своїх попередніх знань і досвіду;

2) акомодація: дитина модифікує своє знання на основі протиріч і недостатності наявних знань (операцій). В результаті дитина отримує новий досвід, тобто відбувається розвиток.

Ідеї конструктивізму виражені і в теорії діяльності, згідно з якою діяльність і дії самої дитини є основою її психічного розвитку (П. Я. Гальперін [90], В. В. Давидов [100] та ін.). Як зазначає М. І. Жалдак, «знання (як і інформацію) передати неможливо: їх набувають у процесі власної пізнавальної діяльності» [362]. Більш того, наголошує С. А. Раков, перехід від наочної дії до дії з моделями освоюваних об'єктів і ситуацій складає невід'ємну умову формування повноцінних розумових здібностей дитини. Основними поняттями конструктивістського підходу С. А. Раков вважає дослідницьке навчання, навчання через діяльність, експериментування, навчання через відкриття та пропонує такі напрями формування дослідницької компетентності:

1) формулювати (ставити) математичні задачі на основі аналізу суспільно та індивідуально значущих задач (ідеалізація, узагальнення, спеціалізація);

2) будувати аналітичні та алгоритмічні (комп'ютерні) моделі задач;

3) висувати та емпірично перевіряти справедливість гіпотез, спираючись на відомі методи (індукція, аналогія, узагальнення, спеціалізація і т. п.), а також на власний досвід досліджень;

4) інтерпретувати результати, що отримані за формальними методами, у термінах вихідної предметної області та інших предметних областях;

5) систематизувати отримані результати: досліджувати межі застосувань отриманих результатів, встановлювати зв'язки з

попередніми результатами, модифікувати вихідну задачу, шукати аналогії в інших розділах і т. п. [216, с. 57].

У [139] зроблено важливе спостереження про те, що власний внесок дитини в своє мислення можна ясно побачити на численних прикладах неправильних уявлень, що висловлюються дітьми і яким діти, природно, не могли навчитися в дорослих. Ці уявлення діти сконструювали самі, намагаючись зрозуміти свої спостереження і враження, пробуючи осмислити свій досвід. Розгляд світу уявлень дітей показує, що мислення маленьких дітей якісно відрізняється від мислення більш старших дітей або дорослих. Відповідно до цього, згідно з конструктивістським підходом, діти мало можуть узяти з повчання дорослих, тобто з прямого навчання. У набагато більшому ступені для свого розвитку вони роблять вирішальні кроки самі і досягають вищих рівнів розвитку шляхом власних зусиль. В результаті зіткнення і активної взаємодії з навколишнім світом діти удосконалюють свої розумові здібності, їх знання стають більш диференційованими і співвіднесеними з реальністю. Але цей прогрес у мисленні є, перш за все, їх власною заслугою, а не заслугою батьків або вихователів. Останні можуть тільки підтримати цей розвиток через створення для них особливо багатого середовища (навколишнього світу), в якому вони (діти) отримують багатообразні враження для подальшого обдумування і в якому вони можуть розгорнути свій багатий творчий потенціал. При цьому мова йде про весь спектр розвитку: когнітивний, емоційний, соціальний, моральний.

Слід зауважити, що за радянських часів у нашій країні теорія Ж. Піаже піддавалася критиці за «біологізм» і відрив від «культури» [157]. В основі цієї критики лежить непорозуміння: під «середовищем» наші автори мали на увазі біологічне середовище, в той час, як у Ж. Піаже «організм», «середовище» і так далі – це метафори. Як для тваринного організму середовище – це природне оточення, так і для людини природним (для неї «природним») є культурне місце існування (мова, устої і т.д.).

Як теорія пізнання і теорія навчання конструктивізм обговорюється у всіх розділах педагогіки [13; 62 та ін.]. В еволюції цієї концепції С. А. Раков [216, с. 108] виділяє три етапи. Перший – це «наївний конструктивізм: якщо знання добре структурувати і систематизувати, то завдяки умілому навчанню учні можуть сформувати свої власні системи знань, які будуть “клонами” або копіями системи знань учителя». Другий – це радикальний конструктивізм, який виходить з тези, що учіння здійснюється у спосіб, що повністю самоорганізується і самокерується. Згідно з такою радикальною позицією педагогічна дія ззовні в процесі придбання знань не є ефективною зовсім. Менш радикальною є позиція прагматичного конструктивізму, що намагається пов'язати поняття конструкції та інструкції. В даний час пануючою є позиція прагматичного конструктивізму, який прагне об'єднати і доповнити ідеї конструктивізму і прагматизму.

Грецьке слово «прагма» означає «дія» або «справа». Згідно прагматизму, основною характеристикою буття людини є дія. Людина може утриматися від пізнання, від занять наукою, – люди жили і в донаукову епоху – але, щоб жити (виживати), людина не може утриматися від дії. Тому практична дія, а не теорія є основним елементом взаємодії людини з дійсністю [38].

Головний прагматичний принцип: «Подумай, які мислимо можливі дії ми можемо приписати в своїй уяві предмету нашого поняття. Тоді наше поняття про дії і є повним поняттям предмету». Це положення служить методом прояснення поняття, згідно з яким зміст значення поняття полягає в його можливих наслідках для дії. Прояснення і можлива корекція понять відбувається завдяки експериментальній взаємодії з дійсністю. Проте для прояснення сенсу поняття зовсім не обов'язково, щоб наслідки настали фактично. Досить мисленого експерименту, в якому розкриваються можливі практичні наслідки і який є широко вживаним у природничо-математичних науках, і зокрема у фізиці.

Результати, здобуті у такий спосіб, повинні себе виправдати в процесі комунікації між людьми, котрі спільно діють і досліджують. За такого підходу істина – це не віддзеркалення дійсності у свідомості і не відповідність мислення наявному положенню справ, а згода, консенсус всіх членів «безмежного дослідницького співтовариства»: те переконання, яке визначене для того, щоб врешті-решт знайти визнання всіх дослідників, є тим, що ми розуміємо під істиною, а предмет, який репрезентує це переконання, є реальним.

Спроба зробити прагматизм дієвим для політики і педагогіки була здійснена Дж. Дьюї [99] в 20-х роках XX століття. Для нього пізнання (поняття) – це інструмент успішної дії (інструменталізм) і служить воно для опанування ситуації або для вирішення практичної проблеми. Мислення і пізнання можна краще за все пояснити, виходячи з того, як вони функціонують у певних зв'язках дії. На цьому філософському фундаменті Дж. Дьюї розробив свої пропозиції для реформи педагогіки (теорії і практики).

Педагогічні ідеї Дж. Дьюї справили великий вплив на загальний характер навчально-виховної роботи в школах США і деяких інших країн, зокрема і на радянську школу 1920-х років, що знайшло своє втілення в так званих комплексних програмах і в методі проектів, які застосовувалися в той час.

Головні питання Дж. Дьюї: як можна так побудувати навчання, щоб учні в майбутньому змогли виконати свою роль громадян у вільному демократичному суспільстві? Як досягти того, щоб учнів не повчали, але щоб учні отримували свій власний досвід? Його відповідь: школа повинна стати життєвим середовищем, ландшафтом учіння, в якому можуть жити і вчитися учні і вчителі: вчитися в процесі життя і жити в процесі навчання.

Звідси важливий принцип подолання ізоляції традиційної школи від досвіду життя, внесення до простору школи позашкільного досвіду і використання життєвих реалій повсякденності в цілях навчання. Підставою такого учення, що розкриває його механізм, є теорія рефлексивного досвіду, яка передбачає в цілісному понятті досвіду дві взаємообумовлені сторони: активний і пасивний досвід. Людина активно впливає на природне й соціальне середовище (активний досвід) і отримує відповідь (пасивний досвід). На основі цієї теорії досвіду конструюється середовище, в якому дитина може вчитися, здійснюючи на елементи середовища якусь дію і отримуючи від нього відповіді, тобто шляхом активної взаємодії з ним.

Обґрунтовуючи значущість демократії для освіти, Дж. Дьюї говорить: «Чи можна знайти хоч одне спростування тези про те, що демократичні суспільні інститути покращують якість людського досвіду, який доступніший і приносить більше задоволення, ніж антидемократичні форми суспільного життя? Чи не є правдою те, що принципи вшанування індивідуальної свободи і гуманності і добродіяння людських взаємин, врешті-решт, зводяться до переконання, що ці категорії є показниками досвіду більш високої якості, ніж методи придушення, примусу і силового тиску? Чи не є це причиною нашого вибору на користь віри в те, що взаємні консультації і консенсус, досягнутий через переконання, роблять можливим досвід кращої якості, який інакше не був би глобально досяжний?» [99, с. 25-26]. Дж. Дьюї відповідає позитивно на поставлені питання, підкреслюючи при цьому, що учні є повноправними членами суспільства, і в шкільних стінах їм має надаватися така ж свобода, як і поза ними.

Критикуючи діяльність тих шкіл, які не справляються із завданням прищеплення культури гуртожитку і поведінки своїм учням, Дж. Дьюї говорить: «Не викласти дітям один з найважливіших життєвих уроків – взаємної співпраці й адаптації – означає потерпіти педагогічну невдачу. Така педагогіка має однобічний характер, оскільки сформовані нею етичні і поведінкові установки мають фундаментальне значення для навчальної діяльності, заснованої на безпосередньому і прямому контакті і спілкуванні з іншими» [99, с. 68].

У конструктивістській педагогіці роль учителя – у співпраці з учнем, у спрямуванні учня в процесі інтелектуального пошуку. Така роль учителя відповідає парадигмі проблемного навчання як сукупності проблемних життєвих ситуацій, що актуалізують потреби учнів у знаннях, необхідних для розв'язання цих ситуацій. Організація навчання у формі проблемних ситуацій (фрагментів дослідницької діяльності) відображає принципи інтегративності і міждисциплінарності.

### *1.4.2 Еволюція та сучасний стан концепції соціального конструктивізму в навчанні*

Колективні навчальні дослідження (спільні дослідження, дослідницькі спільноти) С. А. Раков виділяє як особливо плідні у процесі вдосконалення природничо-математичної освіти. Виділяючи важливість спілкування, взаємодій, які стимулюють плідні обговорення, що сприяють розвиткові знань у навчанні, він пропонує додати слово соціальний до слова конструктивізм, що підкреслює важливість мови та обговорень у комунікаційному середовищі навчання: «дискусії, узгодження та аргументація у дослідницькому підході у навчанні, як і у професійних дослідженнях, сприяють розвиткові знань як у навчанні, так і у підготовці учителя» [216, с. 47]. Розглянемо більш детально теоретичні основи його застосування у процесі навчання.

У соціальних науках конструктивістський підхід до буденного і наукового пізнання, що отримав назву «соціальний конструкціонізм» (або «соціальний конструктивізм»), робить акцент на опосередкованості пізнання комунікацією, дискурсивними практиками і культурними конструктами. Аналіз дискурсивних практик у французькому структуралізмі (перш за все, в роботах М. Фуко (Michel Foucault) [371]), ідеї Л. С. Виготського [86] про культурно-історичну обумовленість людської психіки, соціології науки Т. Куна (Thomas Samuel Kuhn) [149] та інших ідей показує, що всі вони пов'язані з рефлексією соціокультурних чинників породження знань.

У роботі П. Бергера (Peter L. Berger) і Т. Лукмана (Thomas Luckmann) «Соціальне конструювання реальності», що вийшла в 1966 р., були заявлені основні тези соціального конструкціонізму як дослідницької програми, спрямованої на вивчення того, як «...загальноприйнята в тій або іншій культурі «консенсусна реальність» формується в результаті соціальних взаємодій, включаючи значення, класифікації, цінності, соціальні інститути, що сприймаються як частини «об'єктивного» світопорядку» [75].

Соціальний конструкціонізм об'єднує різні напрями на перетині психології, соціології, лінгвістики, антропології, що характеризуються критичним підходом до «саме собою зрозумілого» знання, інтересом до механізмів культурно-історичної обумовленості сприйманої реальності, увагою до соціальних процесів, що опосередковують пізнання, і аналізом зв'язку між знанням і поведінкою (соціальними наслідками знання). З точки зору соціальних конструкціоністів реальність конструюється не індивідом, а суспільством: будь-які ментальні конструкції виступають як продукт комунікації і спільної діяльності людей. У дослідженнях, виконаних у цій парадигмі, міститься критичний аналіз ідеології і домінуючих уявлень про реальність. Як зазначає В. Барр (Vivien Burr), у цих дослідженнях підкреслюється відносність будь-якої теорії, її неповнота і доповнюваність різних точок зору [12].

Американський соціолог К. Дж. Джерджен [103] бачить завдання соціальних наук, теорії яких неминуче містять замасковані цінності, не в описі реального світу, а в конструюванні нових форм стосунків і поведінки. Він критикує «образотворчу» концепцію мови науки, згідно з якою мова функціонує як карта або картина світу, він показує, що мова ученого завжди затверджує певні цінності і стосунки, служить досягненню якогось ефекту. За К. Дж. Джердженом, терміни психології культурно та історично обумовлені (наприклад, такі поняття соціальної психології, як «конформізм» або «когнітивний дисонанс» несуть негативне оцінне навантаження і містять приховані уявлення про належну поведінку).

Один із засновників дискурс-аналізу (дискурс – міркування, направлені на встановлення істини) Р. Харре (Rom Harré) [32] запропонував перемикати увагу з пошуку «Я» як суті на методи конструювання «Я» в ході комунікації і «розповіді про себе». Основна функція дискурсу – «приводити об'єкти в буття, створювати структуру реальності» шляхом встановлення і узгодження значень, зрозумілих учасникам комунікації. «Ментальна реальність» конструюється в «реальності розмови»; замість психологічних феноменів слід вивчати мовні акти, чим і займається дискурс-аналіз. Близькі за сенсом концепції конструювання «Я» були розвинені в теорії соціальної ідентичності [59] і в нарративній психології [46].

Конструкціонізм у дидактиці – філософія навчання, розвинена С. Пейпертом на основі конструктивізму [15]. До активної позиції конструктивізму конструкціонізм додає ідею того, що люди створюють нове знання особливо ефективно, коли вони залучені до створення продуктів, наділених особистісним сенсом. Головне те, що люди в процесі конструктивної діяльності створюють щось важливе для них самих або для тих, хто їх оточує.

У рамках сучасного конструкціонізму навчання людей (зокрема, дітей) проходить набагато ефективніше за умови, що вони самостійно роблять власні відкриття і моделюють свої знання, ніж коли готові знання «вливаються» їм через «крапельницю» традиційного процесу навчання [15].

В основі конструкціонізму лежать дві ідеї розуміння «конструкції» знання. Перша з них заснована на тому, що діти виучуються, активно конструюючи нове знання, не маючи жодної інформації, заздалегідь вкладеною в їхні голови. Друга стверджує, що ефективне навчання відбувається, коли учень залучає до конструювання нових знань особисто значимі артефакти, що представляють його власне знання. Так, створення комп'ютерної мультиплікації, роботів, ігор або віршів краще сприяє процесу навчання, ніж механічне виконання вправ.

Ідеї спеціальної організації учбових взаємодій з розвивальними цілями беруть свій початок у теорії розвитку вищих психічних функцій (мислення, свідомості, довільної поведінки й ін.) Л. С. Виготського, за яким ці функції формуються в ході спілкування і різноманітних соціальних взаємодій. Спілкування з дорослим або більш просунутим однолітком задає для дитини так звану «зону найближчого розвитку» [87, с. 264]. Це те, що дитина поки що не вміє сама, але чого може навчитися за допомогою іншого – партнера у спілкуванні й навчанні.

Якщо спілкування організоване правильно, стверджує Л. С. Виготський, то один крок в навчанні конкретного матеріалу має супроводжуватися двома кроками в інтелектуальному розвитку [120].

За Є. Д. Патаракіним [189], основними принципами соціально-конструктивістського навчання є:

1. *Організація навчання через дослідження.* Відповідно до даної ідеї, можна якісно поліпшити і кількісно прискорити процес пізнання, якщо організувати його як спілкування учнів зі спеціально розробленими об'єктами і середовищами моделювання.

2. *Конструювання навчально-дослідницьких співтовариств.* Для колективних навчальних досліджень важливим є побудова співтовариств. Для цього, крім навчальних об'єктів різної складності і іншого матеріалу, призначеного для дослідження та експериментування, треба створювати співтовариства учасників навчального процесу, сконструювавши такі правила його внутрішніх соціальних взаємодій, які додадуть процесу навчання нових вимірів і тим самим збагатять його. Матеріал для вивчення має бути сконструйований так, щоб по відношенню до нього був можливий особливий розподіл ролей і дослідницьких дій учасників. Розподіл повинен розкривати сутнісні характеристики виучуваної реальності і створювати можливості спільних змістовних обговорень, з метою більш поглибленого розуміння досліджуваного об’єкту чи процесу.

3. *Принцип орієнтації на особистість*. Сучасний підхід до розуміння змісту освіти визначає її як діяльність, спрямовану на удосконалення системи персональних конструктів учнів, тому зміст освіти є особистісно-орієнтованим і формується вчителем разом з учнями у процесі їх особистого руху вздовж індивідуальних освітніх траєкторій.

4. *Принцип насиченості*. Під насиченістю розуміють насичення освітнього простору носіями знання – наявність різноманітної літератури (не тільки підручників), можливість роботи з експертами (не обов’язково з професійними педагогами), з телекомунікаційними мережами (Інтернет, локальними електронними ресурсами), організація наочно-практичної діяльності (робота з лабораторним устаткуванням, з артефактами культури, реальна продуктивна діяльність) тощо. Насичене освітнє середовище дає можливість кожному учневі набути досвід діяльності, необхідний для розвитку особистісної системи конструктів, та вибудувати власну освітню траєкторію.

5. *Принцип співпраці*. Вчитель є не стільки «носієм знання», скільки рівноправним партнером у навчальній комунікації. Важливою складовою принципу співпраці є наявність у кожного учасника освітнього процесу (включаючи вчителя) особистого статусу – неоднакового і динамічно змінного в різних складових освітнього процесу. Прийнято виділяти чотири рівні такого статусу: відвідувач (гість), клієнт, постійний член групи для занять, експерт (статус не призначається, а природним чином визначається самим освітнім співтовариством, так що в однієї й тієї ж людини можуть бути різні статуси в різних ситуаціях). Ще однією складовою принципу співпраці є моніторинг особистих освітніх досягнень, причому мова йде не про оцінку учня вчителем, а про взаємооцінку досягнень освітнім співтовариством.

Освітня система, що реалізує перераховані принципи, є відкритою і спрямованою на формування системи компетентностей.

На роботах Л. С. Виготського [86; 87] базуються і витоки сучасного мережного підходу до процесу навчання (рис. 1.4).

Пізнавальна діяльність передбачає спільне використання засобів і обговорення результатів. Діяльність, пізнання, творчість і навчання мають потребу в інших людях. Навчання вимагає участі як партнерів-однолітків, з якими можна було б сперечатися й співробітничати, так і старших, які могли б оцінити результати діяльності. Модель пізнання за

Л. С. Виготським у сучасній інтерпретації Є. Д. Патаракіна передбачає освоєння світу через наступні культурні медіатори: дорослий-посередник; знак; слово (рис. 1.4).

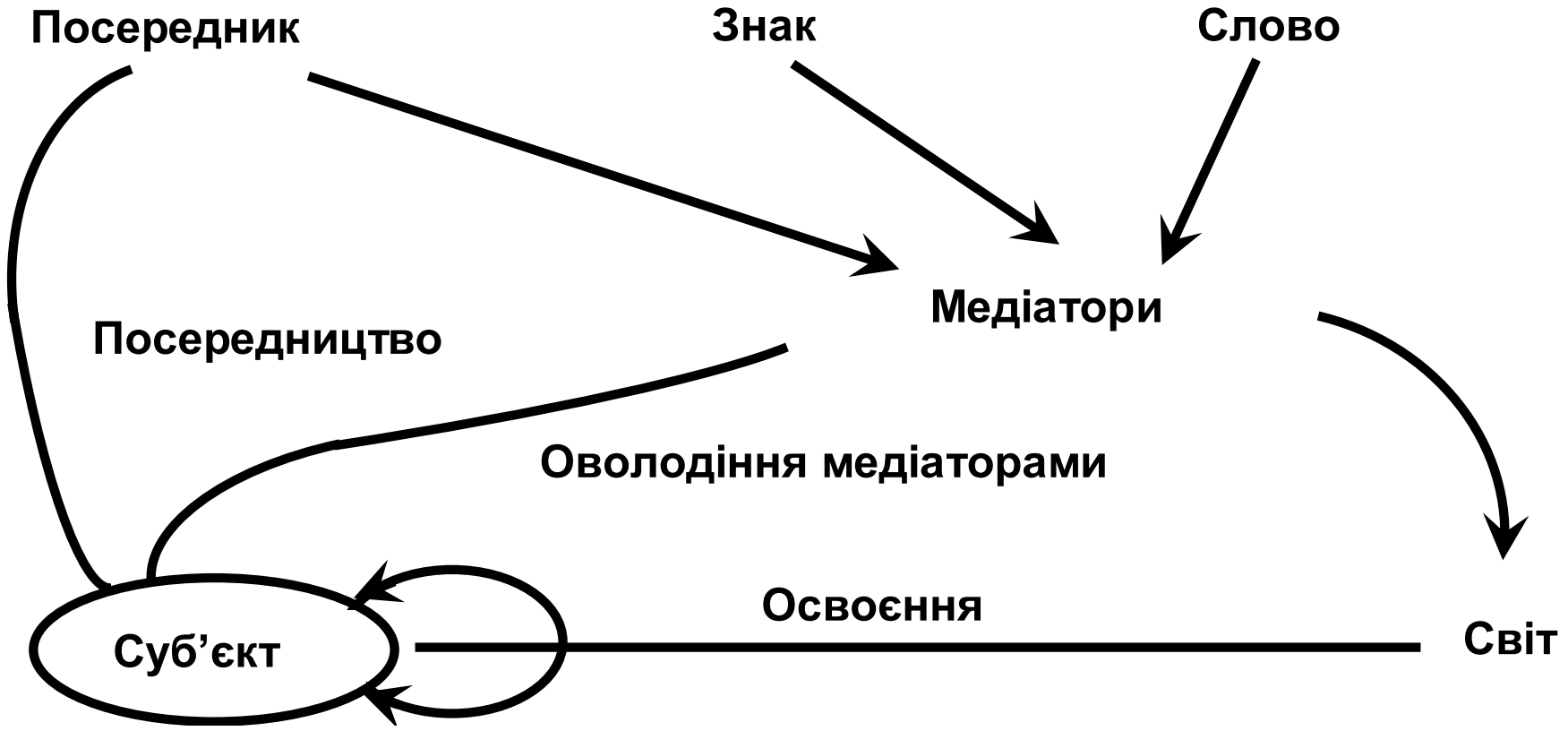

Рис. 1.4. Модель пізнання за Л. С. Виготським у інтерпретації Є. Д. Патаракіна

Пізнання розгортається як процес діяльності, в якому агент, що пізнає, постійно взаємодіє і співробітничає з іншими агентами і використовує різні культурні знаряддя.

Пізніше, розвиваючи теорію діяльності, Ю. Енгестрем (Yrjo Engestrom) [24] включив пізнання в складнішу мережу стосунків, куди, окрім агента, що пізнає, і культурних знарядь, входять:

1) правила – визнані норми, що обмежують дії, здійснювані в рамках системи діяльності;

2) розподіл праці відображає необхідність будувати свою індивідуальну діяльність з урахуванням діяльності інших і «ділитися» діями;

3) співтовариство – являє собою інших людей, які підключаються до індивідуальної дії суб’єкта на рівні діяльності.

Створюючи цей «розширений» трикутник (рис. 1.5), Ю. Енгестрем робив наголос на тому, що людська діяльність завжди соціальна і обов’язково передбачає наявність інших агентів. Спільна діяльність і обговорення цієї діяльності формує співтовариство. Дії над об’єктами вимагають спілкування. Це спілкування з приводу дій і об’єктів має первинне значення для пізнання і навчання.

Уявлення про пізнання як про мережу стосунків з іншими агентами і культурними знаряддями отримало подальший розвиток в теоріях розподіленого пізнання, ситуативного навчання і співтовариствах обміну знаннями – community of practice (суспільство практики) [6; 27].

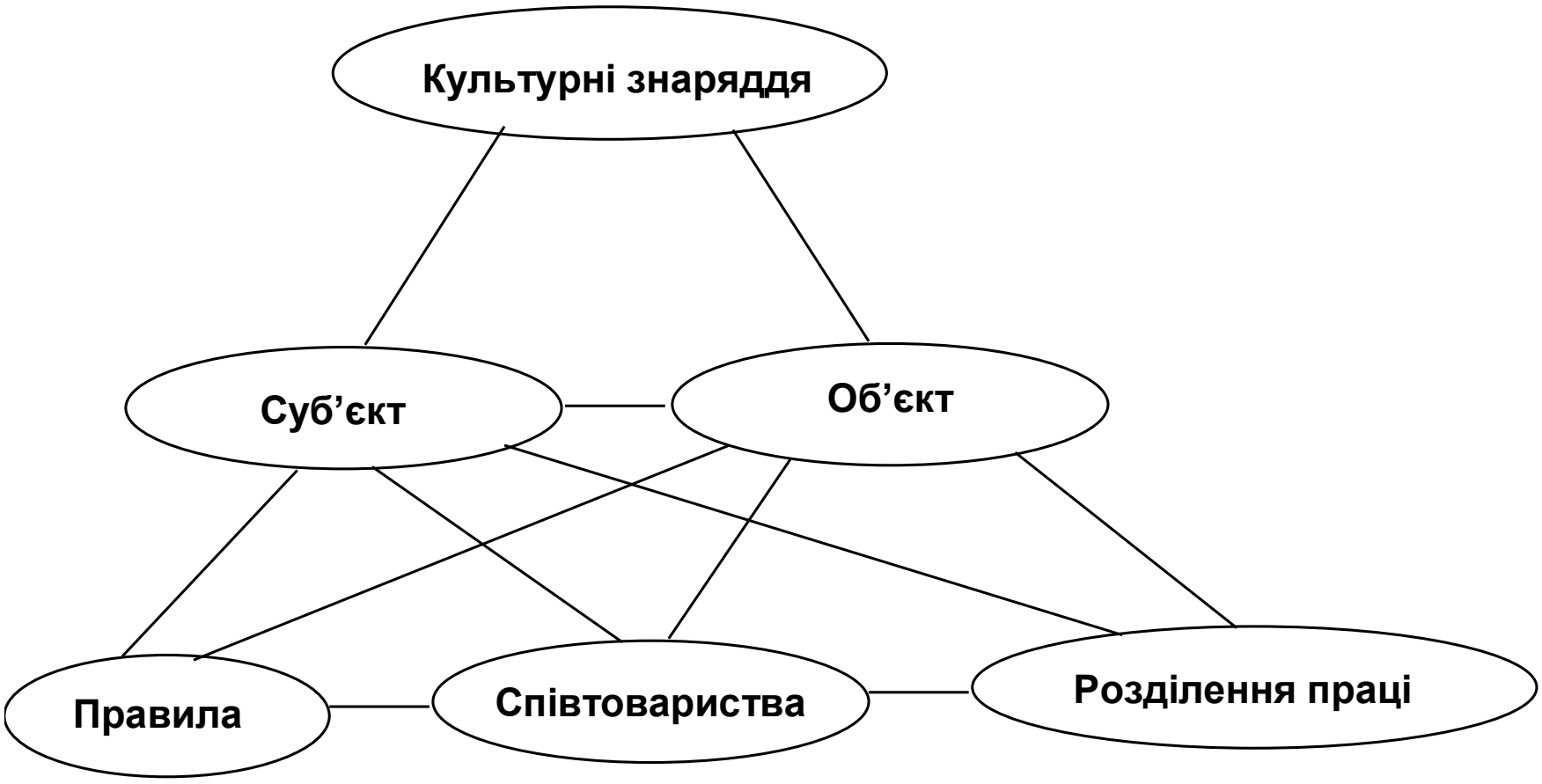

Рис. 1.5. Розширений трикутник пізнання за Ю. Енгестремом

Співтовариство обміну знаннями, або «співтовариство практики», позначає неформальну мережу, яка підтримує зусилля професіоналів у обміні досвідом і побудові спільного знання про предмет їхньої професійної діяльності [394] (рис. 1.6).

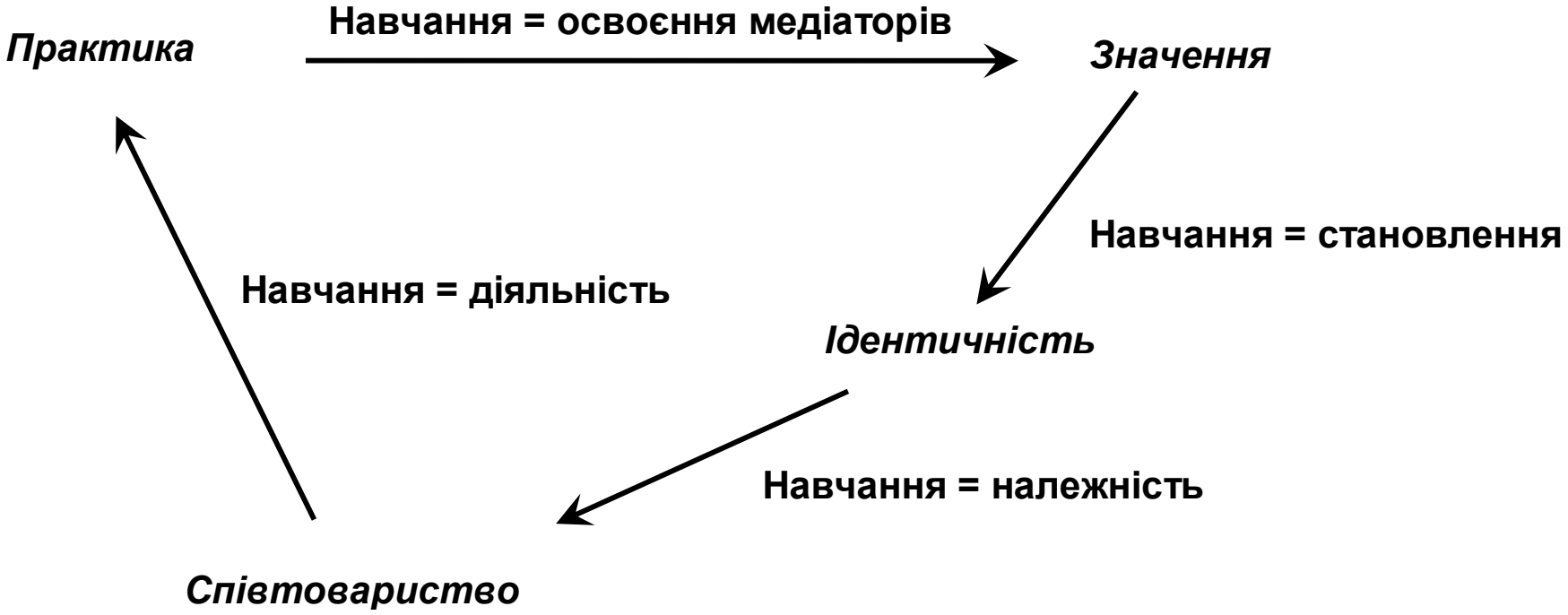

Рис. 1.6. Навчання як просування до центру співтовариства

Як відзначає Д. Норман (Donald A. Norman) [181], знання і пізнання світу не обмежуються головою суб'єкта, що пізнає. Вони розподілені в об'єктах, засобах і інших людях, які нас оточують. Є знання в голові і знання в речах і засобах. У повсякденних ситуаціях наша поведінка визначається комбінацією внутрішніх знань, зовнішньої інформації і обмежень. Люди постійно використовують ці зовнішні можливості, які допомагають учити менше і запам'ятовувати не так точно. Люди організовують свій життєвий простір для того, щоб він підтримував і спрощував їх діяльність. Жодна дитина не народжується школярем, так само, як вона не народжується і скрипалем. Навчання відбувається в співтовариствах, де учбовою практикою є участь в житті суспільства.

Діяльність і мислення людей завжди опосередковані інструментами та об'єктами. Всяка діяльність завжди пов'язана з використанням засобів діяльності і матеріальних об'єктів. Відповідно, всякий об'єкт мислиться і сприймається нами тільки в контексті певної діяльності. Всякий сприйманий об'єкт обов'язково має здатність проводити, здійснювати діяльність і мислення. Ми можемо освоїти тільки ту суть, яку можна представити як матеріальні об'єкти.

Метафора співтовариств із обміну знаннями дуже плідна і дозволяє підкреслити спільний і діяльнісний аспект навчання. В рамках даного підходу всяке навчання мислиться як спільна діяльність, яка обов'язково вимагає зацікавленої участі інших людей, що діють в даній галузі знань або схожих областях. Найбільш відомий приклад організації мережного співтовариства – наукове, в якому поширення і публікація інформації приводить до навчання членів всього співтовариства. Наукові співтовариства підтримують не лише дослідницьку діяльність, поширення інформації про її результати і доступ до цієї інформації, але і доступ до людей, можливість спостерігати за діяльністю експертів, можливість звертатися до них за порадою і допомогою.

Як відзначають Дж. Баукер (Geoffrey C. Bowker) і С. Стар (Susan Leigh Star), усвідомлення не може бути відірване від тих умов, від тієї ситуації, в якій воно відбувається. Для того, щоб освоїти засіб, мало його отримати в своє розпорядження і почати ним користуватися. Необхідно ще сприйняти культуру використання цього засобу. Навчання значною мірою є процес соціалізації, в ході якого люди вчаться говорити, читати, писати, стають школярами, співробітниками офісу, дослідниками тощо [8].

Участь у співтоваристві з обміну знаннями пов'язана з освоєнням тих об'єктів, що їх використовують у своїй діяльності члени співтовариства, Членство в співтоваристві обміну знаннями пов'язане з освоєнням тих об'єктів, які використовують в своїй діяльності члени співтовариства, – текстів, символів, засобів.

Особливу увагу на роль засобів звертають С. Пейперт і його послідовники. У своїх роботах С. Пейперт змістив напрям педагогічних інновацій з пошуку кращих методів викладання на пошук кращих об'єктів, за допомогою яких можна конструктивно діяти і розмірковувати про свою діяльність. С. Пейперт і його колеги передбачили, що побудова власних інтелектуальних структур здійснюється учнем найефективніше в тому випадку, якщо він залучений до створення реального продукту [2].

Модель пізнання за С. Пейпертом представлена на рис. 1.7.

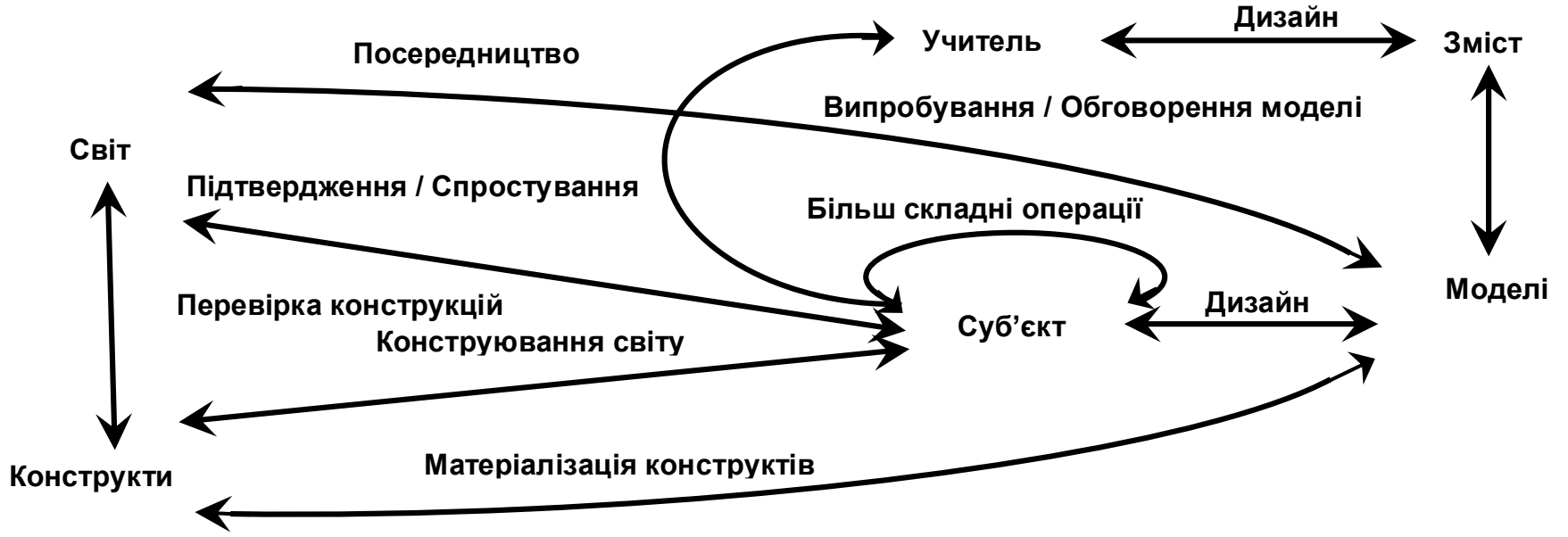

Рис. 1.7. Модель пізнання за С. Пейпертом

У лівій частині схеми відображене ключове для педагогіки положення про те, що знання не може бути передане людині в готовому вигляді. Знання завжди реконструюється самим суб'єктом, котрий пізнає, і це активний процес висунення гіпотез і породження особистих конструктів, які випробовуються суб'єктом на практиці. У правій нижній частині схеми представлене важливе доповнення, зроблене С. Пейпертом, – реконструкція знання здійснюється особливо успішно, якщо гіпотези суб'єкта втілюються в продукти-моделі. В цьому випадку перевірка власних гіпотез пов'язується з процесом випробування і обговорення моделей. Права верхня частина схеми підкреслює роль вчителя як дизайнера багатого інформаційного середовища, в якому відбувається пізнання.

Створення моделей і участь в діяльності всередині співтовариства практики часто зв'язані між собою (рис. 1.8). Діяльність, направлена на конкретний, втілений в матеріалі результат, витягує позитивні якості з понять. Така діяльність контролює розвиток понять, зберігаючи їх природний зв'язок з будовою реальних речей. Учасники багатьох співтовариств створюють моделі, які не лише випробовуються на практиці, але й активно обговорюються всередині співтовариства практики.

Поняття дослідницької спільноти С. А. Раков відрізняє від діяльнісних спільнот («співтовариств практики»), визначаючи, що вчителі та студенти-педагоги, які є членами дослідницької спільноти, презентують та осмислюють свій досвід, опановують теорії та дослідження інших членів спільноти: «поняття дослідницької спільноти може поєднати елементи соціального конструктивізму та елементи соціокультурних теорій: члени спільноти розвиваються та здійснюють свій внесок у безперервний процес реконструювання спільноти на основі критичної рефлексії; дослідницький підхід було сформовано як одну із форм діяльності у спільноті, у якій члени спільноти розвиваються на основі рефлексії дослідницького підходу. Це аналогічно вдосконаленню навчального процесу на основі досліджень самих вчителів їх власної практики навчання математики» [216, с. 51-52].

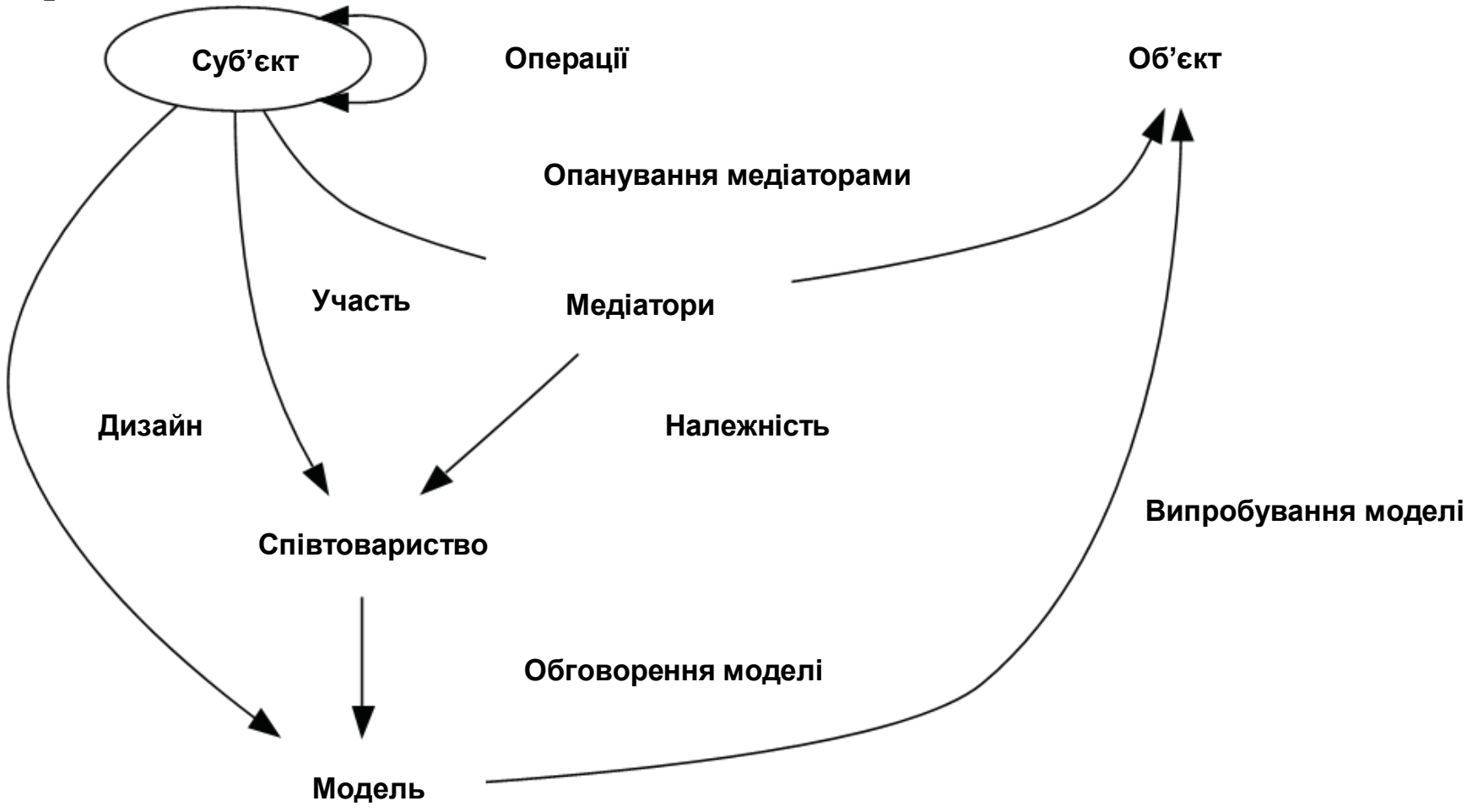

Рис. 1.8. Пізнання в співтоваристві практики

*1.4.3 Соціально-конструктивістські засоби об'єктно-орієнтованого моделювання*

Найбільш видатними представниками соціально-конструктивістського підходу в навчанні є С. Пейперт [192], А. Кей [33], М. Резник [16], В. Данн (Wanda P. Dann) [20], М. Гуздіал (Mark Guzdial) [30].

Конструкціонізм помітно вплинув на педагогічний дизайн і втілився в наступних педагогічних середовищах: Лого [80] і його похідні NetLogo, StarLogo [55] та LibreLogo [48]; Squeak [30] і похідний від нього Scratch [58]; Alice [3; 35] та ін.

Назва Лого (LOGO) походить від грецького «логос», що означає «слово», «сенс», «ідея». У літературі термін «Лого» використовується в двох значеннях:

1) як мова програмування, настільки проста, що її можуть освоїти діти, але настільки потужна й виразна, що й досвідчений програміст знайде в ній багато цікавого;

2) як філософія навчання, система поглядів на процес навчання,

покликана, як вважає С. Пейперт, не частково поліпшити, а корінним чином революціонізувати традиційну організацію навчання [51]. Мова Лого займає в цій системі центральне місце.

Філософія навчання Лого – це новий підхід до організації процесу навчання, що передбачає революційну перебудову системи освіти, яка існує в даний час. Характер і шляхи цієї перебудови трактуються С. Пейпертом відповідно до його уявлення про майбутнє людства. Він вважає, що в найближчі десятиліття відбудеться проникнення комп’ютерів у всі сфери людської діяльності. Це приведе до корінних змін в духовному житті суспільства: комп’ютери принесуть принципово нову, комп’ютерну культуру.

Філософія Лого передбачає перетворення традиційної системи навчання і ґрунтується на ідеї «використання комп’ютера як моделі, яка може вплинути на наш спосіб мислення про самих себе» [192, с. 26].

С. Пейперт протягом п’яти років працював в Міжнародному центрі генетичної епістемології в Женеві, очолюваному Ж. Піаже, виключно високо оцінює його теорію, хоча далеко не зі всім з нею згоден: «Піаже в його теорії стадій страшенно консервативний, майже реакційний стосовно того, що діти не можуть робити. Я прагну розкрити «більш революційного» Піаже, того, чиї епістемологічні ідеї можуть розірвати відомі пута людського розуму» [192, с. 157]. Як і Ж. Піаже, С. Пейперт розглядає дитину як активного конструктора своєї когнітивної структури. Дитина діє в певному соціальному середовищі – культурі, яка є визначальною в психічному розвитку дитини: вона отримує з неї поняття, схеми, способи діяльності і т.д. Культура визначає як специфіку, так і межі розвитку дитини.

Одне з найбільш значних досягнень теорії Ж. Піаже, як вважає С. Пейперт, – розкриття феномену неформального навчання («навчання за Піаже»), тобто засвоєння значного обсягу знань без спеціально організованого навчання. Прикладом «навчання за Піаже» служить засвоєння дитиною рідної мови, що відбувається природним чином – без спеціально організованого навчання, безпосередньо з середовища, що оточує дитину. Навчання рідної мови протікає вкрай ефективно: всі нормальні діти засвоюють рідну мову, чого не можна сказати про предмети, що вивчаються в школі. При «навчанні за Піаже» немає нездібних. Процес навчання не викликає негативного відношення з боку дітей, кожна дитина проявляє активність і інтерес. Значно скороченим виявляється і час навчання – досить порівняти освоєння рідної мови дитиною і вивчення іноземної мови дорослим.

С. Пейперт не згоден з Ж. Піаже щодо вікових меж розвитку форм мислення. Він вважає, що перехід дитини від однієї форми мислення до іншої значною мірою визначається особливостями середовища, в якому діє дитина, а також представленістю понять в культурі; проблема стадій інтелектуального розвитку просто не цікавить його. Так, пізній розвиток формального мислення обумовлений тим, що в нашій культурі не представлені або майже не представлені формальні поняття, схеми і так далі, отже, вони зовсім або майже недоступні дитині. Якби соціальне середовище було організоване так, щоб дитина постійно мала справу з формальними поняттями, то їх засвоєння наставало б у більш ранньому віці.

Об'єкти середовища Лого – це комп'ютер, що «математично говорить за істоту» [192, с. 47] і Черепаха, «кібернетична тварина», керована за допомогою комп'ютера. Навчання в Лого відбувається в процесі «бесід» учня з Черепахою і комп'ютером: на відміну від традиційної організації навчання, в Лого не комп'ютер управляє процесом навчання, а учень «навчає» комп'ютер, «говорячи йому» мовою Лого.

Мова програмування Лого становить основу середовища Лого. Мова Лого вигідно відрізняється від більшості відомих мов програмування в багатьох істотних відношеннях. Вона явно підтверджує думку С. Пейперта, згідно з якою мова програмування може бути дуже простою і в той самий час дуже потужною. Важливими перевагами мови Лого є: діалоговість (інтерактивність), об'єктна організація, розширюваність.

Програмне управління Черепахою є найбільш вражаючим дидактичним досягненням в системі Лого. Черепаха, як і комп'ютер, – об'єкт в середовищі Лого. Надсилаючи команди-повідомлення об'єктові «Черепаха», учні у природний спосіб засвоюють принципи об'єктно-орієнтованого програмування.

У всіх Лого-середовищах Черепаха – це об'єкт, схема або модель, з якою пов'язуються нові знання. Черепаха відкриває можливість персонального конструювання в процесі навчання: роздумуючи над поведінкою цього об'єкту, дитина привчається думати і про саму себе, і навпаки, спостереження над своєю власною поведінкою – важливе джерело ідей для «розмов» з Черепахою.

Черепаха – це мікросвіт, навчальне середовище в середовищі Лого. С. Пейперт дуже високо оцінює мікросвіт Черепахи з точки зору можливості реалізації в ньому неформального навчання. «Робота в мікросвіті Черепахи – це модель вивчення ідей у той самий спосіб, у який ми пізнаємо іншу людину. Учні, котрі працюють в цьому середовищі, безумовно, відкривають в ньому цікаві факти, приходять до узагальнень, засвоюють навики» [192, с. 136]. Діяльність учня в мікросвіті Черепахи нагадує пізнання ним світу в його повсякденній активності. У мікросвіті Черепахи наукові (наприклад, математичні) ідеї стають для учня настільки ж природними, як і його буденний досвід.

У середовищі Лого немає різкого розриву між «науковими» теоріями, що вивчаються в школі, і «ненауковими» теоріями, що створюються самою дитиною (вивчення дитячих теорій займає значне місце в дослідженнях школи Ж. Піаже). У середовищі Лого перехід від «ненаукових» дитячих теорій до «наукових» плавніший; його можна порівняти з процесом налагодження програми. Істотно іншим є у філософії Лого і погляд на помилку учня. Якщо в традиційному шкільному навчанні помилкове рішення оголошується неправильним і повинно бути відкинуто, то в Лого «ми розглядаємо помилки як щось корисне ... діти розуміють, що вчитель теж завжди є учнем і що кожен вчиться на помилках» [192, с. 114]. Будь-яка людина, хоч би трохи знайома з програмуванням, знає, що дуже рідко вдається відразу скласти програму так, щоб вона не містила жодної помилки. Наявність помилок в складеній програмі не означає, що вона має бути відкинута: необхідно виділити помилки, з'ясувати їхні причини і виправити їх. У Лого в процесі виправлення помилок учень поступово удосконалює свою програму, отримуючи при цьому нові знання.

На початку 1990-х років М. Резник запропонував використати мультиагентне співтовариство черепашок для освоєння учнями екологічних стратегій [54]. З безліччю черепашок у мові StarLogo учні могли спостерігати, вивчати й моделювати складні фізичні, хімічні, біологічні й соціальні феномени. Хоча мова створювалася в першу чергу як засіб навчання, у цьому середовищі виявилося можливим ставити й серйозні експерименти із мультагентного моделювання.

Дослідницькі можливості середовища одержали подальший розвиток у мові NetLogo – мережному середовищі колективних навчальних досліджень. Мова була створена Урі Віленським в 1999 році і продовжує активно розвиватися. Середовище програмування NetLogo служать для моделювання ситуацій і феноменів, що відбуваються в природі й суспільстві. NetLogo зручно використати для моделювання складних систем, що розвиваються в часі. Творець моделі може давати вказівки сотням і тисячам незалежних агентів, що діють паралельно. Це відкриває можливість для пояснення та розуміння зв'язків між поводженням окремих індивідуумів і явищами, які відбуваються на макрорівні. Мова NetLogo досить проста, тому учні й учителі можуть створювати в цьому середовищі свої власні навчальні моделі. У той же час NetLogo – це досить потужна мова для побудови дослідницьких моделей і проведення досліджень. У середовищі NetLogo в останні роки були побудовані різні дослідницькі моделі, які використовувалися в наукових статтях і обговорювалися в книгах з мультиагентного моделювання й соціології. У російській освіті середовище NetLogo використовувалося для демонстрації мережних феноменів і для моделювання соціальних феноменів у навчальних курсах з менеджменту [55].

Бібліотека моделей, створених у середовищі NetLogo, велика й поповнюється не тільки розроблювачами, але й членами співтовариства – http://ccl.northwestern.edu/netlogo/models/community.

У цьому співтоваристві можна:

– прочитати опис моделі, її призначення, покладені в основу принципи;

– подивитися виконання програми в мережі (для цього досить просто запустити програму в браузері, Java-applet відпрацює і покаже в окремому вікні, як працює модель;

– скачати модель і запустити на своєму комп’ютері;

– внести до моделі зміни й використати готові процедури, узяті з чужої моделі для своїх власних потреб;

– завантажити свою модель на загальнодоступний сервер і запропонувати її до обговорення й спільного використання.

Можливість збирати програму з набору готових «цеглинок» була в Лого із самого початку. У 1968 році А. Кей, надиханий ідеями С. Пейперта, придумав Dynabook – прообраз персонального комп’ютера. А. Кей бачив роль персонального комп’ютера як особистісного динамічного середовища (метамедіа), що об’єднувало в собі всі інші середовища: текст, графіку, анімацію і навіть те, що ще не винайдено [34].

У 1972 р. А. Кей перейшов у відомий науковий центр Xerox PARC, де й реалізував ці ідеї в новій мові Smalltalk. Саме тоді він запропонував знаменитий термін «об’єктно-орієнтоване програмування». В процесі роботи над Smalltalk А. Кей створив нову концепцію розробки програмного забезпечення – багатовимірне середовище взаємодії об’єктів з асинхронним обміном повідомленнями. В результаті з’явилась можливість підтримки такого середовища не одним, а багатьма комп’ютерами, об’єднаними в мережу. Працюючи над апаратною реалізацією ООП-системи (проект FLEX – повноцінний персональний комп’ютер, що базувався на об’єктах), А. Кей вивчав роботи С. Пейперта з навчання дітей програмування мовою Лого.

Подальшим розвитком FLEX став проект Dynabook – компактний комп’ютер, легко керований, оснащений клавіатурою і пером, безпровідною мережею тощо (в сучасних термінах ми можемо назвати

Dynabook планшетним портативним комп’ютером). У статті 1972 р. А. Кей визначив ціль проекту як «персональний комп’ютер для дітей будь-якого віку» [33]. Smalltalk увібрав у себе багато з даного проекту – в ньому вперше були використані вікна, меню, іконки та маніпулятор «миша». В Smalltalk містяться витоки Microsoft Windows, X Window та MacOS. Інакше кажучи, сучасні інтерфейси користувача еволюціонували паралельно з ООП, а їх формування відбувалося під впливом ідей дослідницького підходу в навчанні.

Сьогодні А. Кей – активний учасник проекту OLPC (One Laptom Per Child – «Кожній дитині – по ноутбуку») [28]. Незважаючи на високу технологічну досконалість ідей проекту Dynabook – «батька» сучасних мобільних пристроїв, головним в ньому є все ж таки ідея «комп’ютера для навчання», основою якого є особистісна зорієнтованість, висока інтерактивність, навчання через гру, спільне навчання, динамічне моделювання, навчання завжди та всюди.

Так само, як і Лого, мова Smalltalk розроблена як програмна частина проекту Dynabook. Вона є одночасно і мовою програмування, і середовищем розробки програм. Це чисто об’єктно-орієнтована мова, у якій абсолютно все розглядається як об’єкти. Як зазначає один з її розробників Д. Інгаллс (Daniel Henry Holmes Ingalls, Jr.), «мета проекту Smalltalk – зробити світ інформації доступним для дітей будь-якого віку. Всі труднощі полягають у тому, щоб знайти й застосувати досить прості й ефективні метафори, які дозволять людині вільно оперувати найрізноманітнішою інформацією від чисел і текстів до звукових і зорових образів» [41]. В основу мови покладені дві прості ідеї: 1) усе є об’єктами; 2) об’єкти взаємодіють, обмінюючись повідомленнями.

Більш глибокий аналіз Smalltalk показує, що це ретельно продумана фундаментальна розробка, яка не має прямих аналогів у традиційній практиці виробництва програмної продукції, оскільки охоплює на єдиній концептуальній основі всі відомі програмно-апаратні рівні віртуальної машини користувача. При цьому мікропроцесорна (апаратна) реалізація основних системних класів може не тільки значно випередити сучасний рівень розвитку комп’ютерних систем, але й забезпечити ефективну реалізацію подальших їх поколінь.

У 1995 р. А. Кей, Д. Інгаллс і Т. Кьохлер (Ted Kaehler) працювали в Apple, залишаючись усе ще зацікавленими у своєму баченні Dynabook у якості середовища розробки для побудови освітнього програмного забезпечення, яке зможуть використати (і навіть програмувати) не лише технічні фахівці. На жаль, у комерційних реалізаціях Smalltalk, що одержали поширення на той час, зникли багато ідей проекту Dynabook, тому, вирішивши, що «правильного» Smalltalk не існує, А. Кей з колегами почали створення Squeak – відкритого вільно поширюваного середовища розробки. У вересні 1996 р. Squeak став доступний в Інтернет. За минулі роки він був успішно перенесений на різні варіанти UNIX, Windows і навіть Windows CE [339].

Сьогодні розробка Squeak продовжується тією ж групою в Walt Disney Imagineering – дане середовище використовується в багатьох диснеївських проектах.

У мультимедійному об'єктно-орієнтованому середовищі Squeak з'являється все більше властивостей проекту Dynabook – потужна 2D- і 3D-графіка, багатоголосний і синтезований звук, підтримка анімації й відео, засоби для роботи з різноманітними медіаформатами тощо. Squeak сьогодні – практичний Smalltalk, у якому дослідник, викладач або зацікавлений студент може переглянути вихідний код для будь-якої частини системи, включаючи графічні примітиви й саму віртуальну машину, і виконати будь-які зміни без необхідності використання мови, відмінної від Smalltalk [156; додаток А].

Squeak включає в себе ряд інтерфейсів користувача: Morphic (основний інтерфейс), eToys (мова візуальних сценаріїв, що базується на Morphic), новий експериментальний інтерфейс Tweak та MVC (наслідуваний від початкового інтерфейсу користувача Smalltalk-80).

Squeak використовується як основний компонент в новій операційній системі Es. Багато розробників Squeak співпрацюють у проекті Croquet (Крокет) – надбудовою Squeak, метою якої є створення мережної операційної системи реального часу, що утворює спільний робочий простір з 2D- та 3D- можливостями між декількома користувачами. Він також забезпечує гнучку структуру, в якій більшість концепцій інтерфейсу користувача можуть бути швидко прототиповані і розгорнуті в потужне середовище моделювання. Додатки, створені за допомогою програмного забезпечення для розробників (SDK), можуть бути використані для підтримки високомасштабованої спільної візуалізації даних, віртуального навчання та вирішення проблем навколишнього середовища, 3D-Wiki, онлайнових ігрових середовищ (MMORPG), взаємопов'язані багатокористувацькі віртуальні середовища та ін.

Таким чином, Squeak є розвиненим об'єктно-орієнтованим мультимедіа-середовищем мови Smalltalk, в якому реалізовані основні концепції конструктивістського підходу та об'єктно-орієнтованого програмування [31]. Вибір Squeak у якості основного компоненту проекту OLPC дозволяє нам говорити про його високу стабільність, наявність мобільних реалізацій – про мобільність, а реалізація об'єктного підходу – про фундаментальність даного середовища.

Scratch – середовище програмування, створене під керівництвом ще одного співробітника С. Пейперта – М. Резника [58]. Scratch дозволяє дітям створювати власні анімовані й інтерактивні історії, ігри й інші творіння. Основне завдання проекту – стати часткою освітньої програми для дітей і підлітків, розвинути в них творчі здібності, логічне мислення і свободу у використанні інформаційних технологій. Все це пропонується розвинути шляхом залучення учнів до процесу конструювання інтерактивних презентацій, мультфільмів, ігор. Діти можуть складати свої програми з блоків команд («цеглинок») так само, як вони будували будиночки і машинки з деталей конструктора LEGO. Основні особливості соціально-конструктивістського середовища Scratch:

1. Блокове програмування. Створення програм у Scratch – це просте поєднання графічних блоків разом в стеках. Блоки розроблені так, щоб їх можна було зібрати тільки в синтаксично правильних конструкціях, що виключає принаймні синтаксичн помилки. Різні типи даних мають різні форми, підкреслюючи несумісність. Зміни в стеках можна робити, навіть коли програма виконується, що дозволяє експериментувати з новими ідеями знов і знов.

2. Маніпуляції даними. Зі Scratch можна створювати програми, які поєднують графіку, анімацію, музику та різні звуки і управляють цими засобами. Scratch розширює можливості управління візуальними даними, що популярні в сьогоденній культурі, – наприклад, додаючи можливість користуватися Photoshop-подібними фільтрами.

3. Спільна робота й обмін. Сайт проекту Scratch надає можливість переглянути проекти інших людей, використовувати і змінювати їхні картинки і скрипти, додати власний проект. Найбільше досягнення – це соціальне середовище і культура, створена довкола самого проекту.

Де і як можна використовувати дане середовище у навчанні майбутніх учителів природничо-математичних дисциплін?

По-перше, при вивченні теми «Алгоритми і виконавці». Чи багато цікавих завдань можна придумати для цього виконавця? Чи всі алгоритмічні структури можна наочно показати? Найскладніше підібрати завдання на використання розгалуження в таких алгоритмах. Дане середовище можна використовувати для створення графічних зображень і анімації. Для цього існують команди малювання і команди руху.

По-друге, при вивченні програмування. Більшість людей розглядають програмування на комп’ютері як нудне заняття, доступне тільки тим, хто має гарну технічну підготовку. І, справді, традиційні мови програмування, такі як Java і C++, складні для вивчення. Завдання

Scratch, як нової мови програмування – змінити це становище. При викладанні програмування мало просто показати і пояснити роботу різних операторів, циклів, умов і т. п. – потрібно вчити дітей мислити особливим чином, розуміти суть команд і алгоритмів. Отже, викладання повинне вестися максимально наочно, а учні – мати можливість негайно бачити результат своїх дій.

Таким чином, Scratch можна розглядати як інструмент для творчості, залишаючи програмування на другому плані. Діти можуть складати історії, малювати й оживляти на екрані придуманих ними персонажів, вчитися працювати з графікою і звуком. Застосувань можливостей Scratch можна знайти безліч: у цьому середовищі легко створювати анімовані листівки, презентації, ігри, мультфільми. Завдяки простоті мови й ідеології в цілому Scratch дозволяє легко навчитися основам програмування. Задаючи поведінку своїх персонажів в програмі, дитина вивчає такі фундаментальні поняття, як цикли та умови.

За своєю внутрішньою архітектурою Scratch базується на Squeak, тому при «вичерпанні» можливостей Scratch у міру розвитку навичок програмування можна перейти до батьківського мультимедійного середовища об’єктно-орієнтованого моделювання Squeak, використовуючи потужні засоби ООП мови Smalltalk.

Так само, як Лого і Squeak, Scratch є стандартним програмним забезпеченням проекту OLPC, метою якого є подолання цифрової нерівності й формування навичок колективних навчальних досліджень.

Scratch цікавий і сам по собі, і тим дослідницьким співтовариством, що склалося довкола нього. У цьому співтоваристві учасники обмінюються й обговорюють результати своєї діяльності – конкретні об’єкти, програми, малюнки тощо. Співтовариство носить відкритий характер. Будь-хто бажаючий може подивитися всі матеріали. Реєстрація необхідна тільки в тому випадку, якщо потрібно завантажити на сервер свій готовий проект.

Усі члени співтовариства діляться своїми проектами та їхніми «рецептами». Немає ніяких секретних рецептів, всі рецепти відкриті для всіх. Кожний рецепт можна використати й видозмінювати. Кожний учасник, котрий зареєструвався у співтоваристві, може опублікувати на сервері співтовариства свій проект. При публікації автор додає до проекту короткий опис і ключові слова – теги. Інші учасники співтовариства можуть відзначити проект який сподобався, відзначити його своїми тегами, залишити коментар, додати проект до галереї з подібною тематикою. Учасники можуть скачати проект, якщо хочуть подивитися його сценарій на своєму комп’ютері, внести деякі зміни й доповнення. Після цього можна знову опублікувати проект на сервері як свій власний. При цьому програма сервера розпізнає проект як похідний і додасть в опис посилання «Цей проект зроблений на основі проекту» → «Посилання на батьківський проект».

Адміністратори співтовариства проводять експертну оцінку проектів і виділяють найбільш складні й несподівані проекти. Ці проекти розміщаються в розділах «Обрані проекти» і «Несподівані проекти» на першій сторінці сайту. У такий спосіб адміністратори можуть привертати увагу учасників співтовариства до проектів, які їм уявляються найбільш цікавими або важливими (рис. 1.9).

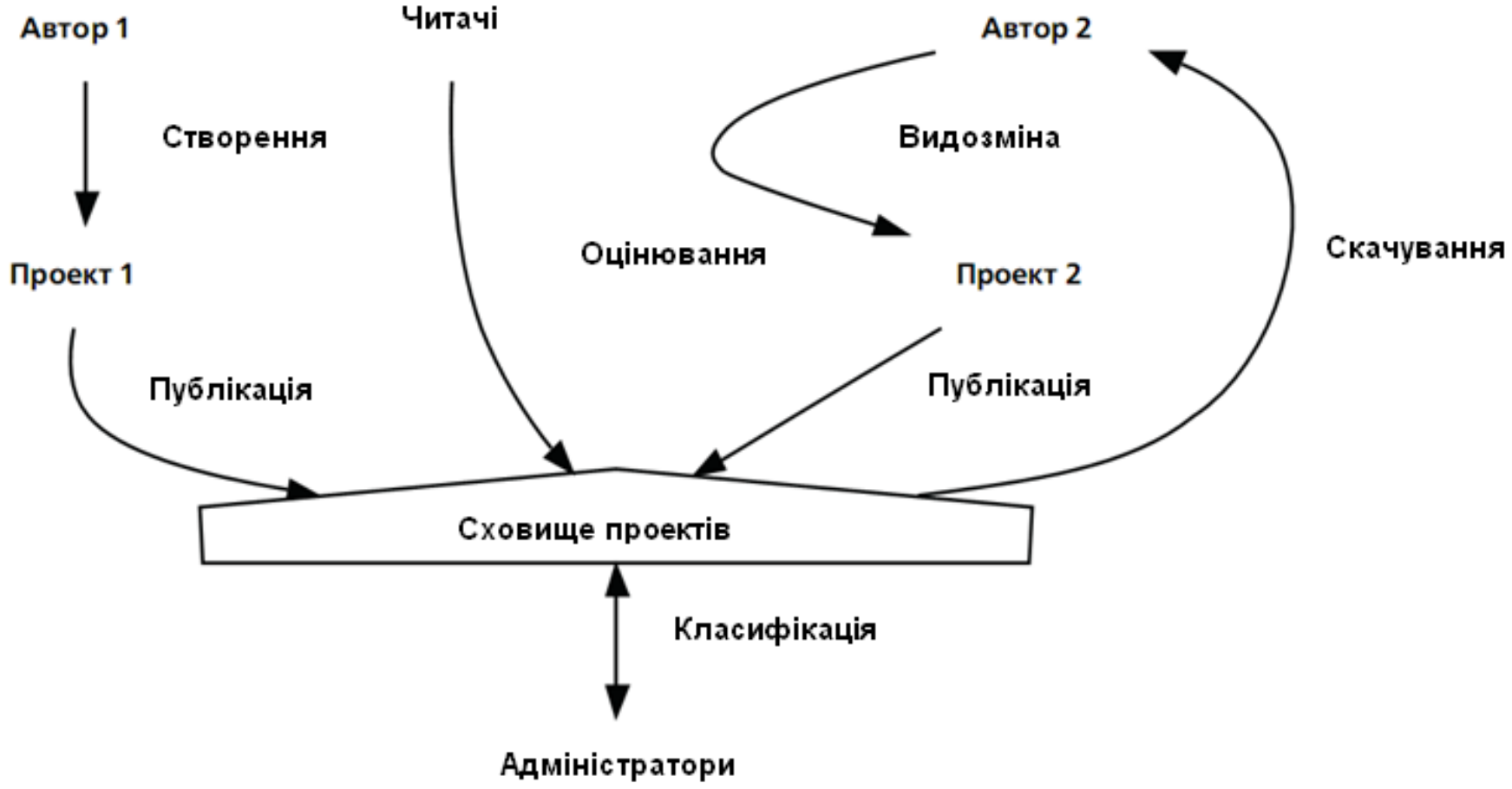

Рис. 1.9. Схема взаємодії учасників у співтоваристві Scratch

Якщо проект виконаний з порушенням правил, прийнятих на освітньому сайті, то учасники можуть відзначити цей проект як неприпустимий за змістом. Причини такої оцінки обов'язково повинні бути пояснені в додатковому повідомленні. Такі випадки розглядаються модераторами сайту, котрі приймають рішення про видалення проекту й позбавленні його автора права розміщення проектів у майбутньому. У співтоваристві Scratch піклуються про безпеку учасників. Діти молодше 13 років не записують при реєстрації адресу своєї електронної пошти. На сайті не публікуються ніякі особисті дані про учасника, крім країни, у якій він перебуває. Учасники співтовариства не можуть обмінюватися приватними повідомленнями. У співтоваристві Scratch припустимі тільки відкриті коментарі до опублікованих проектів.

Scratch привчає нас збирати проект із цеглинок і ділитися результатами своїх дій з іншими людьми. Ці навички важливі не тільки усередині спеціальних середовищ програмування, але й у сучасних мережних співтовариствах. Єдність процесів створення, пошуку й зберігання інформаційних цеглинок всі частіше можна спостерігати на сторінках сучасних сайтів, що використають концепцію Web 2.0. Метафора будівельних блоків, з яких діти й дорослі можуть зібрати прості й дуже складні конструкції, є присутнім не тільки в навчальних проектах, але й у більшості сучасних мережних сервісів форматів Web 2.0, призначених для підтримки організацій і мережних співтовариств обміну знаннями. М. Резник, описуючи педагогічні можливості мови Scratch, використовує метафору спіралі творчого розвитку, представлену на рис. 1.10 [58]:

– люди в уяві представляють, що саме вони хочуть зробити й одержати в результаті;

– люди створюють проект, заснований на своїх уявленнях;

– люди грають із результатами своєї діяльності;

– люди діляться результатами своєї діяльності з іншими людьми;

– люди обмірковують і обговорюють свої результати.

– обговорення й обмірковування приводить до наступних версій нових проектів.

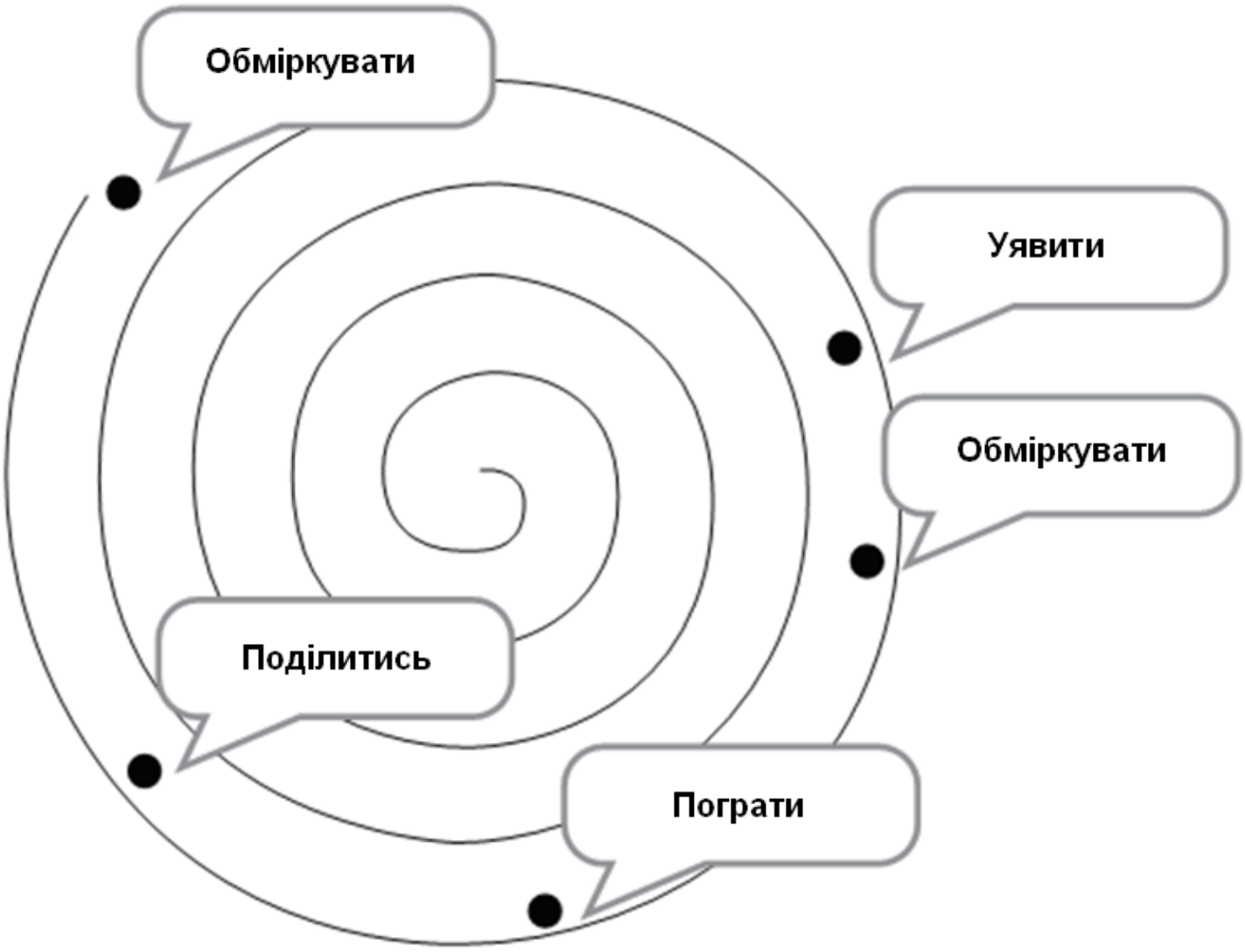

Рис. 1.10. Спіраль творчого розвитку

Середовище розробки Alice [17], так само, як і Scratch – відносно новий проект, що розробляється в університеті Карнегі-Меллона. На відміну від своїх попередників, Alice – повністю тривимірне середовище моделювання. Alice 2.2 позиціонується розробниками як засіб навчання об'єктно-орієнтованого програмування, а Alice 3 – як засіб об'єктно-орієнтованого моделювання. Існує також спрощена версія Alice (Storytelling Alice), що може бути ефективним засобом конструювання навчальних 3D-моделей у середній школі.

Alice є середовищем, в якому можна маніпулювати 3D-об'єктами (рухати, обертати, міняти колір тощо) і створювати програми, що генерують анімацію у віртуальних світах. Вона не просто схожа на сучасні професійні IDE – використовуючи IDE NetBeans, створені в Alice моделі легко перетворити на проекти ООП-мовою Java. Робоча площина Alice розбита на декілька вікон: у одному в реальному часі відображується віртуальний світ; у іншому присутнє дерево об'єктів, і для кожного з них надається набір доступних методів, функцій та властивостей; центральна частина відведена під редактор вихідного коду.

Запропонована в Alice концепція навчання фактично є зануренням у світ об'єктно-орієнтованого моделювання без будь-яких істотних і часто непотрібних спрощень цієї парадигми – маніпулювати об'єктами можна тільки за допомогою їх властивостей, функцій та методів, як убудованих, так і сконструйованих користувачем. В Alice максимально скорочений обсяг уведення з клавіатури: для переважної більшості дій досить миші. Код програми не є текстом в звичному розумінні: в межах одного методу він є набором вкладених блоків, виділених кольором залежно від типу (цикли, умовні переходи й ін.), їх можна згортати, перетягувати, змінюючи порядок, тощо. Незважаючи на таке полегшення, мовою програмування в Alice є не навчальна, а професійна мова Java. Проте середовище розробки Alice надає можливість відображувати створену програму як у Java-стилі (**об'єкт**.метод(*параметри*)), так і в стилі мови Smalltalk (**об'єкт** метод *параметри*).

Проведені творцями проекту дослідження показали, що застосування Alice для навчання програмування початківців сприяє глибшому розумінню концепції ООП – адже якщо студенти відразу починають оперувати поняттями ООП, їм немає необхідності перенавчатися, що неминуче відбувається при переході від однієї парадигми програмування до іншої.

Визначальними особливостями Alice, у порівнянні з іншими середовищами об'єктно-орієнтованого моделювання, є:

– близькість середовища моделювання і мови програмування за інструментами, використовуваними сучасними програмістами – із самого початку навчання освоюються високорівневі концепції;

– скорочення клавіатурного введення до мінімуму, що помітно

спрощує створення програм людьми, котрі не володіють розвиненими навичками набору тексту, та початківцями;

– використання оригінальної ідеї побудови анімацій у віртуальному світі, що дозволяє наочно представляти процес виконання програми і спрощує пошук помилок.

Різні електронні середовища, всередині яких учасники можуть створювати свої власні цифрові об'єкти, обмінюватися такими об'єктами, видозмінювати електронні об'єкти, є соціально-конструктивістськими. До таких середовищ можна віднести багатокористувацькі світи, системи управління знаннями (наприклад, Moodle), різні сервіси Web 2.0. Чим складніші та цікавіші об'єкти, якими може обмінюватися співтовариство, тим більші можливості для навчання воно відкриває для своїх учасників. Як зауважує Є. Д. Патаракін, «... пізнавальна, творча і навчальна діяльність з самого початку мають мережний і колективний характер. Перехід від егоцентричної позиції до розуміння ролі та значення других людей, інших способів конструювання реальності є важливим етапом психологічного розвитку особистості» [189, с. 6].

### 1.5 Умови професійної підготовки майбутніх учителів природничо-математичних дисциплін засобами комп'ютерного моделювання

У сучасних педагогічних дослідженнях, пов'язаних з проблемами вдосконалення функціонування педагогічних систем, підвищення ефективності процесу професійної підготовки, одним з аспектів, що викликають найбільший інтерес, є виявлення, обґрунтування і перевірка умов, що забезпечують успішність здійснюваної діяльності.

Головною метою системи вищої педагогічної освіти є професійна підготовка вчителів високої кваліфікації згідно із соціальним замовленням. Сучасні науковці розглядають підготовку педагога у ЗВО з позиції суб'єкта пізнавальної та навчально-професійної діяльності, здатного цілеспрямовано регулювати власні дії [64].

На сьогодні існує потреба у вирішенні питання щодо належної підготовки у ЗВО майбутніх учителів природничо-математичних дисциплін, удосконалення якої можна досягти за рахунок розробки адекватних умов, що в свою чергу повинно підвищити ефективність навчання. Зауважимо, що головною метою у створенні умов буде вдосконалення професійної підготовки майбутніх учителів природничо-математичних дисциплін на основі фундаменталізації процесу їх підготовки.

*Умова* – категорія філософії, що позначає відношення предмета до навколишньої дійсності, явищ об’єктивної реальності, а також до себе і свого внутрішнього світу. Предмет виступає як дещо обумовлене, а умова – як відносно зовнішнє щодо предмету різноманіття об’єктивного світу.

Умову слід відрізняти від поняття причини, так як на відміну від причини, що безпосередньо породжує те чи інше явище або процес, умова визначає те середовище, в якому останні виникають, існують і розвиваються.

Таким чином, умова – це: 1) стан системи, за якого настає можливість події; 2) обставина, від якої дещо залежить; 3) правила, встановлені в певній сфері діяльності.

У психології досліджуване поняття, як правило, представлено в контексті психічного розвитку і розкривається через сукупність внутрішніх і зовнішніх причин, що визначають психологічний розвиток людини, прискорюють або уповільнюють його, роблять вплив на процес розвитку, його динаміку і кінцеві результати [175, с. 270-271]. Педагоги займають схожу з психологами позицію, розглядаючи умову як сукупність змінних природних, соціальних, зовнішніх і внутрішніх впливів, що впливають на фізичне, моральне, психічне, розвиток людини, її поведінка, виховання і навчання, формування особистості (В. М. Полонський) [206, с. 36]

Ю. К. Бабанський за сферою впливу виділяє дві групи умов функціонування педагогічної системи: зовнішні (природно-географічні, суспільні, виробничі, культурні) і внутрішні (навчально-матеріальні, шкільно-гігієнічні, морально-психологічні, естетичні) [191].

За характером впливу виділяють об’єктивні і суб’єктивні умови. Об’єктивні умови, що забезпечують функціонування педагогічної системи, включають нормативно-правову базу сфери освіти, засоби інформації тощо і виступають в якості однієї з причин, що спонукають учасників освіти до адекватних проявів себе в ньому. Ці умови можуть змінюватися. Суб’єктивні умови, що впливають на функціонування і розвиток педагогічної системи, відображають потенціали суб’єктів педагогічної діяльності, рівень узгодженості їх дій, ступінь особистісної значущості цільових пріоритетів і провідних задумів освіти для учнів тощо.

За специфікою об’єкта впливу виділяють загальні і специфічні умови, які сприяють функціонуванню та розвитку педагогічної системи. До загальних умов належать соціальні, економічні, культурні, національні, географічні та інші умови, до специфічних – особливості соціально-демографічного складу учнів; місцезнаходження освітньої установи, її матеріальні можливості, обладнання освітнього процесу;

виховні можливості навколишнього середовища та ін.

Важливу роль у забезпеченні функціонування і розвитку педагогічної системи відіграють також такі специфічні умови, як: характер морально-психологічної атмосфери в педагогічному та учнівському колективах, рівень педагогічної культури педагогів та ін. Важливу роль при визначенні напрямів розвитку педагогічної системи відіграє урахування просторових умов, в яких існує педагогічна система, тому її функціонування обумовлюється особливостями регіональних, місцевих умов, специфікою навчального закладу, конкретної педагогічної середовища, рівнем кваліфікації необхідних педагогічних кадрів, ступенем оснащеності освітнього процесу (кабінети, навчальні посібники, обладнання та ін.). Необхідність урахування просторових умов, що складають середовище функціонування педагогічної системи, обумовлена реалізацією принципу єдності загального, одиничного та особливого в наукових дослідженнях.

Словник-довідник з професійної педагогіки визначає педагогічні умови як «обставини, від яких залежить та відбувається цілісний продуктивний педагогічний процес підготовки фахівців, що опосередковується активністю особистості, групою людей» [277, с. 243]. Розглядаючи дане поняття, вчені дотримуються кількох позицій. Першої позиції дотримуються дослідники, для яких педагогічні умови є сукупність деяких заходів педагогічного впливу і можливостей матеріально-просторового середовища: комплекс заходів, зміст, методи (прийоми) і організаційні форми навчання і виховання. Другу позицію займають дослідники, що пов’язують педагогічні умови з конструюванням педагогічної системи, в якій вони виступають одним з компонентів: змістовна характеристика одного з компонентів педагогічної системи, в якості якого виступають зміст, організаційні форми, засоби навчання і характер взаємин між учителем та учнями.

Розглядаючи інформаційно-технологічну підготовку вчителів природничо-математичного циклу в системі додаткової професійної освіти, Р. Г. Хамітов визначає наступні її умови: 1) адекватне урахування в змісті інформаційно-технологічної підготовки двох підходів – предметно-орієнтованого, для розвитку педагога; постановки і вирішення професійних проблем і особистісно-орієнтованого підходу – для врахування індивідуальних особливостей педагога і використання його досвіду; 2) створення позитивної мотивації у педагогів, учнів у межах системи додаткової професійної освіти; 3) зацікавленість педагогів у результатах інформаційно-технологічної підготовки; 4) проектування змісту інформаційно-технологічної підготовки в цілому і кожного окремого заняття окремо; 5) включення педагогів у формування середовища навчально-інформаційної взаємодії [373].

До основних завдань підготовки майбутніх учителів природничо-математичних дисциплін О. І. Ордановська відносить створення умов, які сприяють відображенню профільного навчання у змісті навчальних предметів циклів природничо-математичної та професійно-орієнтованої підготовки в освітньому процесі педагогічного ЗВО, а саме: 1) узгодження і систематизація навчальних курсів природничо-математичного циклу і методики їх викладання з позицій профільного навчання; 2) формування у студентів спеціальних знань міжпредметного змісту; 3) створення спеціальних курсів з підготовки студентів до роботи у профільній школі [185].

А. О. Прокубовська виділяє наступні умови розвитку самостійної пізнавальної діяльності студентів ЗВО:

1) самостійна пізнавальна діяльність має розглядатися як компонент професійно-педагогічної підготовки, що розвивається у процесі планування, регулювання і виконання студентами самостійної роботи з використанням інформаційних технологій;

2) теоретико-експериментальною основою розвитку самостійної пізнавальної діяльності студентів виступає самостійна робота, яка є одночасно організаційною формою, засобом і методом навчання;

3) методичне забезпечення має бути побудоване на основі системності, наочності, індивідуальності і включати робочу програму, навчальний посібник, моделюючий пакет, дидактичні матеріали [213].

У рамках дослідження під педагогічними умовами будемо розуміти сукупність необхідних заходів, спрямованих на фундаменталізацію підготовки майбутніх учителів природничо-математичних дисциплін.

Аналіз досліджень останніх років дає можливість зробити висновок, що в контексті нашого дослідження реалізація педагогічних умов має на меті:

– забезпечення фундаментальності природничо-математичної підготовки засобами комп’ютерного моделювання;

– забезпечення організаційно-педагогічного й психолого-педагогічного супроводу підготовки майбутніх учителів природничо-математичних дисциплін;

– визначення форм організації, методів та засобів процесу професійної підготовки студентів природничих, математичних та інформатичних спеціальностей соціально-конструктивістськими засобами комп’ютерного моделювання.

Підготовка майбутніх учителів природничо-математичних дисциплін засобами комп’ютерного моделювання ґрунтується на таких принципах:

– варіативності, альтернативності й доступності освітніх програм, технологій навчання і навчально-методичного забезпечення;

– гнучкості, свободи вибору змісту і форм організації підготовки майбутніх учителів природничо-математичних дисциплін, зокрема засобами технологій соціального конструктивізму;

– упровадження технологій соціального конструктивізму в процес навчання, використання цих технологій у подальшій професійній діяльності;

– зворотного зв'язку: контроль і корекція процесу підготовки за допомогою мережних засобів технологій соціального конструктивізму.

У результаті аналізу філософської та психолого-педагогічної літератури вітчизняних та зарубіжних учених сформульовані основні умови підготовки майбутніх учителів природничо-математичних дисциплін засобами комп'ютерного моделювання:

1) застосування педагогічної технології соціального конструктивізму в процесі підготовки майбутніх учителів природничо-математичних дисциплін;

2) упровадження об'єктно-орієнтованого моделювання у процес навчання інформатичних дисциплін;

3) використання соціально-конструктивістських засобів ІКТ навчання об'єктно-орієнтованого моделювання.

### Висновки до розділу 1

Перспективним напрямом фундаменталізації професійної підготовки майбутніх учителів природничо-математичних дисциплін є посилення ролі основного методу дослідження в природничих науках – методу моделювання, що одночасно виступає в якості провідного методу навчання. При цьому урахування психологічних особливостей відображення свідомістю людини об'єктів оточуючої дійсності вимагає відповідної їх інтерпретації у комп'ютерних моделях, тому розв'язання проблеми фундаменталізації навчання природничо-математичних дисциплін в умовах швидкої зміни засобів ІКТ потребує об'єднання методу моделювання та об'єктно-орієнтованої технології програмування, що разом утворюють якісно нову концепцію – об'єктно-орієнтоване моделювання.

Об'єктно-орієнтоване моделювання – це: 1) вид комп'ютерного моделювання, за якого середовищем моделювання є деяке середовище програмування, що надає можливість конструювання об'єктів, їх використання та обміну повідомленнями між ними; 2) навчальна дисципліна, в якій вивчаються способи конструювання та дослідження об'єктно-орієнтованих моделей. Вид програмування, що використовується при побудові таких моделей, називають об'єктно-орієнтованим програмуванням – це технологія програмування, заснована на поданні програми у вигляді сукупності об'єктів, кожен з яких є реалізацією деякого класу, а класи утворюють ієрархію за принципами наслідування.

Найважливішими етапами об'єктно-орієнтованого моделювання є: 1) об'єктно-орієнтований аналіз (вид аналізу, за якого вимоги до моделі системи сприймаються з точки зору класів та об'єктів, виявлених у предметній галузі); 2) об'єктно-орієнтоване проектування (вид проектування, що відображає процес конструювання об'єктно-орієнтованої моделі та поєднує у собі процес об'єктної декомпозиції та прийоми подання логічної та фізичної, а також статичної та динамічної моделей проектованої системи); 3) обчислювальний експеримент та аналіз його результатів.

Теоретичною основою перебудови професійної підготовки майбутніх учителів природничо-математичних дисциплін на основі об'єктно-орієнтованого моделювання є педагогічна технологія соціального конструктивізму. Ґрунтуючись на засадах вітчизняної педагогічної психології, вона втілює в собі демократичний підхід до освіти, особистісну зорієнтованість, компетентнісний прагматизм, розвиток дивергентного критичного мислення, навчання у спільноті та через спільноту. Перспективним напрямом її реалізації є індивідуальні та колективні навчальні дослідження, а їх проведення вимагає використання таких засобів комп'ютерного моделювання, які забезпечують спільну навчальну діяльність у мережному середовищі.

Умовами професійної підготовки майбутніх учителів природничо-математичних дисциплін засобами комп'ютерного моделювання є: 1) застосування педагогічної технології соціального конструктивізму в процесі підготовки майбутніх учителів природничо-математичних дисциплін; 2) упровадження об'єктно-орієнтованого моделювання в процес навчання інформатичних дисциплін; 3) використання соціально-конструктивістських засобів ІКТ навчання об'єктно-орієнтованого моделювання.

# РОЗДІЛ 2
# СИСТЕМА РЕАЛІЗАЦІЇ УМОВ ПРОФЕСІЙНОЇ ПІДГОТОВКИ МАЙБУТНІХ УЧИТЕЛІВ ПРИРОДНИЧО-МАТЕМАТИЧНИХ ДИСЦИПЛІН ЗАСОБАМИ КОМП'ЮТЕРНОГО МОДЕЛЮВАННЯ

## 2.1 Модель підготовки майбутніх учителів природничо-математичних дисциплін засобами комп'ютерного моделювання

Для теорії та методики професійної освіти розробка моделі професійної підготовки фахівця є незмінно актуальним завданням: зміна соціально-економічних умов, поява та зникнення професій, розвиток технологій та багато інших зовнішніх чинників приводять до розуміння того, що модель фахівця є історичною категорією. Саме тому необхідним є аналіз тих факторів зовнішнього середовища, що визначають професійну підготовку фахівця на поточному етапі розвитку системи освіти в Україні.

У професійній педагогіці моделювання широко використовується в процесі підготовки фахівців та передбачає системний розгляд, з одного боку, професійної діяльності, до якої готують студентів, а з іншого, особливостей організації підготовки студентів. Такі моделі формуються на основі освітньо-кваліфікаційних характеристик та навчальних планів і програм. Водночас для ефективної організації процесу підготовки майбутніх фахівців доцільним є розгляд питання цілепокладання вибору професії, тому до моделей часто включають мотиваційну складову.

Принципи та методи формування моделі фахівця розглядав авторський колектив під керівництвом О. Я. Савельєва [228], який визначив наступну послідовність операцій з їх побудови:

– визначення мети та конкретних задач моделювання;

– аналіз та синтез інформації, що відноситься до сформульованих задач;

– виокремлення основних факторів, що впливають на зміну тенденцій та закономірностей досліджуваного об'єкта чи явища;

– побудова моделі, що базується на задачах, на розв'язання яких спрямована модель [228, с. 25-26].

Складовою моделі підготовки фахівця є зміст освіти, що визначається метою підготовки фахівця. Сама ж мета підготовки визначається вимогами суспільства до підготовки фахівця, вимоги державних та галузевих стандартів вищої освіти, сучасними тенденціями розвитку технологій та суспільства. Зауважимо, що саме така складова галузевих стандартів вищої освіти, як освітньо-кваліфікаційна характеристика фахівця, має бути основою визначення мети підготовки.

Для побудови моделей фахівця дослідники використовують кілька основних підходів: компонентний, технологічний, ступеневий, структурний, модельний, компетентнісний. Розглянемо окремо особливості кожного з них.

За компонентного підходу в моделі виділяються наступні складові:

1) система вимог до професійної підготовки, що включає суспільно, державно, соціально-економічно, соціокультурно та технологічно зумовлені фактори підготовки;

2) мета підготовки, що визначається функціонально-цільовим блоком;

3) методологія підготовки;

4) компоненти підготовки.

Компоненти підготовки поділяють на:

– цільові: цілі класифікуються за часом їх досягнення від найближчих до перспективних;

– змістові містять управлінський, психолого-педагогічний та технологічний блоки;

– організаційні (організаційно-виконавчі);

– оцінно-результативні (контроль та моніторинг якості підготовки).

Змістові та організаційно-виконавчі компоненти часто об’єднують у інтегровану змістово-операційну систему.

За технологічного підходу [114] у змісті моделі виділяють наступні складові:

1) система вимог до професійної підготовки;

2) мета підготовки;

3) зміст навчання;

4) технологія навчання, що містить: вимоги до професійної підготовки фахівця; методи підготовки; засоби підготовки; форми організації навчання;

5) запланований результат підготовки.

За ступеневого підходу на кожному етапі процесу навчання, що відповідає освітньо-професійній програмі підготовки фахівця (навчальний курс, семестр, модуль, тиждень тощо), проектується зміст навчання, що може бути деталізованим відповідними методами та засобами навчання.

За структурного підходу в моделі виділяються складові зовнішнього та освітнього середовищ:

1) заклади системи управління освітою;

2) структура освітньої установи;

3) суб’єкти навчання та управління навчанням;

4) інформаційно-технологічне забезпечення процесу підготовки

фахівця.

За модельного підходу модель підготовки фахівця є складовою системи, що містить також:

1) модель навчального закладу;

2) модель фахівця, що визначається його освітньо-кваліфікаційною характеристикою;

3) модель процесу навчання, деталізована за блоками підготовки та навчальними дисциплінами;

4) модель діяльності фахівця – прогноз його подальшого професійного розвитку;

5) модель перепідготовки фахівця, що відображає соціально, державно та технологічно зумовлені зміни в моделі фахівця.

За компетентнісного підходу, що сьогодні є провідним в Україні, метою підготовки є компетентний фахівець. Нормативною основою підготовки виступає система компетенцій (соціально-особистісних, загальнонаукових, інструментальних, загально-професійних та спеціалізовано-професійних), на основі якої мають бути сформовані виробничі функції майбутнього фахівця (дослідницька, контрольна, проектувальна, прогностична, організаційна, управлінська, технологічна, технічна). Саме за такого підходу до побудови моделі в ній відображаються всі компоненти майбутньої професійної діяльності фахівця: профіль діяльності, узагальнені виробничі функції, структура праці, особливості професійної діяльності, прогноз професійного розвитку тощо.

О. Я. Савельєв, підкреслюючи важливість урахування специфіки педагогічної освіти при розробці моделі фахівця, наголошує на тому, що майбутній учитель повинен володіти технологією проектування власної професійної діяльності, бути здатним до розробки та застосування інноваційних педагогічних технологій. У зв’язку з цим особливу увагу при розробці моделі підготовки майбутнього вчителя природничо-математичних дисциплін слід приділити, з одного боку, перспективним напрямам побудови розвитку освітніх систем (технологічний аспект), а з іншого – інтеграційним основам навчання фізики, математики, хімії, біології, географії та інформатики (фундаментальний аспект). За такого підходу:

1) «той, хто навчається, стає не просто студентом, а формується і розвивається фахівцем, а ... накопичений ним потенціал забезпечує поступальний саморозвиток професійної компетентності в умовах модельованої, імітованої або реальної професійної діяльності» [228, с. 43];

2) студент в інтегративному навчальному курсі оволодіває

соціально-конструктивістськими технологіями перетворення змісту навчання на способи професійної діяльності в швидкозмінних умовах.

До основних груп професійно орієнтованих умінь, якими студенти повинні в процесі навчання в педагогічному ЗВО, відносять:

– психолого-педагогічні: аналітичні, проектувальні, конструктивні, організаторські, комунікативні, контролювальні;

– частинно-методичні – специфічні, пов'язані з навчанням і організацією предметної діяльності;

– спеціальні – уміння в тій області діяльності, якої навчають дітей.

Провідними *соціально-конструктивістськими уміннями* майбутнього вчителя є такі:

– аналізувати власну педагогічну діяльність;

– проектувати розвиток особистості кожної дитини і колективу в цілому;

– прогнозувати результати навчання і виховання;

– планувати свою роботу із керівництва різними видами діяльності учнів на тривалий період часу;

– теоретично обґрунтовано добирати засоби, методи і форми організації навчальної діяльності, щоб забезпечити проектування розвитку особистості та колективу;

– реально представляти і знаходити найбільш раціональні рішення, пов'язані з розміщенням учнів під час різних видів діяльності, з розподілом між ними обов'язків у спільній діяльності, з одночасною організацією учнів для виконання різних видів діяльності;

– обґрунтовано, з урахуванням психологічних особливостей учнів, визначати логічну структуру уроків та інших форм роботи з учнями;

– управляти поведінкою і активністю учнів, захоплювати їх освітнім процесом;

– групувати учнів у процесі діяльності з урахуванням їх взаємин та індивідуальних особливостей;

– знаходити найкращу форму вимог і варіювати їх залежно від індивідуальних особливостей дітей і конкретних педагогічних умов;

– знаходити контакт і правильний тон з різними людьми в різних обставинах;

– визначати по зовнішніх проявах і вчинках дітей зміну їх психологічного стану, корегувати його в конкретних життєвих ситуаціях;

– своєчасно орієнтуватися і обґрунтовано корегувати поставлені педагогічні цілі і педагогічні завдання з урахуванням відповідних реакцій учнів на педагогічні впливи і конкретних педагогічних умов;

– здійснювати поточне конструювання (в тому числі у виборі

методів і засобів педагогічного впливу) з метою досягнення скоригованих завдань;

– стимулювати хід діяльності, враховуючи успіхи і досягнення дітей;

Оволодіння кожним з названих умінь вимагає включення учнів у відповідну спільну навчальну діяльність.

Основними показниками якості підготовки майбутнього вчителя природничо-математичних дисциплін вважатимемо: 1) рівень фундаментальності освіти, що забезпечується посиленням ролі методу моделювання; 2) рівень професіоналізму освіти, що забезпечується посиленням ролі технологій соціального конструктивізму; 3) рівень сформованості компетентності у комп'ютерному моделюванні.

Компетентність з комп'ютерного моделювання, як і будь-які компетентності, мають такі взаємопов'язані складові (згідно нашого визначення):

– *когнітивно-змістову* (гносеологічну);

– *операційно-технологічну* (праксеологічну);

– *ціннісно-мотиваційну* (аксіологічну) – емоційно-ціннісне ставлення до процесу моделювання та аналізу його результатів; уміння знаходити нові, нестандартні рішення задачі; внутрішня мотивація до опанування комп'ютерного моделювання; готовність до активного застосування гносеологічних та праксеологічних складових у практичній діяльності; прагнення до самовдосконалення, потреба у саморозвитку гносеологічних та праксеологічних складових; уміння самостійно приймати рішення, критично ставитись до чужих впливів, здатності за власним почином організовувати діяльність, ставити мету, в разі необхідності вносити в поведінку зміни; вміння постійно і тривало домагатися мети; наполегливість у досягненні мети, прагнення до поліпшення отриманих результатів, незадоволеність досягнутим, намагання домогтися успіху;

– *соціально-поведінкову* – здатність до співпраці у процесі розробки, опису, налагодження, тестування комп'ютерних моделей та аналізу результатів їх роботи, використання засобів для організації спільної роботи над проектом; відповідальність за власну поведінку, за виконання завдань; комунікабельність; здатність до адаптації; схильність до дискусії.

Систематичне і цілеспрямоване формування гносеологічної та праксеологічної складових компетентності з комп'ютерного моделювання майбутнього вчителя природничо-математичних дисциплін виконуватимемо у рамках спецкурсу.

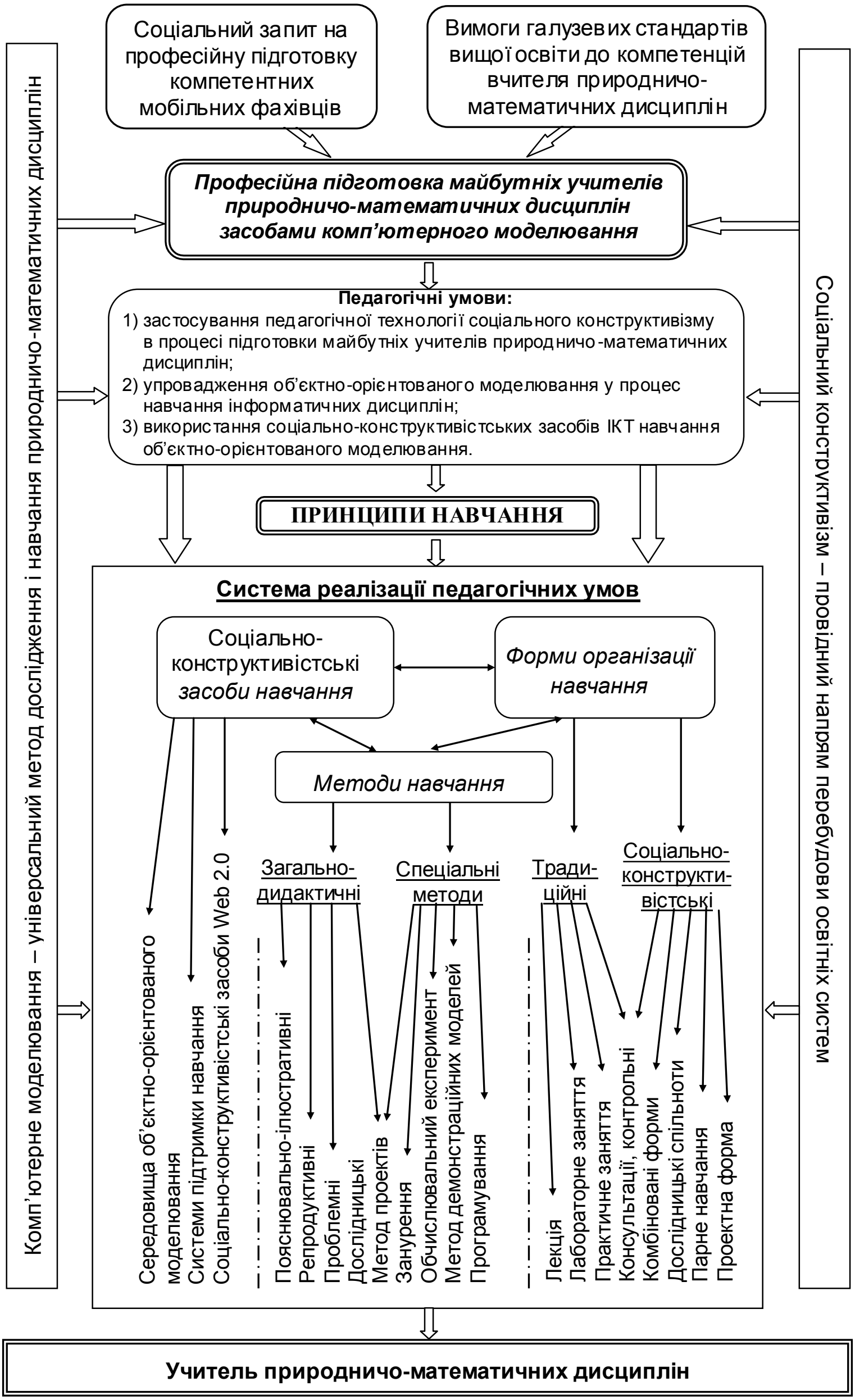

Рис. 2.1. Структурно-функціональна модель підготовки майбутніх учителів природничо-математичних дисциплін засобами комп'ютерного моделювання

Для розробки моделі професійної підготовки майбутніх вчителів природничо-математичних дисциплін (рис. 2.1) було використано інтегрований підхід на основі оптимального поєднання різних типів моделей.

Розроблена структурно-функціональна модель відповідає основним принципам формування моделі фахівця, забезпечуючи:

– відповідність змісту сучасним потребам держави, суспільства і особистості;

– відповідність підходів до формування змісту принципам розробки державних освітніх стандартів;

– відповідність розроблюваного змісту вимогам до рівня вищої освіти як одного із ступенів професійної освіти;

– використання методу моделювання змісту підготовки у відповідності з моделлю діяльності та особистості підготовлюваного фахівця.

Розроблена модель передбачає створення умов для формування активної навчально-пізнавальної діяльності студентів, розвитку спільної навчально-дослідницької діяльності, умінь використання засобів комп’ютерного моделювання і соціально-конструктивістських технологій.

Підготовка майбутніх учителів природничо-математичних дисциплін засобами комп’ютерного моделювання виконується на основі системи вимог, що об’єднуються у дві групи.

До першої групи належать вимоги, що висуваються до майбутнього фахівця зовнішнім та професійним середовищем. Це, насамперед, соціальний запит на підготовку компетентних мобільних фахівців, здатних до навчання протягом всього життя, конкурентоспроможних на ринку освітніх послуг, які вільно володіють засобами розв’язання професійних задач, здатні до ефективної роботи за фахом на рівні світових стандартів.

Друга група вимог визначається освітнім середовищем та насамперед формується на основі освітньо-кваліфікаційної характеристик та освітньо-професійних програм за напрямами підготовки 6.040101 «Хімія», 6.040102 «Біологія», 6.040104 «Географія», 6.040106 «Екологія, охорона навколишнього середовища та збалансоване природокористування», 6.040201 «Математика», 6.040203 «Фізика», 6.040302 «Інформатика». Ці вимоги визначають якісну підготовку фахівця та передбачають:

– оптимізацію методів навчання, інформатизацію навчального процесу та активне використання технологій відкритої освіти;

– розробку інтегрованих та міждисциплінарних курсів та програм;

– формування умов для неперервного професійного зростання кадрів, забезпечення наступності різних рівнів професійної освіти та створення ефективної системи неперервної професійної освіти.

Важливими складовими моделі є теоретично обґрунтовані умови:

1) застосування педагогічної технології соціального конструктивізму в процесі підготовки майбутніх учителів природничо-математичних дисциплін;

2) упровадження об'єктно-орієнтованого моделювання в процес навчання інформатичних дисциплін;

3) використання соціально-конструктивістських засобів ІКТ навчання об'єктно-орієнтованого моделювання.

Розроблена модель забезпечує реалізацію підготовки сучасного вчителя природничо-математичних дисциплін.

Система реалізації виокремлених умов включає в себе соціально-конструктивістські засоби навчання, методи навчання (загальнодидактичні та спеціальні) та форми організації навчання (традиційні та соціально-конструктивістські), підпорядковані загальній меті професійної підготовки засобами комп'ютерного моделювання.

## 2.2 Цілі та зміст навчання спецкурсу «Об'єктно-орієнтоване моделювання»

Реалізація моделі професійної підготовки майбутніх учителів природничо-математичних дисциплін засобами комп'ютерного моделювання вимагає конкретизації змісту навчання, що не є складовою моделі. Як було показано в розділі 1, для цього необхідним є впровадження в процес професійної підготовки спецкурсу «Об'єктно-орієнтоване моделювання».

Вибір об'єктів моделювання в інформатиці як науці залежить від предметної області, а в інформатиці як навчальний дисципліні – від фахової орієнтації майбутнього вчителя. Сьогодні в Україні професійна інформатична підготовка здійснюється переважно на природничо-математичних спеціальностях педагогічних ЗВО, в яких метод моделювання є провідним методом дослідження, тому в процесі проектування змісту курсу об'єктно-орієнтованого моделювання ми виходили з професійно-орієнтовної функції фундаменталізації природничо-математичної освіти, що має наступні структурні компоненти: цільовий, змістовий, технологічний та підсумковий.

Враховуючи, що головною метою професійної підготовки студентів є формування професійних компетентностей, основою *цільової компоненти* обрано суспільне замовлення, державні стандарти вищої освіти та особистий вибір студента. Змістова компонента містить специфічну теорію, що відображає професіоналізацію обраної спеціальності.

Зміст навчання є тим стрижнем, який з'єднує всі рівні системи освіти, визначаючи їхню послідовність та наступність. При формуванні змісту важливо встановити баланс між фундаментальністю та професійною спрямованістю підготовки, реалізувавши виділений Г. О. Михаліним принцип диференційованої фундаментальності [165].

Так, для майбутніх вчителів математики доцільним є комп'ютерне моделювання математичних об'єктів, для майбутніх вчителів фізики доцільним є комп'ютерне моделювання фізичних об'єктів та процесів, для майбутніх вчителів біології та екології доцільним є комп'ютерне моделювання екологічних процесів, а при навчанні майбутніх вчителів хімії більше уваги слід приділити квантово-механічним моделям атомів та молекул тощо. Спільним у всіх випадках є використання таких засобів об'єктно-орієнтованого моделювання, що надають можливість конструювати об'єкти та встановлювати зв'язки між ними, досліджувати явища, процеси, динаміку об'єктів, важкодоступних для спостереження в реальному світі, візуалізуючи рухомі елементи, найбільш важливі з погляду навчальних цілей і завдань характеристики досліджуваних об'єктів і процесів.

*Технологічна компонента* реалізується як відбір засобів, форм та методів професійної підготовки майбутніх учителів природничо-математичних дисциплін. *Підсумкова компонента* для системи професійної підготовки майбутніх учителів природничо-математичних дисциплін засобами комп'ютерного моделювання є діагностичною та вказує на рівень сформованості професійних компетентностей студентів. Вона набуває свого специфічного вираження в *модельному стилі мислення*. Будемо говорити, що студент має модельний стиль мислення, якщо він може:

а) структурувати інформацію про об'єкт у просторі та часі;

б) визначати логічну структуру моделі, створювати графічні образи елементарних явищ, що становлять процес;

в) виявляти основні зміни стану об'єкта або процесу;

г) представляти взаємодію об'єктів і процесів у просторі й часі.

Реалізація розробленої моделі професійної підготовки в навчанні спецкурсу «Об'єктно-орієнтоване моделювання» вимагає побудови відповідної методичної системи навчання. Традиційною моделлю методичної системи навчання є п'ятикомпонентна модель, запропонована А. М. Пишкало [214], в якій використовується системний підхід стосовно компонентів процесу навчання (всі компоненти утворюють єдине ціле із визначеними внутрішніми зв'язками). Згідно з цією моделлю, методична система навчання – це сукупність ієрархічно пов'язаних компонентів: цілей навчання, змісту, методів, засобів і форм організації навчання.

«Можна говорити про те, що поява принципово нових засобів навчання, що якісно змінюють можливості передавання інформації і розширюють можливості організації навчального процесу, приводить до перегляду змісту, форм і методів навчання і може опосередковано позначитися на цілях навчання» [207, с. 7]. Це зауваження майже на 10 років випередило появу комп'ютерів у масовій школі, але з позицій сьогодення можна стверджувати, що в ньому сконцентровані всі основні ідеї створення й обґрунтування методичної системи навчання об'єктно-орієнтованого моделювання: комп'ютер як провідний засіб навчання в значній мірі обумовлює цілі, зміст, методи й форми організації навчання в сучасній вищій школі.

Розробка повноцінної методичної системи навчання спецкурсу «Об'єктно-орієнтоване моделювання» відіграє ключову роль у її функціонуванні як суттєвої складової системи професійної підготовки майбутніх учителів природничо-математичних дисциплін засобами комп'ютерного моделювання. Тому актуальним є аналіз її компонентів, виявлення найбільш слабких місць і проблем, що здатні помітно погіршити її якості і без подолання яких неможливий її подальший розвиток.

Враховуючи особливості природничо-математичних наук (а, відповідно, і спецкурсу «Об'єктно-орієнтоване моделювання»), в структурі методичної системи навчання було виділено технологічну підсистему, що надало можливість максимально відобразити взаємовпливи всіх її компонентів: цільового, змістового та технологічного.

Створюючи методичну систему навчання спецкурсу «Об'єктно-орієнтоване моделювання», ми намагалися:

– урахувати професійну спрямованість підготовки студентів природничо-математичних спеціальностей шляхом диференціації змісту навчання;

– спрогнозувати результати педагогічного впливу, передбачаючи, які компетентності з комп'ютерного моделювання має набути студент і який розвиваючий вплив на нього повинен здійснити зміст навчання;

– забезпечити варіативність форм організації, методів і засобів навчання з опорою на соціально-конструктивістський підхід у навчанні.

Отже, виходячи з визначеної структури, було виділено цільовий, змістовий та технологічний компоненти методичної системи навчання спецкурсу «Об'єктно-орієнтоване моделювання».

*Мета* (*ціль*) навчання – ідеалізоване передбачення кінцевих результатів навчання; те, до чого прагнуть учасники навчального процесу – студенти і викладачі. За традиційним підходом до визначення процесу навчання через знання, уміння та навички, він переслідує три основні групи взаємопов'язаних цілей: 1) освітня – формування у студентів наукових знань, спеціальних і загальнонавчальних умінь і навичок; 2) розвивальна – розвиток мови, мислення, пам'яті, творчих здібностей, рухової та сенсорної систем; 3) виховна – формування світогляду, моралі, естетичної культури тощо.

*Головною метою* нашої методичної системи є формування компетентності в об'єктно-орієнтованому моделюванні через сукупність спеціальних знань, умінь та навичок, що забезпечують студентам можливість застосовувати засоби обчислювальної техніки спочатку в навчальній, а в перспективі й у професійній дослідницькій діяльності. Зміст курсу містить сукупність двох взаємопов'язаних складових: теоретичної та практичної. Теоретична складова спрямована на формування у студентів модельного мислення, навичок об'єктно-орієнтованого аналізу предметної області та проектування об'єктно-орієнтованої моделі, на ознайомлення з методологією об'єктно-орієнтованого моделювання і особливостями її комп'ютерної реалізації мовою об'єктно-орієнтованого програмування. Практичний аспект пов'язується з набуттям умінь добору середовища об'єктно-орієнтованого моделювання, адекватного розв'язуваній задачі та формі організації навчального дослідження (індивідуального чи колективного), навичок роботи в різних середовищах, проведення обчислювального експерименту та прийняття рішення про адекватність моделі об'єктові дослідження. У загальній структурі курсу об'єктно-орієнтованого моделювання обсяг практичних занять має переважати над обсягом теоретичних.

Цілі навчання спецкурсу «Об'єктно-орієнтоване моделювання»:

– формування навичок об'єктно-орієнтованого моделювання як найбільш природного способу дослідження систем різної складності;

– ознайомлення з основними принципами конструювання та дослідження об'єктно-орієнтованих моделей;

– формування навичок індивідуальних та колективних навчальних досліджень засобами технологій соціального конструктивізму.

Обговорюючи питання змістового компоненту навчання, дамо визначення цьому поняттю згідно Закону України «Про вищу освіту» [119].

*Зміст навчання* – структура, зміст і обсяг навчальної інформації, засвоєння якої забезпечує особі можливість здобуття вищої освіти і певної кваліфікації.

Зміст навчання на рівні певної навчальної дисципліни – обумовлена цілями та потребами суспільства система знань, умінь і навичок, професійних, світоглядних і громадянських якостей, що має бути сформована в процесі навчання з урахуванням перспектив розвитку суспільства, науки, техніки, технологій, культури та мистецтва.

Зміст освіти є одним із факторів економічного і соціального прогресу суспільства і повинен бути орієнтований: на забезпечення самовизначення особистості, створення умов для її самореалізації; розвиток суспільства; посилення та вдосконалення засад правової держави.

Визначення змісту навчання об'єктно-орієнтованого моделювання необхідно здійснювати з урахуванням принципів, спільних для будь-яких навчальних дисциплін, так і властивих насамперед для інформатичних [152, с. 70]:

1. *Принцип відповідності навчальним цілям.* Цілі навчання об'єктно-орієнтованого моделювання визначаються, виходячи із загальних цілей освіти – формування компетентності особистості, причому не лише в моделюванні, а й в інших спеціальних професійних, загальнопрофесійних та ключових компетентностей.

2. *Принцип науковості.* Вимога науковості передбачає взаємозв'язок теорії, розробки, аналізу й оцінювання ефективності, реалізації та застосування моделей. Зміст спецкурсу повинен складатися з тих розділів і тем, які важливі для практики моделювання незалежно від обраного підходу до навчання самого об'єктно-орієнтованого моделювання.

3. *Принцип фундаментальності*, реалізація якого передбачає: опанування методологічно важливих та інваріантних знань, що мають довгий термін життя, необхідних для професійної діяльності вчителя природничо-математичних дисциплін; розвиток творчої і пізнавальної активності та самостійності студентів; модернізація систем професійної підготовки з урахуванням перспектив «економіки знань» та суспільства сталого розвитку; системність засвоєння природничо-математичних дисциплін на основі глибокого розуміння сучасного стану та існуючих проблем відповідних наук [260, с. 86].

4. *Принцип відкритості.* Цей принцип передбачає можливість корекції змісту спецкурсу залежно від освітнього напряму підготовки, без порушення цілісності фундаментального ядра дисципліни.

5. *Принцип сучасності.* Швидкий розвиток інформаційно-комунікаційних технологій вимагає регулярного перегляду навчальної програми дисципліни з метою модернізації застарілих компонентів, що для майбутніх учителів є особливо актуальним з огляду на особливості їхньої професійної діяльності в умовах широкого впровадження засобів ІКТ у навчальний процес.

6. *Принцип перспективності.* Цей принцип передбачає формування в студентів готовності до подальшого навчання протягом усього життя, що надасть їм можливість бути здатними до вирішення професійних проблем у майбутньому.

7. *Принцип вирівнювання знань.* Зміст спецкурсу «Об’єктно-орієнтоване моделювання» повинен включати пропедевтичний модуль, вивчення якого забезпечить початкове опанування середовища та мови об’єктно-орієнтованого моделювання.

Окрім цього, добір змісту навчального матеріалу має здійснюватися з урахуванням основних дидактичних принципів навчання: посильної складності, системності, послідовності і систематичності навчання, наочності змісту і діяльності, активності і самостійності, свідомості, індивідуалізації і колективності навчання тощо.

*Змістова частина* спецкурсу (навчальний матеріал) включає широкий спектр задач із різних предметних галузей і передбачає опанування технології об’єктно-орієнтованого моделювання в соціально-конструктивістських середовищах. Як зазначалось у розділі 1, основний зміст навчання об’єктно-орієнтованого моделювання складає конструювання та дослідження динамічних та імітаційних моделей. З урахування принципу вирівнювання знань, зміст курсу складають 3 модулі (додаток Г).

У *першому модулі* «Вступ до об’єктно-орієнтованого моделювання» розглядаються базові поняття та уявлення спецкурсу (поняття про моделювання, види моделей, об’єктно-орієнтоване моделювання; об’єктно-орієнтоване програмування та об’єктно-орієнтовані мови; абстракція, інкапсуляція, спадкування, поліморфізм як основа об’єктно-орієнтованої методології; етапи об’єктно-орієнтованого моделювання: об’єктно-орієнтований аналіз, проектування, обчислювальний експеримент та аналіз його результатів) і виконується огляд середовищ об’єктно-орієнтованого моделювання (зокрема, виділяються універсальні середовища моделювання, середовища для конструювання динамічних моделей та середовища для конструювання імітаційних моделей). На підставі аналізу придатності середовищ для дослідження різних класів моделей пропонується в навчанні студентів природничо-математичних спеціальностей:

а) при розгляді динамічних моделей послугуватися середовищами об’єктно-орієнтованого моделювання VPython та Squeak як основними, а Sage, Alice та NetLogo – як додатковими;

б) при розгляді імітаційних моделей скористатися середовищами об'єктно-орієнтованого моделювання Alice та NetLogo як основними, а Sage, VPython та Squeak – як додатковими.

У процес навчання спецкурсу «Об'єктно-орієнтоване моделювання» студентів спеціальностей «Математика», «Інформатика» середовище Sage із додаткового переноситься в основні. Це не вимагає додаткових витрат часу через спільність мови ООП в середовищах VPython та Sage.

*Другий модуль* «Об'єктно-орієнтовані динамічні моделі» присвячений розгляду динамічних моделей математичної екології (динаміка одновидової популяції, модель «Хижак-жертва», вікові моделі Леслі (Patrick H. Leslie)), класичної механіки (динаміка коливних систем, рух тіл в полі сили тяжіння, моделювання аеродинамічних об'єктів та явищ), явищ молекулярної фізики й фізики твердого тіла (атомна та молекулярна динаміка) та електродинаміки (рух заряду в електричному та магнітному полях).

У *третій модуль* «Об'єктно-орієнтовані імітаційні моделі» включені моделі кліткових автоматів (модель поширення чуток, модель «Хижак-жертва», модель поширення пожежі, гра «Життя»), стохастичні моделі (модель броунівського руху, модель відмов обладнання, модель росту кристалу), моделі фрактальних об'єктів та процесів (моделі регулярних фракталів, задача перколяції, моделі електролізу, модель утворення берегової лінії).

Підсумковий контроль знань за спецкурсом – екзамен у формі захисту індивідуальних та колективних дослідницьких проектів.

### 2.3 Соціально-конструктивістські форми організації та методи навчання

Як зазначалося, окрім цільового та змістового компонентів, методична система навчання містить ще технологічний, до складу якого входять: форми організації, методи та засоби навчання. Враховуючи, що засоби навчання об'єктно-орієнтованого моделювання реалізують соціально-конструктивістський підхід у навчанні, вони обговорюються окремо в наступному підрозділі.

*Форма організації навчання* – це: а) обмежена в просторі та часі взаємозумовлена діяльність педагога й учня, викладача й студента [148, с. 247]; б) цілеспрямована, чітко організована, змістовно насичена й методично забезпечена система пізнавального та виховного спілкування, взаємодії, співпраці викладачів та студентів [144, с. 316]. Виходячи із проаналізованих у першому розділі дисертації підходів до організації соціально-конструктивістського навчання, пропонуємо наступне уточнення: *соціально-конструктивістська форма організації навчання* – це цілеспрямована, чітко організована та технологічно забезпечена система індивідуальної, парної та колективної роботи учасників навчальної спільноти (з викладачів та студентів).

В. Г. Крисько розподіляє форми організації навчання на *навчально-планові* (урок, лекція, семінар, домашня робота, екзамен та ін.), *позапланові* (бригадно-лабораторні заняття, консультації, конференції, гуртки, екскурсії, заняття за поглибленими та допоміжними програмами) і *допоміжні* (групові та індивідуальні заняття, групи вирівнювання, репетиторство) [144]. При організації соціально-конструктивістського навчання провідними навчально-плановими формами стають лекція (в т. ч. лекція-семінар) та екзамен (у формі захисту індивідуальних та колективних дослідницьких проектів), провідними позаплановими формами стають робота над колективним дослідницьким проектом, очні та розподілені консультації, Інтернет-конференції, заняття за поглибленою програмою (при підготовці конкурсних робіт).

Взаємодія учасників навчального процесу є основою поділу форм організації навчання на три групи: 1) індивідуальні заняття, в тому числі – самонавчання; 2) колективно-групові заняття; 3) індивідуально-колективні заняття.

Найпоширенішою в навчанні майбутніх учителів природничо-математичних дисциплін є *лекційно-лабораторна* форма, характерними ознаками якої є: постійний склад навчальних груп; строге визначення змісту навчання; певний розклад навчальних занять; поєднання індивідуальної та колективної форм роботи; провідна роль викладача; систематична перевірка й оцінювання знань [152]. При використанні соціально-конструктивістського підходу в навчанні роль викладача змінюється з власне викладання на фасилітацію.

Як зазначає Н. М. Бібік [106, с. 953-954], фасилітація (англ. faciletate – полегшувати, сприяти) – стиль педагогічного спілкування, який передбачає полегшення взаємодії під час спільної діяльності; ненав’язлива допомога групі чи окремій людині в пошуку способів виявлення і розв’язання проблем, налагодженні комунікативної взаємодії між суб’єктами діяльності. Поняття «фасилітатор» введене К. Р. Роджерсом (Carl Ransom Rogers), який вчителя називає фасилітатором спілкування, вважає, що він має допомогти учневі вчитися, увиразнити себе як особистість, зацікавити, підтримати під час пошуку знань [56].

За соціально-конструктивістського навчання викладач-фасилітатор, крім налагодження зв’язку студентів між собою та студентів з викладачем, виконує також функцію управління процесом навчання з активним застосуванням ІКТ. Перехід до ролі фасилітатора вимагає від викладачів вищої школи набуття нових та посилення сформованих навичок: 1) письмового та аудіального спілкування; 2) тайм-менеджменту в синхронному та асинхронному режимі спілкування; 3) гнучкої організації індивідуального та інклюзивного навчання засобами ІКТ.

На рис. 2.2 показано, як на прикладі соціально-конструктивістського навчання, зокрема спецкурсу «Об'єктно-орієнтоване моделювання», із зростанням частки самостійної роботи відбувається зміна ролі викладача, студента та технології навчання.

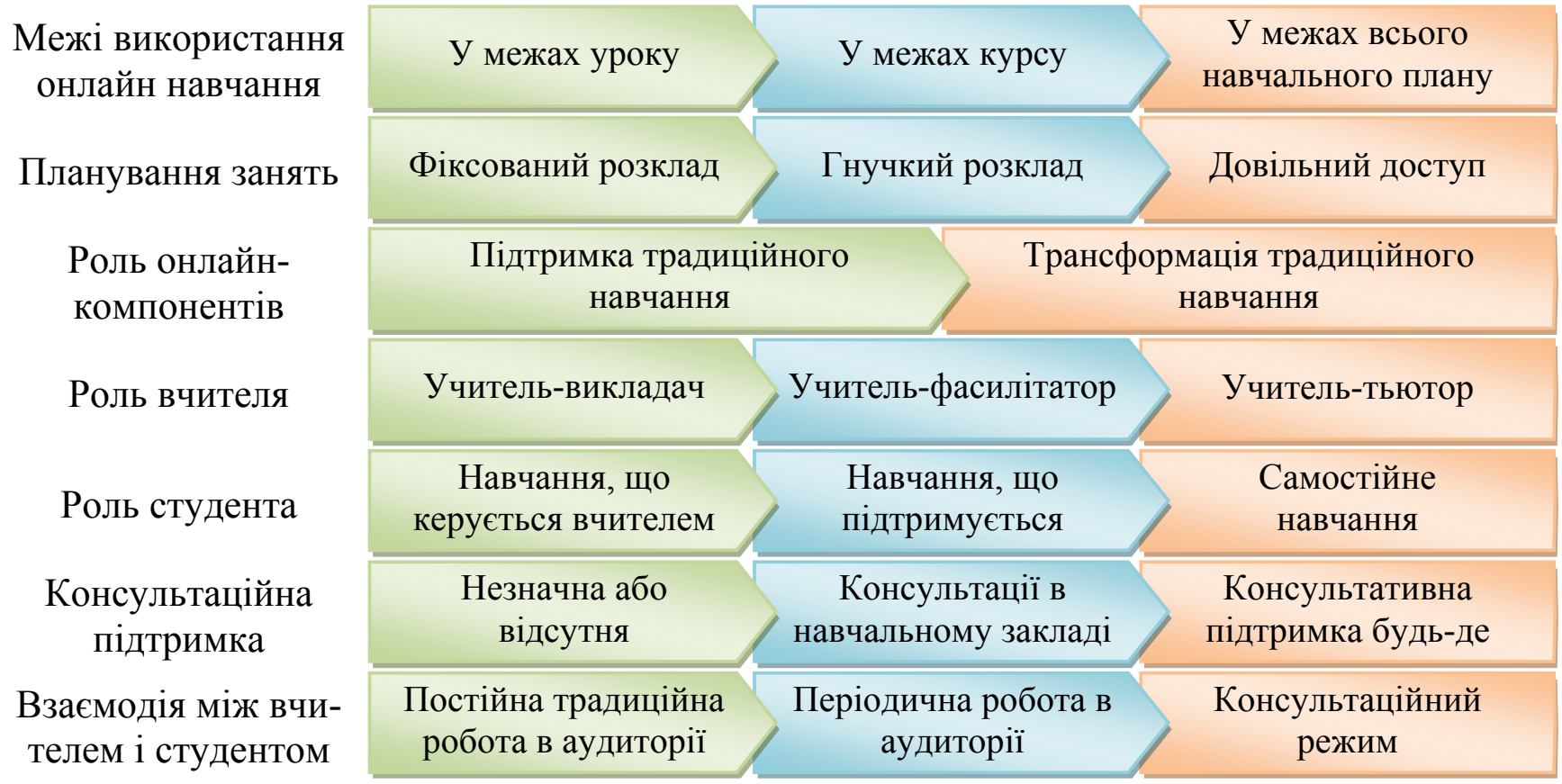

Рис. 2.2. Зміни, що відбуваються в процесі переходу до соціально-конструктивістського навчання (за Дж. Уотсоном (John Watson) [63, с. 12])

Ураховуючи характеристики особливостей комунікативної взаємодії як між викладачем і студентами, так і між самими студентами, серед загальних форм організації навчання виділяють фронтальні, колективні, групові, парні, індивідуальні, а також зі змінним складом студентів.

*Фронтальне* навчання застосовується при роботі всіх студентів над одним і тим самим змістом або при засвоєнні одного й того самого виду діяльності та передбачає роботу викладача з усією групою (потоком, підгрупою) в єдиному темпі, із спільними завданнями. Ця форма організації навчання широко використовується на лабораторних заняттях, на початку вивчення предмету (теми) при реалізації словесного, наочного й практичного методів, а також у процесі контролю знань.

Слід зазначити, що в міру засвоєння загальних способів навчальних дій робота студентів стає все більш індивідуальною та незалежною від зовнішньої допомоги та вказівок викладача, що є одним із чинників зміни його ролі.

Так, у процесі навчання спецкурсу «Об'єктно-орієнтоване моделювання» майбутніх учителів природничо-математичних дисциплін фронтальне навчання є основною формою навчання першого модулю, коли виконуються фронтальні лабораторні заняття з опанування основ роботи в середовищах моделювання Alice (мова Java), Squeak (мова Smalltalk), VPython (мова Python)

*Колективна* форма навчання відрізняється від фронтальної тим, що студентська група розглядаються як цілісний колектив зі своїми лідерами й особливостями соціальної взаємодії. При реалізації взаємодії в мережному середовищі, зокрема, на Інтернет-форумах, студентських лідерів доцільно призначати модераторами відповідних розділів сайту.

У *групових* формах навчання студенти працюють у групах, створюваних на різній основі та на різний термін. Це досить типова форма організації соціально-конструктивістського навчання при *роботі над проектами*, що відображає реальний поділ праці в колективі фахівців, які працюють над одним завданням. Ураховуючи, що однією із складових перевірки ефективності розроблених педагогічних умов є захист результатів виконання проекту з об'єктно-орієнтованого моделювання, виконаного в соціально-конструктивістському середовищі, дана форма стає однією з провідних.

При навчанні в складі групи в ній виникає інтенсивний обмін різноманітними повідомленнями, тому групові форми можуть бути ефективними в групах з учасниками різного рівня підготовки й мотивації. Це зумовлює доцільність при розподілі студентів за різними дослідницькими проектами до складу груп включати студентів з різними рівнями навчальних досягнень з природничо-математичних дисциплін.

У *парному* навчанні основна взаємодія відбувається між двома студентами, котрі можуть обговорювати завдання, здійснювати взаємонавчання або взаємоконтроль. Парні форми організації навчання, так само, як і групові, відносяться до *гнучких форм*, конкретизацією яких в процесі навчання об'єктно-орієнтованого моделювання є групове та парне програмування [260, с. 201].

*Індивідуальна* форма навчання передбачає взаємодію викладача з одним студентом. Особливого поширення ця форма набуває при використанні мережних засобів ІКТ (текстовий, голосовий та відеозв'язок).

В умовах комп'ютерної аудиторії управляти індивідуальною діяльністю студентів досить складно: ситуація за кожним комп'ютером практично унікальна. Вихід полягає в тому, щоб залучити до навчання сильних студентів (у тому числі в рамках виконання індивідуальних та колективних дослідницьких проектів) та, за висловом А. П. Єршова, «автоформалізувати власний педагогічний досвід». Сучасна реалізація цієї форми знайшла своє відображення в методі *учіння через навчання* [39].

В навчанні інформатики можна говорити про індивідуальне навчання при контакті з колективним знанням, що реалізується у формі «студент і комп'ютер». Працюючи один на один з комп'ютером (точніше, з навчальною програмою), студент у своєму темпі здобуває знання, сам вибирає індивідуальний маршрут вивчення навчального матеріалу в рамках заданої теми. За О. І. Бочкіним, «радикальна відмінність цієї форми від класичної самостійної форми роботи в тім, що програма є зручним для використання «зліпком» інтелекту й досвіду її автора» [78].

У програмне забезпечення спецкурсу «Об'єктно-орієнтоване моделювання» засоби зв'язку та організації колективної роботи є вбудованими, тому індивідуальне навчання об'єктно-орієнтованого моделювання можливе лише за умови ігнорування засобів організації спільної роботи у відповідних середовищах об'єктно-орієнтованого моделювання (спільне редагування, обмін об'єктами та моделями, оприлюднення проектів у мережі). Тому індивідуальне навчання об'єктно-орієнтованого моделювання не є провідною формою організації соціально-конструктивістського навчання.

Застосування мережних ІКТ сприяє інтеграції кращих сторін індивідуальної та фронтальної форм навчання – так, за рахунок розміщення середовищ моделювання в Інтернет (Sage), тиражування середовищ моделювання через Інтернет, розміщення навчальних курсів у мережі, використання комунікаційних ресурсів Інтернет зберігається й перевага фронтальних форм: можливість вчитися у кращих викладачів, використовувати адаптивні джерела навчальних матеріалів. Це допомагає реалізувати одне з найважливіших завдань викладача вищої школи – розвиток у студентів самостійної пізнавальної активності та навичок спільної навчальної діяльності.

Зовнішні форми організації навчання майбутніх учителів природничо-математичних дисциплін позначають певний вид заняття: лекція, семінар, практичне заняття, лабораторне заняття, практикум, факультативне заняття, екзамен, предметні гуртки, студентські наукові товариства й т. д.

*Лекція* – усне систематичне та послідовне подання матеріалу з певної проблеми, методу, теми, питання й т. д. У вищій школі ця форма є основною в процесі навчання і має два змісти: це і форма, і метод. Лекція завжди фронтальна. При застосуванні мережних ІКТ (зокрема, засобів для проведення вебінарів, Інтернет-трансляцій тощо) лекція може відбуватися не тільки за традиційною формою; використання відеозапису лекції та її конспекту (наприклад, у вигляді гіпертексту або презентації) підсилює самоуправління пізнавальною діяльністю. Виходячи із поділу аудиторних годин на лекційні та лабораторні заняття у відношенні 1:2, у процесі навчання об’єктно-орієнтованого моделювання перевага віддається узагальнюючим лекціям та лекціям-семінарам.

*Додаткові* (*консультаційні*) форми організації навчання розраховані на окремих студентів або групу з метою задоволення підвищеного інтересу до навчального предмета, вироблення вмінь і навичок, заповнення пробілів у знаннях. На консультаціях можуть бути роз’яснені окремі питання, організоване повторне пояснення теми і т. п. Так, на консультаціях з об’єктно-орієнтованого моделювання можуть бути виконані об’єктно-орієнтований аналіз предметної області, об’єктно-орієнтоване проектування та надані рекомендації з реалізації моделі мовою ООП.

Для задоволення пізнавального інтересу та поглибленого вивчення об’єктно-орієнтованого моделювання з окремими студентами передбачені заняття, на яких досліджуються моделі підвищеної складності, обговорюються наукові проблеми, що виходять за рамки програми, даються рекомендації із самостійного опанування проблем, що цікавлять студентів. Результати цієї роботи можуть бути використані в процесі виконання конкурсних, курсових та кваліфікаційних робіт.

Розрізняють поточні, тематичні й узагальнюючі (наприклад, при підготовці до екзаменів або заліків) консультації. Консультації найчастіше є груповими (від 5 студентів), що однак не виключає й індивідуальних консультацій.

Залучення мережних ІКТ до навчального процесу з об’єктно-орієнтованого моделювання надає можливість організації *дистанційних консультацій* у формі електронного листування, чату чи форуму, аудіо-, відео-конференції тощо. Їх запровадження дозволяє: будь-якому студенту отримувати консультацію в зручний для нього час; надати процесу консультування публічності через розміщення питань та відповідей на спеціалізованому форумі; як наслідок, – підвищити ефективність консультування шляхом уникнення однотипних питань; накопичувати банк типових питань та відповідей на них, доступних різним поколінням студентів; оперативно доводити до відома студентів завдання контрольних та практичних робіт, робити оголошення тощо.

Додатковими перевагами даної форми консультування є її прозорість, сучасність та відповідність вимогам Болонського процесу.

Для впровадження дистанційної форми консультування необхідними є наступні першочергові заходи: 1) створення та розміщення в мережі Інтернет консультаційного пункту у вигляді форуму з розділами, що ведуться окремими викладачами; 2) розробка та затвердження положення про дистанційне консультування студентів; 3) забезпечення студентів та викладачів засобами мережі Інтернет для проведення дистанційного консультування.

Навіть у найпершій програмі курсу «Основи інформатики та обчислювальної техніки» [212] передбачалися три основних види організаційного використання кабінету обчислювальної техніки на уроках – демонстрація, фронтальна лабораторна робота й практикум. Ці ж форми застосовуються й у вищій школі.

*Демонстрація.* Використовуючи демонстраційний екран (мультимедійні дошку, проектор тощо), викладач показує різні навчальні елементи змісту курсу (елементи інтерфейсу, фрагменти програм, схеми, тексти й т. п.). При цьому викладач сам працює на комп’ютері, а студенти спостерігають за його діями або відтворюють їх. У деяких випадках викладач пересилає демонстрації на студентські комп’ютери (мобільні пристрої), а студенти працюють з ними самостійно. Зростання ролі й дидактичних можливостей використання комп’ютерних демонстрацій пояснюється покращенням мультимедійних характеристик комп’ютерів. Основна дидактична функція демонстрації – унаочнення нового навчального матеріалу.

У процесі навчання об’єктно-орієнтованого моделювання з’являється нова форма демонстрації: *демонстрація, керована викладачем*. За такої форми демонстрації викладач проводить студентів через усі етапи конструювання моделі: вибір об’єктів, визначення реакції об’єктів на повідомлення, встановлення зв’язку між різними об’єктами та вибір початкового повідомлення, з якого розпочинається функціонування моделі. При цьому на будь-якому етапі конструювання моделі може бути виконана перевірка її на адекватність предметній області, а результати виконання розміщені в мережі. Запис дій, що їх виконує викладач під час такої демонстрації, зокрема, захоплення екранних кадрів, запис голосових та текстових коментарів, може бути використаний для створення навчальних відеоматеріалів (рис. 2.3).

*Лабораторна робота* (фронтальна) є основною формою роботи в комп’ютерному класі. Діяльність студентів може бути як синхронна, так і асинхронна. Нерідко відбувається швидке «розтікання» фронтальної діяльності навіть при спільному вихідному завданні. Роль викладача під час фронтальної лабораторної роботи – фасилітативне спостереження за роботою студентів (у тому числі через мережу) та надання їм оперативної допомоги. У процесі навчання об’єктно-орієнтованого моделювання фронтальна лабораторна робота є провідною формою організації навчання за першим модулем.

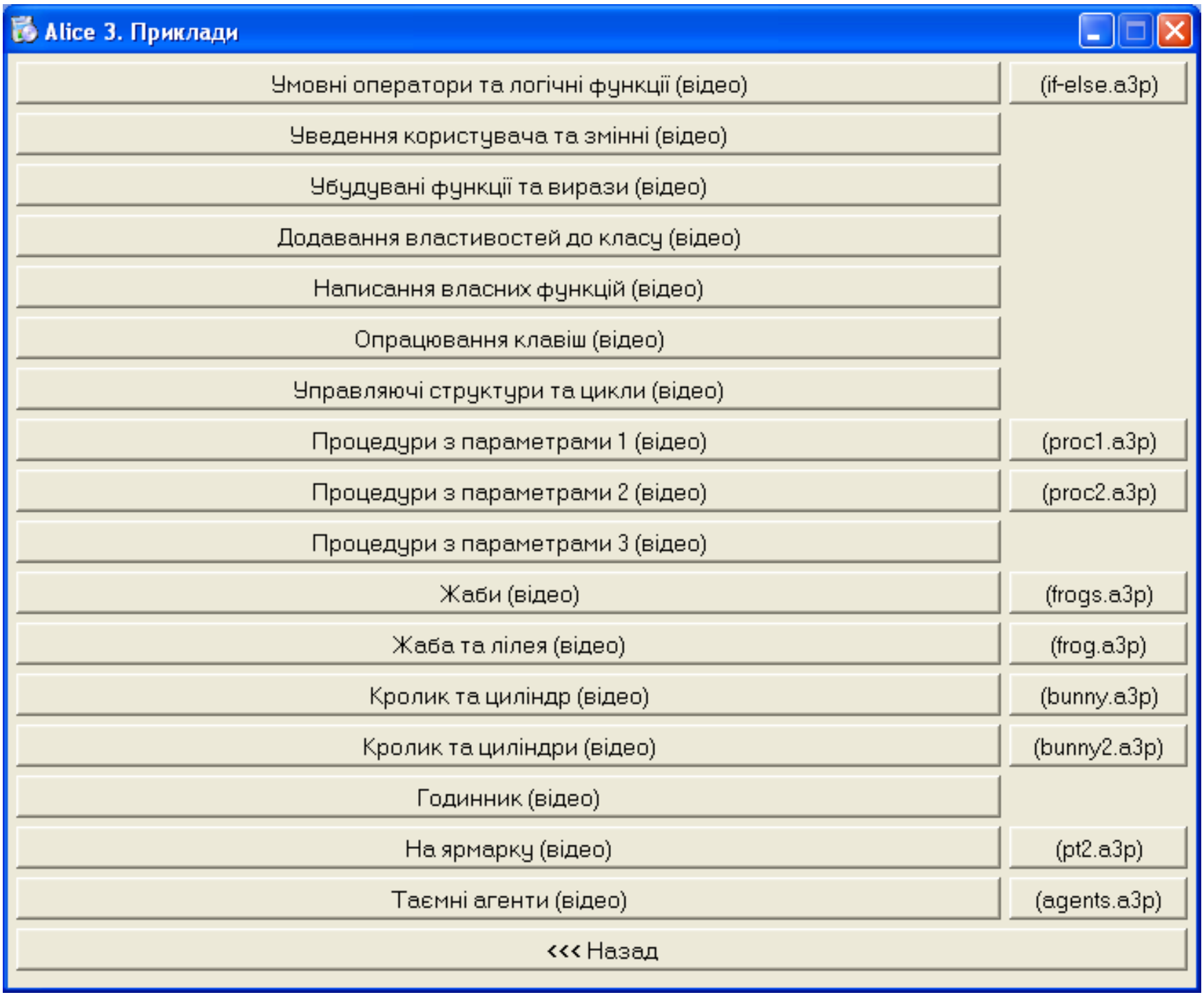

Рис. 2.3. Приклади відео-демонстрацій до навчального посібника [348]

*Індивідуальний практикум* – більш високорівнева форма роботи в порівнянні із фронтальними лабораторними роботами, що характеризується різнотипністю завдань як за рівнем складності, так і за рівнем самостійності; більшою опорою на підручники, довідковий матеріал, ресурси Інтернет тощо. Лабораторні заняття за другим та третім модулем проводяться у формі індивідуального практикуму. Студенти одержують індивідуальні завдання на початку спецкурсу. Як правило, такі завдання видаються для відпрацьовування навичок побудови динамічних та імітаційних моделей.

*Лабораторно-обчислювальний практикум* (*за типом «занурення»*) – форма, за якою передбачається інтенсивна концентрована робота студентів у комп’ютерному класі з відривом від інших занять протягом 1-2 тижнів. У ході занурення може бути опрацьований матеріал з окремого курсу або сукупності тем.

*Семінари* та *практичні заняття* є перехідною формою від фронтальної до індивідуальної роботи. В навчанні інформатичних дисциплін необхідно виробляти ряд немашинних та домашинних навичок і вмінь (наприклад, розв’язування завдань з теоретичних основ інформатики, розробка та обговорення алгоритму, моделі тощо). Практичне заняття – найбільш адекватна форма роботи для колективного осмислення того, що треба зробити або вже зроблено на комп’ютері, і чому саме такі результати отримані.

Важливим інтелектуальним умінням є здатність до розгорнутого коментованого прогнозу (передбачення) результатів, отриманих за допомогою комп’ютера на основі нагромадженого досвіду роботи з ним. Для його формування доцільно застосовувати семінарські заняття.

Студентам слід знати, що саме зараховується як результат роботи на семінарі, адже при вивченні суспільно-гуманітарних дисциплін це є лише виступи, доповнення та участь у дискусії. На семінарах з комп’ютерного моделювання можливі контрольовані результати: текст алгоритму роботи моделі, готовий для введення; таблиця виконання алгоритму, складена без застосування комп’ютера; проект роботи із програмною реалізацією моделі; відповіді на питання інструкції; інструкція до власної або чужої програми; коментарі до своєї або чужої програми; опис очікуваних результатів роботи з програмною реалізацією моделі. Незважаючи на те, що навчальною програмою зі спецкурсу «Об’єктно-орієнтоване моделювання» практичні та семінарські заняття не передбачені, їх елементи використовуються на лекціях-семінарах, демонстраціях та вебінарах (при виконанні індивідуальних та колективних дослідницьких проектів).

*Проектна форма навчання.* Основою проектної форми навчання є соціально-конструктивістський підхід у навчанні. Ознаками проектної форми навчання є: 1) наявність організаційного етапу підготовки до проекту – самостійний вибір і розробка варіанту виконання, вибір програмних і технічних засобів, вибір джерел потрібних відомостей; 2) вибір із числа учасників проекту лідера (організатора, координатора), розподіл ролей; 3) наявність етапу самоекспертизи й самооцінки (рефлексії), захисту результату та оцінювання рівня виконання; 4) кожна група може займатися розробкою окремого проекту або брати участь у втіленні колективного проекту. Реалізація проектної форми навчання об’єктно-орієнтованого моделювання можлива, зокрема, на лабораторно-обчислювальному практикумі за типом «занурення».

*Методи навчання* – це впорядковані способи взаємопов'язаної діяльності викладача та студента (їх взаємосприяння), спрямовані на досягнення цілей навчання та ефективне розв'язання навчально-виховних завдань [148, с. 87].

У дидактиці вищої школи існують різні трактування цього поняття. В. Ю. Биков, Ю. І. Машбиць, М. Л. Смульсон, М. І. Жалдак та інші автори посібника [186] підходять до навчання як до управління навчальною діяльністю і розглядають метод навчання як спосіб управління та істотну детермінанту навчальної діяльності, що реалізується в системі навчальних впливів, у способі включення студентів у процес відтворення педагогом фрагменту навчальної діяльності, в «полі самостійності» студентів (характеризується відхиленням від нормативного способу розв'язання навчальних задач, при яких учням не надається допомога), у формах організації навчання і у модальності обміну інформацією між студентом і викладачем. А. М. Алексюк також визначає метод навчання як спосіб організації і управління з боку викладача пізнавальною діяльністю студентів [68, с. 46]. Р. А. Нізамов зазначає, що навчання не можна редукувати до управління. Воно включає в себе такі функціональні види діяльності: а) викладача – організацію діяльності студентів із засвоєння знань, формування навичок і вмінь; виклад сутності наукових знань, складних теоретичних положень; контроль знань і вмінь; стимулювання пізнавальної діяльності студентів; б) студента – засвоєння знань; формування вмінь; добування нових знань [177, с. 124]. З. І. Слєпкань під методом навчання розуміє способи роботи викладача і студентів, за допомогою яких досягається оволодіння знаннями, навичками й уміннями, формується світогляд студентів, розвиваються їхні здібності [274, с. 105].

Об'єктно-орієнтований підхід до методів навчання передбачає визначення того, що і як саме студенти мають робити з навчальним матеріалом, які властивості і зв'язки між об'єктами необхідно розкривати. Метод є центральною ланкою детермінації процесу навчання зовнішніми обставинами.

Поряд з поняттям «метод навчання» в теорії та педагогічній практиці використовуються поняття «прийом навчання», «методичний прийом». Прийнято вважати, що метод як спосіб діяльності складається із прийомів або окремих дій, спрямованих на розв'язування педагогічних завдань.

У методах навчання можна виділити змістову і формальну сторони. Змістова сторона включає такі компоненти: 1) зміст, різні моделі, аналогії, алгоритми, використання яких дає змогу засвоїти сутність навчальних предметів; 2) розумові, передусім мислительні, дії, потрібні для засвоєння змісту навчальних предметів і додаткового змісту (загальнологічні дії, а також дії, через які розкриваються принципи побудови навчального матеріалу тощо); 3) співвідношення між цілями навчання, з одного боку, та прямими і непрямими його продуктами, з іншого. Формальна сторона методів навчання характеризується співвідношенням активності викладача та студентів, характером поєднання колективних та індивідуальних форм навчальної роботи, співвідношенням зорових та слухових форм подання навчального матеріалу, кількістю і складністю завдань, що стоять перед студентами, мірою допомоги, що надається їм тощо. При цьому діяльність викладача, з одного боку, обумовлена метою навчання, закономірностями засвоєння й характером навчальної діяльності студентів, а з іншого боку – вона сама обумовлює діяльність студентів, реалізацію закономірностей засвоєння й розвитку.

Існують різні підходи до класифікації методів навчання.

Методи навчання … «є категорією історичною, … змінюються зі зміною цілей та змісту навчання» [151, с. 86]). Якщо за критерій класифікації обрати джерело знань, то розрізняють такі методи: словесні (розповідь, пояснення, бесіда, лекція, інструктаж), наочні (ілюстрація, демонстрація, спостереження), практичні (лабораторна робота, практична робота, вправи, робота на виробництві).

Оскільки *загальні методи навчання* численні і мають багато характеристик, їх можна класифікувати за кількома напрямами:

1. *За характером навчання* (ступенем самостійності та творчості) – система загально-дидактичних методів навчання І. Я. Лернера та М. М. Скаткіна [104]: пояснювально-ілюстративні (студенти сприймають знання в «готовому» вигляді), репродуктивні методи (застосування вивченого матеріалу на основі зразка або правила), метод проблемного викладання (педагог ставить проблему, формулює пізнавальне завдання, а потім, розкриваючи систему доведень, порівнюючи погляди, різноманітні підходи, показує способи розв’язання поставленого завдання, при цьому студенти стають свідками і співучасниками наукового пошуку), частково-пошуковий (організація активного пошуку розв’язання поставлених або самостійно сформульованих пізнавальних завдань, над якими студенти працюють самостійно або під керівництвом викладача на основі евристичних програм та вказівок; такий метод – надійний спосіб активізації мислення, пробудження інтересу до пізнання), дослідницький (після аналізу матеріалу, постановки проблеми і визначення завдань студенти самостійно опрацьовують наукові джерела, проводять спостереження і виміри, виконують інші дії пошукового характеру; найбільш повно ініціатива, самостійність та творчий пошук виявляються в дослідницькій діяльності).

2. *За основними компонентами діяльності викладача* – система методів Ю. К. Бабанського [72], що включає три великі групи методів навчання:

1) методи організації та здійснення навчальної діяльності (словесні, наочні, практичні репродуктивні й проблемні, індуктивні й дедуктивні, самостійної роботи та роботи під керівництвом викладача);

2) методи стимулювання й мотивації навчання (методи формування інтересу: пізнавальні ігри, аналіз життєвих ситуацій, створення ситуацій успіху; методи формування обов'язковості й відповідальності в навчанні: роз'яснення суспільної й особистісної значимості навчання, пред'явлення педагогічних вимог);

3) методи контролю й самоконтролю (усний і письмовий контроль, лабораторні й практичні роботи, машинний і безмашинний програмований контроль, фронтальний і диференційований, поточний і підсумковий).

Частково-дидактичні методи навчання можна класифікувати:

– за особливостями подання та характером сприймання матеріалу – система традиційних методів: словесні методи (розповідь, бесіда, лекція та ін.); наочні (показ, демонстрація та ін.); практичні (лабораторні роботи, твори та ін.);

– за ступенем взаємодії викладача та студентів: подання матеріалу, бесіда, самостійна робота;

– в залежності від конкретних дидактичних завдань: підготовка до сприймання, пояснення, закріплення матеріалу й т. д.;

– за принципом розчленовування або з'єднання знань: аналітичний, синтетичний, порівняльний, узагальнюючий, класифікаційний;

– за характером руху думки від незнання до знання: індуктивний, дедуктивний.

М. П. Лапчик [152], О. І. Бочкін [78] та Н. В. Морзе [171], крім загально-дидактичних та частково-дидактичних, виділяють ще *спеціальні методи навчання* інформатичних дисциплін, до яких відносять *метод доцільно дібраних задач* та *метод демонстраційних прикладів*.

До спеціальних методів навчання об'єктно-орієнтованого моделювання відносяться обчислювальний експеримент та програмування. Це пов'язано з наступними обставинами:

1) обчислювальний експеримент є методологією інформатики як науки, тому його можна віднести до методів її навчання [151, с. 91];

2) обчислювальний експеримент є складовою методу моделювання, тому його можна віднести до методів дослідження природничо-математичних наук;

3) цілі навчання у вищій школі включають необхідність засвоєння як певної сукупності наукових фактів, так і методів отримання цих фактів, що використовуються в самій науці, а програмування відображає метод пізнання, що застосовується в інформатиці. При цьому під терміном «програмування» розуміють діяльність, яка у вузькому сенсі зводиться до простого кодування відомого алгоритму, а в широкому – співпадає з методологією інформатики, тобто є тотожною обчислювальному експерименту [151, с. 92].

У визначенні об’єктно-орієнтованого моделювання термін «програмування» присутній в характеристиці середовища моделювання, мовою якого має бути мова об’єктно-орієнтованого програмування.

Частина назв форм організації навчання об’єктно-орієнтованого моделювання виступають і як назви методів навчання: це, насамперед, лекція, метод проектів та лабораторно-обчислювальний практикум (за методом «занурення»).

*Лекція* як метод навчання відноситься до словесних методів. У професійній підготовці майбутніх учителів природничо-математичних дисциплін лекції звичайно використовуються при поданні нового досить об’ємного і складного матеріалу з використанням прийомів активізації навчально-пізнавальної діяльності студентів, у тому числі через привчання їх до конспектування лекційного матеріалу. В ході лекції студентом сприймається навчальний матеріал, потім у свідомості відбувається його аналіз, після чого цей матеріал знову виражається словами (реконструюється у вигляді конспекту лекції). Конспект є вже фіксацією продуктів мислення студента, що вимагає від нього значної розумової напруги, тому уміння слухати й конспектувати лекцію виробляється поступово.

Якщо рівень підготовки студентів групи досить високий, то можна провести *лекцію-прес-конференцію*. Перед лекцією студенти опрацьовують навчальний матеріал, використовуючи дидактичні матеріали, на початку лекції впродовж декількох хвилин студенти передають запитання в письмовій формі викладачу. Викладач, переглянувши і відсортувавши запитання, починає лекцію, розглядаючи лише ті питання, що викликали труднощі у студентів. Це дає змогу інтенсифікувати процес навчання, залучити студентів до активної діяльності щодо оволодіння змістом курсу, процесом здобуття нових знань, ознайомити з новою для студентів формою організації навчальних занять [218, с. 145].

Одним із основних сучасних інноваційних методів активного навчання є *метод проектів.* У професійній підготовці майбутніх учителів цей метод широко впроваджується за всіма напрямами підготовки (теоретичні основи впровадження методу проектів розроблені в працях Є. С. Полат [180]). Типи проектів: індивідуальні й групові, локальні й телекомунікаційні. Згідно з Є. С. Полат, навчальний телекомунікаційний проект – це соціально-конструктивістський метод навчання, який передбачає спільну навчально-розвивальну діяльність учасників, які можуть бути територіально віддаленими, для досягнення значущої для них мети (результату) узгодженими методами, що вимагають застосування засобів комп'ютерних телекомунікацій. Характерними ознаками навчальних телекомунікаційних проектів є самостійна дослідницька діяльність їх учасників, пов'язана із розв'язанням навчальної проблеми, що має на меті отримання практичного результату та спирається на більшості або на кожному своєму етапі на використання засобів комп'ютерних телекомунікацій [180, с. 168]. У процесі навчання спецкурсу «Об'єктно-орієнтоване моделювання» метод проектів застосовується на етапі підготовки до заключної атестації з навчальної дисципліни.

*Занурення* відноситься до методів концентрованого навчання природничо-математичних дисциплін. Розглядаючи педагогічну технологію концентрованого навчання, А. О. Остапенко виділяє такі види занурення [187]:

1) «занурення» як модель інтенсивного навчання із застосуванням сугестивного впливу;

2) занурення як модель тривалого заняття одним або кількома предметами: занурення в предмет (однопредметне занурення); двопредметна система занурення; тематичне занурення (занурення в образ); евристичне (метапредметне) занурення; занурення в порівняння (міжпредметне занурення); занурення в культуру («діалог культур»); занурення як компонент колективного способу навчання; виїзне занурення; циклова (конвеєрна) система навчання.

Найпоширеніша форма реалізації занурення в навчанні об'єктно-орієнтованого моделювання майбутніх учителів природничо-математичних дисциплін – лабораторно-обчислювальний практикум.

Прикладом нового соціально-конструктивістського методу навчання є *учіння через навчання* (з німецької *Lernen durch Lehren*), що активно пропагується Ж.-П. Мартаном (Jean-Pol Martin) [39] (Католицький університет Айхштетт-Інгольштадт, Німеччина). Це метод навчання, при якому студенти самі – за допомогою викладача – готовлять і проводять заняття (це може стосуватися і його окремих частин). Основа методу не є новою: ще в Давньому Римі існувала приказка «Docendo discimus» – «навчаючи, учимося самі». В XIX столітті ця ідея стала частиною Белл-Ланкастерської системи взаємного навчання. Широкого поширення цей метод набув завдяки заснованій в 1987 році Ж.-П. Мартаном мережі, що охоплює кілька тисяч учителів, а з 2001 року «Учіння через навчання» переживає особливий підйом у зв'язку зі шкільними реформами в Німеччині.

Для професійної освіти цей метод цікавий насамперед своїм кібернетичним трактуванням, згідно якого навчальні комунікації моделюються нейронною мережею. У природних нейронних мережах навчання відбувається в головному мозку, при цьому нейрони утворюють стабільні, тривалі з'єднання. В нейронних мережах продукуються знання, інтегруючись і створюючи в рамках цих взаємозв'язків нові більш ефективні з'єднання (емергенції).

Як можна перенести цю модель на організацію й проведення заняття? Викладач повинен подбати про те, щоб студенти інтенсивно спілкувалися й створювали довгострокові, пов'язані з матеріалом контакти, тобто викладач повинен піклуватися про те, щоб студенти колективно продукували знання. Це відбувається найкраще в рамках невеликих дослідницьких проектів, в тому числі – телекомунікаційних.

Які необхідні умови для функціонування колективу людей? Ж.-П. Мартан вказує, що «учні повинні отримувати радість від свого завдання й відчувати, що ця колективна робота відбувається з наміром поліпшити світ» [39] (етична мотивація). Комунікації повинні бути вільними: комунікативні бар'єри мають бути усунуті (чим простіші й швидші комунікації, тим краще). Викладач повинен добре знати кожного студента для того, щоб сприяти їх активній і продуктивній навчально-соціальній взаємодії один з одним (орієнтація на ресурси). Чим компетентніші окремі члени колективу, чим компетентніші викладачі, тим краще функціонує колектив.

Учіння розглядається як органічна продуктивність мозку й ґрунтується на погодженості молекулярних, клітинних і системних нейронних процесів у суміжних підсистемах моторики, сенсорики та асоціації. Встановлено, що: 1) учіння відбувається в контурах регулювання, які селективно стабілізуються шляхом структурного й функціонального узгодження; 2) учіння відбувається за певними правилами, що відповідають індивідуальній мотиваційній та емоційній динаміці й якими обумовлюється успіх в навчанні; 3) сенсомоторні та асоціативні контури регулювання включаються, підсилюючись, у навчальний процес, що й приводить до учіння через навчання.

Метод учіння через навчання ґрунтується на конструюванні комунікативних умінь студентів та вимагає від них відкритості, дружелюбності, концентрації, для чого, зокрема, заохочується демократична поведінка.

У процесі структурування групи комунікації стають усе інтенсивнішими, тому викладач повинен звикнути до того, щоб з кожного повідомлення відразу ж розпізнавати основне висловлення й співвідносити його з іншими повідомленнями. Він стає організатором колективного міркування й повинен обережно направляти розумові потоки, не втручаючись занадто часто. Він не повинен випустити з уваги зміст, втручаючись насамперед у сам процес, щоб комунікація між студентами відбувалася безупинно. Як зазначає М. Ю. Кондратьєв, здатність до комунікації в колективі стає основною якістю студентів [138].

Викладач як організатор колективного самоаналізу повинен піклуватися про те, щоб він вів до однієї мети, а саме до доведення нового матеріалу до всіх студентів. На початку заняття ще панує змістова невизначеність (відсутність лінійності), проте шляхом спільної роботи, крок за кроком повинна виникнути ясність (лінійність на основі досвіду). Базою підготовки до впровадження методу викладачем може бути його діяльність як модератора форумів, де з хаотично поступаючих повідомлень конструюються знання.

Існує паралель між процесом конструювання за методом учіння через навчання і способом наповнення Інтернет-енциклопедії. Той факт, що знання за методом учіння через навчання презентуються студентами, які не мають статусу експертів, привертає увагу одногрупників. У такий спосіб всі студенти закликаються працювати над поліпшенням ще незавершеного знання. Так само і з Інтернет-енциклопедією: користувачі тільки тому готові критично працювати спільно над текстами, що вони не визнають переваг в знаннях авторів статей. Тільки через рівноправність всіх користувачів стає можливим, що наявне – можливо, спочатку дилетантське – знання буде занесене до енциклопедії. Ця нова форма конструювання знання позначає перехід до суспільства знань, у якому всі рівноправно беруть участь у соціальному конструюванні знань.

Учіння через навчання ґрунтується на трьох компонентах: педагогічно-антропологічному, навчально-теоретичному та систематичному, предметно-спрямованому та змістовому.

З точки зору *педагогічно-антропологічного аспекту* учіння через навчання посилається в основному на піраміду потреб А. Маслоу (Abraham Harold Maslow). Завдання формування в інших знань повинно задовольняти потреби в надійності (за А. Маслоу – в структурі самосвідомості [164]), соціальному контакті й соціальному визнанні, а також у самореалізації й змісті (трансцендентність).

*Навчально-теоретичний та систематичний аспект* протипоставлений традиційному способу подання навчального матеріалу. В той час, як на занятті, центром якого є викладач, відбувається, як правило, рецептивне сприйняття навчального матеріалу, знання за методом учіння через навчання затребувані самими студентами. Виходячи з підготовленого, але ще не систематизованого на занятті матеріалу, перед студентами постає завдання здобути із цього матеріалу шляхом оцінювання, зважування та систематизації відповідні знання (Linearitaet a posteriori). Цей процес може відбуватися лише за інтенсивної комунікації.

З погляду *предметно-спрямованого та змістового аспекту* (в оригінальній методиці Ж.-П. Мартана спрямованого на навчання іноземних мов) цей метод повинен усунути віддавна існуюче протиріччя між звиканням (біхевіористичний компонент), співвідношенням матеріалу (когнітивний компонент) та аутентичною взаємодією (комунікативний компонент). У змістовому плані застосування методу вимагає, щоб навчальний матеріал став основою для міркувань. При роботі з підручником його зміст подається студентам. Якщо робота з підручником закінчена, то передбачається, що вони самі в рамках методу проектів виробляють нові знання й формують їх у всіх інших. На цьому етапі мотивація студентів дуже сильно залежить від якості змісту: вони повинні відчувати, що таке обговорення є для них професійно значущим (трансцендентне відношення: потреба в змісті).

Перед розглядом нової теми викладач розподіляє матеріал малими дозами між групами студентів (максимально три студенти); кожна група одержує окрему частину матеріалу, а також завдання повідомити цей зміст всім іншим. Студенти, котрі одержали завдання, дидактично підготовляють матеріал. Під час такої підготовки, що відбувається на занятті, викладач підтримує окремі групи, надає імпульси та поради. Швидко виявляється, що студенти без проблем справляються із цим завданням, адже вони могли спостерігати, які прийоми застосовує сам викладач.

Відразу ж потрібно звернути увагу на те, що учіння через навчання в жодному разі не повинно розумітися як фронтальне заняття, проведене студентами: вони повинні постійно відповідними засобами переконуватися, що матеріал зрозумілий тим, кому він адресований (коротко запитувати, узагальнювати, залучати до партнерської роботи). Тут викладач повинен втручатися, якщо він бачить, що комунікація не вдається або що застосовувані студентами прийоми мотивації не спрацьовують.

Переваги методу:

1. Матеріал опрацьовується інтенсивніше, а студенти виявляються істотно активнішими.

2. Студенти набувають додатково до предметних знань таких ключових умінь: здатність працювати в команді; здатність до планування; надійність; презентація й коментування; самосвідомість.

До недоліків методу відносять більші часові витрати (у порівнянні з іншими методами навчання).

Метод учіння через навчання знаходить своє застосування у всіх предметах. Так, у Німеччині він рекомендується як відкритий метод активізації навчально-пізнавальної діяльності, він може бути застосований і як метод підвищення кваліфікації за позааудиторною формою.

Студентам цей метод дає можливість тренувати мислення, щоб самим продукувати знання, гармонійно поєднуючи дослідження й навчання. Цей метод виявився особливо ефективним для стимулювання та обмежування традиційно багаторазової деталізації матеріалу. Учіння через навчання можна застосовувати і у великих групах (з кількістю учасників від 15 до 35).

При виборі та поєднанні методів навчання об'єктно-орієнтованого моделювання необхідно керуватися наступними *критеріями*:

– відповідність цілям і завданням навчання, виховання й розвитку;

– відповідність змісту досліджуваного матеріалу (складність, новизна, характер, можливість наочного подання матеріалу);

– відповідність реальним навчальним можливостям студентів: рівню підготовленості (навченості, розвиненості, вихованості, ступінь володіння інформаційно-комунікаційними технологіями), особливостям групи;

– відповідність наявним технічним умовам та відведеному для навчання часу;

– відповідність ергономічним умовам (час за розкладом, наповнюваність аудиторії, тривалість роботи за комп'ютером і т. д.);

– відповідність індивідуальним особливостям і можливостям самих викладачів (риси характеру, рівень володіння тим чи іншим методом, стосунки з групою, попередній досвід, рівень психолого-педагогічної, методичної та інформаційно-технологічної підготовки).

У табл. 2.1 наведено узагальнення класифікації традиційних та комп'ютерно-орієнтованих форм організації та методів навчання.

*Таблиця 2.1*

**Традиційні та комп'ютерно-орієнтовані форми організації та методи навчання у ЗВО (за Ю. В. Триусом [367])**

| Компоненти технології навчання | Традиційні | Комп'ютерно-орієнтовані |
|---|---|---|
| **Форми організації навчання** | Лекції, практичні заняття, семінари, лабораторні роботи, навчальні дискусії, самостійна позааудиторна робота, індивідуальна або групова науково-дослідна робота, поточні та підсумкові форми контролю:<br>– контрольні роботи,<br>– тестування,<br>– колоквіуми,<br>– модульний контроль,<br>– заліки, екзамени | Комп'ютерно-орієнтовані лекції, семінари, практичні і лабораторні заняття, контрольні роботи тощо; комп'ютерно-орієнтована науково-дослідна робота; комп'ютерне тестування; дистанційні форми:<br>– трансляція;<br>– чат (текстовий, графічний);<br>– відео- і телеконференції,<br>– інтерактивні форми проведення лекцій, семінарів, практичних й лабораторних занять, навчальних дискусій та ін.;<br>– комп'ютерно-орієнтовані екзамени й заліки |
| **Методи навчання (за джерелом здобуття знань)** | **Вербальні методи навчання** | |
| | Лекція; розповідь; пояснення; бесіда; робота з підручником, довідковою, науково-популярною та навчальною літературою | Робота з електронними підручниками, довідковим матеріалом комп'ютерних програм; робота з відомостями, що отримуються через глобальну мережу Internet |
| | **Наочні методи навчання** | |
| | Демонстраційний експеримент; самостійне спостереження | Робота з навчаючими та навчально-контролюючими програмами |
| | **Практичні методи навчання** | |
| | Виконання лабораторних робіт; виконання практикумів; розв'язування доцільно дібраних задач | Дослідницька робота в комп'ютерних лабораторіях; обчислювальні експерименти; телекомунікаційні проекти |

Таким чином, у системі професійної підготовки майбутніх учителів природничо-математичних дисциплін:

– провідними *формами організації навчання* є: комп'ютерно-орієнтована лекція, лабораторні роботи; самостійна позааудиторна робота; індивідуальна та колективна навчально-дослідницька робота; модульний контроль та підсумковий контроль у формі захисту результатів виконання телекомунікаційного проекту; дистанційні форми (текстова, аудіо та відеотрансляція, вебінари тощо);

– провідними *методами навчання* є: лекція; робота з підручником, довідковою, науково-популярною та навчальною літературою, з відомостями, що отримуються через глобальну мережу Internet; обчислювальний експеримент; виконання лабораторних робіт; виконання телекомунікаційних проектів.

### 2.4 Соціально-конструктивістські засоби навчання комп'ютерного моделювання майбутніх учителів природничо-математичних дисциплін

*Засоби навчання* – матеріальні й ідеальні об'єкти, які використовуються в освітньому процесі як носії відомостей (інформаційних ресурсів) та інструменти діяльності вчителя (викладача) й учнів (студентів), що застосовуються ними як окремо, так і спільно [367, с. 230].

Автори [186] під засобом навчання розуміють будь-який об'єкт, який використовується для досягнення певної цілі: поза неї засобу не існує, бо не існує засобу взагалі – він існує лише в системі засіб–ціль.

До засобів навчання належать: природне і соціальне оточення, обладнання, підручники, книги, наукові видання, комп'ютери і комп'ютерні мережі з відповідним програмним забезпеченням та інформаційними ресурсами, зокрема електронні підручники, довідники, енциклопедії, електронні бібліотеки.

Існують різні класифікації засобів навчання. Одна з них – класифікація за дидактичною функцією [172, с. 158]:

– інформаційні засоби (підручники, навчальні посібники);

– дидактичні засоби (таблиці, плакати, відеофільми, програмні засоби навчального призначення, демонстраційні приклади);

– технічні засоби навчання (аудіовізуальні засоби, комп'ютери, засоби телекомунікацій, відеокомп'ютерні системи, мультимедіа, віртуальна реальність).

Інша – класифікація за способом моделювання: матеріальні (технічні) засоби, зокрема комп'ютер, матеріалізовані (знакові об'єкти, малюнки, схеми тощо), а також ідеальні (це різноманітні знання) [186].

Програмні засоби навчання є матеріалізованими ідеальними засобами, що реалізуються за допомогою технічних засобів.

Дидактичне призначення використовуваних програмних засобів може бути різним: опанування нового матеріалу (наприклад, за допомогою програми навчального призначення), закріплення нового матеріалу (наприклад, за допомогою програми-тренажера), перевірка рівня засвоєння навчального матеріалу або операційних навичок (наприклад, за допомогою програм автоматизованого контролю або тестування).

Серед засобів навчання виділяють (умовно) традиційні і нові інформаційно-комунікаційні технології, які слід гармонійно поєднувати і взаємодоповнювати в процесі навчально-пізнавальної діяльності. Під засобами нових інформаційно-комунікаційних технологій розуміють програмно-апаратні засоби й пристрої, що функціонують на базі комп'ютерної техніки, а також сучасних засобів і систем інформаційного обміну, забезпечення операцій щодо пошуку, збирання, накопичення, зберігання, опрацювання, подання і передавання повідомлень [172, с. 162].

Таким чином, засоби навчання об'єктно-орієнтованого моделювання майбутніх учителів природничо-математичних дисциплін – це система програмних засобів, спрямованих на реалізацію: 1) цілей навчання об'єктно-орієнтованого моделювання (різні середовища моделювання); 2) системи основних та допоміжних навчальних впливів (системи управління навчанням); 3) навчальної комунікації в мережному середовищі (зокрема, засобами систем управління навчання); 4) колективних навчальних досліджень (мережні засоби спільної роботи середовищ об'єктно-орієнтованого моделювання та засоби Web 2.0).

Розглянемо їх більш докладно.

### *2.4.1 PyGeo – соціально-конструктивістське середовище об'єктно-орієнтованого моделювання геометричних об'єктів*

Інтерактивне геометричне програмне забезпечення (середовища динамічної геометрії) надає можливість створювати геометричні побудови (переважно планіметричні). Побудова починається з кількох базових точок, що використовуються для визначення ліній, кіл чи інших точок. Після завершення побудови можна її змінити простим переміщенням однієї з базових точок.

Найбільш відомими планіметричними середовищами динамічної геометрії є C.a.R., Cabri Geometry, Euklid DynaGeo, DG, Dr Genius, Dr. Geo, GRAN-2D, The Geometer's Sketchpad, GeoProof, GRACE, Kgeo, OpenEuclide. Серед тривимірних середовищ можна виділити Cabri 3D, Euler 3D, Geomview, GRAN-3D, SingSurf, PyGeo.

Як показав С. А. Раков, середовища динамічної геометрії є ефективними засобами реалізації соціально-конструктивістського підходу в навчанні [216], тому в процесі професійної підготовки студентів спеціальності «Математика» доцільним є використання такого засобу динамічної геометрії, що одночасно є і середовищем об'єктно-орієнтованого моделювання. Виходячи з даного нами означення, основою такого середовища має бути об'єктно-орієнтована мова програмування.

Середовище об'єктно-орієнтованого моделювання планіметричних та стереометричних об'єктів PyGeo реалізовано мовою Python – інтерпретованою об'єктно-орієнтованою мовою програмування високого рівня з динамічною семантикою, розроблена Гвідо ван Россумом (Guido van Rossum) у 1990 р. [57]. Структури даних високого рівня разом із динамічною семантикою та динамічним зв'язуванням роблять її привабливою для швидкої розробки програм, а також як засіб інтеграції існуючих компонент. У Python підтримуються модулі та пакети модулів, що сприяє модульності та повторному використанню коду. Інтерпретатор Python та стандартні бібліотеки доступні на всіх основних платформах, тобто сам інтерпретатор є мобільним. У Python підтримується кілька парадигм програмування, зокрема, об'єктно-орієнтована, процедурна, функціональна та аспектно-орієнтована.

Серед основних *переваг* Python можна назвати такі: «чистий» синтаксис (для виділення блоків використовуються відступи); мобільність програм (внаслідок інтерпретованої природи мови); стандартний дистрибутив має велику кількість корисних модулів (включаючи модулі для розробки графічного інтерфейсу); можливість використання в діалоговому режимі (корисне для експериментування та розв'язування простих задач); стандартний дистрибутив має просте, але разом із тим досить потужне середовище розробки; зручність для розв'язування математичних задач (містить засоби роботи з комплексними числами, цілими числами довільної довжини, в режимі безпосереднього введення може використовуватись як потужний символьний калькулятор).

Елегантний синтаксис Python, динамічне опрацювання типів, ефективні структури даних високого рівня, простий, але ефективний підхід до об'єктно-орієнтованого програмування, а також те, що це інтерпретована мова, роблять її ідеальною для швидкої розробки та прототипування програм.

Суттєвою відмінністю середовища PyGeo від інших засобів динамічної геометрії є те, що динамічні побудови PyGeo за внутрішньою структурою є програмами мовою Python, що обробляються безпосередньо її інтерпретатором (рис. 2.4).

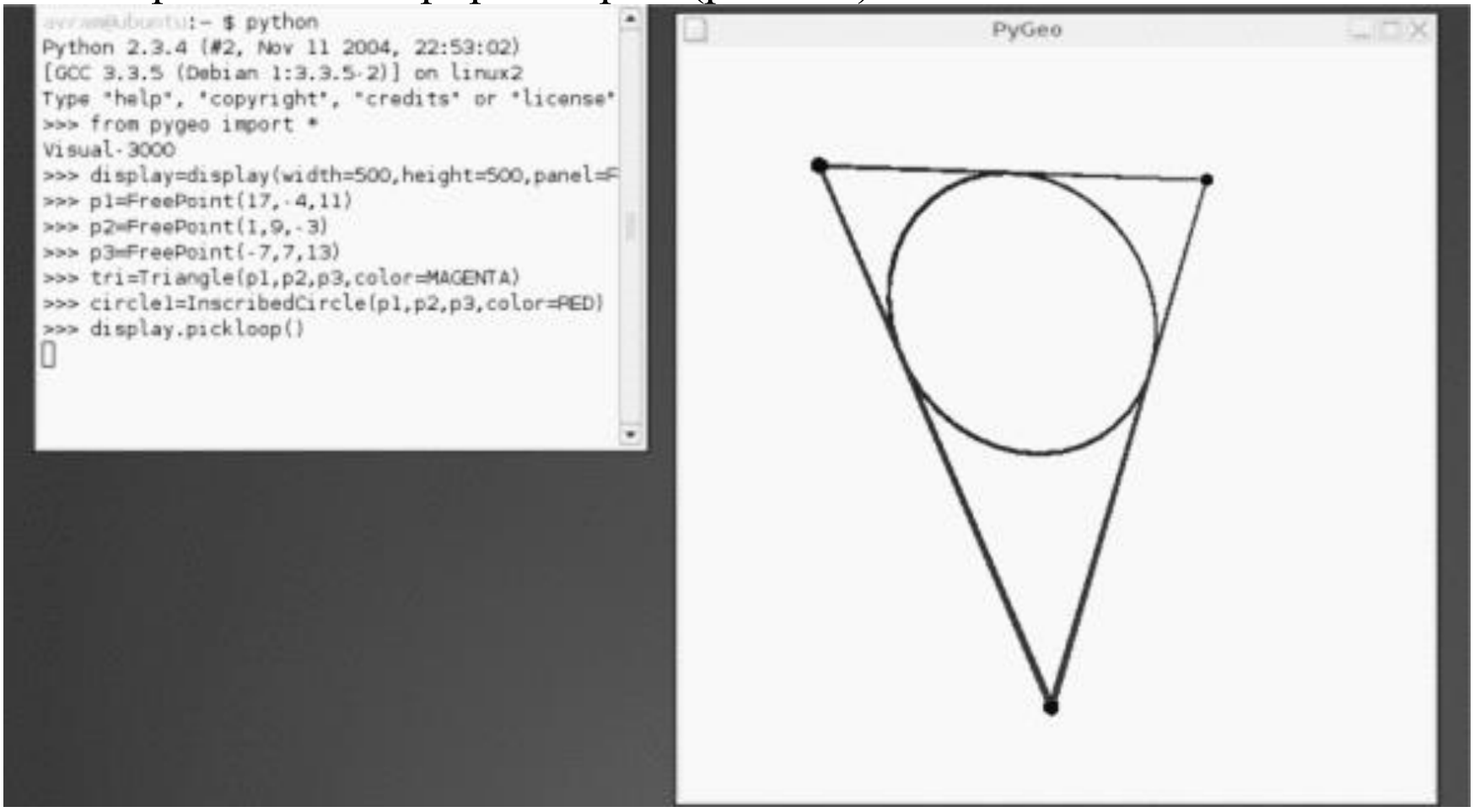

Рис. 2.4. Побудова в командному режимі Python

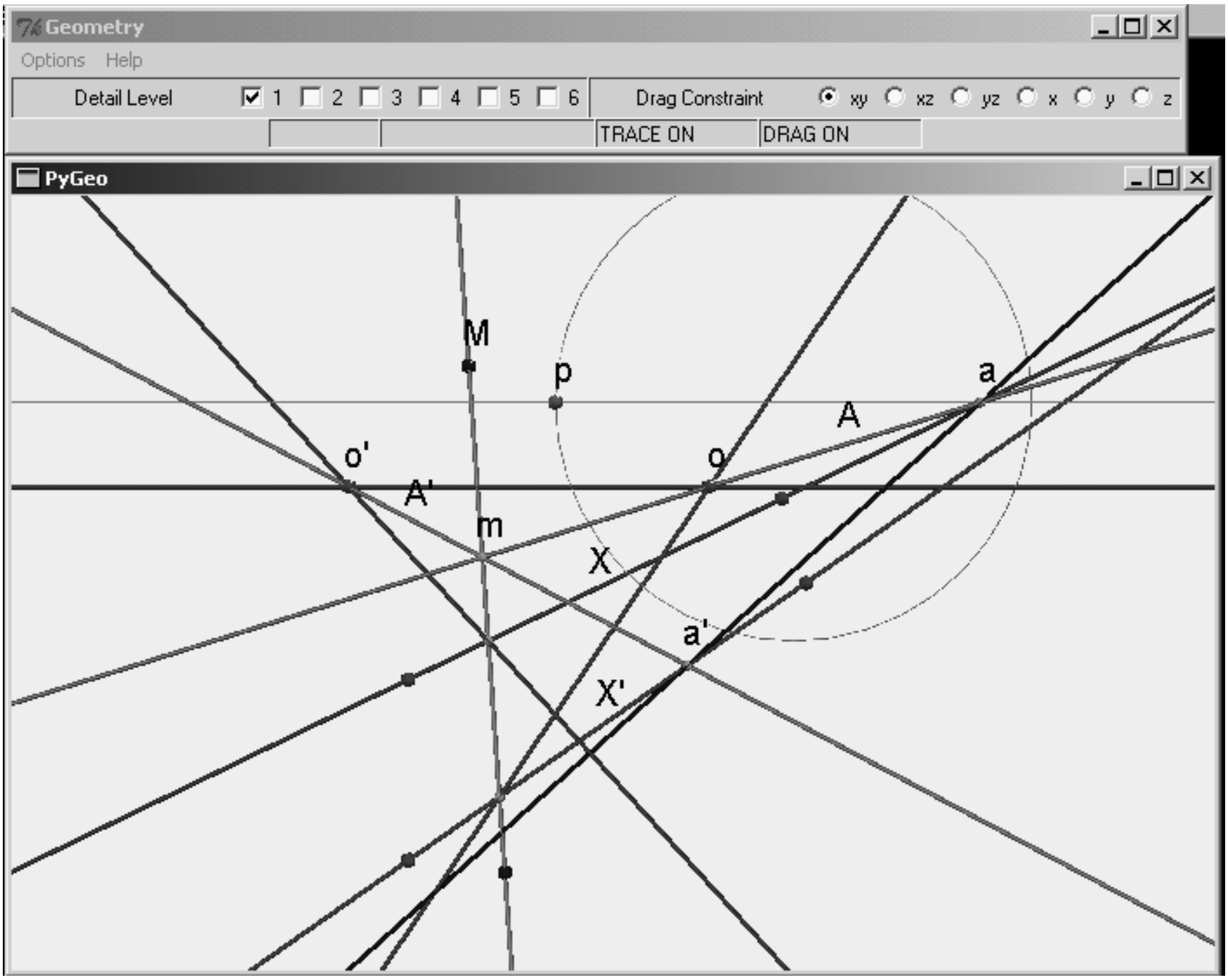

Рис. 2.5. Робочий сеанс середовища PyGeo з панеллю керування

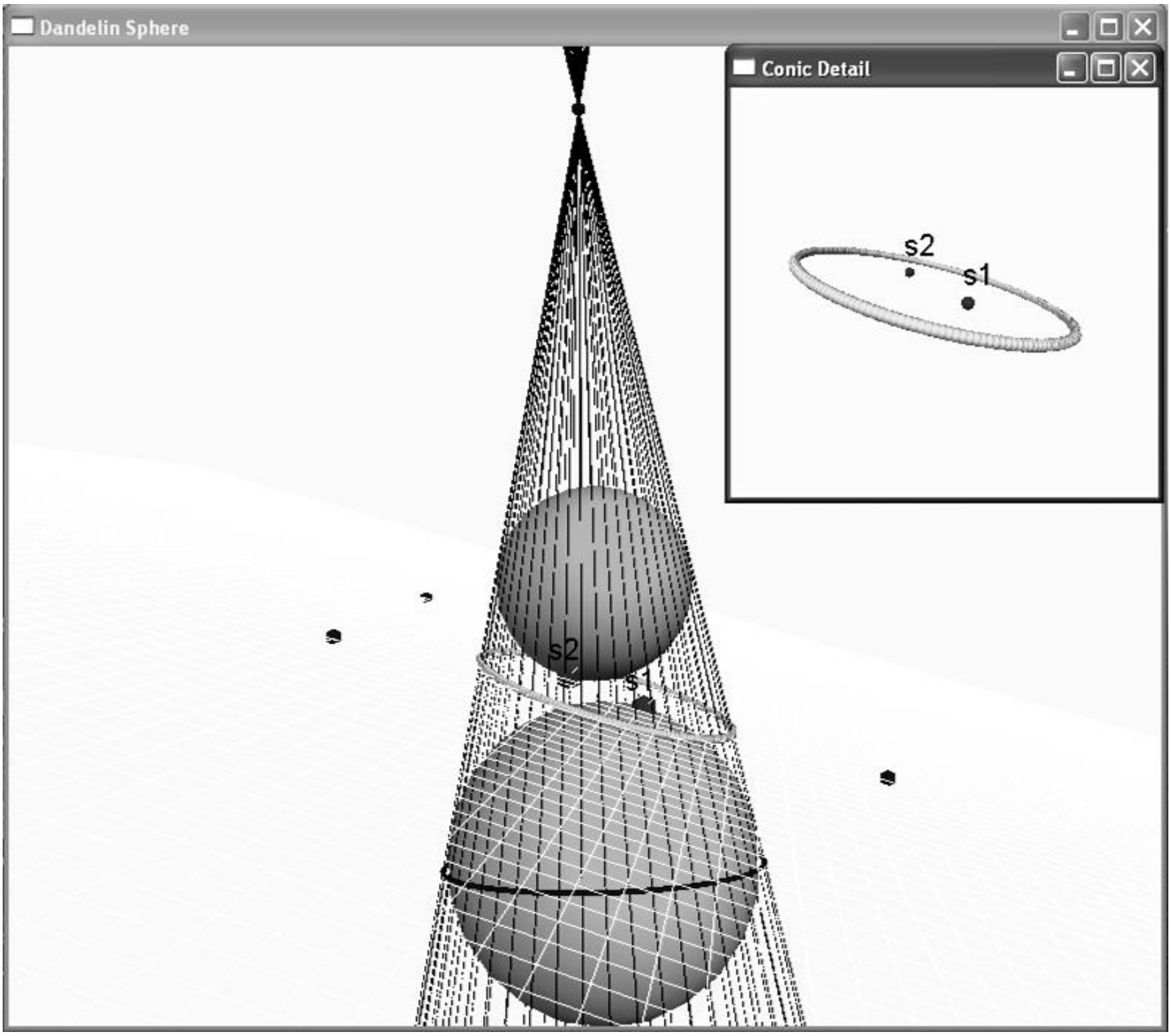

Рис. 2.6. Робочий сеанс PyGeo з головним вікном та вікном деталізації

Таким чином, динамічні побудови в PyGeo будуються програмно, що суттєво відрізняє його від інших програмних засобів, орієнтованих на візуальні побудови.

Визначальною рисою PyGeo є те, що його застосування надає можливість природним чином поєднати засоби ООП з алгоритмами геометричних побудов, що є неможливим засобами візуальних середовищ.

Реалізуючи ідею С. Пейперта «навчатися, навчаючи», PyGeo надає можливості не лише використання вбудованих об'єктів, а й побудови на їх основі нових, реалізуючи концепцію наслідування. З точки зору геометрії, об'єкти PyGeo поділяються на дві групи: об'єкти класичного евклідового простору разом із тісно пов'язаними з ними поняттями проективних простору та площини (на рис. 2.5 показано розв'язання задачі на проективній площині) і об'єкти, пов'язані з геометрією комплексних чисел, зокрема об'єкти на проективній комплексній площині та об’єкти на сфері Рімана.

Убудованими об’єктами PyGeo є:

– прості об’єкти (точка, лінія, площина тощо);

– складені об’єкти, наприклад, об’єкт-набір ліній, що проходять через задану точку заданої площини;

– перетворення об’єктів, наприклад, об’єкт, створений проектуваннями об’єктів на задану площину через певну точку в просторі (рис. 2.6 ілюструє побудову перерізів у середовищі PyGeo).

Всі об’єкти PyGeo – прості, складені, перетворені – можуть бути передані іншим об’єктам у якості параметру, що дозволяє виконати дослідження таких явищ, як ітеративне перетворення.

З точки зору поведінки об’єкти поділяються на: об’єкти, що можуть бути інтерактивно переміщені в межах області (евклідового простору, комплексної площини) чи обмежені конкретним об’єктом цієї області (наприклад, конкретним колом, лінією); анімовані об’єкти, що переміщуються уздовж визначеного шляху, наприклад, точка, що рухається вздовж кола, яке котиться по іншому колу; фіксовані об’єкти, наприклад, точка з визначеними координатами; залежні об’єкти, що змінюють власну позицію лише при зміні позиції об’єкта, від якого вони залежать.

### *2.4.2 VPython – середовище об’єктно-орієнтованого моделювання фізичних процесів*

Програма мовою Python може бути розширена функціями та типами даних, створеними іншими мовами (C, C++, Pascal тощо). Одним із таких розширень є VPython, за допомогою якого було побудоване середовище PyGeo.

Основою VPython є модуль Visual, орієнтований на подання візуальних об’єктів у тривимірному просторі. Процес створення програм із застосуванням VPython суттєво спрощений у порівнянні з традиційними засобами програмування тривимірної графіки. Під час роботи програми користувач має можливість змінювати позицію об’єкта, розмір сцени, кут огляду і т.п.

У процесі навчання спецкурсу «Об’єктно-орієнтоване моделювання» VPython може використовуватись як середовище, яке надає студентам можливість зосередитись на процесі розробки динамічних моделей з другого модулю курсу, а не на деталях їх візуалізації [303].

Динамічне моделювання відтворює процес функціонування й розвитку об’єктів у часі й у просторі. Зокрема, виходячи з основної задачі механіки, змістовою складовою динамічних моделей може бути багато задач з курсу фізики, тому в процесі навчання об’єктно-орієнтованого моделювання середовище VPython було обрано провідним засобом навчання студентів спеціальності «Фізика» (спеціалізація «Інформатика»).

3D-природа VPython допомагає в процесі навчальних досліджень побачити нові дані й шляхи розв'язання задачі, адже динамічні моделі надають можливість вивчати не тільки готові результати, але й розглядати процес їх отримання і дослідження, формуючи в студентів здатність продукувати нестандартні ідеї й рішення, сприяючи розвитку модельного мислення.

Таким чином, особливістю реалізації динамічних моделей у середовищі VPython є: 1) мультимедійність – в моделях можуть використовуватися такі елементи мультимедійних технологій, як графіка, анімація, звук, текст, відео; 2) інтерактивність – кожний об'єкт VPython на екрані доступний для вивчення, видозміни та комбінування з іншими об'єктами з метою надання кінцевому користувачеві можливості активного експериментування з ними; 3) компактність – файли моделей, створені засобами мови ООП Python, мають малий розмір (до декількох кілобайт, що забезпечується зберіганням моделей у вигляді вихідних текстів програм), що відіграє істотну роль при передаванні по мережі, а також при розміщенні в системі підтримки навчання та на сервері мережної дослідницької спільноти.

У процесі розробки динамічних моделей в середовищі об'єктно-орієнтованого моделювання VPython студенти набувають *умінь*:

1) реалізувати окремі елементи модельованого об'єкту (процесу, явища) у вигляді статичної графіки;

2) реалізувати інтерактивну взаємодію окремих елементів модельованого об'єкту (процесу, явища);

3) використовувати глобальну мережу для пошуку й розміщення моделей;

4) конструювати, досліджувати та удосконалювати об'єктно-орієнтовану модель процесу чи явища;

5) використовувати набуті навички в майбутній професійній діяльності.

Набуті уміння можуть надалі бути використані в процесі навчальних досліджень в конкурсних, курсових та кваліфікаційних роботах з фізики, зокрема – для реалістичної анімації модельованих об'єктів (процесів, явищ):

1) побудова реалістичних анімацій вимагає гарного розуміння природи руху (адже людський мозок дуже добре виявляє «неприродний» рух), тому будь-яка така комп'ютерна анімація має спиратися на закони фізики;

2) реалістична фізична анімація має бути тривимірною, надавати користувачу можливість оглядати об'єкти з різних точок зору, наближувати чи віддаляти їх тощо;

3) середовище для розробки анімацій має надавати можливість у швидкий спосіб, зосереджуючись, насамперед, на фізичній природі модельованого явища, побудувати та анімувати комп'ютерну модель.

Ураховуючи, що екзамен зі спецкурсу об'єктно-орієнтованого моделювання проводиться у формі захисту індивідуальних та колективних дослідницьких проектів, нами були розроблені методичні рекомендації для самостійної роботи з об'єктно-орієнтованого моделювання фізичних об'єктів (процесів, явищ) у середовищі VPython, сутність яких полягає в наступному:

1. Спочатку вводяться поняття координатної системи, позиції матеріальної точки, векторів та скалярів, зміщень, сталої, миттєвої та середньої швидкості. Замість координатної площини вводиться координатний простір (рис. 2.7а), в якому позиція матеріальної точки задається тривимірним вектором, що є вбудованим об'єктом VPython. Для зображення матеріальних точок використовується графічний об'єкт «сфера», для зображення векторів – «стрілка». Дії над векторами ілюструються графічно (рис. 2.7б). Зміщення трактується як векторна величина, залежна від часу.

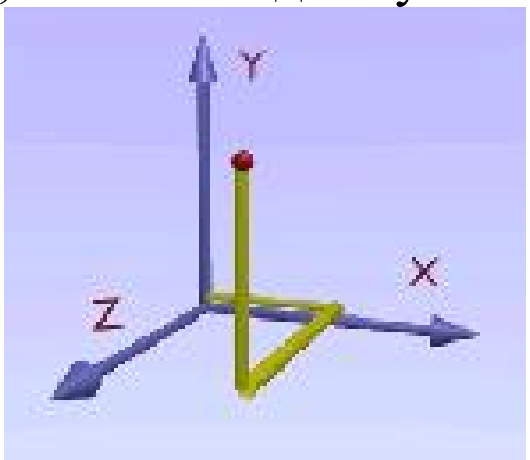

а) координатний простір VPython

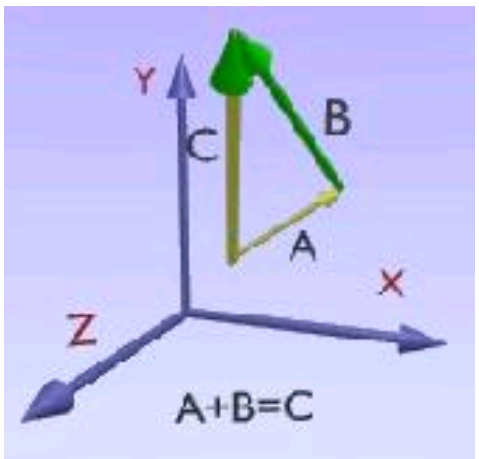

б) векторний простір VPython

Рис. 2.7. Дії над векторами у VPython

2. Далі обговорюється поняття прискорення та визначається загальновідомий алгоритм обчислень:

1) знайти прискорення $\boldsymbol{a}_i$;

2) змінити швидкість: $\upsilon_{i+1}=\upsilon_i+\boldsymbol{a}_i\Delta t$;

3) визначити нове положення: $\boldsymbol{r}_{i+1}=\boldsymbol{r}_i+\upsilon_i\Delta t$.

Для ілюстрації застосування алгоритму розглядаються моделі рівномірного та рівноприскореного руху по прямій, параболічного руху (тіла, кинутого під кутом до горизонту), колового руху зі сталою швидкістю.

3. Розглядаються специфічні засоби VPython – робота з файлами,

списками, клавіатурою, мишею.

4. Обговорюються способи виявлення зіткнень тіл: зокрема, якщо при розрахунку для двох куль модуль різниці відстаней між ними менше або дорівнює сумі їх радіусів, то вважатимемо, що тіла зіткнулись.

5. Далі обговорюється рух тіл за наявності та відсутності тертя, пружна та непружна деформації, вводяться поняття центру мас, імпульсу та законів збереження імпульсу, моменту імпульсу і кінетичної енергії.

6. Обговорюються поняття ООП: об'єкт, клас, конструктор, наслідування тощо. Тут доцільно розглянути комплексний приклад «Вишневе дерево» (рис. 2.8), для реалізації якого пропонується створити два класи: «лист» та «плід». Різні дерева будуються за допомогою генератора псевдовипадкових чисел. На цьому прикладі демонструється ефект коливання листя під дією вітру.

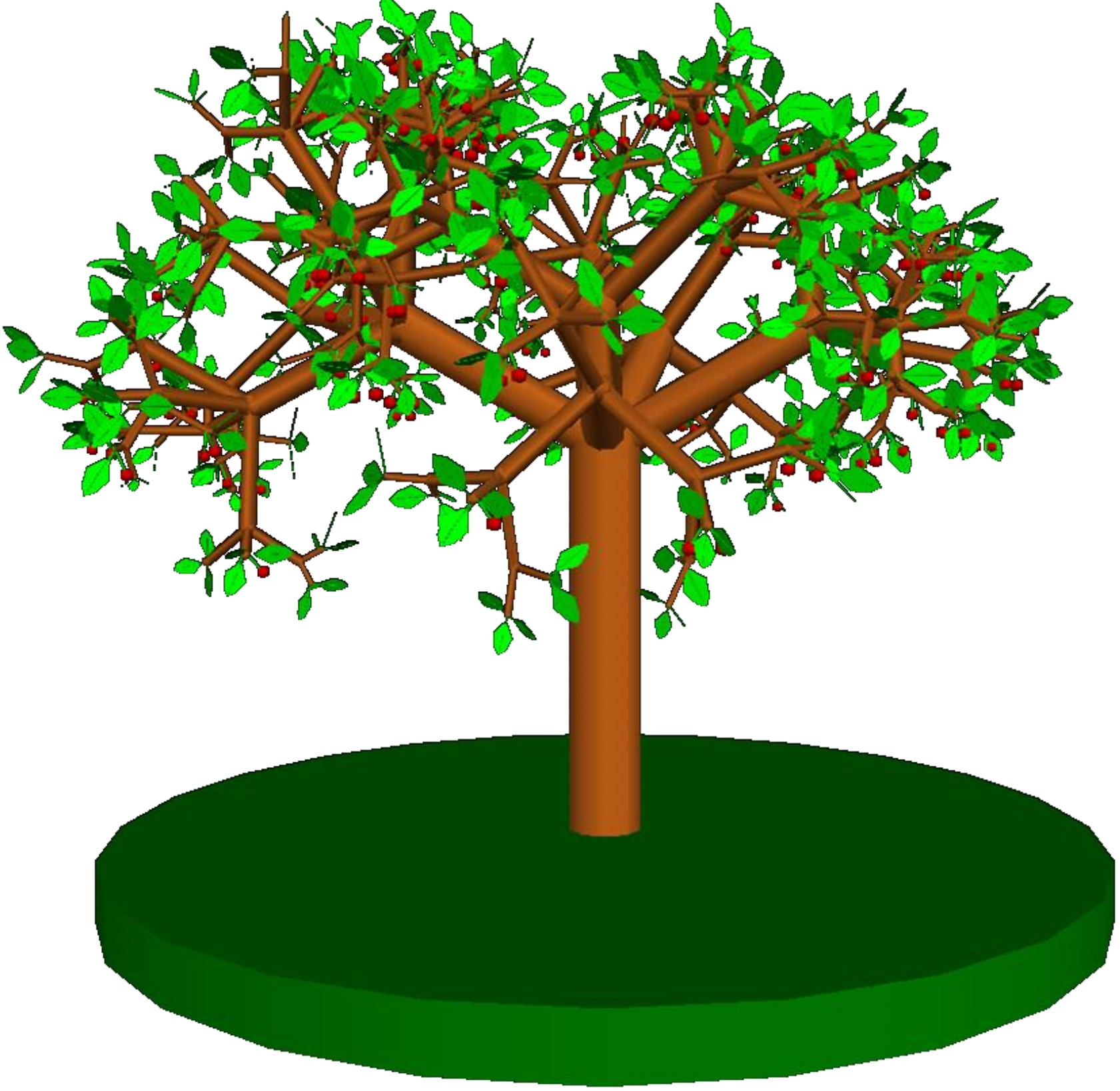

Рис. 2.8. Об'єктно-орієнтована модель коливання листя під дією вітру

7. Уводяться поняття полярних координат, кутової швидкості, кутового прискорення, моменту інерції. Завершуються методичні рекомендації найпростішою моделлю ходьби людини.

Не зважаючи на те, що середовище VPython призначено насамперед для побудови динамічних моделей, його можна використати і для побудови імітаційних моделей з третього модулю курсу (рис. 2.9).

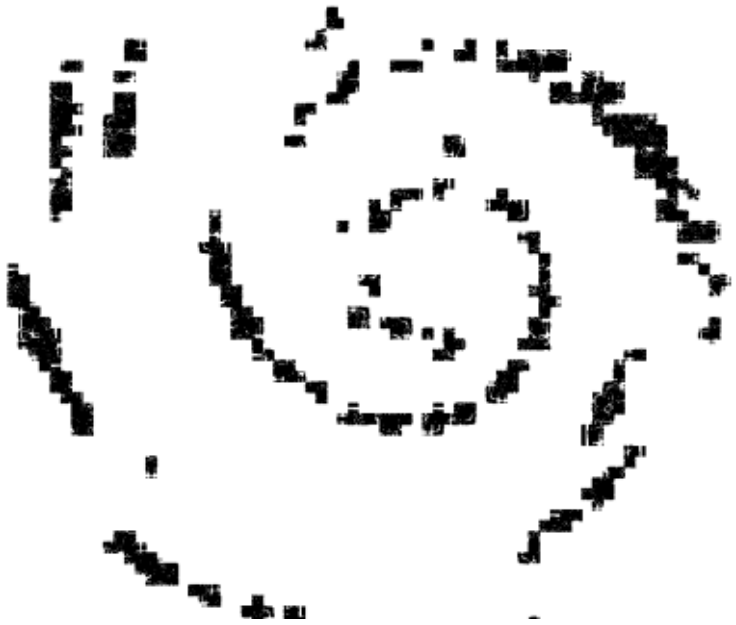

а) перколяційна модель спіральної галактики

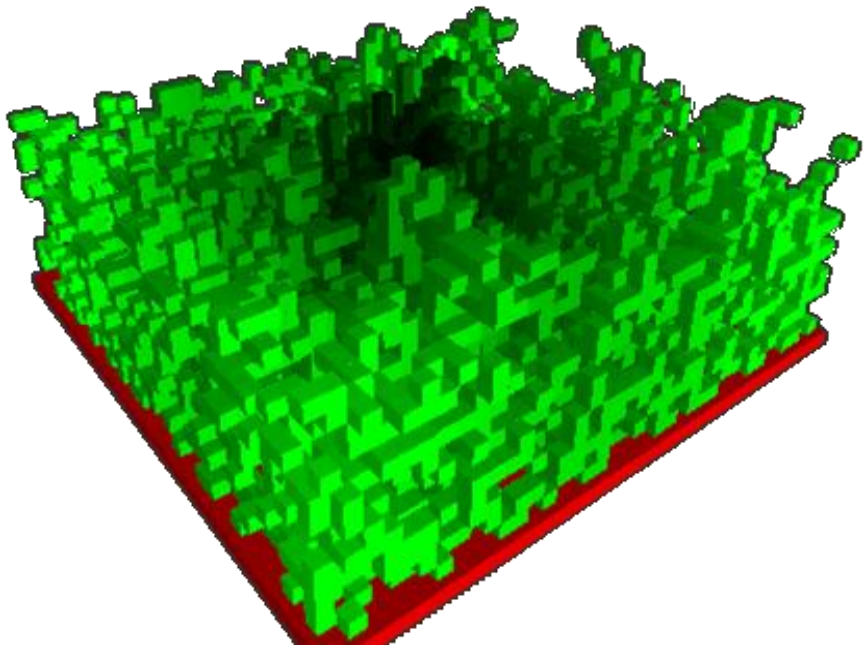

б) модель електролізу на пласкому катоді

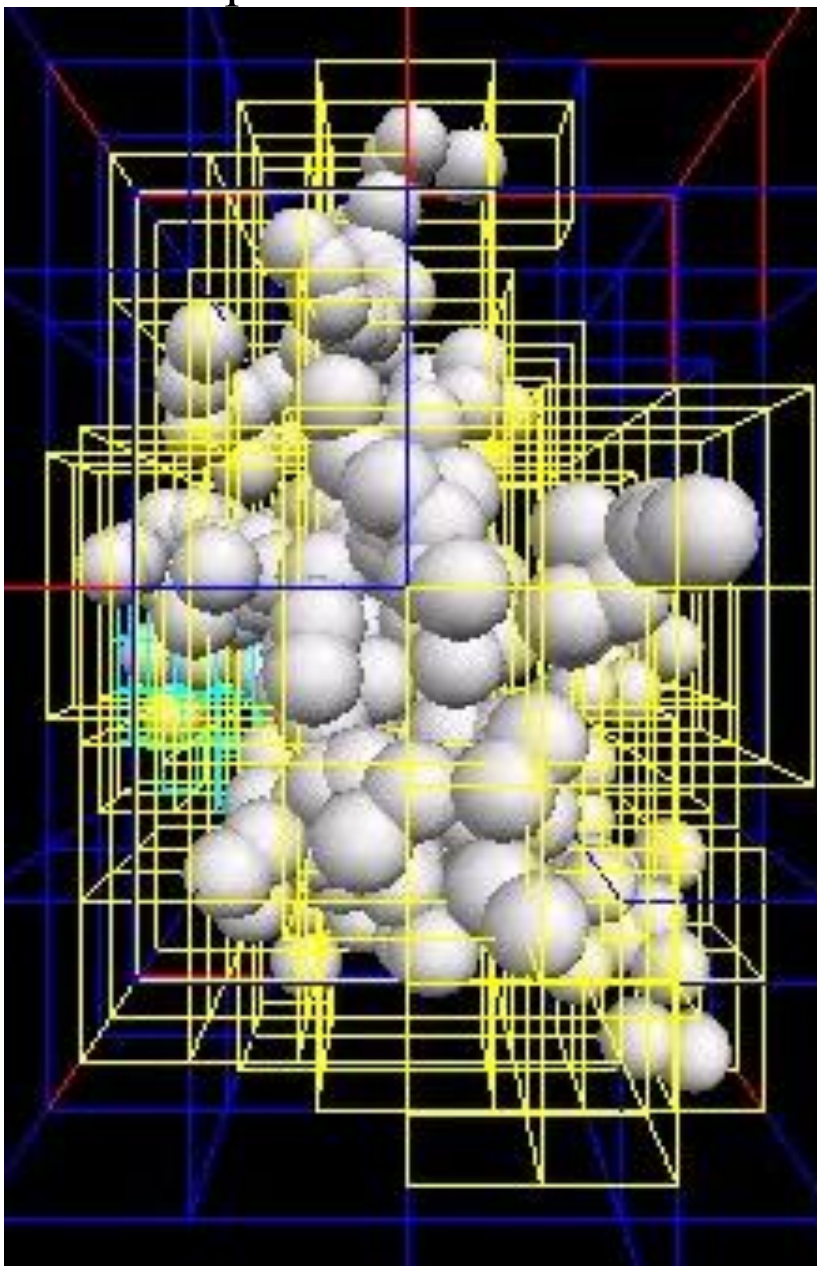

в) модель агрегації з обмеженням дифузії

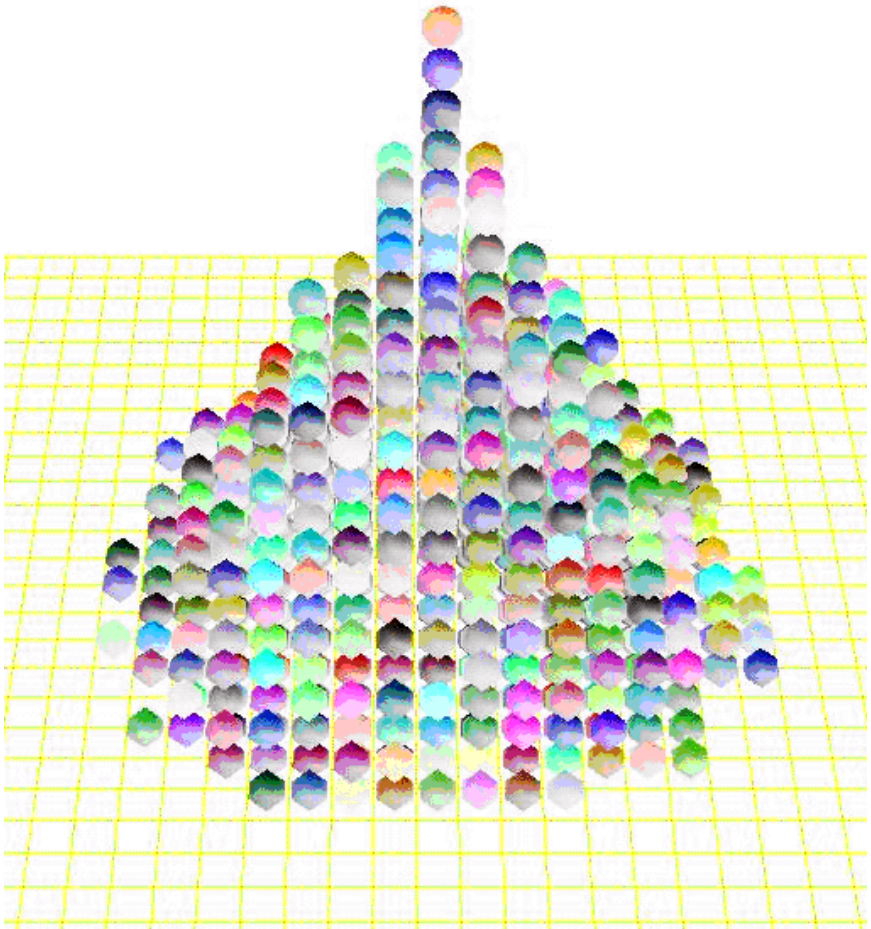

г) модель самоорганізованої критичності – виникнення лавини осипань у купі піску

Рис. 2.9. Приклади побудови імітаційних моделей у середовищі VPython

Тематика індивідуальних та колективних дослідницьких проектів, що пропонується у навчальній програмі спецкурсу (додаток Г), є орієнтовною: вибір теми проекту часто пов'язується з тематикою курсових робіт із фахових дисциплін. Найцікавішими серед тем проектів, обраних студентами, є «Моделювання визначальних фізичних експериментів», «Модель стабільності орбіт в точках Лагранжа», «Моделювання процесу електролізу на точковому катоді», «Зіткнення галактик», «Динаміка кристалічної решітки», «Нерелятивістське розсіювання електронів», «Моделювання електричного та магнітного полів», «Вибрані моделі класичної механіки», «Модель стабільності атмосфери», «Дифракція та інтерференція світла».

У додатку В наведено тематику індивідуальних та колективних дослідницьких проектів, що виконувались студентами 3 курсу фізико-математичного факультету Криворізького державного педагогічного університету в середовищі об'єктно-орієнтованого моделювання Squeak, розглянутого у розділі 1.

Комплексне використання в процесі навчання об'єктно-орієнтованого моделювання різних середовищ можливе шляхом їх інтеграції. Як було показано в [281], вдосконалення техніки моделювання вимагає переходу від застосування електронних таблиць до розгляду інформаційних моделей, що забезпечується об'єктно-орієнтованим середовищем. Обмеження на динаміку зображень та обсяг обчислень, що накладають відомі середовища електронних таблиць [306], виступають тією межею, подолання якої вимагає зміни середовища – наприклад, на VPython [303]. На жаль, поряд із отриманням нової можливості – побудови динамічних об'єктно-орієнтованих моделей, – ми втрачаємо стару: зручний інтерфейс користувача, що його надає середовище електронних таблиць.

Для подолання цієї проблеми Ч. Праяга (Chandra Prayaga) пропонує простий інтерфейс між VPython та Excel, що базується на використанні VBA та дозволяє об'єднати 3D-властивості VPython з інтерфейсом користувача Excel для введення даних. Сценарій VBA читає дані з комірок таблиці та зберігає у текстовому файлі, що читається VPython та застосовується при візуалізації моделі. У своїй статті [53] автор ілюструє цю ідею на прикладі руху заряду в магнітному полі.

Для перевірки цієї ідеї було створено ряд комбінованих інтерфейсів:

1. Так, раніше для роботи з програмним комплексом VPNBody [126; 23; додаток Б] необхідно було створювати та редагувати текстові файли з параметрами системи, інколи – досить об'ємні. У розробленому інтерфейсі всі дані вводяться та зберігаються в електронній таблиці, а відповідний макрос VBA генерує файл та запускає VPNBody.

2. У Excel можна було побудувати фазову площину, проте за побудованим зображенням знайти початок та кінець фазової траєкторії недосвідченому оку було важко [323]. Імпорт результатів розрахунку до VPython надав можливість анімувати процес побудови фазової траєкторії.

Перший з розроблених інтерфейсів розв'язує проблему відсутності інтерфейсу до відомого програмного забезпечення, другий – відсутності адекватного відображення результатів розрахунків. Таким чином, з'являється можливість не тільки надати середовищу VPython інтерфейс електронних таблиць (як це пропонує Ч. Праяга), а й розширити межі застосування самих електронних таблиць шляхом підвищення наочності результатів обчислень.

Наступним кроком з інтеграції середовищ моделювання є створення COM-сервера, що використовує VPython, та макросу VBA (з Excel) для керування дисплейним об'єктом VPython. У такий спосіб користувач електронних таблиць отримує всі можливості VPython без необхідності опанування нової мови програмування.

Зауважимо, що все описане може бути виконано також у середовищі OpenOffice Calc (зі змінами, зумовленими іншими діалектом Basic та об'єктною моделлю). Застосування OpenOffice замість Microsoft Office надає можливості для ще більш тісної інтеграції середовищ моделювання, адже інтерпретатор Python є вбудованим у OpenOffice. Єдина дія, необхідна для виконання програм на VPython викликом з макросів OpenOffice – встановлення модуля Visual.

### *2.4.3 Sage – середовище об'єктно-орієнтованого моделювання математичних об'єктів*

Широке застосування мови Python у Web-системах керування змістовим наповненням створює умови для конструювання об'єктно-орієнтованих моделей у мережних середовищах. Гарним прикладом такого середовища є Web-СКМ Sage, що надає можливість будь-якому її користувачеві створювати динамічні та імітаційні об'єктно-орієнтовані моделі, співпрацюючи при цьому з іншими користувачами, оприлюднювати результати індивідуальних та колективних дослідницьких проектів у Web тощо.

У роботі [386] виділено також такі характеристики Web-СКМ Sage, як підтримка технологій Web 2.0 та можливість здійснення інтеграції з системами підтримки навчання.

Використання Web-СКМ Sage у процесі навчання об'єктно-орієнтованого моделювання майбутніх учителів математики надає можливість:

1) об'єднати засоби візуалізації динамічних моделей, що їх надає

VPython, із убудованими засобами Sage для здійснення аналітичних та чисельних розрахунків, що дозволяє користувачеві не лише подавати результати обчислень у зручній для сприйняття формі, будувати дво- та тривимірні графіки кривих та поверхонь, гістограми тощо, а й візуалізувати перебіг процесу, що моделюється, засобами 3D-графіки;

2) створювати класи мовою Python, спрямовані на реалізацію нових математичних об'єктів.

Пункт 2.4 «Плану дій щодо поліпшення якості фізико-математичної освіти на 2009–2012 роки», затвердженого Наказом МОН України №1226 від 30.12.2008 р. [211], передбачає удосконалення змісту навчальних програм з базових математичних дисциплін, враховуючи комп'ютеризацію усіх видів інженерної діяльності (дискретна і комп'ютерна математика, нечіткі методи і «м'які» обчислення).

На виконання «Плану ...» на базі Криворізького металургійного інституту ДВНЗ «Криворізький національний університет» створено сервер мобільного математичного середовища (http://korpus21.dyndns.org:8000), що широко використовується у ЗВО м. Кривого Рога. На основі локалізованої версії Sage створено мобільне математичне середовище «Вища математика», до змісту якого включені питання, пов'язані з «м'якими» обчисленнями.

Ураховуючи, що обчислювальне ядро Sage не має безпосередньої підтримки «м'яких» обчислень, було створено кільце нечітких трикутних чисел над множиною Q: клас Fuzzy(Ring) з методами __init__, _repr_ та _element_constructor_. Останній метод заповнює кільце елементами класу FuzzyNumber(RingElement), конструктор якого може мати один з трьох видів:

1) FuzzyNumber(left, median, right) – створює трикутне нечітке число з лівою границею left, правою right та медіаною median;

2) FuzzyNumber(left, right) – створює трикутне нечітке число з лівою границею left, правою right та медіаною median=(left+right)/2;

3) FuzzyNumber(median) – створює трикутне чітке число.

У класі FuzzyNumber визначено як стандартні операції над нечіткими трикутними числами (додавання, віднімання, множення, ділення, піднесення до степеня, порівняння та ін.), так й нестандартні (перетворення нечіткого числа на рядок, вектор, список, графічний об'єкт, чітке число тощо). Приклад різних інтерпретацій моделі трикутного нечіткого числа у мобільному Web-СКМ Sage показано на рис. 2.10.

Розроблене програмне забезпечення повністю інтегроване у Sage завдяки новому типу даних FF (статичному об'єкту кільця Fuzzy), застосування якого дозволяє будувати нечіткі матриці, поліноми та інші стандартні об’єкти Sage.

```
n1=FuzzyNumber(1,2,5)
n2=FuzzyNumber(3,5,10)
n3=n1+n2
html('$n_1=%s, n_2=%s, n_1+n_2=%s$'%(n1,n2,n3))
show(plot(n1)+plot(n2)+plot(n3))
```

evaluate

$$n_1 = (1, 2, 5), n_2 = (3, 5, 10), n_1 + n_2 = (4, 7, 15)$$

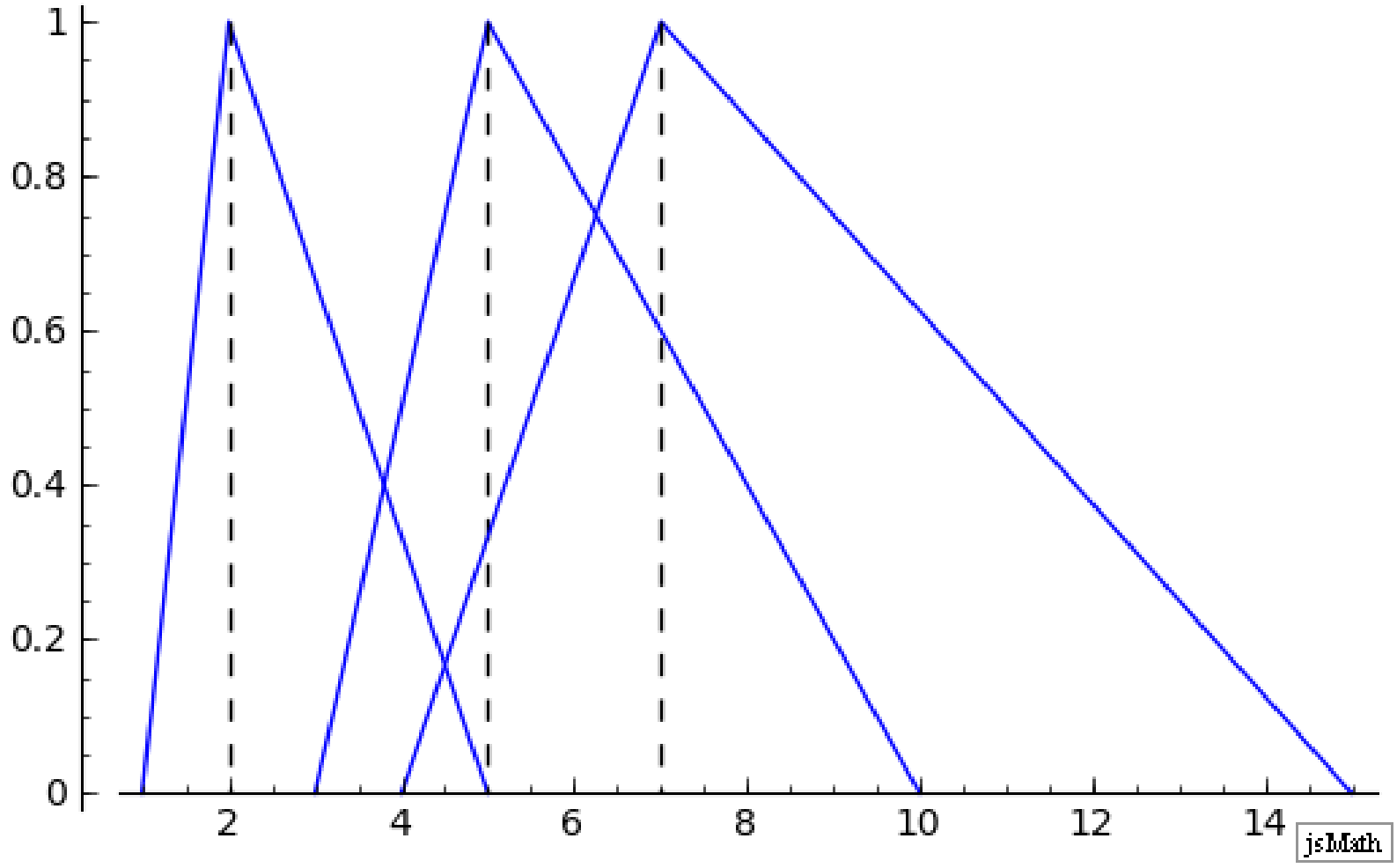

Рис. 2.10. Модель нечіткого трикутного числа

Наприклад:

```
#конструювання матриці 3 чотирьох трикутних чисел
m1=matrix(FF,2,2,[FuzzyNumber(),FuzzyNumber(4),
                  FuzzyNumber(4,9),FuzzyNumber(2,5,8)])
m1[1,0]=m1[1,0]+20
#нечіткий визначник
det(m1)
m2=matrix(FF,2,2) #нечітка нульова матриця
#нечітка одинична матриця
m3=matrix(FF,2,2,[FuzzyNumber(1),FuzzyNumber(0),
                  FuzzyNumber(0),FuzzyNumber(1)])
m1^3 #степінь нечіткої матриці
```

У відповідності до [284], розглянемо методику використання Web-СКМ Sage для побудови фрактальних об’єктів.

Одним із компонентів Sage є СКМ Maxima, що має в своєму складі модуль dynamics для візуалізації динамічних систем та побудови фрактальних об’єктів. Нажаль, в процесі інтеграції Maxima у Sage засоби цього модуля виявились обмежено працездатними у зв’язку з невідповідністю структурі Sage. У зв’язку з цим було виконане розширення можливостей Sage засобами побудови траєкторій руху динамічних систем та фрактальних об’єктів, що відповідають його структурі.

Розроблений Н. А. Хараджян модуль Sage для побудови траєкторій руху динамічних систем фрактальних об’єктів включає наступні функції: evolution, evolution2d_, chaosgame staircase, ifs, orbits.

Функція evolution2d_(fun,state,initial,n) розраховує розташування на площині ($n$+1)-ої точки послідовності, що визначається дискретною динамічною системою з рекурентними співвідношеннями fun($u_{n+1}=F(u_n, v_n)$, $v_{n+1}=G(u_n, v_n)$) з змінними положення state та початковими значеннями initial($u_0$, $v_0$).

*Приклад*. Побудуйте графік динамічної системи:

$$\begin{cases} 10x_{n+1} = 6x_n(1+2x_n)+8y_n(x_n-1)-10y_n^2-9; \\ 10y_{n+1} = x_n(1-6x_n+4y_n)+y_n(1+9y_n)-4. \end{cases}$$

*Розв’язок*. Дві функції, які породжують систему, підставимо в якості аргументів до функції evolution2d_ та отримаємо графік еволюції системи (рис. 2.11):

```
f=0.6*x*(1+2*x)+0.8*y*(x-1)-y^2-0.9
g=0.1*x*(1-6*x+4*y)+0.1*y*(1+9*y)-0.4
a=evolution2d_([f,g], [x,y], [-0.5,0],50000)
show(point(a,pointsize=3))
```

Графіки на рис. 2.11 є фракталами, про що свідчить їх самоподібність (існування частин, які повторюються, при довільному збільшенні будь-якої частини зображення).

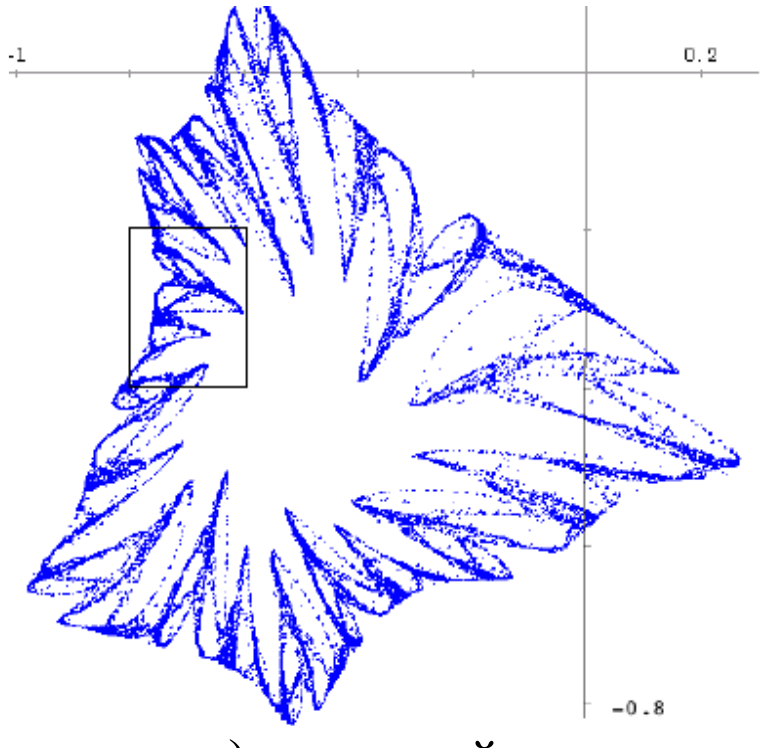

а) загальний вид

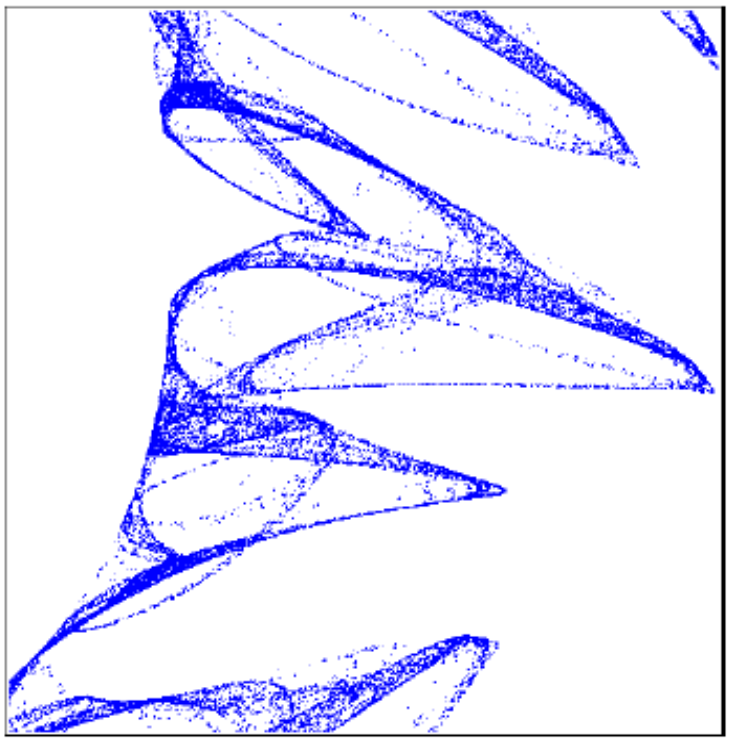

б) фрактал у збільшеній області

Рис. 2.11. Фрактальна динаміка у Sage

Функція chaosgame(point,p0,b,n) реалізує так звану «гру хаосу»: спочатку розраховується точка $p_0=[x_0, y_0]$, що належить списку point=($[x_1, y_1]$, ..., $[x_m, y_m]$), яка обирається у випадковий спосіб. Наступна точка розраховується із відрізку, що з'єднує попередню точку з випадково обраною на відстані, рівній довжині цього відрізка, помноженій на $b$. Процедура повторюється $n$ разів.

Виконавши команди:

```
c1=chaosgame([[0, 0], [1, 0], [0.5,0.866025]], [0.1, 0.1], 
1/2,30000)
show(point(c1))
```

отримаємо фрактальну поверхню − серветку Серпінського (рис. 2.12). На рис. 2.13 зображено дивний атрактор в 5 точках з вершиною в центрі площини та половинним коефіцієнт стиснення.

```
c2=chaosgame([[0,0],[0,1],[1,0],[1,1],
[0.5,0.5]],[0.5,0.5], 1/3,30000)
show(point(c1))
```

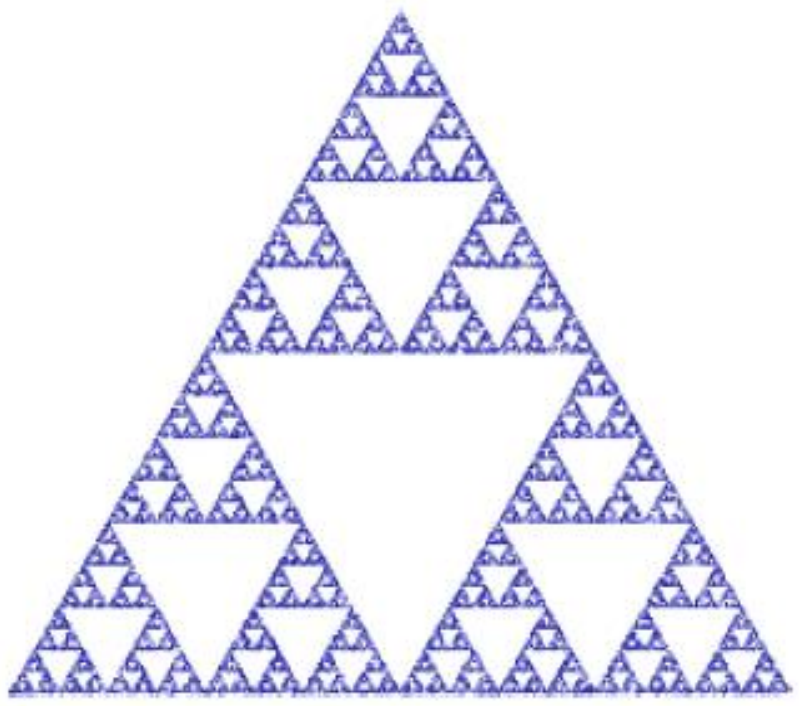

Рис. 2.12. Серветка Серпінського

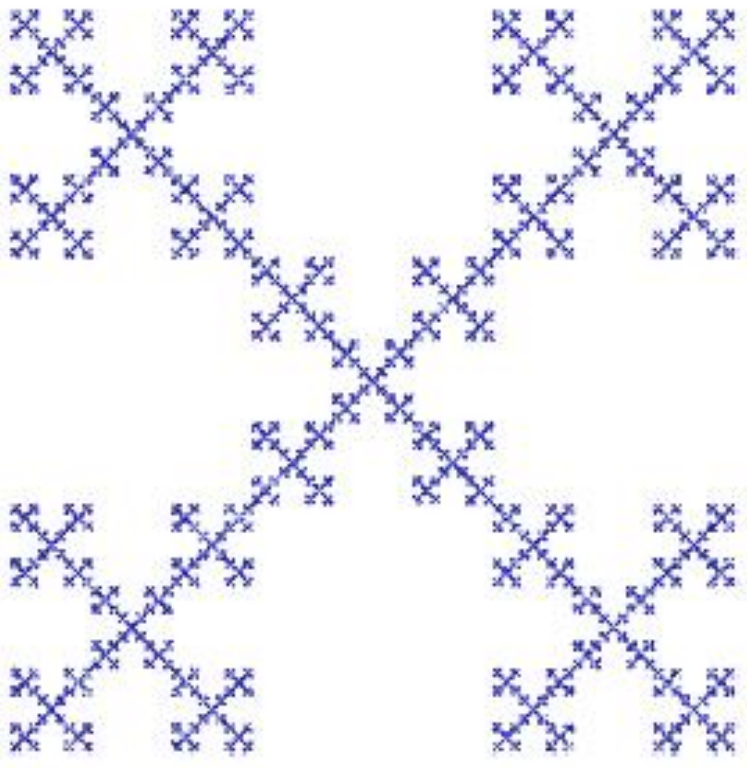

Рис. 2.13. Дивний атрактор

Перед тим, як пояснювати наступну функцію, нагадати студентам форму запису матриць, адже основним параметром наступної функції є список з матриць. Функція ifs(prob,mat,point,p0,n) реалізує метод системи ітеративних функцій. Параметрами функції є перелік ймовірних значень prob із списку $m$ матриць mat, списку $m$ точок point, одна з яких є початковою $p_0$ та кількістю ітерацій $n$, яку треба виконати.

```
m1=matrix([[-0.550, -0.179], [-0.179, 0.550]])
p1=matrix([[-0.438], [0.382]])
m2=matrix([[-0.246, 0.193], [0.275, 0.365]])
p2=matrix([[-0.379], [0.538]])
```

```
m3=matrix([[0.006, -0.014], [-0.147, -0.459]])
p3=matrix([[-0.283], [0.490]])
c=ifs([1,2,3],[m1,m2,m3],[p1,p2,p3],\
[-0.3,0.2],10000)
show(point(c))
```

Для побудови спочатку використовували в якості початкової точки початок координат та починали з малого значення *n*, а для збільшення рівня деталізації використали в якості початкової точки (–0.3, 0.2), що є частиною фракталу, і брали більшу кількість ітерацій.

Засобами функції ifs можна отримати дивну папороть Баренслі (рис. 2.15):

```
a1=matrix([[0.85,0.04],[-0.04,0.85]])
a2=matrix([[0.2,-0.26],[0.23,0.22]])
a3=matrix([[-0.15,0.28],[0.26,0.24]])
a4= matrix([[0,0],[0,0.16]])
p1=matrix([[0],[1.6]])
p2=matrix([[0],[1.6]])
p3=matrix([[0],[0.44]])
p4=matrix([[0],[0]])
c=ifs([85,92,99,100],[a1,a2,a3,a4],
[p1,p1,p3,p4],[5,0],50000)
show(point(c,rgbcolor=(0,1,0)))
```

Для дослідження моделей кліткових автоматів у Sage слід скористатися відповідним доповненням системи – модулем cellular.

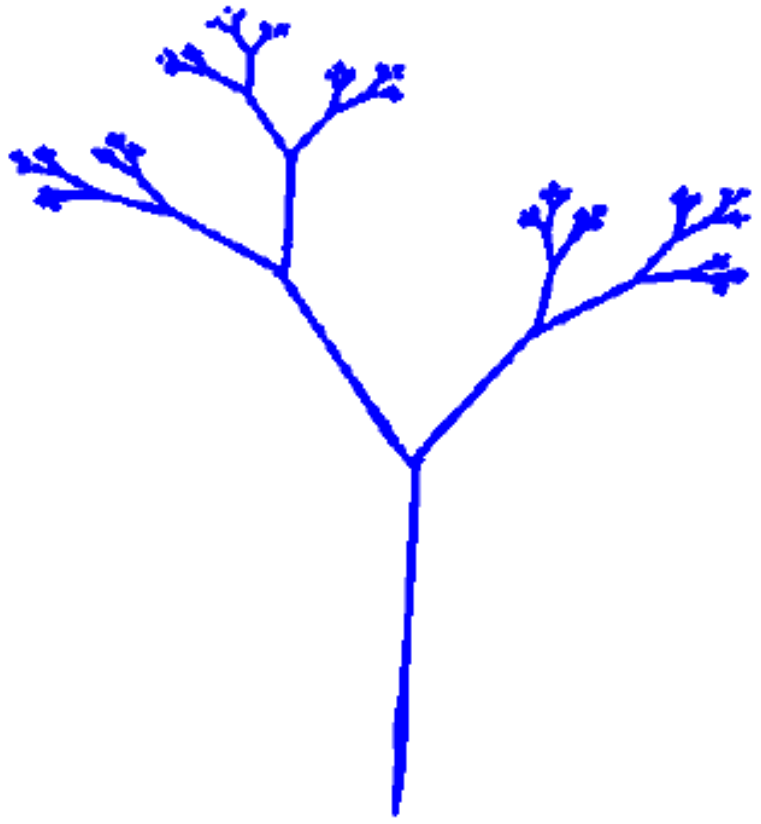

Рис. 2.14. Фрактальне дерево, що утворюється системою з трьох ітераційних функцій

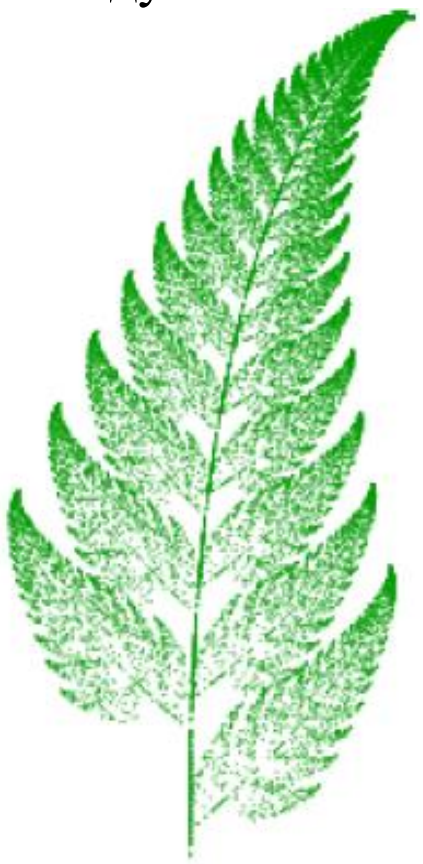

Рис. 2.15. Фрактальний об’єкт «дивна папороть Баренслі», що утворюється системою з чотирьох ітераційних функцій

### *2.4.4 Moodle – соціально-конструктивістське середовище підтримки навчання об'єктно-орієнтованого моделювання*

Moodle – вільно поширювана система підтримки навчання зі зручним інтерфейсом та системою допомоги, засобами підтримки всіх етапів процесу навчання, що виділяє її з переліку інших систем цього ж класу [279, с. 335].

Є. М. Смирновою-Трибульскою визначені вимоги до системи підтримки навчання майбутніх вчителів природничо-математичних дисциплін та показано, що Moodle відповідає цим вимогам [280, с. 333]: наявність інтерфейсу, допомоги та документації рідною мовою; урахування реальних можливостей користувача (просте, інтуїтивне обслуговування на довільному комп'ютері, в будь-якій операційній системі і довільним підключенням до мережі, без необхідності інсталяції спеціального програмного забезпечення та обладнання); урахування реальних технічних і фінансових умов навчального закладу (безкоштовність, невисокі вимоги до обладнання та пропускної здатності мережі); урахування потреб і можливостей викладача (просте керування змістом та навчально-пізнавальною діяльністю користувачів, легка комунікація з ними, можливість швидкого створення документів, простого надання доступу, впорядкування та опису різних типів даних, у тому числі мультимедійних); функціональна еластичність (нескладний початок роботи, можливість розширення наявних компонентів); доступність інструментів, що забезпечують можливість співпраці між користувачами (запис розмов, спілкування між групами); урахування педагогічних вимог (наявність інструментів та засобів для підтримки всіх етапів і компонентів процесу навчання).

Саме тому серед існуючих систем підтримки навчання для розробки спецкурсу «Об'єктно-орієнтоване моделювання» для майбутніх учителів природничо-математичних дисипцлін нами було обрано систему Moodle. Як відзначає автор системи М. Дугіямас (Martin Dougiamas) [22], вона була розроблена на засадах соціального конструктивізму:

1. У справжньому середовищі навчання всі ми одночасно є потенційними вчителями та учнями. Для реалізації цього принципу в Moodle передбачена велика кількість інструментів, що надають широкі можливості студентам зі створення та модифікації навчальних матеріалів (форуми, Wiki, глосарії, бази даних, семінари, блоги, особисті повідомлення та ін.).

2. Ми вчимося особливо гарно, коли створюємо чи намагаємося щось пояснити іншим людям (вчимося, навчаючи). Для реалізації цього принципу у Moodle передбачені наступні інструменти: а) форуми та блоги, що надають можливість обговорення процесу та результатів навчальної діяльності (зокрема, індивідуальними та колективними дослідницькими проектами); б) Wiki, що надає можливість організації спільної роботи над дослідницькими проектами; в) глосарії, що надають можливість організації спільної роботи над переліком термінів; г) бази даних, що є розширенням ідеї глосаріїв до роботи над будь-якими структурованими записами; д) семінари, що надають можливість багатокритеріального обговорення та оцінювання результатів навчальної діяльності студентів. У процесі навчання майбутніх учителів природничо-математичних дисциплін даний принцип є особливо важливим, оскільки «навчання учити» є основою змісту їх психолого-педагогічної та методичної підготовки.

3. Великий вклад у навчання вносить спостереження за діяльністю колег. Для реалізації цього принципу у Moodle передбачено широкий спектр інструментів для спостереження за активністю студентів та викладачів у курсі.

4. Розуміння інших людей дозволить навчати їх більш індивідуально. Для реалізації цього принципу в Moodle передбачено широкий спектр комунікаційних інструментів: форуми, чати, особисті повідомлення, блоги, анкети, опитування, інструменти для доступу до відомостей про активність студентів та викладачів у курсі.

5. Навчальне середовище має бути гнучким, надаючи учасникам освітнього процесу простий інструмент для реалізації їх навчальних потреб. Із урахуванням цього принципу реалізуються усі інструменти Moodle: комунікаційні, навчальні та адміністративні.

Систему Moodle широко використовують у багатьох навчальних закладах України, Росії, Польщі та інших країн для організації дистанційного навчання і підтримки традиційного процесу навчання різних дисциплін. Викладачі використовують широкий спектр реалізованих у системі засобів для надання студентам доступу до навчальних матеріалів, комунікації між суб’єктами навчального процесу, здійснення контролю рівня навчальних досягнень студентів тощо.

Використання Moodle у процесі професійної підготовки майбутніх учителів природничо-математичних дисциплін засобами комп’ютерного моделювання передбачає застосування засобів цієї системи для досягнення бажаних результатів навчання. Зазначимо, що мова йде про гармонійне поєднання засобів традиційного та дистанційного навчання, формування на їх основі відкритого інформаційно-навчального середовища (системи інформаційно-комунікаційних та традиційних засобів, спрямованих на організацію навчальної діяльності студентів). У рішеннях Всесвітньої конференції ЮНЕСКО з вищої освіти 2009 року вказується, що формування компетентностей XXI століття можливе при комплексному застосуванні відкритої та дистанційної освіти і засобів ІКТ, що створюють умови широкого доступу до якісної освіти, зокрема – на основі відкритих освітніх ресурсів [1, с. 3].

Застосування відкритої системи підтримки навчання Moodle створює умови для надання процесу навчання якості неперервності шляхом технологічної інтеграції аудиторної та позааудиторної роботи студентів у систему комбінованого навчання [4]. А. М. Стрюк у [295] визначає комбіноване навчання як педагогічно виважене поєднання технологій традиційного, електронного, дистанційного та мобільного навчання, спрямоване на інтеграцію аудиторного та позааудиторного навчання. У даному визначенні підкреслюється проміжна роль комбінованого навчання між традиційним (переважно аудиторним) і дистанційним (переважно позааудиторним) навчанням, його відповідність системним принципам відкритої освіти (мобільності учасників навчального процесу, рівного доступу до освітніх систем, надання якісної освіти, формування структури та реалізації освітніх послуг) та провідна роль систем підтримки навчання (зокрема, Moodle) в організації навчальної діяльності.

Як відзначає К. Р. Колос, під час конструювання змісту навчання засобами системи Moodle можна забезпечити створення продуктивного освітнього поля, можливостей для творчості, активності, самостійності [136, с. 88], тому в основу створення спецкурсу «Об’єктно-орієнтоване моделювання» у системі Moodle покладені принципи наочності, науковості, індивідуалізації та диференціації навчання.

До навчального курсу у системі Moodle включено відомості стосовно призначення курсу, відомості про можливості консультацій та зустрічей з викладачем, словник основних термінів з навчальної дисципліни, перелік тем індивідуальних та колективних дослідницьких проектів, що пропонуються студентам та ін.

Навчальний матеріал спецкурсу «Об’єктно-орієнтоване моделювання» у системі Moodle структурований за трьома навчальними модулями, до кожного з яких включено:

– методичні вказівки щодо опанування змістом модуля;

– конспекти лекцій з тем модуля (рис. 2.16);

– глосарій (рис. 2.17);

– дидактичні матеріали до тем модуля;

– посилання на ресурси Інтернету, текстові і мультимедійні файли (графічні файли, відео-, аудіофайли) навчального призначення;

– тести (рис. 2.18);

– моделі;

– перелік тем індивідуальних та колективних дослідницьких проектів.

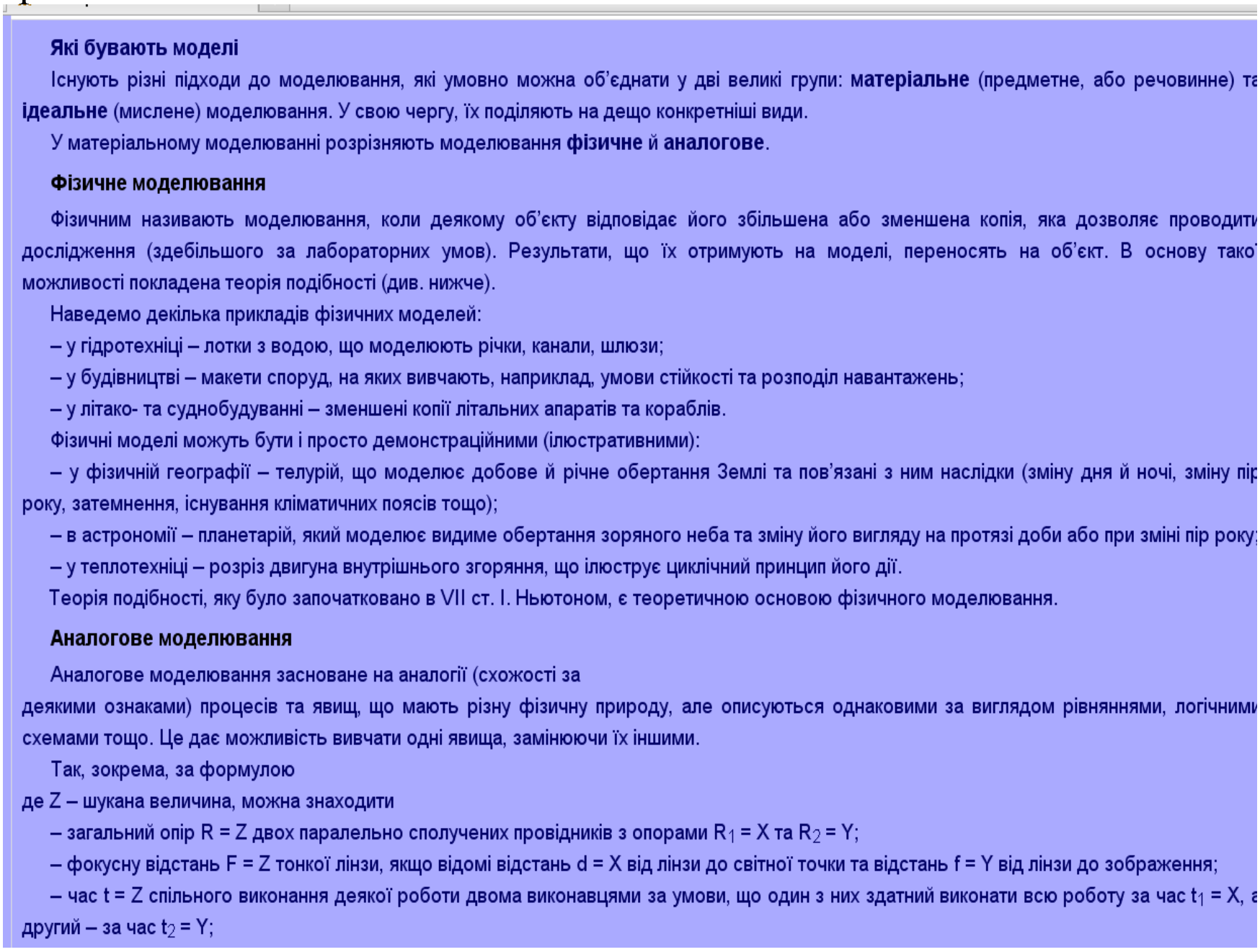

**Які бувають моделі**

Існують різні підходи до моделювання, які умовно можна об’єднати у дві великі групи: **матеріальне** (предметне, або речовинне) та **ідеальне** (мислене) моделювання. У свою чергу, їх поділяють на дещо конкретніші види.

У матеріальному моделюванні розрізняють моделювання **фізичне** й **аналогове**.

**Фізичне моделювання**

Фізичним називають моделювання, коли деякому об’єкту відповідає його збільшена або зменшена копія, яка дозволяє проводити дослідження (здебільшого за лабораторних умов). Результати, що їх отримують на моделі, переносять на об’єкт. В основу такої можливості покладена теорія подібності (див. нижче).

Наведемо декілька прикладів фізичних моделей:

– у гідротехніці – лотки з водою, що моделюють річки, канали, шлюзи;

– у будівництві – макети споруд, на яких вивчають, наприклад, умови стійкості та розподіл навантажень;

– у літако- та суднобудуванні – зменшені копії літальних апаратів та кораблів.

Фізичні моделі можуть бути і просто демонстраційними (ілюстративними):

– у фізичній географії – телурій, що моделює добове й річне обертання Землі та пов’язані з ним наслідки (зміну дня й ночі, зміну пір року, затемнення, існування кліматичних поясів тощо);

– в астрономії – планетарій, який моделює видиме обертання зоряного неба та зміну його вигляду на протязі доби або при зміні пір року;

– у теплотехніці – розріз двигуна внутрішнього згоряння, що ілюструє циклічний принцип його дії.

Теорія подібності, яку було започатковано в VII ст. І. Ньютоном, є теоретичною основою фізичного моделювання.

**Аналогове моделювання**

Аналогове моделювання засноване на аналогії (схожості за деякими ознаками) процесів та явищ, що мають різну фізичну природу, але описуються однаковими за виглядом рівняннями, логічними схемами тощо. Це дає можливість вивчати одні явища, замінюючи їх іншими.

Так, зокрема, за формулою

де Z – шукана величина, можна знаходити

– загальний опір R = Z двох паралельно сполучених провідників з опорами $R_1$ = X та $R_2$ = Y;

– фокусну відстань F = Z тонкої лінзи, якщо відомі відстань d = X від лінзи до світної точки та відстань f = Y від лінзи до зображення;

– час t = Z спільного виконання деякої роботи двома виконавцями за умови, що один з них здатний виконати всю роботу за час $t_1$ = X, а другий – за час $t_2$ = Y;

Рис. 2.16. Фрагмент конспекту вступної лекції до курсу

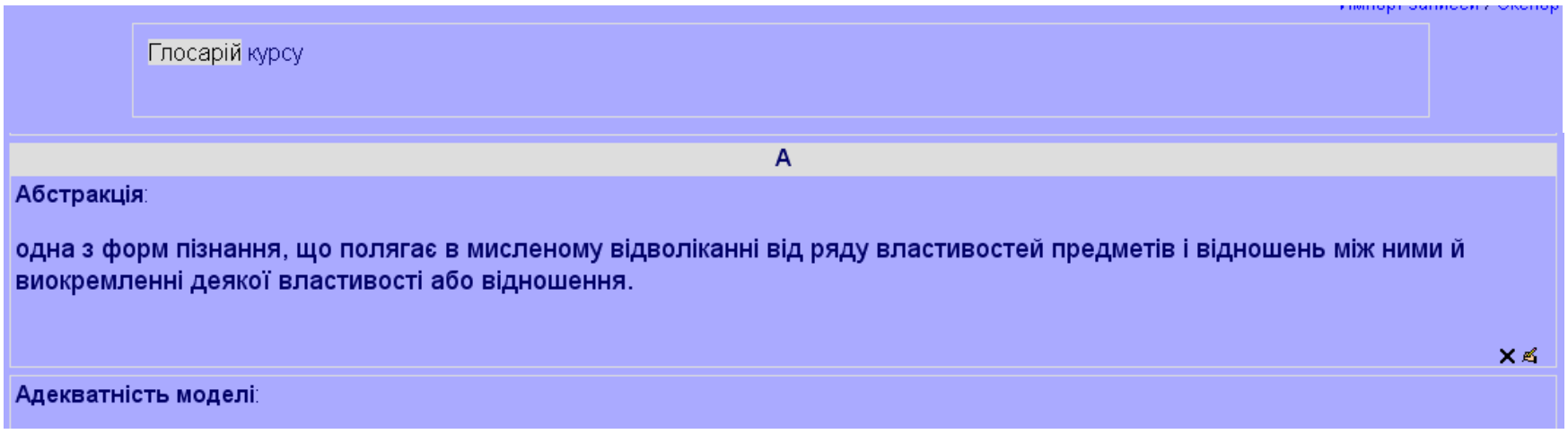

Глосарій курсу

А

**Абстракція**:

**одна з форм пізнання, що полягає в мисленому відволіканні від ряду властивостей предметів і відношень між ними й виокремленні деякої властивості або відношення.**

**Адекватність моделі**:

Рис. 2.17. Фрагмент глосарію

Використовуючи розроблений таким чином курс, студенти мають змогу: обирати довільний навчальний модуль, тему курсу; переглядати конспект лекції з теми, зміст основних понять і фактів; опановувати навчальний матеріал та переглядати приклади моделей, завантажуючи файли з дидактичними матеріалами; ознайомлюватися з мультимедійними, зокрема мережними, ресурсами до тем курсу, користуючись відповідними гіперпосиланнями; переглядати протоколи лабораторних робіт, методичні вказівки щодо їх виконання; проходити тестування за обраною темою або за змістом декількох тем (у навчальному чи контролюючому режимі); розміщувати в Moodle свої індивідуальні та колективні дослідницькі проекти, власні портфоліо тощо.

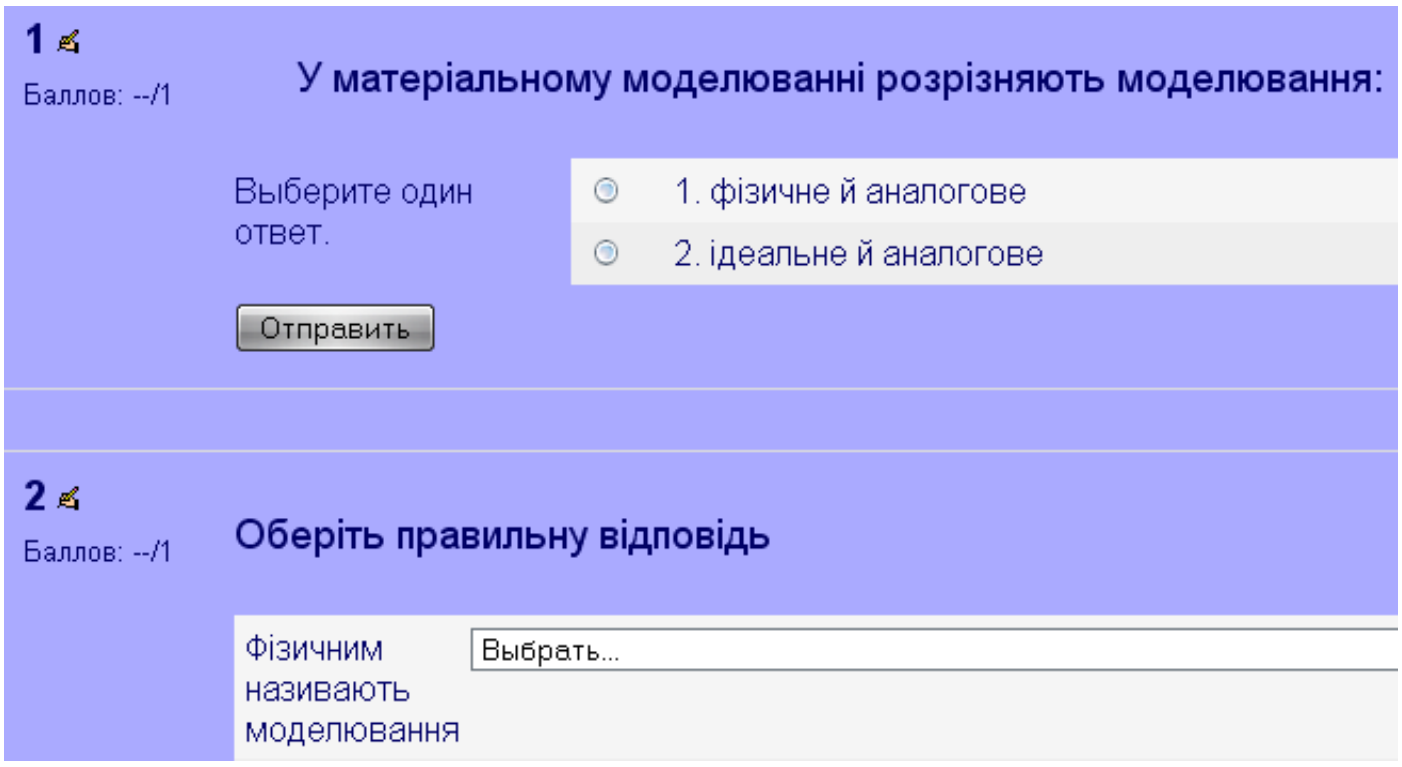

Рис. 2.18. Фрагмент тесту до першого модуля

До дидактичних матеріалів курсу належать моделі, створені у середовищах об’єктно-орієнтованого моделювання, текстові, відео файли, HTML-сторінки.

Поняття, що зустрічаються в курсі, є елементами соціального досвіду, який студенти мають зробити своїм індивідуальним досвідом: означення поняття виступає як деяка точка зору, основа для оцінки об’єктів, які досліджують студенти. Так, наприклад, поняття «модель» може бути означене так: 1) модель – довільний образ, аналог (мислений чи умовний: зображення, опис, схема, креслення, графік, план, карта і т. п.) певного об’єкта, процесу або явища («оригіналу», «прообразу» даної моделі), який використовується як його «замісник», «представник»; 2) модель – реально існуюча або уявна система, яка замінює і відображає у пізнавальних процесах іншу систему – оригінал, знаходиться з нею у відношенні подібності, завдяки чому, досліджуючи модель можна одержати відомості про оригінал; 3) модель – деякий матеріальний чи уявний об’єкт, який у процесі вивчення замінює об’єкт-оригінал, зберігаючи певні важливі для дослідження типові характеристики і властивості оригіналу.

Аналізуючи різні означення поняття «модель», виділяючи в них суттєві ознаки, студенти можуть вже на етапі ознайомлення з новим поняттям зробити висновок щодо того, чи можна вважати той або інший об’єкт моделлю виучуваного явища чи процесу. Оцінювання студентами різноманітних об’єктів щодо належності їх до певного класу об’єктів сприяє формуванню поняття у свідомості студентів як абстрактного, узагальненого образу.

Після формулювання означення воно має бути включене до складу навчальних дій студентів, які вони виконують з відповідними об'єктами (дії розпізнавання, виведення наслідків, класифікації; дії, пов'язані зі встановленням ієрархічних відношень у системі понять тощо). Здійснення зазначених дій, у свою чергу, передбачає розвиток у студентів прийомів наукового мислення (аналізу, синтезу, порівняння, оцінювання, узагальнення та ін.), конструювання системи базових знань та умінь з навчальної дисципліни.

У процесі навчання об'єктно-орієнтованого моделювання студенти створюють записи у глосаріях до курсу в середовищі Moodle, причому це відбувається як на етапі першого знайомства студентів з поняттями та об'єктами, так і в процесі подальшого оволодіння навчальним матеріалом (доповнення записів новими відомостями про класи, об'єкти, зв'язки з іншими об'єктами, процесами, явищами тощо). Наприклад, сформульоване на етапі ознайомлення з новим матеріалом означення поняття «модель» студенти можуть доповнити відомостями про види моделей, ієрархічні зв'язки між ними, прикладами типових моделей тощо. Таке поетапне створення записів у глосаріях сприяє кращому засвоєнню понять, міцності здобутих знань, встановленню внутрішньопредметних та міжпредметних зв'язків, формуванню у студентів здатності до рефлексії своєї навчально-пізнавальної діяльності, запобігає формалізму при засвоєнні понять.

За допомогою засобів Moodle для роботи зі словниками викладач може вибрати зі списку зручний формат перегляду глосарію; переглядати прізвища авторів записів; за необхідності встановити режим, при якому всі записи студентів спочатку передаються на розгляд викладачу і лише у випадку схвалення стають доступними для перегляду всім; дозволити або заборонити студентам коментувати, друкувати, оцінювати записи колег, редагувати власні записи, визначати термін більше одного разу; вибрати шкалу оцінювання записів; встановити автоматичне посилання на записи у глосарії тощо.

Послуга автоматичного посилання на записи у глосарії дає змогу студентам швидко переглянути означення понять, що зустрічаються в процесі опанування змісту спецкурсу з використанням Moodle. Це сприяє відновленню в пам'яті змісту понять, спонукає студентів до діяльності щодо доповнення глосаріїв новими записами.

Спільна робота з конструювання глосаріїв, що передбачає додавання студентами коментарів та оцінок до запропонованих їхніми колегами означень понять, є елементом навчального дослідження, що сприяє підвищенню інтересу студентів до предмета, рівня їх навчальної та професійної мотивації. Студенти активно включаються в таку роботу, знаходять помилки в міркуваннях свої колег, здійснюють взаємне оцінювання, редагують свої записи. При цьому викладач виконує роль експерта, переглядає та оцінює записи студентів, вказує на їхні помилки, спрямовує дискусію у потрібному напрямку, встановлює терміни роботи студентів над створенням глосарію, додає схвалені ним означення до «глобального» глосарію.

Використання системи гіперпосилань між елементами спецкурсу в системі Moodle надає студентам можливість обирати розділи, теми спецкурсу, рівень складності опанування матеріалом, працювати у зручному для них темпі. Диференціації навчання сприяє те, що в системі Moodle викладач може організувати колективну роботу студентів у групах при виконанні ними завдань дослідницьких проектів тощо.

Розроблений курс відповідає виділеним Н. В. Морзе критеріям повноти структури курсу (рис. 2.19) та навчальним матеріалам з модулів курсу (рис. 2.20).

Результати індивідуальних та колективних навчальних досліджень можуть бути представлені у системі Moodle її засобами:

1) проекти, виконані у середовищі об’єктно-орієнтованого моделювання Alice, можуть бути включені до системи віртуальної реальності Wonderland, доступ до якої можливий засобами відповідного модуля системи Moodle (проекти, розроблені у Alice 2.0, 2.2 та похідної від них системи Rebeca, можуть бути представлені на сторінках Moodle за допомогою Java-аплету aliceapplet.jar та бібліотеки Java3D);

2) результати роботи проектів, виконаних у середовищі об’єктно-орієнтованого моделювання Alice, можуть бути засобами середовища розміщені на відеосервері YouTube, а відповідні посилання на відеоролики – на сторінках Moodle;

3) проекти, виконані у середовищі об’єктно-орієнтованого моделювання NetLogo, можуть бути розміщені на сторінках Moodle за допомогою відповідного модуля системи;

4) проекти, виконані у середовищі об’єктно-орієнтованого моделювання Sage, можуть бути розміщені на сторінках Moodle на рівні фреймової інтеграції Web-інтерфейсів [386, с. 85];

5) проекти, виконані у середовищі об’єктно-орієнтованого моделювання Squeak, можуть бути збережені разом з образом віртуальної машини середовища, а відповідні посилання – розміщені на сторінках Moodle;

6) проекти, виконані за допомогою засобів модуля Visual (у середовищах VPython, VPNBody, PyGeo), можуть бути представлені у Moodle вихідними програмними текстами мовою Python.

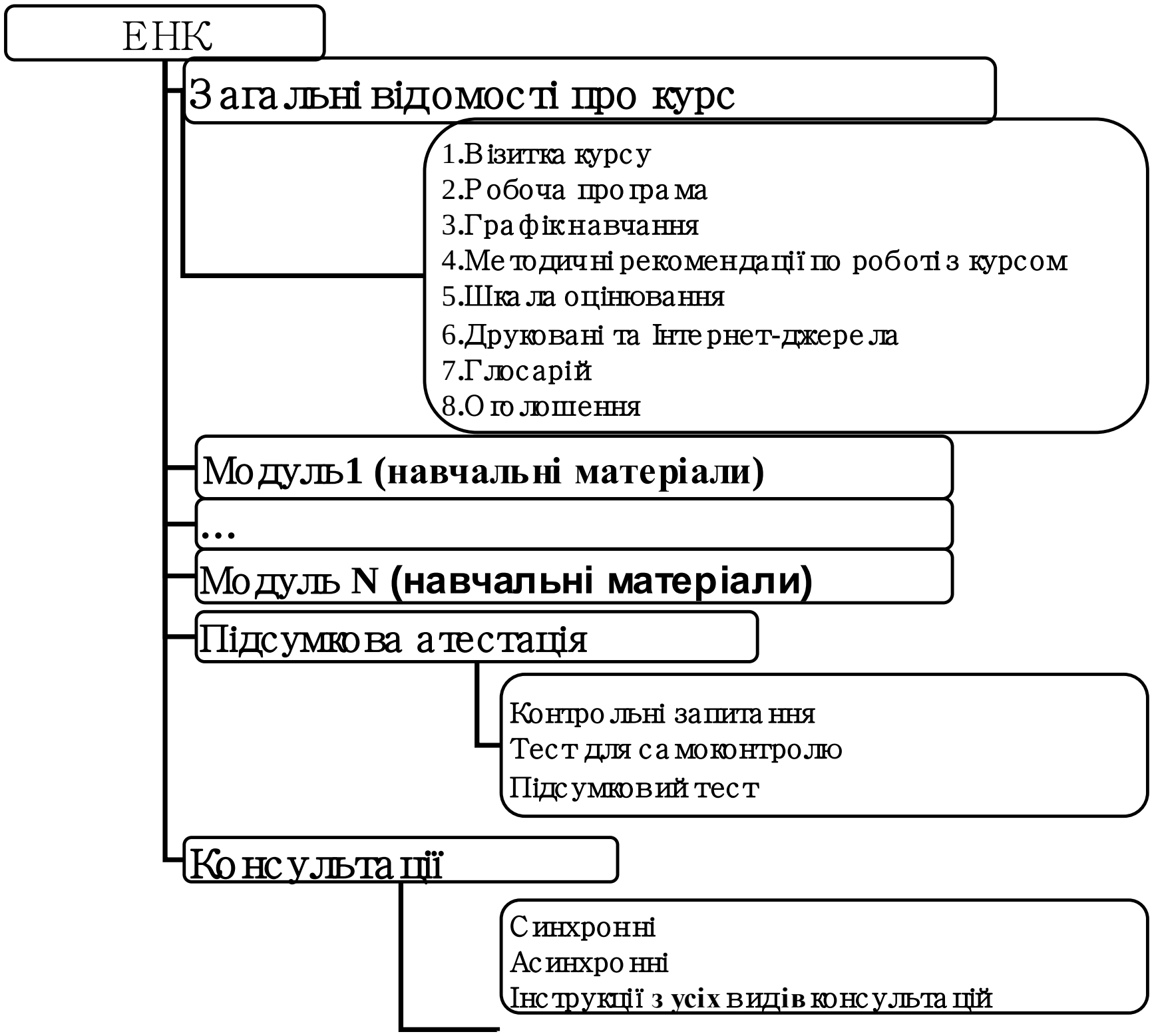

Рис. 2.19. Критерії повноти структури курсу (за Н. В. Морзе)

Таким чином, використання розробленого у системі Moodle електронного навчального курсу з об'єктно-орієнтованого моделювання надає можливість організувати індивідуальну та колективну роботу студентів з оволодіння навчальним матеріалом курсу у процесі виконання навчальних досліджень. Соціально-конструктивістський підхід до навчання об'єктно-орієнтованого моделювання сприяє інтенсифікації, індивідуалізації, диференціації та соціалізації навчання. Крім того, як зазначає М. В. Рафальська, в процесі роботи в середовищі розробленого курсу майбутні вчителі оволодівають прийомами роботи з Moodle, освоюють технологію розробки навчальних матеріалів, що є необхідною умовою формування їх інформатичних компетентностей [218].

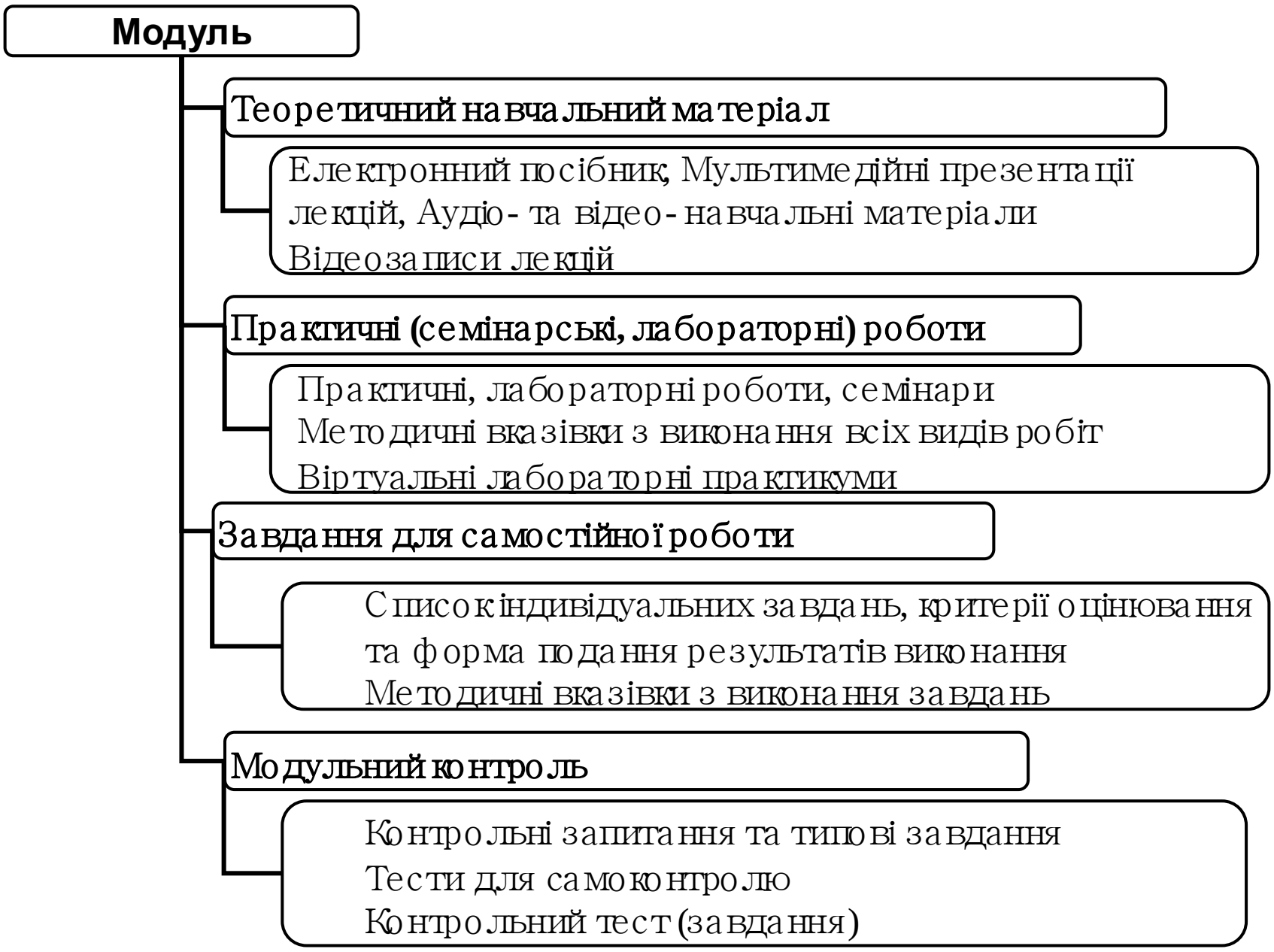

Рис. 2.20. Критерії повноти навчальних матеріалів з модулів курсу (за Н. В. Морзе)

### *2.4.5 Соціально-конструктивістські засоби Web 2.0*

Саме ці засоби, вичерпно описані у роботах Є. Д. Патаракіна [188; 189], утворюють технологічну основу комбінованого навчання об'єктно-орієнтованого моделювання.

Мережна інфраструктура, завдяки якій студенти одержують доступ до Web, відкриває можливості для використання в педагогічній практиці соціальних сервісів.

Загальні складові соціальних сервісів такі:

– використання соціальних сервісів підштовхують людей до участі в колективній діяльності. Сервіси надають прості рішення для повсякденних завдань – поширення фотографій та відеороліків, пошук потрібних відомостей;

– беручи участь у колективній діяльності з використанням соціальних сервісів, люди змінюють свою позицію зі споживчої на творчу (дослідницьку);

– соціальні сервіси допомагають людям взаємодіяти між собою, задаючи прості правила такої взаємодії.

Мережні сервіси перетворюються на засоби, за допомогою яких користувачам можна співпрацювати. Сучасні мережні організації ґрунтуються на участі й співробітництві людей, вони надають людям засоби й можливості вкладатися в розвиток загального змісту через створення Web-сайтів, блоггінг, оцінювання книг, розміщення в мережі фотографій і відео, участь у колективному редагуванні енциклопедій тощо.

Дж. Фішер (G. Fischer) пише про зміни, що відбуваються в освіті, використовуючи поняття мережної співучасті та метадизайну [21]. В епоху телебачення й Web 1.0 розроблювачі навчального середовища створювали й фільтрували навчальні потоки, кінцевими споживачами яких були учні (студенти). На сучасному етапі розробники навчальних систем і дизайнери навчальних мереж повинні планувати діяльність таким чином, щоб учні (студенти) могли не тільки знайомитися зі змістом, але й виступати в ролі його авторів (співтворців).

Серед різноманіття способів класифікації сервісів Web 2.0 найбільш простою є класифікація за принципом «що можна робити за допомогою цього засобу»:

– Wiki – сфера діяльності, у якій автори працюють над Wiki-сторінками колективних гіпертекстів;

– блоги – сфера діяльності, у якій окремі автори залишають свої записи, а інші їх коментують;

– пошукова сфера, у якій учасники шукають, зберігають і класифікують знайдені відомості;

– соціальні мережі – сфера діяльності, у якій люди встановлюють зв’язки один з одним;

– карти;

– логосфера – сфера діяльності, у якій автори створюють і обмінюються своїми програмами та їхніми фрагментами;

– хмари сервісів, у яких учасники використовують все різноманіття сервісів, зібраних «під парасолькою» якоїсь однієї корпорації.

Соціальні сервіси й діяльності усередині мережних співтовариств у процесі навчання об’єктно-орієнтованого моделювання надають наступні можливості:

1) застосування відкритих, безкоштовних та вільних електронних ресурсів;

2) самостійна та колективна розробка мережних навчальних матеріалів;

3) опанування навичок пошуку, використання та інтеграції різноманітних начальних матеріалів;

4) спостереження за діяльністю учасників співтовариства;

5) виконання індивідуальних та колективних мережних досліджень.

Так, система Moodle, починаючи з версії 2.0, надає можливість інтеграції навчальних курсів із соціальними сервісами Інтернет. Організацію колективної роботи над студентськими дослідницькими проектами доцільно координувати за допомогою Документів Google – хмари сервісів, використання яких надає можливість втілити в життя проектну форму роботи на всіх етапах співпраці – від постановки задач до оформлення звітів та їх подання [166, с. 50]. Серед можливих способів обміну та спільного використання Документів Google І. С. Мінтій виокремлює такі: 1) спільне використання документу (надання іншому користувачу прав читача чи співавтора); 2) надсилання електронною поштою як вкладення; 3) публікація в Інтернет.

Таким чином, Документи Google надають можливість організувати як традиційні способи обміну документами (надсилання електронною поштою як вкладення), так і нові: спільна робота з документами в он-лайн режимі дозволяє реалізувати повноцінну співпрацю всіх учасників проекту, а публікація документів в мережі Інтернет сприяє ознайомленню з ними широкого загалу користувачів.

Структура логосфери суттєво різниться для різних середовищ об’єктно-орієнтованого моделювання: від розміщення результатів моделювання у вигляді відеоролику у Alice та обміну вихідними текстами програм через форум до організації колективної роботи над дослідницькими проектами у Web-середовищі.

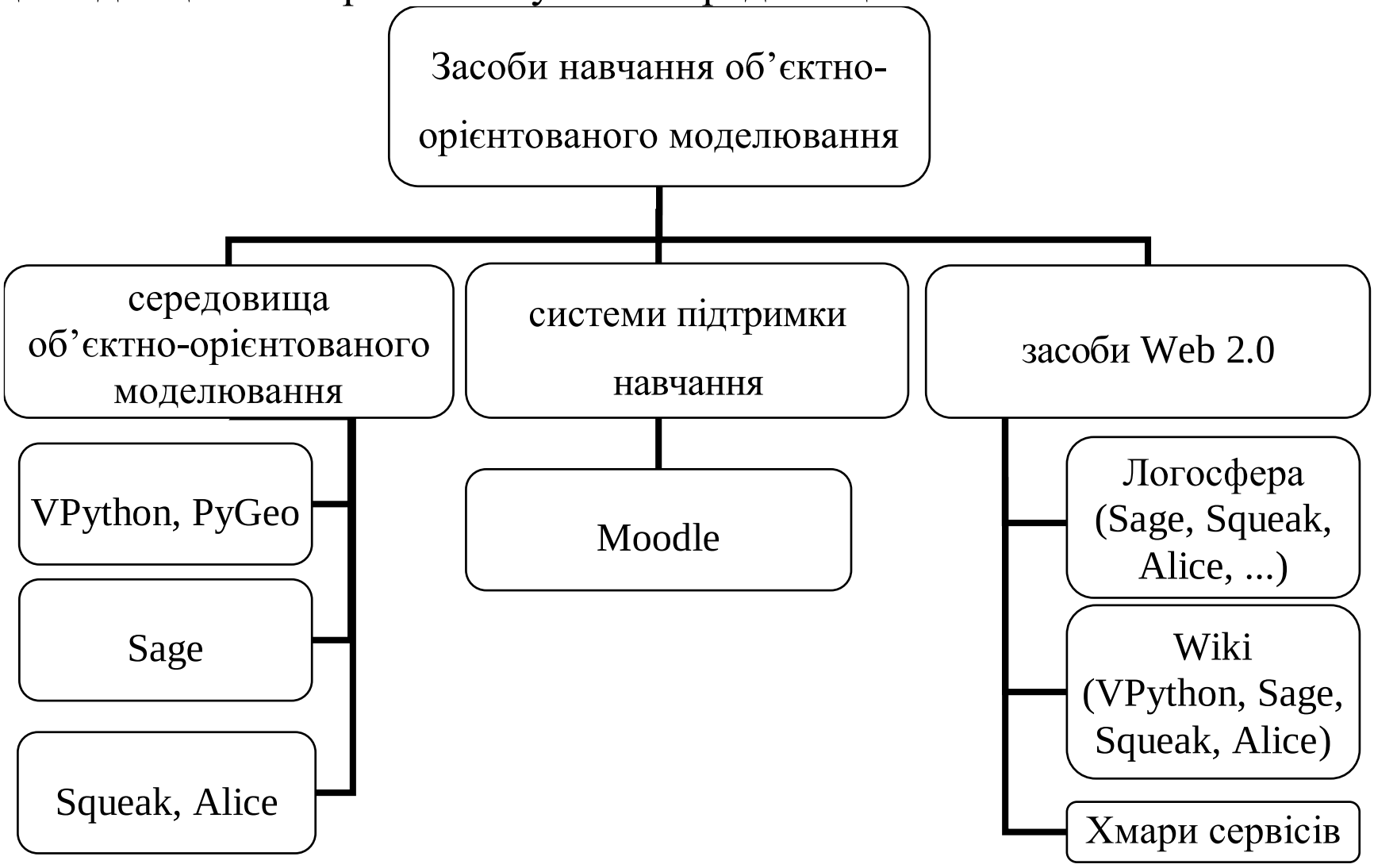

Рис. 2.21. Система засобів навчання об’єктно-орієнтованого моделювання

На рис. 2.21 показано засоби навчання об'єктно-орієнтованого моделювання: середовища об'єктно-орієнтованого моделювання (насамперед, VPython, PyGeo, Sage, Squeak та Alice), системи підтримки навчання (Moodle) та засоби Web 2.0 (логосферу, Wiki та хмари сервісів).

### 2.5 Критерії, рівні та показники сформованості компетентності з комп'ютерного моделювання майбутніх учителів природничо-математичних дисциплін

Процес формування тієї чи іншої компетентності може бути досить тривалий і здійснюватися під впливом різних факторів (навчання у закладах освіти, професійна діяльність, міжособистісне спілкування), тому говорити про наявність у студентів певної компетентності некоректно. Далі, говорячи про набуття студентами певних компетентностей, будемо розуміти їх сформованість на певному рівні [218, с. 24].

Таким чином, процес формування компетентності з комп'ютерного моделювання є перехід від одного рівня (нижчого) до іншого (вищого), а для оцінювання рівня її сформованості необхідно виокремити самі рівні її сформованості та критерії, які б надали можливість здійснювати перевірку рівня сформованості компетентності.

Аналіз літератури надав можливість зробити висновок, що не існує одностайної думки стосовно кількості рівнів сформованості компетентності.

Так, О. М. Спірін виокремлює шість основних рівнів (початковий, мінімально-базовий, базовий, підвищений, поглиблений та дослідницький) [292]. Є. М. Смирнова-Трибульска [280] виділяє дев'ять рівнів, що згруповані у три блоки: 0-2 – базовий (елементарний) рівень, 3-5 – середній (функціональний) рівень, 6-8 – просунутий (системний) рівень.

Таким чином, у формуванні компетентності студентів можна виділити три рівні (низький, середній, високий) та три стадії (становлення, активного розвитку, саморозвитку). Стадії «горизонтального» просування (становлення, активного розвитку, саморозвитку) відбивають кількісне накопичення «критичної маси» суб'єктивних характеристик компетентності в кожного студента. «Вертикальне» просування – це якісний стрибок як перехід на вищий рівень розвитку.

На стадії становлення відбувається засвоєння студентами знань, вироблення умінь на репродуктивному рівні, формування мотивації до вивчення предмету та позитивного ставлення до програмування. На стадії активного розвитку студенти осмислено оперують уміннями та знаннями, мають потребу в особистій самореалізації в інформаційному середовищі, мають такі розвинені якості, як рефлективність, креативність, критичність мислення, мають сформовані навички саморегуляції діяльності. Основна мета стадії саморозвитку – розвиток самостійності, творчої активності, самоорганізації та самоуправління, актуалізація потреби у саморозвитку.

Я. Б. Сікора розглядає такі рівні сформованості компетентності: адаптивний (низький), алгоритмічний (середній), частково-пошуковий (достатній), творчий (високий) [272, с. 11].

Адаптивний рівень характеризується недостатньою сформованістю професійних намірів, відсутністю необхідних знань та вмінь, репродуктивним виконанням діяльності, неадекватною самооцінкою.

Для алгоритмічного рівня характерним є епізодичний інтерес до професії, недостатнє вміння використовувати наявні знання, нестійка потреба у самовдосконаленні.

Частково-пошуковий рівень відрізняється розвиненою суб'єктною позицією, яка виявляється в усвідомленості своїх дій та можливостей, прагненні до прийняття рішень, внесенні змін при використанні запозиченого досвіду, наявністю інтересу до професії, розумінням її значущості, проте недостатньою чіткістю у визначенні цілей формування компетентності.

Творчий рівень характеризується сформованістю стійкого інтересу до професії; здатністю до нестандартного розв'язання завдань, поглиблення знань та прийняттям усвідомлених рішень з урахуванням прогнозування наслідків своїх дій; прагненням до самовираження, самовдосконалення, об'єктивної самооцінки в професійній діяльності; володінням способами самодіагностики і саморозвитку.

С. О. Дружилов у загальному випадку в моделі навчання виділяє чотири стадії сформованості компетентності, що характеризують процес навчання, починаючи від стадії початкового знайомства з новим матеріалом (знаннями, концепціями, навичками) і закінчуючи стадією сформованої компетентності [105, с. 33].

Перша стадія: несвідома некомпетентність – у людини немає необхідних знань, умінь, навичок, і вона не знає про їх відсутність або взагалі про можливі вимоги щодо них для успішної діяльності. Ця стадія характеризується наступною професійної самооцінкою: «Я не знаю, що я не знаю». Коли людина усвідомлює відсутність знань, умінь, навичок, необхідних для даної діяльності, вона переходить на другу стадію.

Друга стадія: свідома некомпетентність – людина усвідомлює, що їй не вистачає професійних знань, умінь, навичок. Тут можливі два результати а) конструктивний (як форма прояву особистісної та професійної активності) і б) деструктивний (форма соціальної пасивності). Конструктивний шлях означає, що усвідомлення суб'єктом своєї професійної некомпетентності сприяє підвищенню його мотивації на здобуття відсутніх професійних знань, умінь, навичок. Деструктивний результат може призводити до виникнення почуття невпевненості у своїх силах, психологічного дискомфорту, підвищеної тривожності та ін., що заважає подальшому професійному навчанню. Для другої стадії характерна наступна професійна рефлексія суб'єкта: «Я знаю, що я не знаю».

Третя стадія: свідома компетентність – людина знає, що входить в структуру і становить зміст її професійних знань, умінь і навичок і може їх ефективно застосовувати. Для третьої стадії характерна професійна самооцінка суб'єкта в такій формі: «Я знаю, що я знаю».

Четверта стадія: несвідома компетентність – коли професійні навички повністю інтегровані, вбудовані в поведінку; професіоналізм є частиною особистості. Несвідома компетентність характеризує рівень майстерності.

Таким чином, дослідники виділяють, як правило, 3 рівні сформованості компетентності: низький, достатній та високий. У дослідженні ми розглядатимемо процес формування компетентності з комп'ютерного моделювання як перехід між цими рівнями: низький → достатній → високий.

*Низький рівень* характеризується негативним або індиферентним ставленням до процесу розробки, опису, налагодження, тестування комп'ютерних моделей та аналізу результатів їх роботи; поверхневими, несистемними знаннями з моделювання, наявністю окремих, розрізнених вмінь; слабкою мотивацією до опанування моделювання. *Достатній рівень* передбачає виявлення інтересу до процесу розробки, опису, налагодження, тестування комп'ютерних моделей та аналізу результатів їх роботи; упорядкованими, структурованими знаннями, достатніми вміннями; проявленням здатності до співпраці у процесі моделювання, використанням засобів для організації спільної роботи над проектом; здатністю до самонавчання. *Високий рівень* характеризується позитивним ставленням до процесу розробки, опису, налагодження, тестування комп'ютерних моделей та аналізу результатів їх роботи; стійкими, ґрунтовними знаннями, творчим підходом, уміннями до нестандартного розв'язання завдань, умінням відстоювати власну думку, постійною здатністю до співпраці у процесі моделювання, використанням засобів для організації спільної роботи; здатністю до самонавчання.

Діагностику рівнів сформованості компетентності з комп'ютерного моделювання виконуватимемо за показниками з табл. 2.2.

*Таблиця 2.2*

**Оцінювання рівнів сформованості компетентності з комп'ютерного моделювання та вага складових у загальній сформованості компетентності**

| *i* | Складова ($S_i, i=\overline{1,4}$) | Рівні сформованості складової | | | вага складової ($p_i, i=\overline{1,4}$) |
|---|---|---|---|---|---|
| | | низький | достатній | високий | |
| | | 0 | 1 | 2 | |
| 1 | гносеологічна | має низький рівень знань про особливості комп'ютерного моделювання, види моделей та етапи моделювання | має певні знання про особливості комп'ютерного моделювання, види моделей та етапи моделювання | має високий рівень знань про особливості комп'ютерного моделювання, види моделей та етапи моделювання | 0,4 |
| 2 | праксеологічна | не вміє пояснити призначення та функції побудованої моделі, описати етапи розробки моделі у середовищі моделювання, пояснити відмінність реалізацій моделі, спроектувати, описати, перевірити та проаналізувати результати комп'ютерного моделювання | частково вміє пояснити призначення та функції побудованої моделі, описати етапи розробки моделі у середовищі моделювання, пояснити відмінність реалізацій моделі, спроектувати, описати, перевірити та проаналізувати результати комп'ютерного моделювання | вміє пояснити призначення та функції побудованої моделі, описати етапи розробки моделі у середовищі моделювання, пояснити відмінність реалізацій моделі, спроектувати, описати, перевірити та проаналізувати результати комп'ютерного моделювання | 0,3 |

| $i$ | **Складова** $(S_i, i=\overline{1,4})$ | **Рівні сформованості складової** | | | **вага складової** $(p_i, i=\overline{1,4})$ |
|---|---|---|---|---|---|
| | | **низький** | **достатній** | **високий** | |
| | | **0** | **1** | **2** | |
| 3 | аксіологічна | процес проектування, опису, перевірки та аналізу результатів моделювання не становить для студента жодної цінності; внутрішня мотивація до опанування комп'ютерного моделювання відсутня | низький рівень зацікавленості процесами проектування, опису, перевірки та аналізу результатів моделювання; присутня певна внутрішня мотивація до опанування комп'ютерного моделювання | висока зацікавленість та внутрішня мотивація до опанування комп'ютерного моделювання, готовність застосовувати набуті знання та вміння у практичній діяльності | 0,2 |
| 4 | соціально-поведінкова | відсутня здатність до співпраці у процесі комп'ютерного моделювання; невміння використовувати засоби спільної роботи в процесі розробки моделей | інколи має здатність до співпраці у процесі комп'ютерного моделювання; частково вміє використовувати засоби спільної роботи в процесі розробки моделей | постійна готовність до співпраці з комп'ютерного моделювання; раціональне та творче використання засобів спільної роботи в процесі розробки моделей | 0,1 |

Кожна складова компетентності з комп'ютерного моделювання оцінюватиметься за трьохбальною шкалою (від 0 до 2), що відповідає низькому, достатньому та високому рівню сформованості відповідної складової. Показники сформованості складової діагностуватимуться методами педагогічного спостереження, в процесі контролю знань, захисту проектів. Вага кожної складової у загальній сформованості компетентності з комп'ютерного моделювання визначалась методом експертних оцінок. Таким чином, числове значення рівня сформованості компетентності з комп'ютерного моделювання може бути визначено за формулою:

$$B = k\sum_{i=1}^{4} S_i p_i \ .$$

Множник *k* перед сумою (в нашому випадку – 50) обирається для того, щоб отримане в результаті числове значення рівня сформованості компетентності було цілим числом. Таким чином, у відповідності до експертних оцінок ваги всіх складових компетентності з комп'ютерного моделювання, значення *B* може бути в межах від 0 до 100. Отримані числові значення розподілятимемо за рівнями сформованості компетентності з комп'ютерного моделювання у такий спосіб: 0-39 балів – низький рівень, 40-74 бали – достатній, 75-100 балів – високий.

**Висновки до розділу 2**

На основі теоретично обґрунтованих у розділі 1 педагогічних умов була розроблена модель професійної підготовки майбутніх учителів природничо-математичних дисциплін засобами комп'ютерного моделювання, спрямована на формування активної навчально-пізнавальної діяльності студентів, розвитку їх інтелектуальних здібностей, пізнавальної самостійності, навичок спільної навчально-дослідницької діяльності, умінь використання засобів комп'ютерного моделювання і соціально-конструктивістських технологій. Основними складовими моделі є система вимог до професійної підготовки майбутніх учителів природничо-математичних дисциплін, умови їх підготовки засобами комп'ютерного моделювання та система реалізації педагогічних умов, що включає в себе соціально-конструктивістські засоби навчання, методи навчання (загальнодидактичні та спеціальні) та форми організації навчання (традиційні та соціально-конструктивістські), підпорядковані загальній меті професійної підготовки засобами комп'ютерного моделювання. Реалізація першої умови потребувала переходу до соціально-конструктивістських форм організації та методів навчання.

Реалізацією соціально-конструктивістського підходу до навчання об'єктно-орієнтованого моделювання є середовища моделювання, що надають можливість використання убудованих об'єктно-орієнтованих мов програмування, візуального конструювання інтерфейсу користувача, іменування об'єктів рідною мовою, об'єктно-орієнтованого, подієво-орієнтованого та візуального підходів до програмування в межах одного середовища, виконання індивідуальних та колективних дослідницьких проектів у мережному середовищі. З метою скорочення терміну початкового опанування середовища об’єктно-орієнтованого моделювання доцільним є вибір синтаксично компактних, розширюваних мов програмування, інтегрованих з середовищем розробки, таких як Python, Smalltalk та Java.

Реалізація другої умови потребувала розробки та впровадження спецкурсу «Об’єктно-орієнтоване моделювання» для майбутніх учителів природничо-математичних дисциплін, провідною ціллю якого є формування навичок об’єктно-орієнтованого моделювання як найбільш природного способу дослідження систем різної складності; ознайомлення з основними принципами конструювання та дослідження об’єктно-орієнтованих моделей; формування навичок індивідуальних та колективних навчальних досліджень.

Основний зміст навчання об’єктно-орієнтованого моделювання складає конструювання та дослідження динамічних та імітаційних моделей. У першому модулі «Вступ до об’єктно-орієнтованого моделювання» розглядаються базові поняття та уявлення курсу і виконується огляд середовищ об’єктно-орієнтованого моделювання (зокрема, виділяються універсальні середовища моделювання, середовища для конструювання динамічних моделей та середовища для конструювання імітаційних моделей). На підставі аналізу придатності середовищ для моделювання різних класів моделей пропонується у навчанні майбутніх учителів природничо-математичних дисциплін: а) при розгляді динамічних моделей послугуватися соціально-конструктивістськими середовищами об’єктно-орієнтованого моделювання VPython та Squeak як основними, а Sage, Alice та NetLogo – як додатковими; б) при розгляді імітаційних моделей скористатися соціально-конструктивістськими середовищами об’єктно-орієнтованого моделювання Alice та NetLogo як основними, а Sage, VPython та Squeak – як додатковими. Другий модуль «Об’єктно-орієнтовані динамічні моделі» присвячений розгляду динамічних моделей математичної екології, класичної механіки, молекулярної фізики і фізики твердого тіла та електродинаміки. У третій модуль «Об’єктно-орієнтовані імітаційні моделі» включені моделі кліткових автоматів, стохастичні моделі, моделі фрактальних об’єктів та процесів.

Зміст навчання, відображений у навчальному посібнику, передбачає застосування соціально-конструктивістського середовища об’єктно-орієнтованого моделювання Alice, локалізованого автором українською та російською мовами. Представлені у посібнику проекти розроблені у такий спосіб, щоб проілюструвати основні концепції об’єктно-орієнтованого, подієво-орієнтованого та візуального підходів до розробки програмного забезпечення та створити умови для опанування різних підходів у межах єдиного середовища моделювання. Розробка навчальних моделей, створених у Alice, може бути продовжена у професійному середовищі програмування NetBeans.

Реалізація третьої умови вимагає застосування трьох груп соціально-конструктивістських засобів ІКТ навчання об'єктно-орієнтованого моделювання: 1) середовищ об'єктно-орієнтованого моделювання, доопрацьованих та локалізованих автором (VPython, PyGeo, Sage, Squeak та Alice); 2) систем підтримки навчання (Moodle) та 3) засобів Web 2.0 (насамперед, це логосфера, Wiki та хмари сервісів).

# РОЗДІЛ 3
# ЕКСПЕРИМЕНТАЛЬНЕ ДОСЛІДЖЕННЯ ЕФЕКТИВНОСТІ ПРОФЕСІЙНОЇ ПІДГОТОВКИ МАЙБУТНІХ УЧИТЕЛІВ ПРИРОДНИЧО-МАТЕМАТИЧНИХ ДИСЦИПЛІН ЗАСОБАМИ КОМП'ЮТЕРНОГО МОДЕЛЮВАННЯ

## 3.1 Завдання та зміст дослідно-експериментальної роботи

Наукове дослідження у будь-якій галузі, зокрема і в галузі теорії та методики професійної освіти майбутніх учителів природничо-математичних дисциплін, виступає як специфічна форма пізнавальної діяльності, яка передбачає відображення педагогічної дійсності в емпіричному процесі. «Діяльність у галузі науки – наукове дослідження – особлива форма процесу пізнання, таке систематичне і цілеспрямоване вивчення об'єктів, у якому використовуються засоби і методи наук та яке завершується формуванням знань про об'єкти, що вивчаються» [182, с. 27].

З метою перевірки гіпотези дослідження та ефективності умов підготовки майбутніх учителів природничо-математичних дисциплін засобами комп'ютерного моделювання в процесі навчання спецкурсу з об'єктно-орієнтованого моделювання було проведено педагогічний експеримент.

Окреслена тема дослідження – умови професійної підготовки майбутніх учителів природничо-математичних дисциплін засобами комп'ютерного моделювання – своїми методологічними засадами включає принципи єдності теорії та практики, об'єктивності, всебічності, комплексності, системності охоплення всіх сторін та опосередкувань об'єкта дослідження. Як складний вид діяльності, дослідно-експериментальна робота потребує забезпечення стану упорядкованості та спрямованості діяльності об'єктів і суб'єктів управління, рівно як і відносин між ними.

Структура організації експерименту включає: цілепокладання, планування, розроблювання інструментарію, безпосереднє проведення дослідження, збір і опрацювання результатів, аналіз та інтерпретацію експериментальних даних [182, с. 66-67].

За С. У. Гончаренком, педагогічний (психолого-педагогічний) експеримент – це комплексний метод дослідження, який забезпечує науково-об'єктивну і доказову перевірку правильності обґрунтованої на початку дослідження гіпотези. Він надає можливість глибше, ніж інші методи, перевірити ефективність тих або інших новацій в галузі навчання і виховання, порівняти значущість різних факторів у структурі педагогічного процесу і обрати найкраще (оптимальне) для відповідних ситуацій їх поєднання, виявити необхідні умови реалізації певних педагогічних завдань [106, с. 253].

Експеримент є складовою наукового дослідження. «Головною метою експерименту може бути виявлення властивостей об'єктів, що досліджуються, перевірка справедливості гіпотез і на цій основі всебічне і глибоке вивчення теми наукового дослідження» [269, с. 201].

Мета експерименту з окресленої теми дослідження полягає у перевірці ефективності реалізації у практичній діяльності ЗВО з підготовки майбутніх учителів природничо-математичних дисциплін умов їх підготовки засобами комп'ютерного моделювання, а саме у виявленні її педагогічної доцільності, достовірності окреслених параметрів ефективності основних компонентів педагогічної системи: цілей та завдань, змістового ресурсу, форм організації, технологічного забезпечення, критеріїв оцінювання якості за її кінцевим результатом – оволодіння студентами проектувально-технологічними вміннями реалізовувати в практиці природничо-математичної освіти соціально-конструктивістські засобами об'єктно-орієнтованого моделювання.

Логіка етапів педагогічного експерименту в цілому відображала послідовність наступних дій [275, с. 193]:

– підготовка педагогічного дослідження:

а) вибір теми;

б) визначення її актуальності;

в) визначення ступеня вивченості;

– розробка програми дослідження:

а) окреслення об'єкта та предмета дослідження;

б) визначення мети;

в) постановка завдань;

г) розроблення робочої гіпотези;

д) визначення методів дослідження;

е) опрацювання даних;

ж) розроблення календарного плану;

– збір емпіричних відомостей, їх кількісне та якісне опрацювання;

– оформлення результатів, висновків і рекомендацій наукового дослідження;

– упровадження результатів дослідження у процес професійної підготовки майбутніх учителів природничо-математичних дисциплін.

На кожному етапі було використано комплекс методів науково-педагогічного дослідження:

– теоретичний аналіз джерел з проблеми дослідження;

– вивчення та узагальнення досвіду роботи викладачів ЗВО та аналіз конкретних експериментальних досліджень;

– цілеспрямоване педагогічне спостереження;

– бесіда, анкетування студентів та викладачів;

– теоретичний аналіз дидактичних можливостей застосування засобів технологій соціального конструктивізму у процесі навчання майбутніх учителів природничо-математичних дисциплін;

– метод статистичного опрацювання результатів педагогічного експерименту;

– вивчення та аналіз результатів діяльності студентів та викладачів.

Результати теоретичного дослідження дозволили виявити зміст, структуру, функції комп'ютерного моделювання у професійній підготовці майбутніх учителів природничо-математичних дисциплін, умови професійної підготовки майбутніх учителів природничо-математичних дисциплін засобами комп'ютерного моделювання на засадах об'єктно-орієнтованого та соціально-конструктивістського підходів, виконати розробку програмно-методичного комплексу з об'єктно-орієнтованого моделювання та його впровадження у навчальний процес з моделювання.

З урахуванням цих результатів розроблено програму експериментального дослідження.

У процесі розробки програми дослідження були виділені такі кроки її реалізації:

1) *підготовчий крок* передбачав:

– аналіз системи професійної підготовки майбутніх учителів природничо-математичних дисциплін у ЗВО;

– визначення бази та завдань дослідно-експериментальної частини дослідження;

– формулювання теми та гіпотези, постановка проблеми, визначення цілей та завдань дослідження;

2) *констатувальний крок* передбачав:

– аналіз базових понять досліджуваної проблеми;

– виявлення та теоретичне обґрунтування сучасних наукових знань про комп'ютерне моделювання;

– дослідження особливостей організації педагогічного процесу, організації соціально-конструктивістської роботи студентів у процесій навчальних досліджень, з'ясування проблем професійної підготовки майбутніх учителів природничо-математичних дисциплін;

– виявлення тенденцій розвитку професійної підготовки майбутніх учителів природничо-математичних дисциплін;

3) *інтуїтивний крок* передбачав:

– отримання вихідних даних;

– відбір навчальних груп для проведення експерименту;

– проведення необхідних діагностичних досліджень;

4) *концептуальний крок* передбачав розробку концептуальної моделі професійної підготовки майбутніх учителів природничо-математичних дисциплін засобами комп’ютерного моделювання;

5) *дослідницький крок* передбачав дослідження умов, що впливають на якість професійної підготовки майбутніх учителів природничо-математичних дисциплін;

6) *змістово-технологічний крок* передбачав:

– створення умов для розкриття та розвитку дослідницьких умінь у процесі спільної навчальної діяльності;

– розробка технології навчання об’єктно-орієнтованого моделювання майбутніх учителів природничо-математичних дисциплін на основі сучасних і перспективних технологій програмування, поєднання традиційних та інноваційних (зокрема, соціально-конструктивістських) форм організації, методів та засобів навчання;

7) *організаційно-педагогічний крок* передбачав:

– застосування умов професійної підготовки майбутніх учителів природничо-математичних дисциплін засобами комп’ютерного моделювання;

– застосування створених навчальних посібників і комп’ютерних моделей;

– використання запропонованих видів спільної дослідницької роботи студента;

– застосування технологій соціального конструктивізму у процесі навчання об’єктно-орієнтованого моделювання;

8) *когнітивно-операційний крок* передбачав впровадження теоретичних і практичних результатів у процес професійної підготовки майбутніх учителів природничо-математичних дисциплін у ЗВО України;

9) *оцінно-результативний крок* передбачав:

– отримання кінцевих даних;

– статистичне опрацювання та аналіз результатів;

– систематизацію та узагальнення результатів дослідження;

– формування прогностичних напрямів розвитку професійної підготовки фахівців майбутніх учителів природничо-математичних дисциплін.

## 3.2 Основні етапи дослідно-експериментальної роботи

Реалізація програми дослідження проходила у три етапи:

1) констатувальний (1999-2003 рр.);

2) пошуковий (2003-2007 рр.);

3) формувальний (2007-2011 рр.).

Завданням першого етапу дослідження було вивчення існуючого стану досліджуваного явища та виділення вихідних положень дослідження. Для реалізації поставленого завдання було виконано аналіз навчальних програм та підручників з шкільних курсі математики, інформатики, фізики, хімії, біології, на підставі якого було визначено основні проблеми, спільні для старших класів середньої школи та молодших курсів педагогічних ЗВО.

Головну увагу на констатувальному етапі дослідження було приділено розробці профільних курсів комп'ютерного моделювання, застосуванню засобів Web 1.0 для конструювання навчальних матеріалів (зокрема, з курсу інформатики середньої школи) та засобам навчальних матеріалів від несанціонованого доступу.

Результати констатувального експерименту виявили наступне:

1. Незважаючи на поширеність методу моделювання у навчанні природничо-математичних дисциплін, у процесі навчання комп'ютерного моделювання в межах курсу інформатики переважають імперативний підхід та відповідні процедурні мови програмування (Basic, Pascal, C). При цьому навчання об'єктно-орієнтованого моделювання із застосуванням об'єктно-орієнтованих середовищ (Visual Basic, Delphi, C++Builder) не відбувається, а головна увага зосереджується на візуальних засобах побудови інтерфейсу користувача.

2. Засоби Web 1.0 надають можливість індивідуальної роботи зі створення навчальних ресурсів, в той час як спільне їх конструювання було надзвичайно утрудненим.

Виявлені невідповідності

– між станом проблеми у спеціальній і методичній літературі та сучасним рівнем розвитку засобів програмних систем;

– між педагогічним потенціалом середовищ об'єктно-орієнтованого моделювання та невикористанням об'єктно-орієнтованого підходу в процесі моделювання;

– між перспективністю застосування технологій соціального конструктивізму в процесі професійної підготовки учителів природничо-математичних дисциплін та наявним станом навчання комп'ютерного моделювання і непристосованістю засобів Web 1.0 для організації колективних навчальних досліджень зумовили вибір мети дослідження, досягнення якої вимагало перебудови системи професійної підготовки майбутніх учителів природничо-математичних дисциплін із урахуванням:

1) процесів інтеграції системи освіти України у світовий освітній простір. Зокрема, приєднання України до Болонського процесу означає

зближення концептуальних (філософських, організаційних і правових) основ вищої освіти України і Європейського співтовариства;

2) інноваційних шляхів підвищення якості педагогічної освіти. Зокрема, Концепція розвитку неперервної педагогічної освіти на основі порівняльного аналізу систем педагогічної освіти України і Європейського співтовариства визначає освітню інноватику одним із пріоритетних напрямів наукового пошуку [140];

3) важливості вищої педагогічної освіти для формування та розвитку відкритого суспільства, створення умов для науково-технічного прогресу та мобільності кадрового забезпечення освіти, науки, економіки.

Другий етап дослідження характеризувався добором засобів реалізації соціально-конструктивістського підходу в навчанні, придатних для застосування в процесі навчання об’єктно-орієнтованого моделювання студентів природничо-математичних та інформатичних спеціальностей педагогічних ЗВО, в результаті якого було виявлено, що переважна більшість програмних засобів не мають підтримки вітчизняних користувачів та орієнтовані на застосування авторських мов програмування (найчастіше візуальних).

Виявлені проблеми дозволили сформулювати вимоги до засобів навчання об’єктно-орієнтованого моделювання на основі технологій соціального конструктивізму:

– локалізованість інтерфейсу користувача та мови програмування;

– наявність можливості створення імен програмних об’єктів рідною мовою;

– можливість створення коротких програм, що мають розвинену функціональність;

– підтримка об’єктно-орієнтованого, подієво-орієнтованого та візуального підходів до програмування в межах одного середовища;

– розширюваність середовища моделювання за рахунок створення власних та імпортування зовнішніх об’єктів;

– інтеграція середовища моделювання з логосферою та засобами Web 2.0;

– наявність можливості імпортування розроблених моделей у професійні середовища програмування.

Огляд існуючих засобів навчання показав, що жоден з них не відповідає сформульованим вимогам, проте найбільш близьким до їх реалізації були середовища моделювання Squeak, VPython, Sage та Alice. Головними їх недоліками були відсутність можливості створення імен програмних об’єктів рідною мовою, нелокалізованість інтерфейсу користувача та слабка інтеграція з логосферою. Додатковим обмеженням на їх застосування була мала кількість джерел українською та російськими мовами, присвячених цим середовищам.

Для усунення виявлених недоліків були використані локалізації середовищ Squeak (рис. 3.1) та Sage (рис. 3.2) російською мовою, що надають можливість застосовувати й українську.

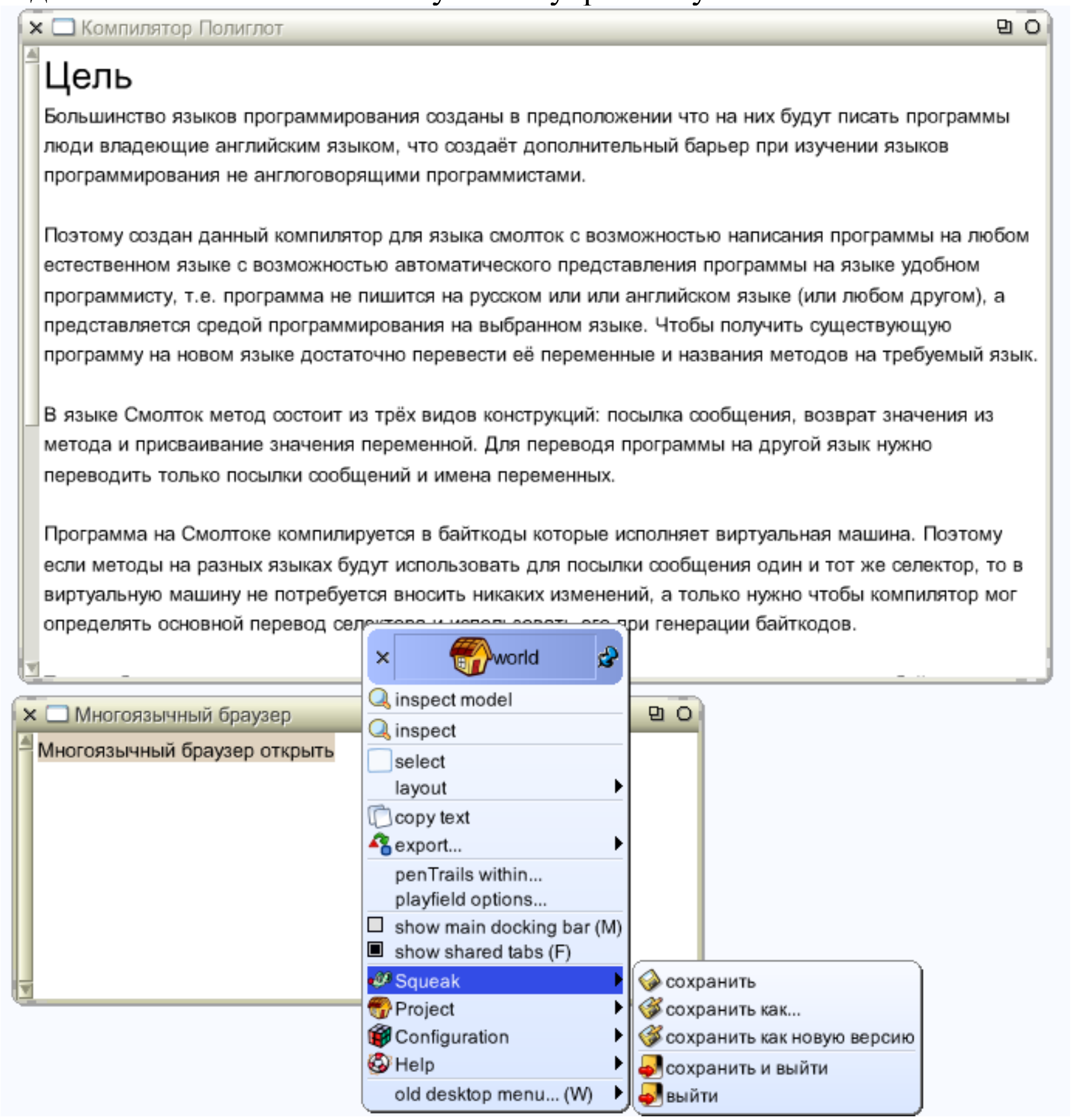

Рис. 3.1. «Поліглот» – локалізація Squeak російською мовою

За результатами дослідної експлуатації вказаних середовищ було створено авторську локалізацію середовища моделювання VPython, що є основою середовища динамічної геометрії PyGeo та середовища моделювання сонячно-подібних систем VPNBody.

На початку третього етапу дослідження з метою інтернаціоналізації середовища об’єктно-орієнтованого моделювання Alice було створено його третю версію. Процедура інтернаціоналізації виконується головним розробником системи – Д. Косгроувом (Dennis Cosgrove), локалізація – міжнародною командою.

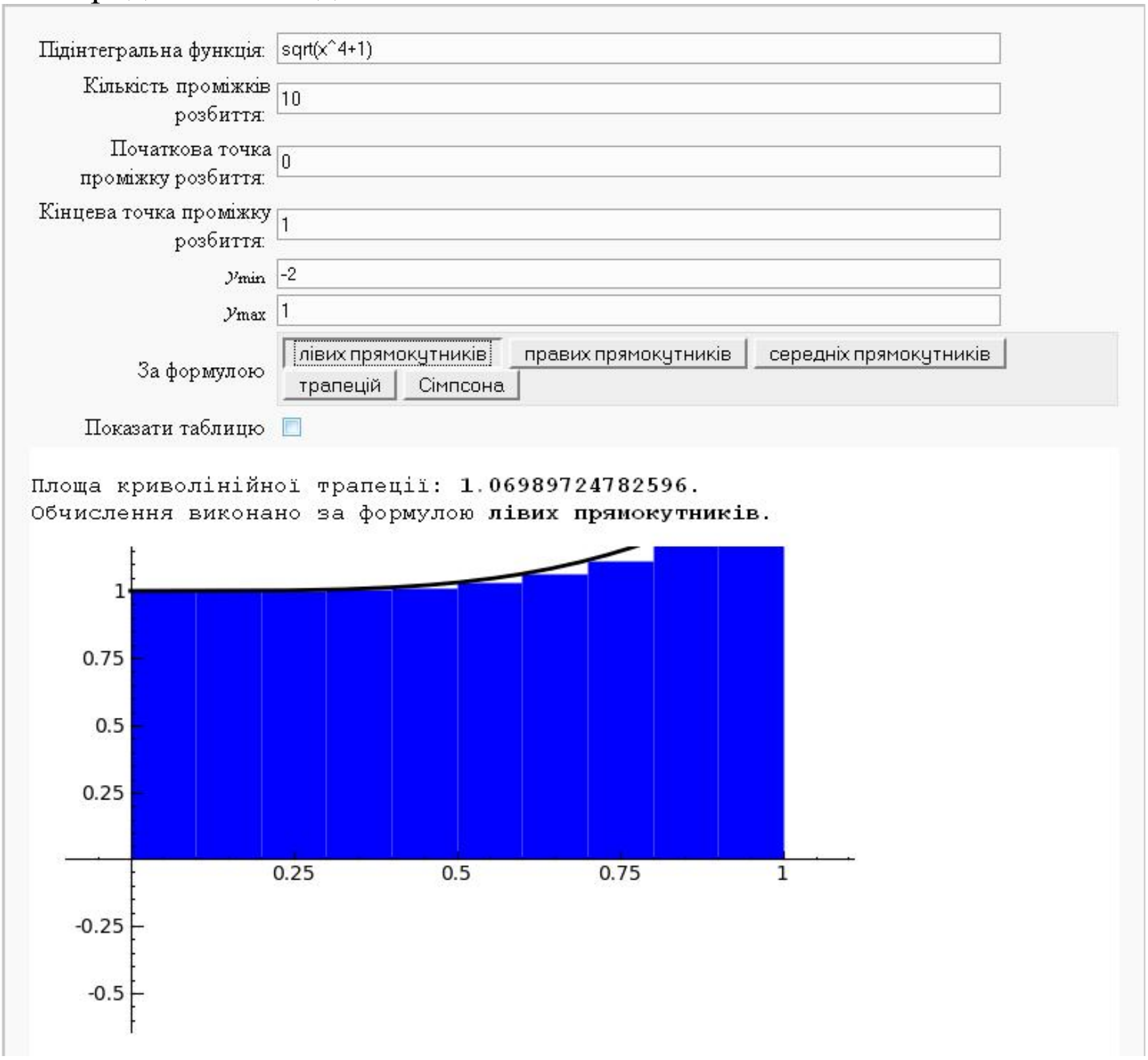

Рис. 3.2. Модель чисельного методу інтегрування, розроблена у локалізованій версії середовища Sage

Сучасні інструменти локалізації передбачають розподілену роботу на потокового сервері (Launchpad, Transifex, Pootle, Narro та ін.). Для локалізації Alice 3 в університеті Карнегі-Меллона (м. Піттсбург) було налаштовано сервер Narro (рис. 3.3).

До переваг Narro відносяться підтримка багатьох форматів файлів перекладу, продуманість системи роботи з перекладачами, перевірка правопису, записів і проста перевірка пунктуації, спільна пам'ять перекладів для проектів, можливість голосування користувачами за варіант перекладу, подання RSS з повідомленнями про появу нових рядків, додавання нових пропозицій та зміну контексту.

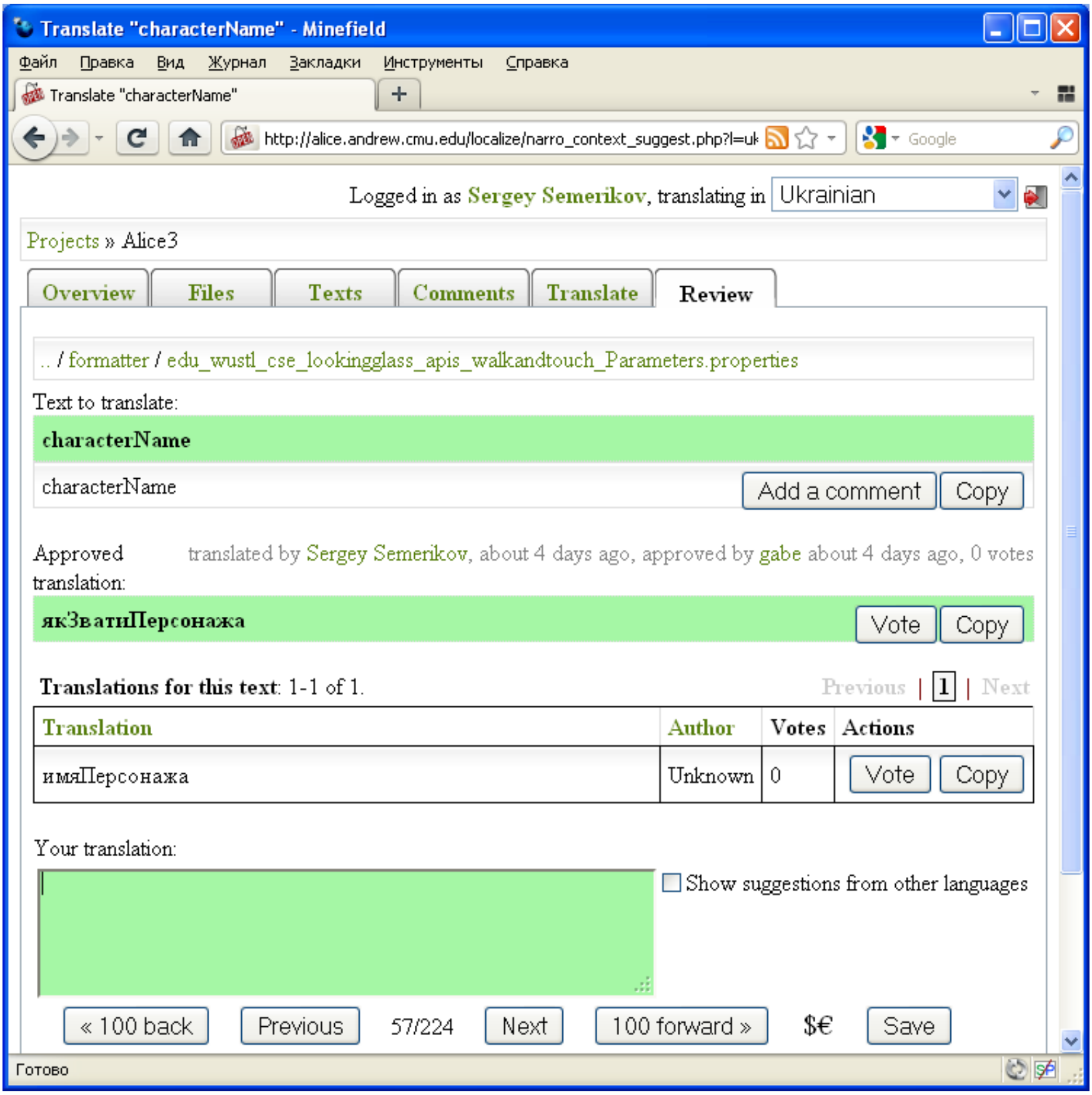

Рис. 3.3. Сервер Narro в університеті Карнегі-Меллона

Створена вітчизняна версія середовища об’єктно-орієнтованого моделювання Alice (рис. 3.4) задовольняє всім поставленим вимогам до засобу навчання об’єктно-орієнтованого моделювання на основі технологій соціального конструктивізму:

1) локалізованість як інтерфейсу користувача, так й ядра системи (мови програмування та засобів конструювання);

2) наявність можливості створення імен програмних об’єктів будь-якою мовою, що підтримується засобами інтернаціоналізації Java-програм (в т. ч. – з ієрогліфічним поданням імен);

3) можливість створення коротких програм, що мають розвинену функціональність: використовують об’єкти 3D-графіки, що переміщуються в автоматичному режимі або керованих користувачем;

4) підтримка об’єктно-орієнтованого, подієво-орієнтованого та візуального підходів до програмування в межах одного середовища;

5) розширюваність середовища моделювання за рахунок створення власних (як візуальних, так і невізуальних) об’єктів та імпортування зовнішніх об’єктів (починаючи з третьої версії, до складу середовища Alice входять усі об’єкти-«персонажі» середовища Sims II);

6) інтеграція середовища моделювання з логосферою забезпечується можливістю розміщення проектів, розроблених у Alice, на Wiki-сервері, та обміном проектами між учасниками форуму (http://alice.org); додаткова інтеграція середовища моделювання із засобами Web 2.0 забезпечується можливістю експорту з нього відеофрагментів, що демонструють роботу моделі, на сервер потокового відео YouTube;

7) можливість імпортування моделей, розроблених у Alice, в професійне середовище програмування NetBeans, забезпечується надбудовою Alice3ProjectWizard, що додає у середовище NetBeans палітру компонентів та подій, опрацьовуваних Alice (рис. 3.5);

8) наявність у середовищі Alice режиму показу програми мовою Java створює умови для переходу від навчального середовища моделювання до професійних середовищ Eclipse, NetBeans та ін.

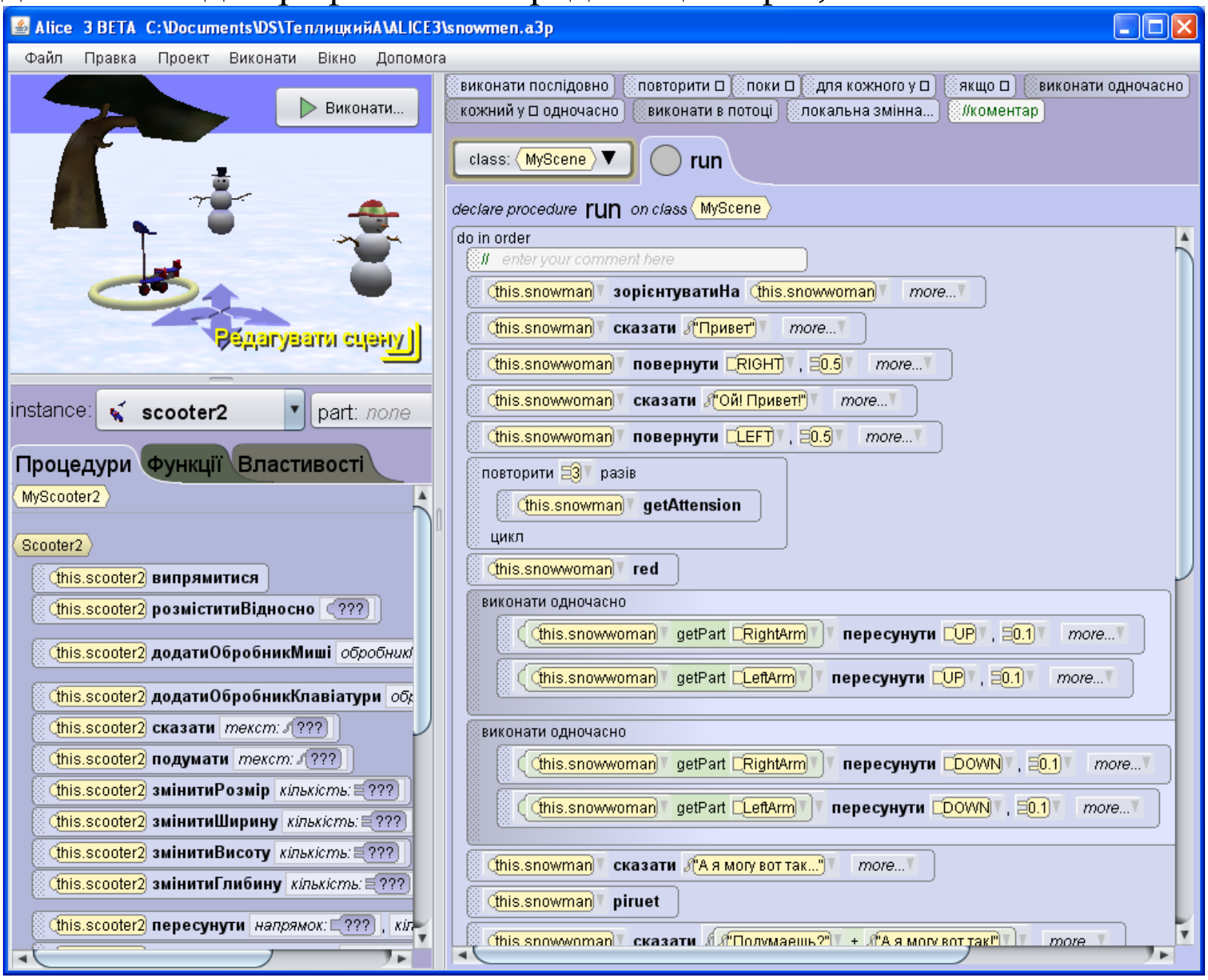

Рис. 3.4. Стан української локалізації Alice 3 Beta на 01.09.2011 р.

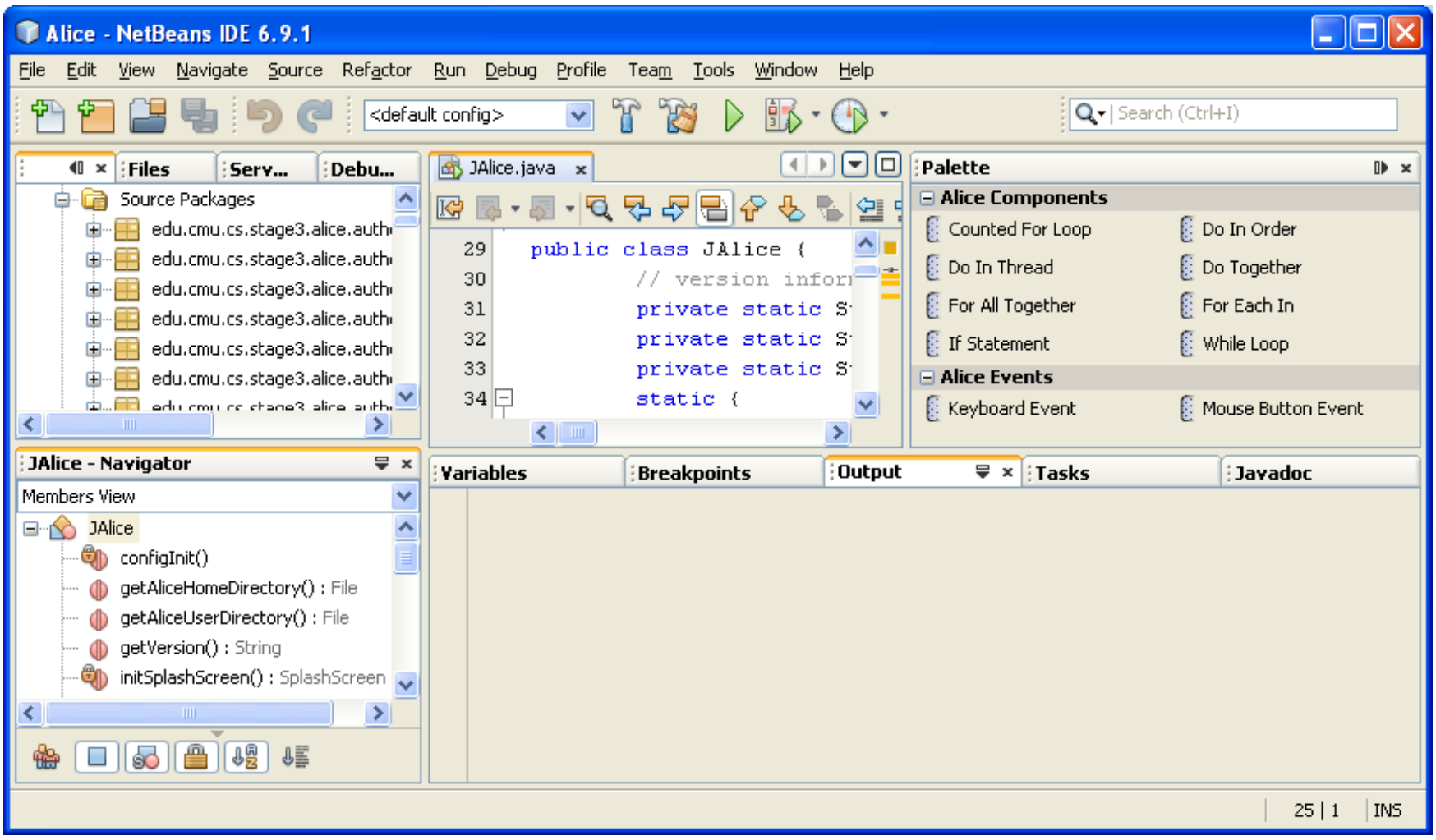

Рис. 3.5. Середовище NetBeans, налаштоване для роботи з моделями Alice

Виконана розробка та локалізація програмного забезпечення створили умови для організації навчання об'єктно-орієнтованого моделювання на основі комбінованого навчання (інтеграції аудиторного, дистанційного та мобільного навчання) із залученням засобів Web 2.0. Це надало можливість вперше розглянути Alice не як середовище навчання початківців програмування на основі сценарного підходу, як це пропонують З. С. Сейдаметова та Ф. В. Шкарбан [384], а як повноцінне середовище моделювання, в якому може бути реалізований весь спектр моделей, пропонованих у розглянутому вище спецкурсі «Об'єктно-орієнтоване моделювання» для студентів природничо-математичних спеціальностей педагогічних університетів.

Упровадження розробленої системи умов професійної підготовки майбутніх вчителів природничо-математичних дисциплін засобами комп'ютерного моделювання та експериментальна перевірка її ефективності була виконана протягом 2007-2008, 2008-2009, 2009-2010 та 2010-2011 н. р. За розробленою методикою навчалися 315 студентів.

### 3.3 Статистичне опрацювання та аналіз результатів формувального етапу педагогічного експерименту

Для перевірки ефективності умов підготовки майбутніх учителів природничо-математичних дисциплін засобами комп'ютерного моделювання було виконано порівняння розподілів студентів за рівнем сформованості компетентності з комп'ютерного моделювання.

Контрольні та експериментальні групи формувалися наступним чином:

– до контрольних груп (КГ) відносилися студенти природничо-математичних спеціальностей, що навчалися за традиційною моделлю підготовки;

– до експериментальних груп (ЕГ) відносилися студенти природничо-математичних спеціальностей, що навчалися за авторською моделлю підготовки на основі розроблених педагогічних умов.

Результати формувального експерименту в контрольних та експериментальних групах наведено у табл. 3.1. Порівняння гістограми розподілів студентів (у відсотках, оскільки маємо різну кількість студентів для контрольної та експериментальної груп) за кількістю набраних згідно табл. 2.2 балів показані на рис. 3.6.

*Таблиця 3.1*

**Порівняльний розподіл студентів за рівнем сформованості компетентності з комп'ютерного моделювання в контрольних та експериментальних групах**

| **Кількість балів** | **Рівень** | **КГ** | | **ЕГ** | |
|---|---|---|---|---|---|
| | | **студ.** | **%** | **студ.** | **%** |
| 1–34 | низький | 0 | 0,00 | 0 | 0,00 |
| 35–39 | | 4 | 11,76 | 1 | 1,92 |
| 40–67 | достатній | 12 | 35,29 | 5 | 9,62 |
| 68–74 | | 8 | 23,53 | 13 | 25,00 |
| 75–81 | високий | 6 | 17,65 | 16 | 30,77 |
| 82–89 | | 4 | 11,76 | 13 | 25,00 |
| 90–100 | | 0 | 0,00 | 4 | 7,69 |

Опрацювання результатів експерименту та оцінка ефективності розроблених педагогічних умов здійснювалась методами математичної статистики [270]. Оскільки задача полягала у виявленні відмінностей в розподілі певної ознаки при порівнянні двох емпіричних розподілів, згідно [270, с. 34] можна скористатись $\chi^2$-критерієм Пірсона (Karl Pearson) або $\lambda$-критерієм Колмогорова-Смирнова.

**$\chi^2$-критерій Пірсона**

У нашому дослідженні вибірки випадкові і незалежні. Враховуючи, що інтервали з нульовими частотами неприпустимі, а не менше 80% частот мають бути більше 5, було виконано поєднання інтервалів 1–34 і 35–39 та 82–89 і 90–100, тому шкалою вимірювань є шкала з $C$=5 категоріями: 1–39, 40–67, 68–74, 75–81, 82–100.

Накладено одну незалежну умову, отже, кількість степенів свободи $v$=$C$–1=4.

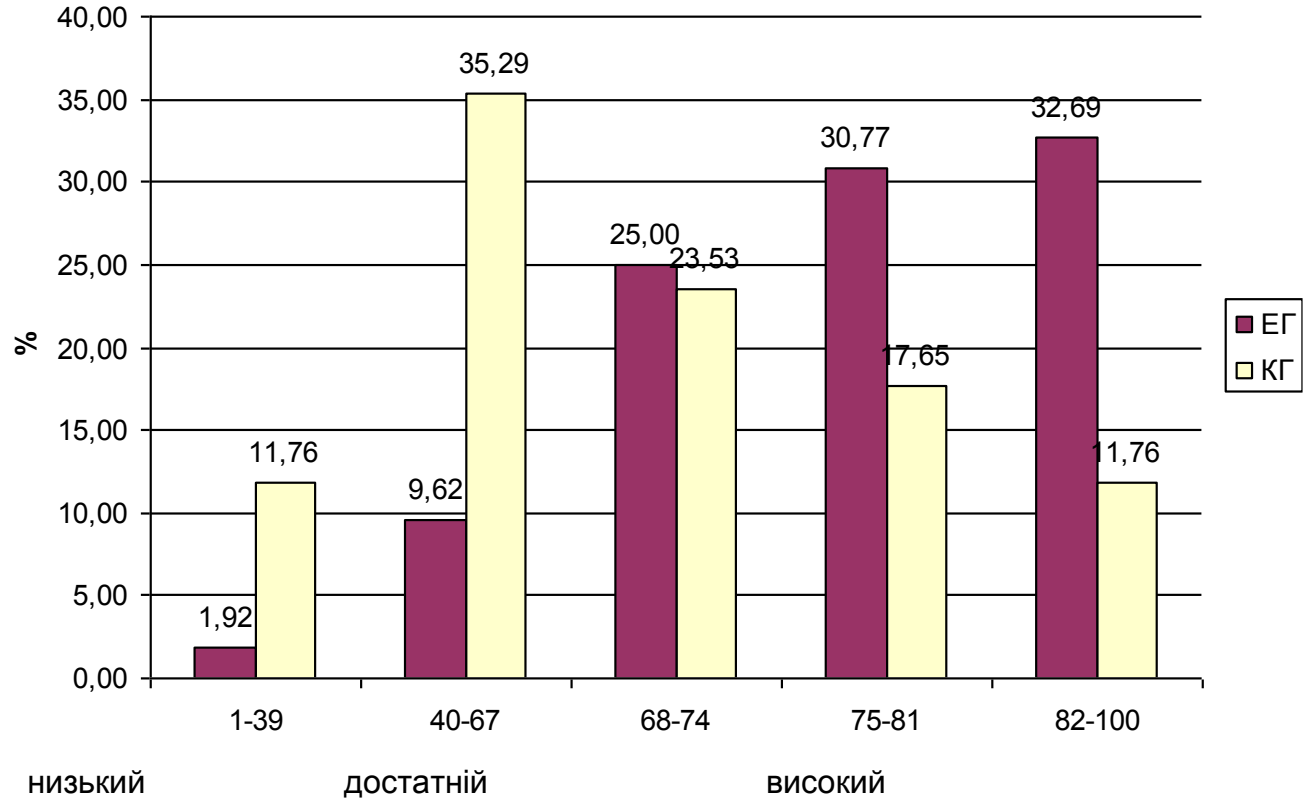

Рис. 3.6. Порівняльний розподіл студентів на етапі формувального експерименту у контрольних та експериментальних групах за рівнем сформованості компетентності з комп'ютерного моделювання

Нульова гіпотеза $H_0$: ймовірність попадання студентів контрольної ($n_1$=34) та експериментальної вибірки ($n_2$=52) в кожну з $i$ ($i$=0, 1, 2, 3, 4) категорій однакова, тобто $H_0$: $p_{1i}=p_{2i}$ ($i$=0, 1, 2, 3, 4), де $p_{1i}$ – ймовірність оцінювання сформованості компетентності з комп'ютерного моделювання учасників контрольної групи на $i$ балів ($i$=0, 1, 2, 3, 4) та $p_{2i}$ – ймовірність оцінювання рівня сформованості компетентності з комп'ютерного моделювання учасників експериментальної групи на $i$ балів ($i$=0, 1, 2, 3, 4);

Альтернативна гіпотеза $H_1$: $p_{1i}\neq p_{2i}$ хоча б для однієї із $C$ категорій.

Значення $\chi^2$ обчислюється за формулою:

$$T=\frac{1}{n_1 n_2}\sum_{i=0}^{C-1}\frac{\left(n_1 Q_{2i}-n_2 Q_{1i}\right)^2}{Q_{1i}+Q_{2i}}.$$

$Q_{1i}$ – кількість учасників контрольної групи, які набрали $i$ балів;

$Q_{2i}$ – кількість учасників експериментальної групи, які набрали $i$ балів.

Із таблиці значень $\chi^2$ для рівня значущості $\alpha$=0,05 і кількості степенів свободи $\nu$=4 визначаємо критичне значення статистики $T_{крит}$=9,488.

Обчислення критерію $\chi^2$ для експериментальної та контрольної вибірки після проведення формувального експерименту показало, що $T>T_{крит}$ (15,372>9,488). Це є основою для відхилення нульової гіпотези. Прийняття альтернативної гіпотези дає підстави стверджувати, що ці вибірки мають статистично значущі відмінності, тобто *розроблена система професійної підготовки майбутніх вчителів природничо-математичних дисциплін засобами комп'ютерного моделювання є* *більш ефективною, ніж традиційна.*

**λ-критерій Колмогорова-Смирнова**

Для більшої переконливості виконаємо перевірку отриманих під час формувального експерименту вибірок за λ-критерієм Колмогорова-Смирнова (табл. 3.2). Цей критерій є непараметричним і застосовується за таких умов:

– вибірки випадкові і незалежні;

– категорії впорядковані за зростанням або спаданням.

Оскільки обидві ці умови для отриманих нами вибірок виконуються, ми можемо застосувати цей критерій для оцінювання відхилення розподілу в експериментальних групах від розподілу в контрольних групах. Позначимо:

$F(x)$ – невідома функція розподілу ймовірностей рівня сформованості компетентності з комп'ютерного моделювання майбутніх учителів природничо-математичних дисциплін в контрольних групах;

$G(x)$ – невідома функція розподілу ймовірностей рівня сформованості компетентності з комп'ютерного моделювання майбутніх учителів природничо-математичних дисциплін в експериментальних групах.

Нульова гіпотеза $H_0$: $F(x)=G(x)$

Альтернативна гіпотеза $H_1$: $F(x)\neq G(x)$

Коли гіпотеза $H_0$: $F(x)=G(x)$ справджується, відхилення

$$D=\sup_x\left|G(x)-F(x)\right|$$

мале, а коли гіпотеза $H_0$ не справджується, це відхилення велике.

Результати опрацювання експериментальних даних наведені в табл. 3.2, з якої видно, що $D$=0,355.

*Таблиця 3.2*

**Обчислення критерію Колмогорова-Смирнова**

| Рівень | Бали | Абсолютна частота | | Накопичена частота | | Відносна накопичена частота | | D |
|---|---|---|---|---|---|---|---|---|
| | | КГ | ЕГ | КГ | ЕГ | КГ | ЕГ | |
| низький | 1–39 | 4 | 1 | 4 | 1 | 0,118 | 0,019 | 0,098 |
| достатній | 40–67 | 12 | 5 | 16 | 6 | 0,471 | 0,115 | ***0,355*** |
| | 68–74 | 8 | 13 | 24 | 19 | 0,706 | 0,365 | 0,340 |
| високий | 75–81 | 6 | 16 | 30 | 35 | 0,882 | 0,673 | 0,209 |
| | 82–100 | 4 | 17 | 34 | 52 | 1,000 | 1,000 | 0,000 |

Граничні значення $\varepsilon_{0,05;\,34}$=0,2480, $\varepsilon_{0,05;\,52}$=0,1921.

Звідси $D > \varepsilon_{\alpha;n}$ (0,355>0,2480 та 0,355>0,1921), тобто у відповідності з λ-критерієм Колмогорова-Смирнова нульова гіпотеза $H_0$: $F(x)=G(x)$ відхиляється і приймається альтернативна гіпотеза $H_1$: $F(x) \neq G(x)$.

На рис. 3.7 подано графічну інтерпретацію розподілів $F(x)$ та $G(x)$.

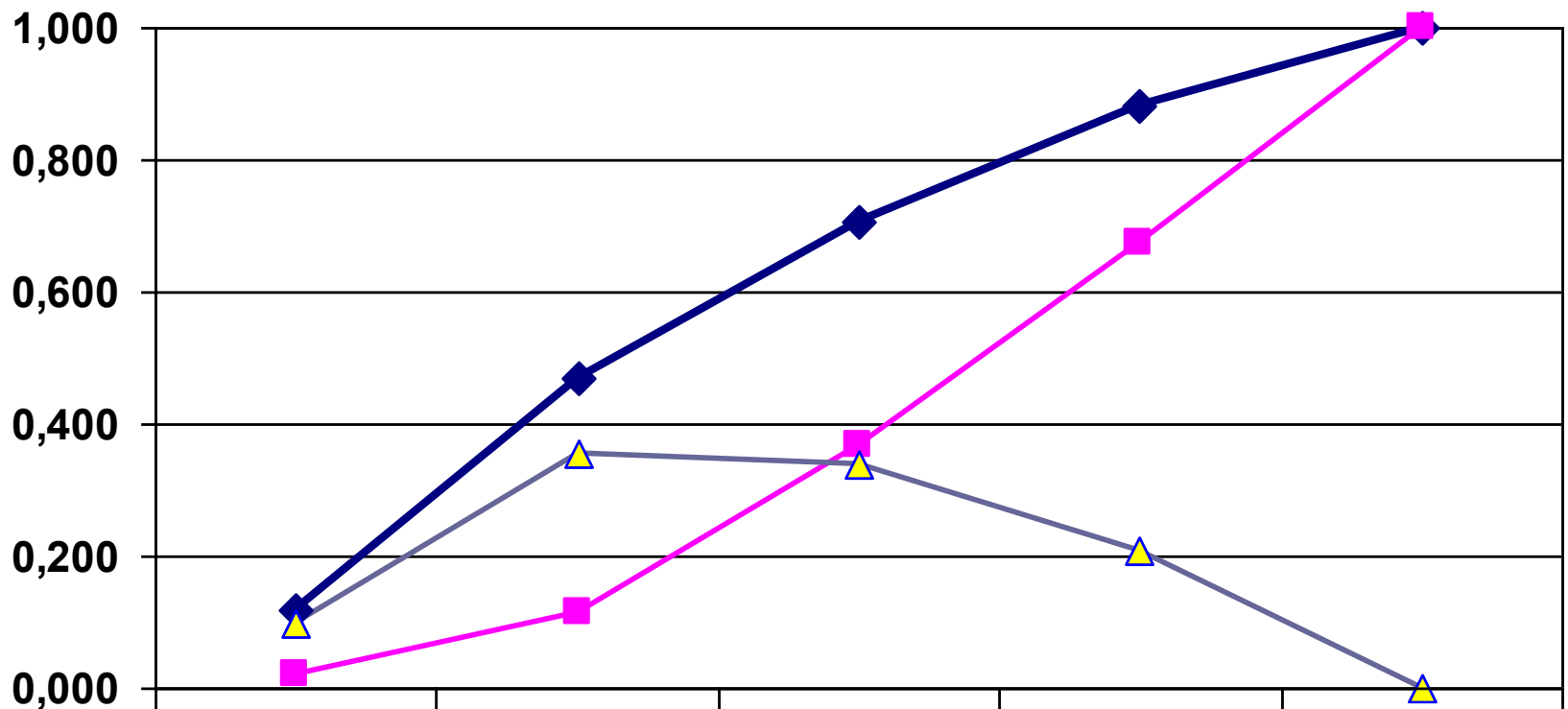

| | 1-39 | 40-67 | 68-74 | 75-81 | 82-100 |
|---|---|---|---|---|---|
| КГ | 0,118 | 0,471 | 0,706 | 0,882 | 1,000 |
| ЕГ | 0,019 | 0,115 | 0,365 | 0,673 | 1,000 |
| D | 0,098 | 0,355 | 0,340 | 0,209 | 0,000 |

Рис. 3.7. Графіки функцій розподілу студентів у контрольних та експериментальних групах за балами, що характеризують рівень сформованості компетентності з комп'ютерного моделювання, та модуля їх різниці (*D*)

Це означає, що існує відмінність розподілу рівня сформованості компетентності з комп'ютерного моделювання студентів, які навчалися за традиційною системою професійної підготовки і розробленої автором. Таким чином, студенти, що навчалися в експериментальних групах, мали більш високі бали, що характеризують рівень сформованості компетентності з комп'ютерного моделювання.

Ураховуючи, що в експериментальних групах навчання студентів відбувалось на основі виокремлених педагогічних умов, можна припустити, що саме це і дало можливість досягти кращих результатів. Отже, можна говорити про експериментальне підтвердження висунутої гіпотези.

Аналіз результатів експериментальної роботи дозволив зробити висновок про те, що уведення в процес професійної підготовки майбутніх учителів природничо-математичних дисциплін об'єктно-орієнтованого моделювання створює умови для:

– *фундаменталізації професійної підготовки*: об'єктно-орієнтоване моделювання стає інтегруючим елементом навчання природничо-математичних дисциплін, що відображає спільну для них технологію досліджень – моделювання і зокрема, математичне комп'ютерне моделювання;

– *посилення міжпредметних зв'язків природничо-математичних дисциплін*: моделі відповідних предметних галузей безпосередньо відображаються у змісті навчання об'єктно-орієнтованого моделювання;

– *дослідницького підходу в навчанні*: у процесі навчання об'єктно-орієнтованого моделювання цілеспрямовано формуються навички організації та проведення індивідуальних та колективних навчальних досліджень;

– *інтеграції різних технологій програмування*: володіння засобами об'єктно-орієнтованого моделювання допомагає об'єднати технології об'єктно-орієнтованого, подієво-орієнтованого та візуального програмування в єдиному середовищі.

## Висновки до розділу 3

Аналіз результатів педагогічного експерименту показав, що впровадження технологій соціального конструктивізму у процес професійної підготовки майбутніх учителів природничо-математичних дисциплін створює умови для формування у них навичок спільної дослідницької діяльності у мережних середовищах, критичного мислення, вмінь застосовувати методи та засоби технологій соціального конструктивізму в професійній діяльності.

Уведення у процес професійної підготовки майбутніх учителів природничо-математичних дисциплін об'єктно-орієнтованого моделювання створює умови для фундаменталізації професійної підготовки, посилення міжпредметних зв'язків природничо-математичних дисциплін, дослідницького підходу в навчанні, інтеграції різних технологій програмування. Вивчення комп'ютерного моделювання в рамках спецкурсу «Об'єктно-орієнтоване моделювання» сприяє розвитку у майбутнього фахівця достатньо широкого погляду на методи та технології моделювання і формуванню у нього міжпредметних знань, які допоможуть йому у подальшій професійній діяльності. Опанування комп'ютерного моделювання у соціально-конструктивістському середовищі сприяє формуванню модельного стилю мислення в процесі виконання дослідницьких навчальних проектів, що дає змогу самостійно конструювати власну систему знань, вмінь та навичок у співтовариствах практики та масових відкритих дистанційних курсах.

Метою розробленої системи реалізації педагогічних умов є забезпечення фундаментальності природничо-математичної підготовки засобами комп'ютерного моделювання, організаційно-педагогічного й психолого-педагогічного супроводу підготовки майбутніх учителів природничо-математичних дисциплін та визначення форм організації, методів і засобів процесу професійної підготовки студентів природничих, математичних та інформатичних спеціальностей соціально-конструктивістськими засобами комп'ютерного моделювання. Для досягнення цієї мети було створено навчально-методичний комплекс, що включає що включає в себе авторський навчальний посібник, відеоуроки, локалізоване українською та російською мовами середовище об'єктно-орієнтованого моделювання Alice, а також інструкції щодо роботи над курсом в цілому та його окремими модулями.

Комплексне забезпечення обґрунтованих умов дає можливість покращити підготовку майбутніх учителів природничо-математичних дисциплін засобами комп'ютерного моделювання. Результати педагогічного експерименту підтверджують ефективність розробленої системи реалізації умов підготовки майбутніх учителів природничо-математичних дисциплін засобами комп'ютерного моделювання.

# ВИСНОВКИ

У монографії запропоновано теоретичне узагальнення та нове бачення розв’язання наукової проблеми, що полягає в окресленні умов підготовки майбутніх учителів природничо-математичних дисциплін засобами комп’ютерного моделювання, перевірці їхньої ефективності в процесі навчання спецкурсу з об’єктно-орієнтованого моделювання. Результати теоретичного дослідження й педагогічного експерименту надали підстави для формулювання висновків наукового пошуку.

1. Унаслідок аналізу сучасних тенденцій професійної підготовки вчителів природничо-математичних дисциплін обґрунтовано перспективний напрям її фундаменталізації через посилення ролі спільного для природничих наук методу дослідження – моделювання, що водночас постає як провідний метод навчання. Урахування психологічних особливостей відображення свідомістю людини об’єктів навколишньої дійсності вимагає їх інтерпретації в комп’ютерних моделях, тому розв’язання проблеми фундаменталізації змісту професійної підготовки вчителів природничо-математичних дисциплін в умовах швидкої зміни засобів ІКТ потребує об’єднання методу моделювання та об’єктно-орієнтованої технології програмування, які разом утворюють якісно нову концепцію – об’єктно-орієнтоване моделювання. Упровадження в процес професійної підготовки вчителів природничо-математичних дисциплін об’єктно-орієнтованого моделювання створює умови для: а) фундаменталізації навчання (об’єктно-орієнтоване моделювання є інтегрувальним елементом навчання природничо-математичних дисциплін, що відображає спільну для них технологію досліджень – моделювання); б) посилення міжпредметних зв’язків природничо-математичних дисциплін (моделі предметних галузей безпосередньо відображені в змісті навчання об’єктно-орієнтованого моделювання); в) реалізації технологій соціального конструктивізму в навчанні (у процесі навчання об’єктно-орієнтованого моделювання цілеспрямовано формуються навички організації та проведення індивідуальних і колективних навчальних досліджень); г) інтеграції різних технологій програмування (володіння засобами об’єктно-орієнтованого моделювання допомагає об’єднати технології об’єктно-орієнтованого, подієво-орієнтованого та візуального програмування в єдиному середовищі).

2. Теоретичною основою професійної підготовки вчителів природничо-математичних дисциплін засобами комп’ютерного моделювання є соціально-конструктивістський підхід. Перспективний напрям його реалізації – індивідуальні та колективні навчальні дослідження, проведення яких вимагає використання таких засобів об'єктно-орієнтованого моделювання, що вможливлюють спільну навчальну діяльність у мережевому середовищі. Побудова освітніх спільнот суттєво полегшується за умови застосування систем підтримки навчання та засобів організації спільної роботи й подання її результатів у Web.

3. До умов професійної підготовки майбутніх учителів природничо-математичних дисциплін засобами комп'ютерного моделювання зараховано: 1) застосування педагогічних технологій соціального конструктивізму в процесі підготовки майбутніх учителів природничо-математичних дисциплін; 2) упровадження об'єктно-орієнтованого моделювання в процес навчання інформатичних дисциплін; 3) використання соціально-конструктивістських засобів ІКТ навчання об'єктно-орієнтованого моделювання. На основі цих умов теоретично обґрунтована та розроблена модель професійної підготовки майбутніх учителів природничо-математичних дисциплін засобами комп'ютерного моделювання, спрямована на формування активної навчально-пізнавальної діяльності студентів, розвиток їхніх інтелектуальних здібностей, пізнавальної самостійності, навичок спільної навчально-дослідницької діяльності, умінь використання засобів комп'ютерного моделювання й соціально-конструктивістських технологій. До основних складників моделі належить система вимог до професійної підготовки майбутніх учителів природничо-математичних дисциплін, умови їх підготовки засобами комп'ютерного моделювання та система реалізації умов, що включає соціально-конструктивістські засоби навчання, методи навчання (загальнодидактичні й спеціальні) і форми організації навчання (традиційні й соціально-конструктивістські), підпорядковані загальній меті професійної підготовки засобами комп'ютерного моделювання.

4. Реалізація виокремлених умов потребувала розроблення й упровадження спецкурсу «Об'єктно-орієнтоване моделювання» для майбутніх учителів природничо-математичних дисциплін, провідною метою якого є формування навичок об'єктно-орієнтованого моделювання як найбільш природного способу дослідження систем різної складності; ознайомлення з основними принципами конструювання та дослідження об'єктно-орієнтованих моделей; формування навичок індивідуальних і колективних навчальних досліджень. Для досягнення цієї мети створено навчально-методичний комплекс, що включає авторський навчальний посібник, відеоуроки, локалізоване українською та російською мовами середовище об'єктно-орієнтованого моделювання Alice, а також інструкції щодо роботи над курсом у цілому та його окремими модулями. Представлені в посібнику навчальні проекти ілюструють різні способи реалізації дослідницького підходу до навчання та основні концепції об'єктно-орієнтованого, подієво-орієнтованого й візуального підходів до розроблення програмного забезпечення.

5. Педагогічний експеримент підтвердив, що впровадження в навчальний процес спецкурсу «Об'єктно-орієнтоване моделювання» покращує підготовку майбутніх учителів природничо-математичних дисциплін засобами комп'ютерного моделювання; впливає на методику навчання природничо-математичних дисциплін на всіх її рівнях: на рівні цілей навчання (з'являється мета навчання природничо-математичних дисциплін як моделей інформаційних процесів та об'єктів навколишнього світу); на рівні змісту навчання (створюються умови для фундаменталізації навчання, посилення міжпредметних зв'язків та інтеграції різних технологій навчання); на рівні методів навчання (вимагає широкого застосування соціально-конструктивістського підходу до навчання); на рівні засобів навчання (виникає необхідність застосування середовищ об'єктно-орієнтованого моделювання (насамперед VPython, Sage, Squeak» та Alice), систем підтримки навчання (насамперед Moodle) і засобів Web 2.0 (насамперед логосфери, Wiki та хмар сервісів)); на рівні форм організації навчання (створює умови для реалізації комбінованого навчання, навчання в малих групах, парного та групового програмування, зокрема у Web-середовищах).

Проведене дослідження умов підготовки майбутніх учителів природничо-математичних дисциплін засобами комп'ютерного моделювання не вичерпує можливостей застосування технологій моделювання та соціального конструктивізму, що дає підстави для окреслення напрямів подальших досліджень: 1) розроблення теоретико-методичних засад підготовки майбутніх учителів на основі дослідницьких підходів до навчання; 2) тенденції розвитку соціально-конструктивістського навчання у вищій освіті США та країн Європейського Союзу; 3) комп'ютерне моделювання як інтеграційна основа навчання інформатичних дисциплін у фаховій підготовці майбутніх учителів.

## СПИСОК ВИКОРИСТАНИХ ДЖЕРЕЛ

# ДОДАТКИ

## Додаток А
## Об'єктно-орієнтоване моделювання в середовищі Squeak

У роботі [339] авторами було розкрито основні підходи до об'єктно-орієнтованого моделювання в мультимедіа-середовищі мобільного навчання Squeak. Розглянемо першу лабораторну роботу з курсу «Об'єктно-орієнтоване моделювання», спрямовану на формування основних навичок програмування мовою Smalltalk в середовищі Squeak.

### Хід роботи
### Практичні завдання

1. Розробити програму, яка буде виводити привітання 'Hello, world!!!'. Зберегти текст програми на диску.
2. Розробити програму, яка виводитиме на екран список чисел від 0 до 25 через кому.
3. Розробити програму, яка виводитиме на екран список чисел, що відповідають наступним умовам:
   – парне число;
   – непарне число;
   – число містить в кінці запису цифру 5;
   – число кратне 2 і 3 одночасно;
   – число кратне 3 або 5.
   – парні числа, які в розряді одиниць мають число більше 3 і менше 7.
4. Розробити програму побудови таблиці sin, cos, tan і обернених до них функцій.
5. Вивести таблицю додавання, віднімання та множення на числа від 2 до 9 у 8 стовпчиків.
6. Вивести матрицю вигляду:

1 2 3 4 5 6 7 8 9
0 1 2 3 4 5 6 7 8
0 0 9 1 2 3 4 5 6
0 0 0 7 8 9 1 2 3
0 0 0 0 4 5 6 7 8
0 0 0 0 0 9 1 2 3
0 0 0 0 0 0 4 5 6
0 0 0 0 0 0 0 7 8
0 0 0 0 0 0 0 0 9

## Теоретичні відомості
## Обчислення елементарних виразів

Для того, щоб перевірити можливість виконання елементарних операцій зі змінними, управляючими структурами, числами, рядками файлами, можна скористатись робочою областю. Роботу з робочою областю слід використовувати в разі необхідності перевірки виконання тих чи інших елементарних дій.

Для відкриття робочої області слід натиснути лівою кнопкою миші в довільному вільному місці робочого столу і з’явиться меню, в якому слід вибрати пункт **Open...** (рис. А.1). У новому меню, яке з’явиться слід вибрати пункт **Workspace (k)**, після чого можна розпочати перевірку виконання елементарних операцій (рис. А.2).

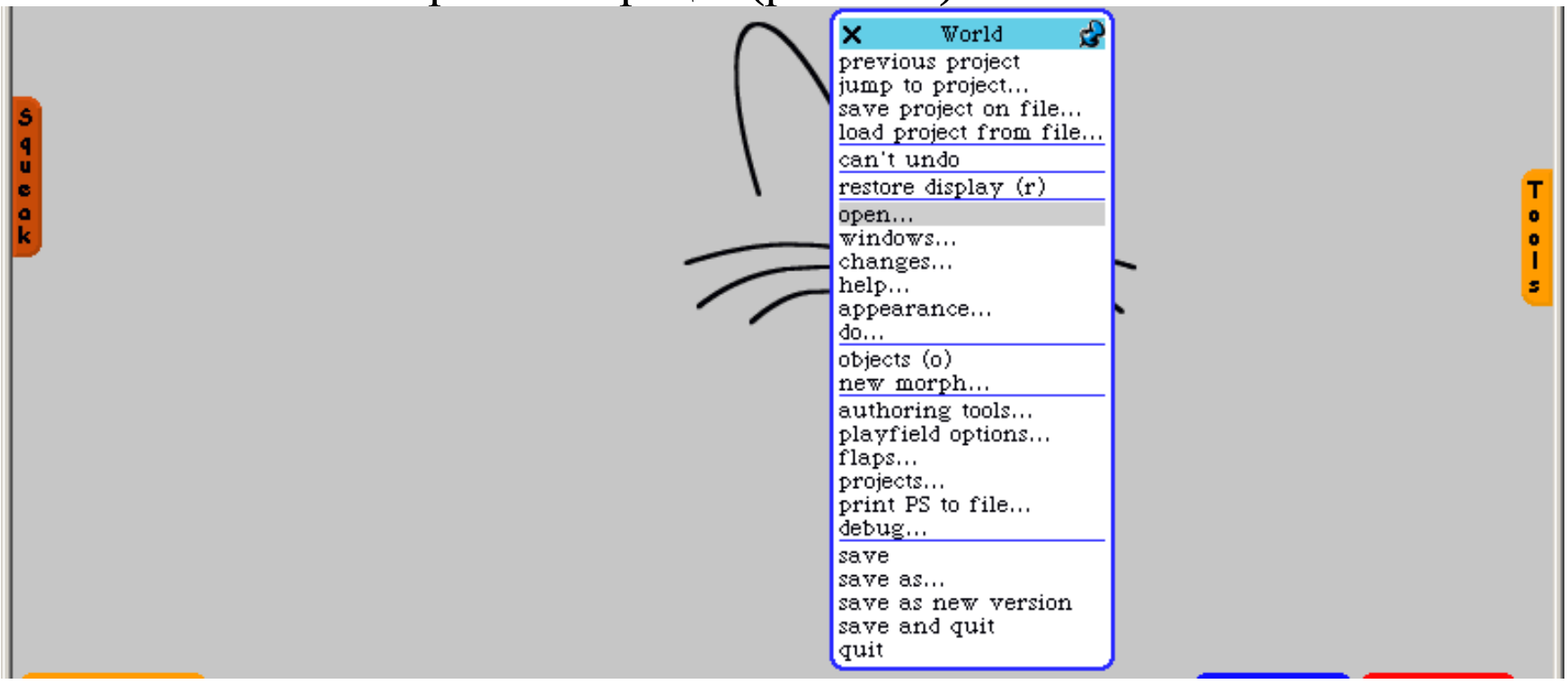

Рис. А.1. Відкриття робочої області

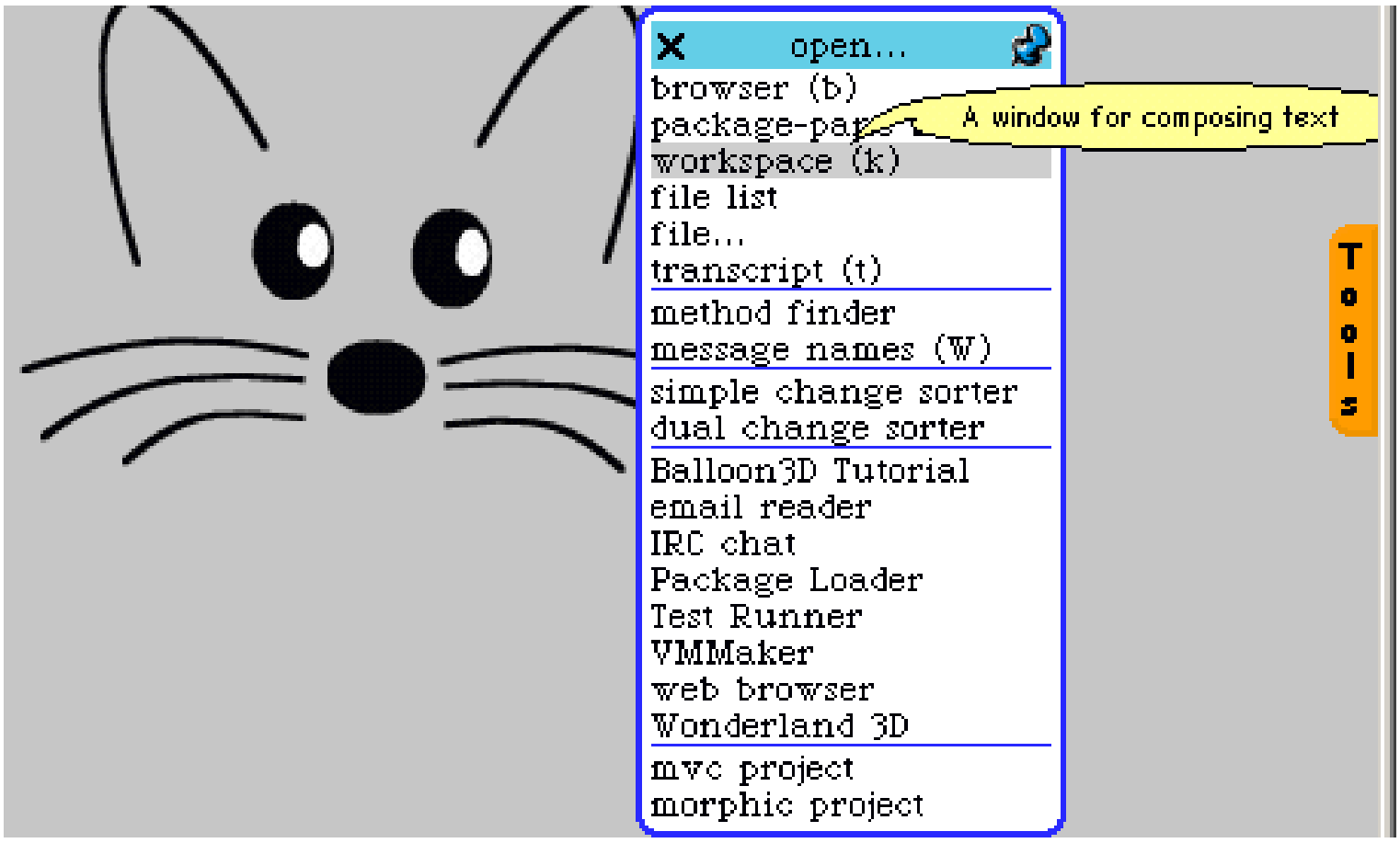

Рис. А.2. Вибір відкриття робочої області.

*Приклад*: Наберіть у робочій області наступний текст: **3+4** та натисніть Alt+p для виконання обчислення даного виразу та його друку в робочій області. Результат зображено на рис. A.3.

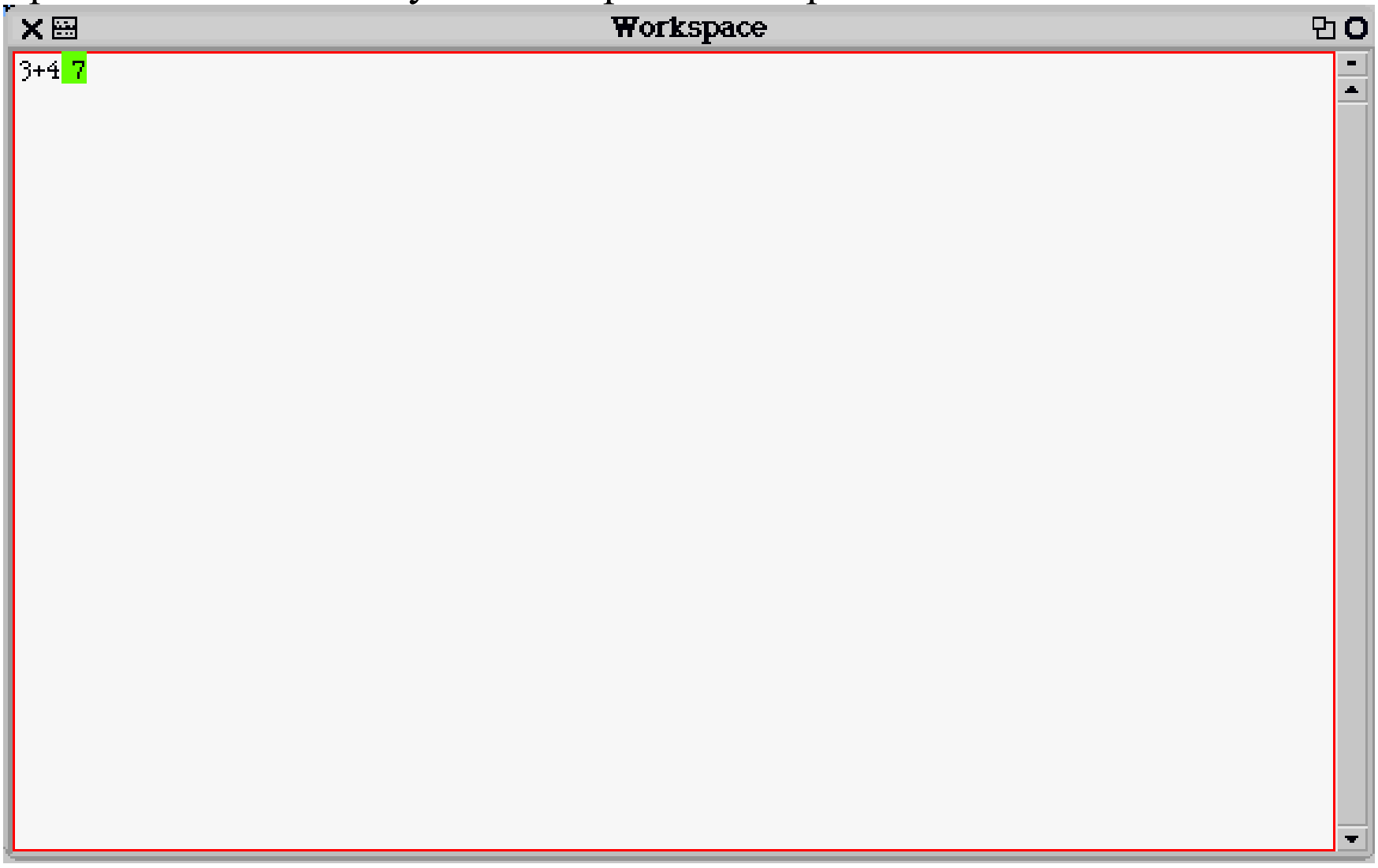

Рис. A.3. Обчислення виразу в робочій області

Іменем змінної може бути одне слово, що починається з літери. Зв’язування змінної зі значенням виконується за допомогою операції **:=** або ←.

```
aVariable := 12.
aVariable ← 'Here is a String'.
```

Кожен завершений вираз відокремлюється точкою. Будь-який вираз у Smalltalk можна сприймати як твердження. Приклади вірних тверджень:

```
1 < 2.
12 positive.
3 + 4.
```

Для змінних є певна область дії. Якщо створено змінну **aVariable**, як вказано в попередньому прикладі, то вона буде діяти в межах робочої області **Workspace**. Створена змінна буде доступна в межах робочої області доти, доки робоча область буде відкрита.

```
myVariable := 34.5.
```

Для оголошення глобальних змінних, першу літеру в імені змінної слід вказувати у верхньому регістрі. Система виведе діалогове вікно, яке вимагатиме підтвердження того, що дана змінна є дійсно глобальною (рис. A.4).

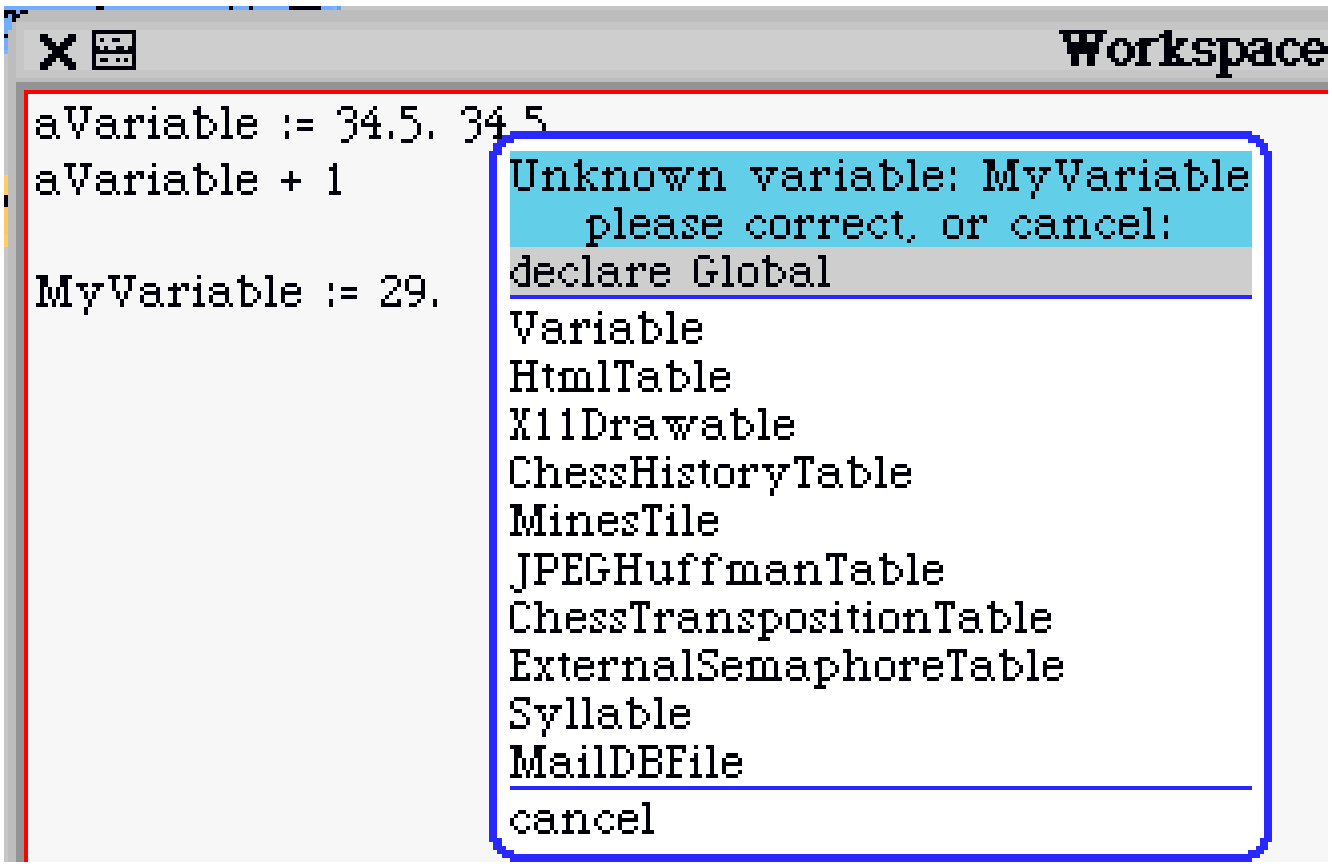

Рис. А.4. Визначення глобальних змінних

### Виведення результатів обчислень

Визначений вище спосіб виведення на екран значення обчисленого виразу (**Alt+p**) не є достатньо зручним. Слушно, коли програміст має бажання виводити результати в окремому вікні. Створимо таке вікно, яке дозволить легко виводити на екран результати виконання різноманітних виразів. Для відкриття вікна **Transcript** необхідно в меню **open…** вибрати пункт **transcript (t)**. Тепер можна виводити дані у вікно **Transcript**, посилаючи відповідні значення у вигляді рядка, використовуючи повідомлення **show** (рис. А.5).

### Синтаксичні конструкції
### Перевірка умови

Синтаксична конструкція перевірки умови дає змогу виконувати операції порівняння і в залежності від їх результатів виконувати різні фрагменти коду.

Формат запису:

(умова) ifTrue: [дія].

Даний формат є аналогом конструкції „if … then …", і дозволяє, в разі істинності умови, виконати фрагмент коду, записаний в квадратних дужка. Якщо є потреба виконати якісь дії в разі хибного значення умови, можна використати іншу конструкцію, що має формат:

(умова) ifFalse: [дія].

Дія цього оператора протилежна попередньому і є аналогом запису „if not … then …". Для повного формату „if … then … else …" слід поєднати ці дві структури в одній.

(умова)

ifTrue: [дія]
ifFalse: [дія].

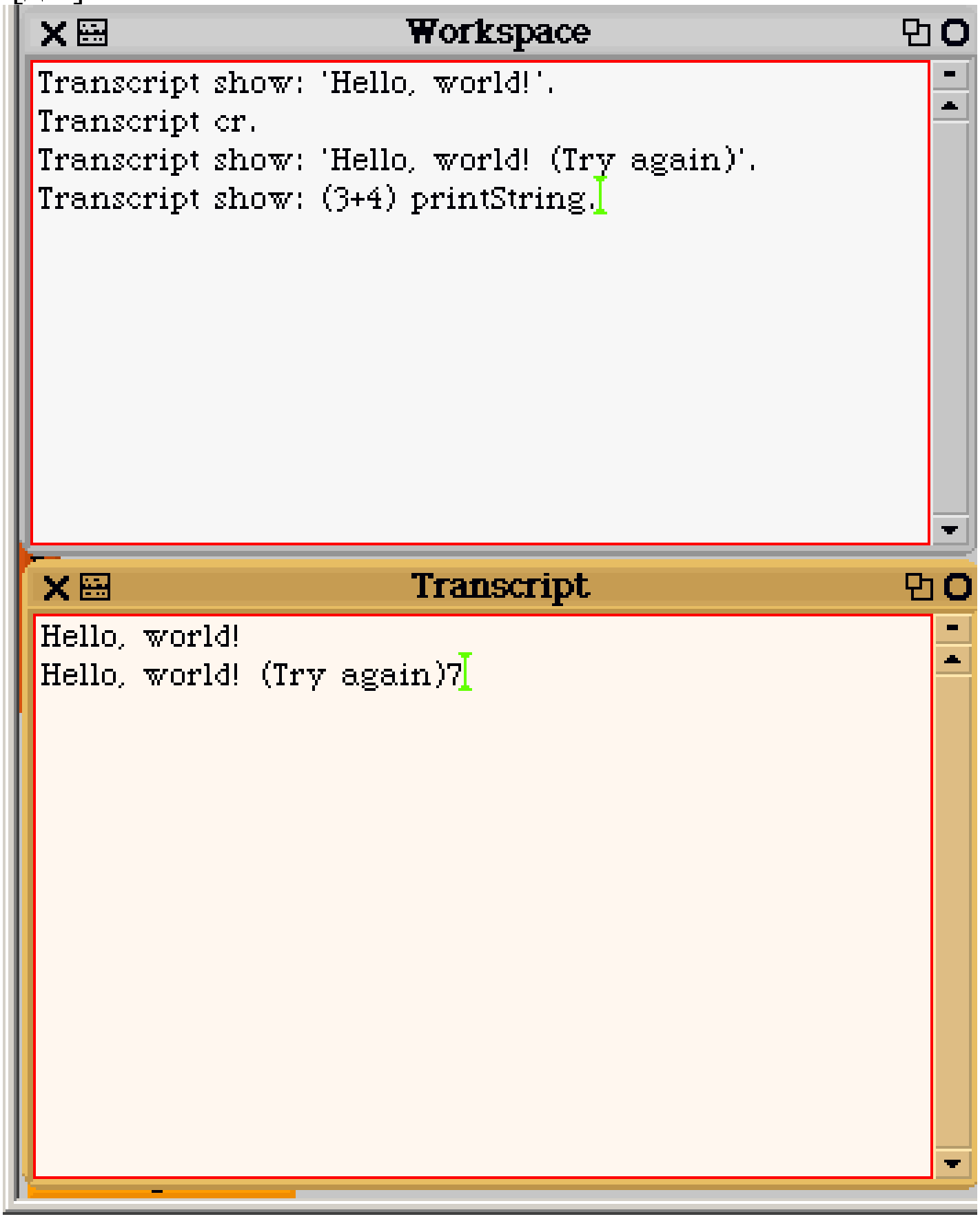

Рис. А.5. Використання Transcript для виведення значень обчислень

Приклади використання даних конструкцій наведено на рис. А.6.

Ще один приклад використання конструкції перевірки умови з використанням логічних операцій:

```
((a < 12) and: [b > 13])
ifTrue: [Transcript show: ‘True!’]
ifFalse: [Transcript show: ‘False!’].
```

Зверніть увагу на запис другої умови в квадратних дужках. Така форма запису символізує те, що and:, як і ifTrue:, ifFalse: є повідомленням для об’єкту логічний вираз. Те ж саме стосується и повідомлення or:.

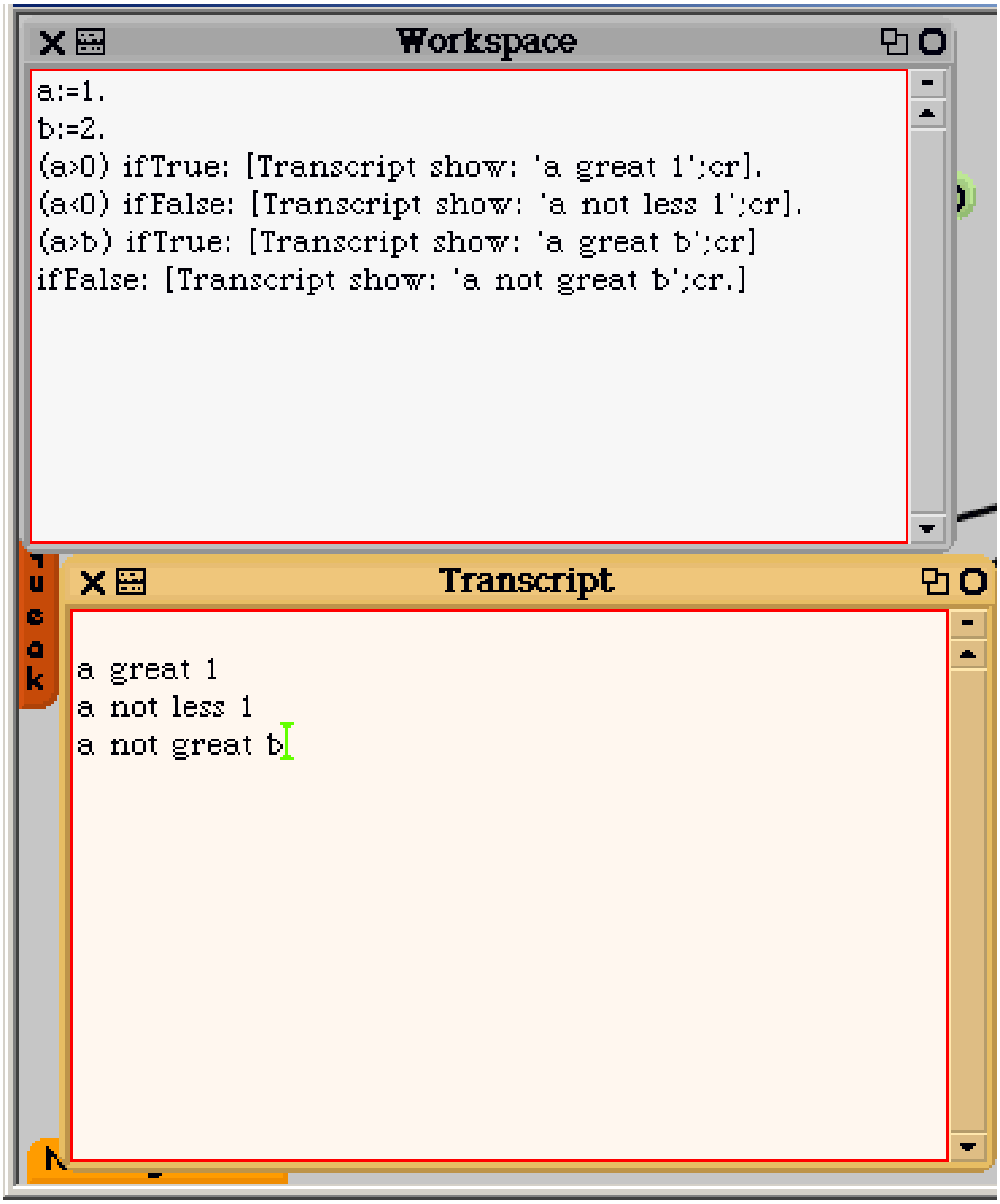

Рис. А.6. Використання синтаксичних конструкцій перевірки умови

## Циклічні конструкції

```
a := 0.
[a < 10] whileTrue:
[a := a + 1. Transcript show: '10 times...'].
```

Даний фрагмент коду демонструє використання циклічної конструкції whileTrue:, яка 10 разів виконує операцію виведення фрази '10 times...' на екран.

Але для розв'язання даної задачі більш лаконічним було б використання конструкції timesRepeat:

```
10 timesRepeat: [Transcript show: '10 times...'].
```

Для багатьох мов програмування дана конструкція мала б використати оператор for. Для нього також є аналог, приклад якого

можна переглянути:

```
0 to: 9 do: [:index | Transcript show:
(index printString),’ times...’].
0 to: 9 do: [:i | Transcript show:
(i printString),’ times...’].
```

Зверніть увагу, що між : і | записується ім’я змінної, що виступає в якості лічильника кількості виконаних операцій.

### Основні операції
### (бінарні)

| | |
|---|---|
| **4 + 3** | Додавання |
| **32.3 – 5** | Віднімання |
| **65 * 32** | Множення |
| **67 / 42** | Ділення (результатом є дійсне число) |
| **10 // 3** | Цілочисельне ділення |
| **10 \\ 3** | Ділення з остачею |

### (унарні)

**(-4) abs** модуль числа.
**90 sin** синус від кута заданого в радіанах
**anArray at: 5** повертає об’єкт масиву, який знаходиться на 5 позиції масиву **anArray**

### (логічні)

**12 positive** перевірка числа на додатність

### Додаток Б

## Об'єктно-орієнтоване моделювання динаміки сонячно-подібних систем у середовищі VPNBody

При вивченні теми 4 «Моделі класичної механіки» («Рух тіл в полі сили тяжіння») курсу об'єктно-орієнтованого моделювання для студентів спеціальності «Фізика» в якості допоміжного програмного забезпечення ми застосовуємо VPNBody – колекцію модулів мовою програмування Python, створену Родні Даннінгом (Rodney Dunning) [23] та локалізовану нами. Назва VPNBody є поєднанням Візуального Python (VP) та N-частковий (N-Body). Комплекс призначений для моделювання систем під дією сили тяжіння, що складаються з гравітаційно домінуючого об'єкта (зірка) і декількох менших об'єктів (планети).

Для використання VPNBody необхідний інтерпретатор Python і модуль VPython. Якщо вони встановлені, запуск системи здійснюється вибором файлу *vpnb.py* (рис. Б.1).

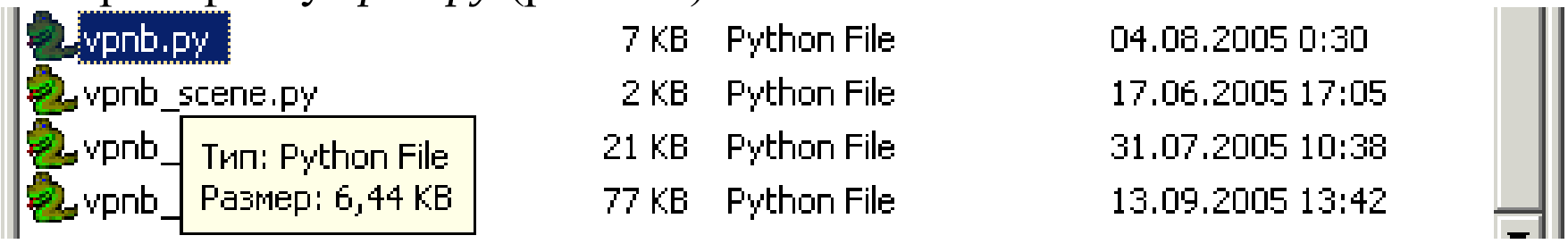

Рис. Б.1. Запуск системи

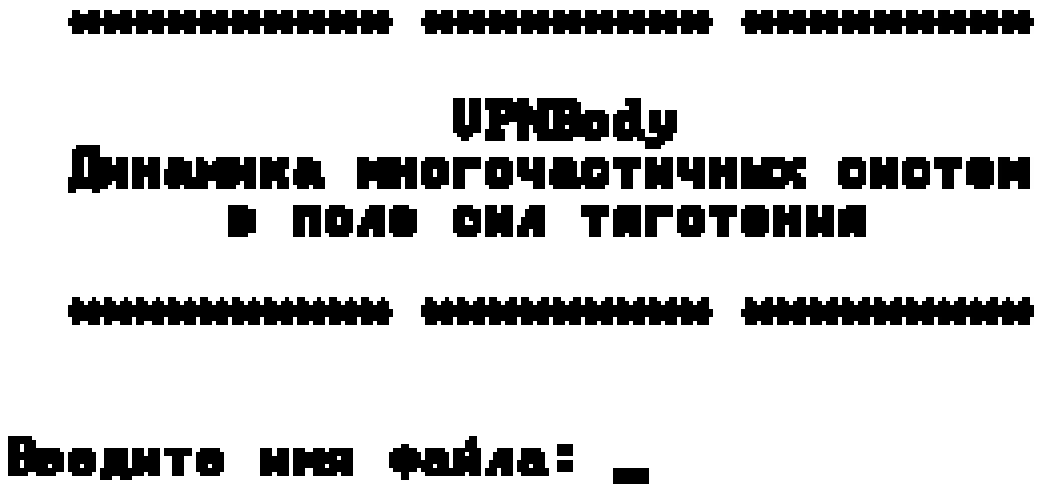

Рис. Б.2. Початковий екран системи

Увівши у запрошенні inner_planets.txt, одержимо наступне вікно (рис. Б.3).

Модуль VPython дає наступні можливості керування вікном анімації:

1. Призупинити і відновити анімацію – натискання лівої кнопки миші.

2. Поворот сцени (системи координат) – переміщення миші при натиснутій правій кнопці.

3. Зміна розміру сцени – утримуючи натиснутими ліву і праву кнопки миші, тягти мишу до верху екрану.

4. Завершення моделювання – натиснути клавішу Escape.

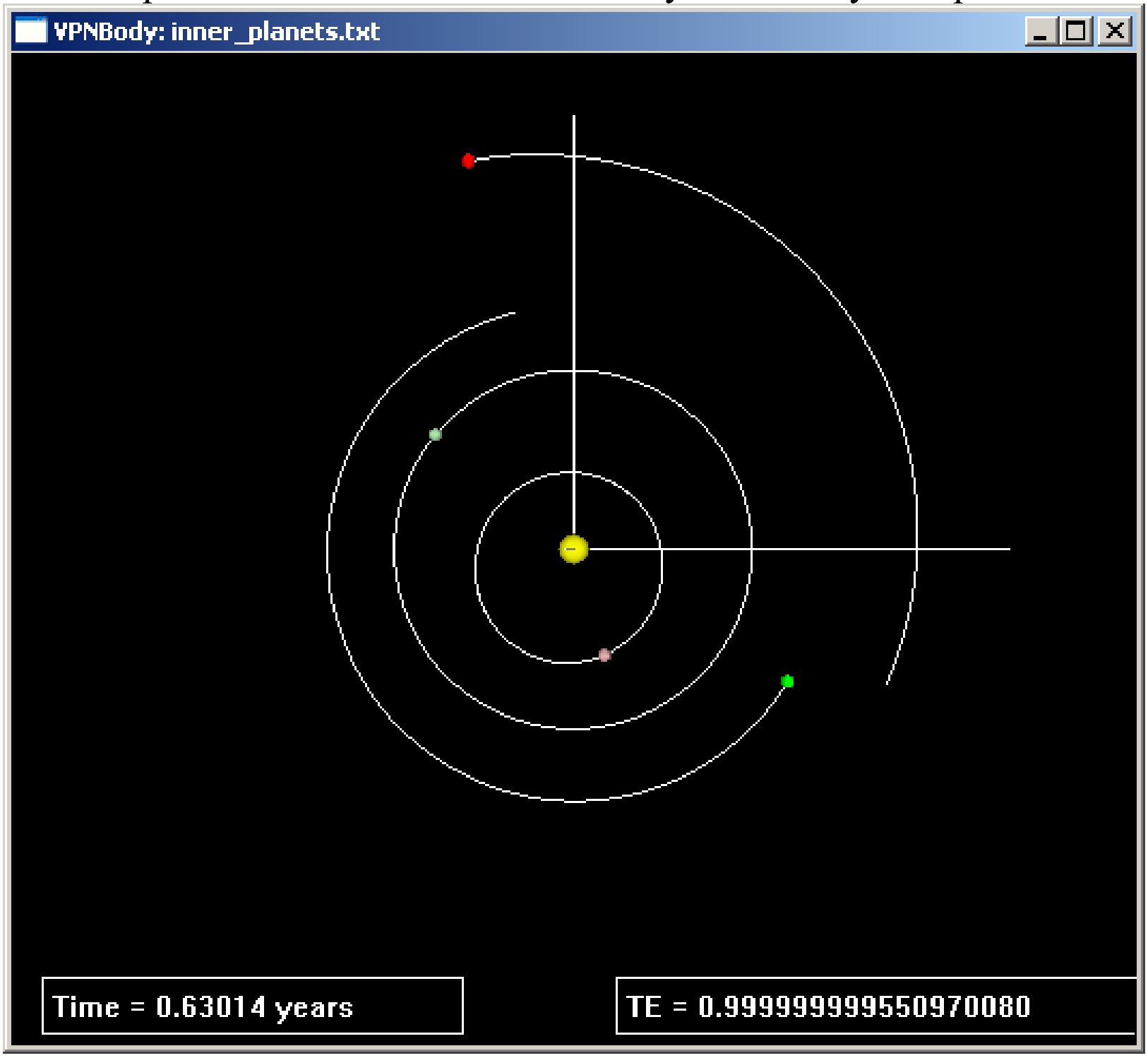

Рис. Б.3. Моделювання руху внутрішніх планет

При масштабуванні та обертанні сцени можна побачити тонкі лінії, що йдуть від Сонця – це необов’язкові координатні вісі (рис. Б.4).

На рис. Б.4 Сонце розташоване в центрі поля зору. Внизу: ліворуч – модельний час (у роках), праворуч – повна механічна енергія системи в одиницях початкової повної механічної енергії. Чим більш це число відрізняється від 1, тим менш точні результати моделювання.

На моделювання динаміки Сонячної системи у пропонованому курсі відведено найбільший час не через складність матеріалу (адже дана задача може бути розв’язана і засобами, доступними допитливому школяреві), а внаслідок багатства та наочності моделей руху. Люди завжди були заворожені небом – колективне людське прагнення зрозуміти рух планет було тією рушійною силою, що сформувала сучасну фізику й астрономію. Сьогодні інтерес до динаміки Сонячної системи залишається сильним і серед професійних учених, і в колі аматорів. Наприклад, недавнє відкриття планет навколо інших зірок підняло важливе питання про те, чи могла б землеподібна планета рухатися по орбіті іншої зірки в межах критичної зони, в якій вода перебуває в рідкому стані на поверхні планети. VPNBody забезпечує зручний та ефективний вступ до цієї області наукових досліджень.

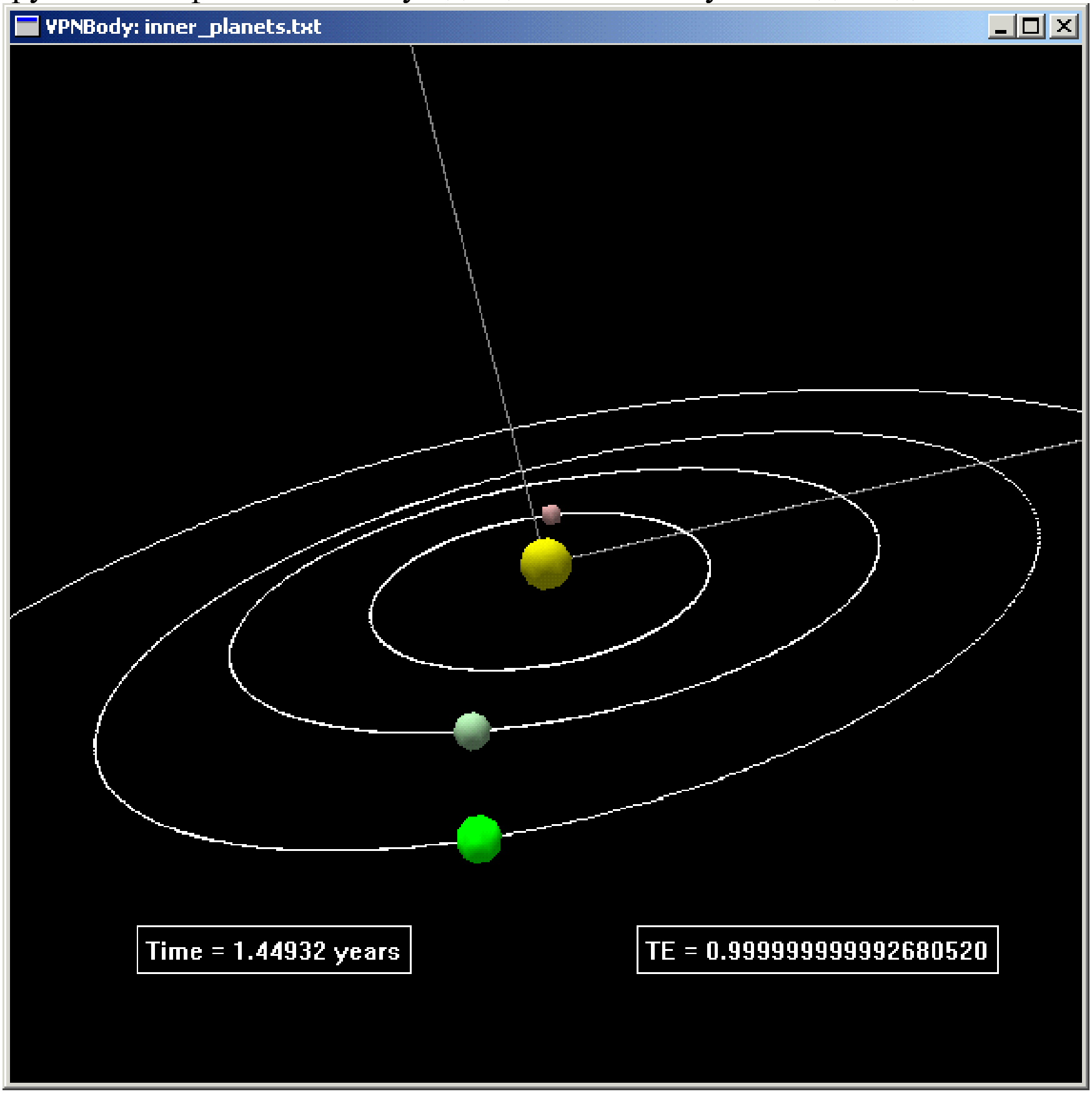

Рис. Б.4. Моделювання внутрішніх планет після обертання і зміни розмірів сцени

Як зазначалося вище, VPNBody призначений для моделювання систем, що складаються з гравітаційно домінантного об'єкта, по орбіті якого рухаються декількома менших об'єктів. Очевидний приклад такої системи – батьківська зірка і дочірні планети, астероїди, комети. Інший приклад – система планета-місяць.

У Сонячній системі зірка і планети взаємодіють за законом всесвітнього тяжіння:

$$F_{ij} = G\frac{m_i m_j}{r_{ij}^2} \qquad (*)$$

В усіх моделях нас цікавить еволюція системи у часі. Взагалі, існує два типи інформації, що ми намагаємося добути з моделі:

1) нас можуть цікавити точні відносні положення планет у даний момент часу,

2) ми можемо бути зацікавлені стабільністю і довгостроковою тривалою еволюцією орбіт – чи може, приміром, землеподібна планета $\varepsilon$-Андромеди залишатися в межах населеної зони настільки довго, щоб там виникло життя?

Тип інформації, що ми його хочемо одержати, визначає методи, які використовуються для дослідження системи.

Еволюція системи визначається обчисленням траєкторії для кожної планети за рівнянням (*). Є багато способів зробити це, проте всі вони можуть бути поділені на дві великі категорії. У першій категорії – алгоритми Рунге-Кутта (РК). Алгоритм РК високого порядку може використовуватися для відстеження руху планет з майже машинною точністю (принаймні, певний час). Помилка, пов’язана з алгоритмами РК, не обмежена, і тому ці алгоритми не можуть звичайно використовуватися для довгострокового моделювання системи. В другій категорії – симплектичні алгоритми. На відміну від РК-методів, симплектичні алгоритми не зберігають повну механічну енергію системи, але енергетична помилка обмежена. Таким чином, симплектичний алгоритм може використовуватися для еволюції системи протягом майже необмеженого часу (можливі обмеження зумовлюються помилками округлення). Взагалі, РК-методи використовуються, коли ми хочемо знати відносні положення планет на деякий момент часу в найближчому майбутньому, а симплектичні методи використовуються, коли ми хочемо знати, чи буде орбіта даної планети стійкою протягом кількох мільйонів років.

Для одержання траєкторії кожного об’єкта системи з рівняння (*) ми повинні знати *початкові умови* системи. Існує кілька способів їхнього задання. Перший – розташування планет на одній з осей («парад планет») і задання для кожної з них швидкості [6–7]. Але більш зручно застосовувати *орбітальні елементи*. Орбітальні елементи планети описують її розміри (одне число), форму орбіти (одне число), орієнтацію (три числа) орбіти в просторі, і поточне положення об’єкта на орбіті.

Перший орбітальний елемент – велика піввісь $a$ еліптичної орбіти планети відносно зірки.

Другий орбітальний елемент – ексцентриситет $e$ планетної орбіти.

Ексцентриситет описує форму орбіти ($e$ = 0 визначає ідеальне коло, граничне значення ексцентриситету $e$ = 1).

Третій орбітальний елемент – нахил $i$, що вимірюється відносно заданої опорної площини. У Сонячній системі опорною вважається площина орбіти Землі і нахили решти планетних орбіт вимірюються відносно неї. Рис. Б.5 показує орбіти Меркурія і Землі. Нахил орбіти Меркурія відносної земної – $i$ орбіти Меркурія.

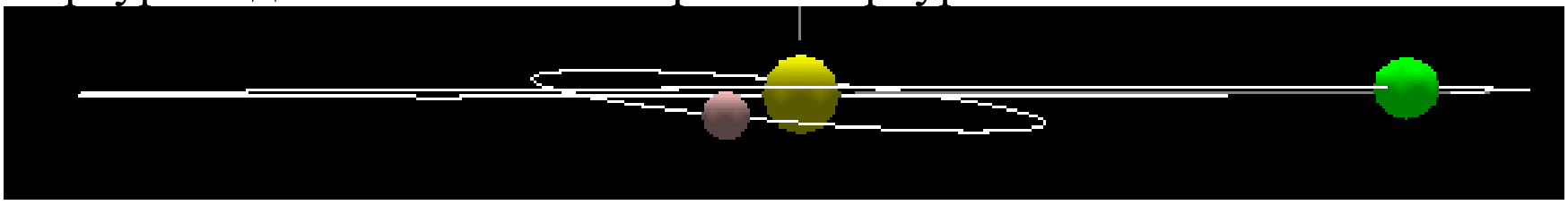

Рис. Б.5. Нахил орбіти Меркурія відносно земної $i$ = 7°

В анімаціях VPNBody опорна площина – завжди початкова площина сцени (площина комп'ютерного монітора).

Четвертий орбітальний елемент – довгота точки сходження $\Omega$. Точка сходження (висхідний вузол) – точка в опорній площині, у якій орбіта планети при русі нагору перетинає опорну площину. Довгота точки сходження – кут, вимірюваний в опорній площині проти годинникової стрілки від осі $x$ до точки сходження. В Сонячній системі вісь $x$ вказує на точку весняного рівнодення – позицію в небі, що її займає Сонце у перший день весни.

П'ятий орбітальний елемент – напрямок на перицентр $\omega$. Перицентр – точка зближення на найкоротшу відстань між планетою і зіркою (відповідно точка максимального віддалення називається *апоцентром*). Напрямок на перицентр – кут $\omega$, що вимірюється проти годинникової стрілки в площині орбіти планети від точки сходження до перицентра.

Шостий орбітальний елемент – середня аномалія $M$. Середня аномалія – кут, вимірюваний проти годинникової стрілки в площині орбіти планети від перицентра до позиції *середньої планети*. Реальні планети рухаються по орбіті зі змінною швидкістю; через збереження кутового моменту ця швидкість найбільша в позиції перицентра і найменша в позиції апоцентра. Середня планета рухається по орбіті з *постійною* швидкістю так, що середня планета і реальна планета збігаються в позиціях перицентра й апоцентра.

Приклади моделей.

**asteroids.txt** – включає Сонце, Марс, Юпітер, Сатурн і 15 астероїдів з випадковими орбітальними елементами, розташовані між орбітами Марса і Юпітера (рис. Б.6).

**comet_inner.txt** – включає Сонце, Меркурій, Венеру, Землю, Марс і малу комету, що рухається по орбіті з великим ексцентриситетом (рис. Б.7*а*).

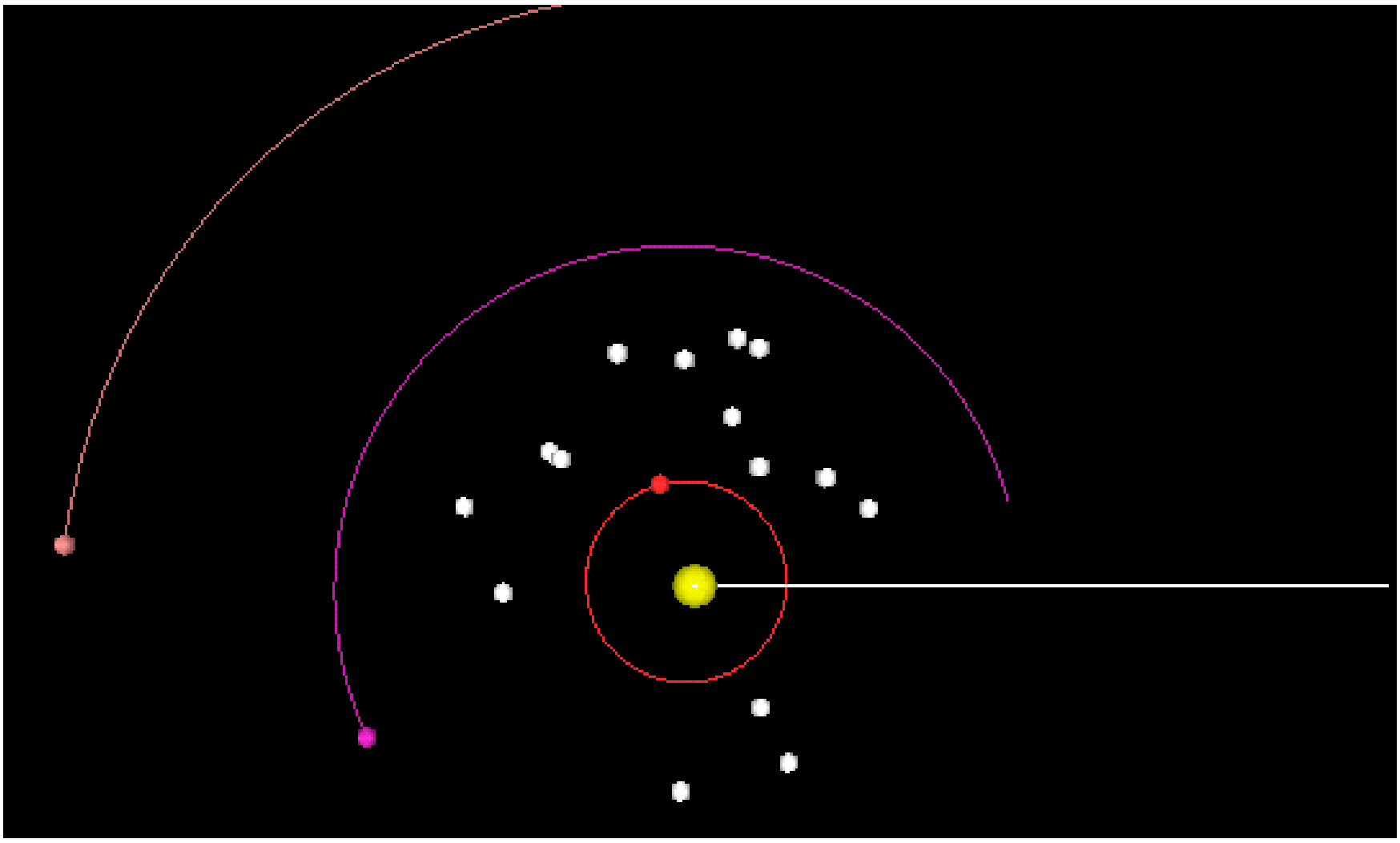

Рис. Б.6

**comet_outer.txt** – включає Сонце, Юпітер, Сатурн, Уран, Нептун, Плутон і малу комету (рис. Б.7*б*).

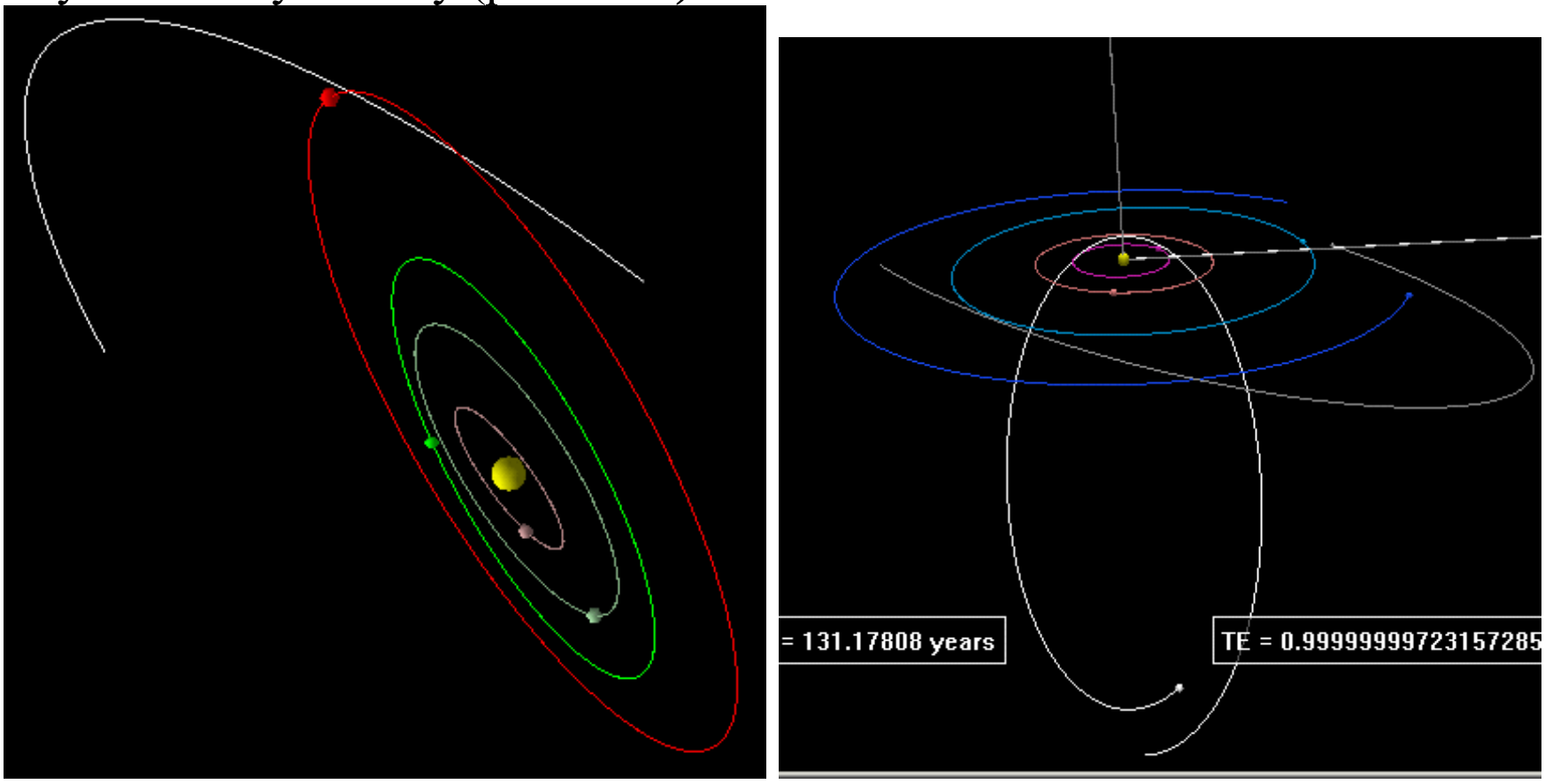

а) б)

Рис. Б.7

**compare_a.txt** – включає зірку і три планети з ідентичними орбітальними елементами, за винятком півосей. Анімація ефективно показує відношення між півосями й орбітальним періодом (рис. Б.8).

**compare_ap.txt** – включає зірку і три планети з ідентичними орбітальними елементами, за винятком напрямку на перицентр

(рис. Б.9).

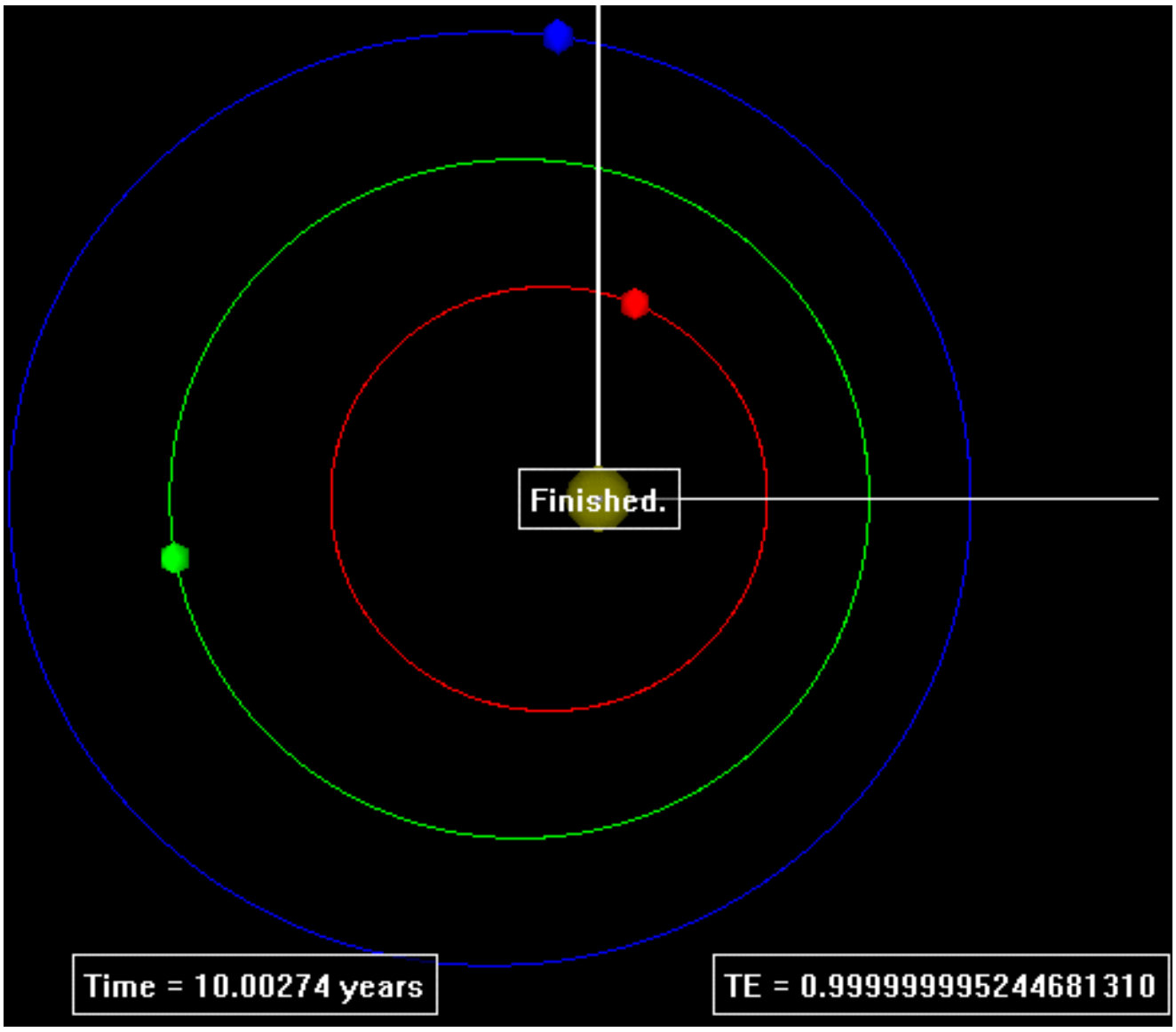

Рис. Б.8

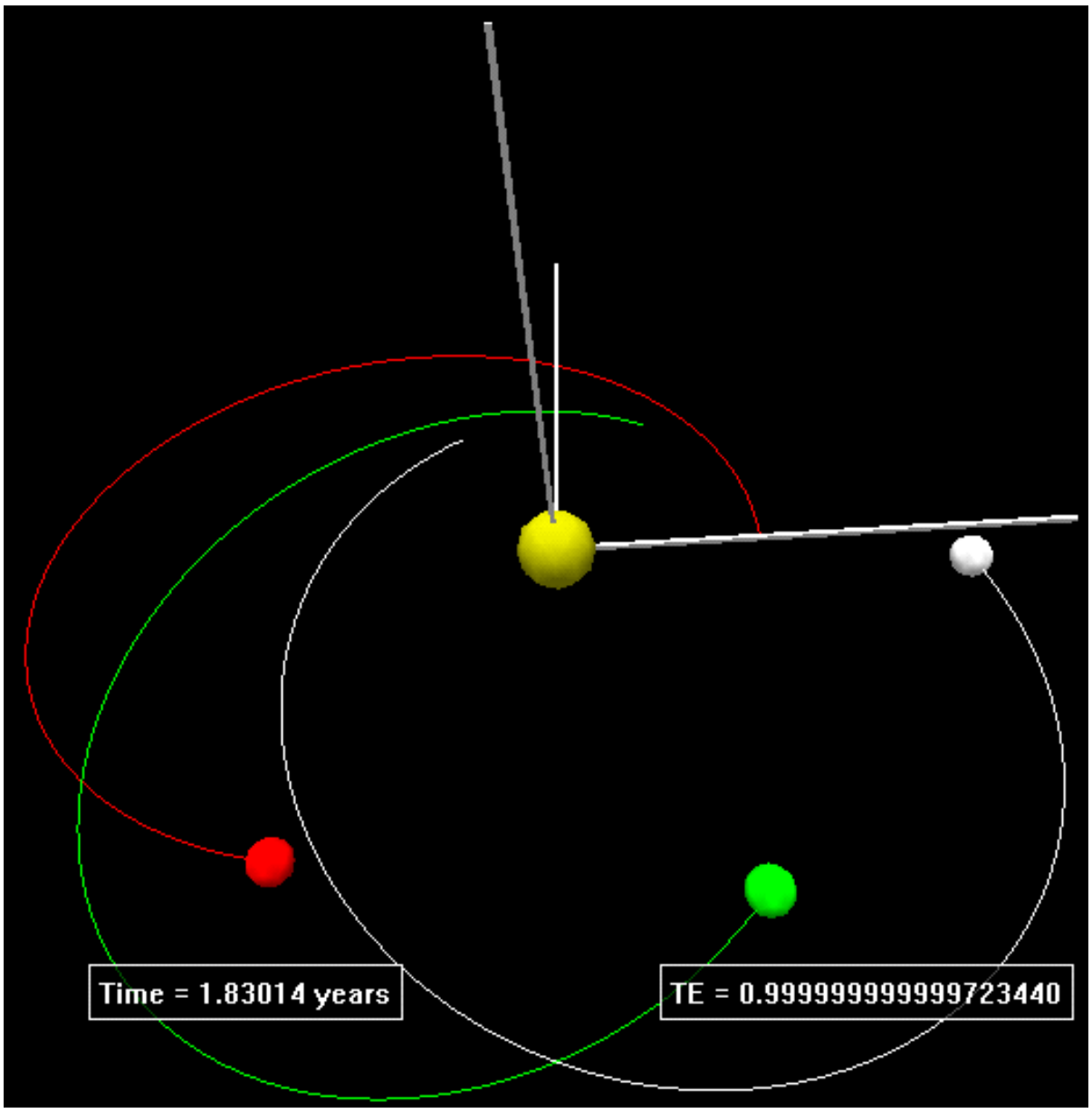

Рис. Б.9

**compare_e.txt** – включає зірку і три планети з ідентичними орбітальними елементами, за винятком ексцентриситету (рис. Б.10). Кожна планета починає рух з напрямку на перицентр, повертаючись за однаковий час. Це модель показує, що між ексцентриситетом і орбітальним періодом немає зв'язку.

**compare_i.txt** – включає зірку і три планети з ідентичними орбітальними елементами, за винятком орбітальних нахилів (рис. Б.11). Два ненульових нахили демонструють геометричний зміст даного поняття. Один з нахилів, більший за 90°, демонструє ретроградну орбіту. Як відомо, в Сонячній системі ретроградну орбіту має Уран.

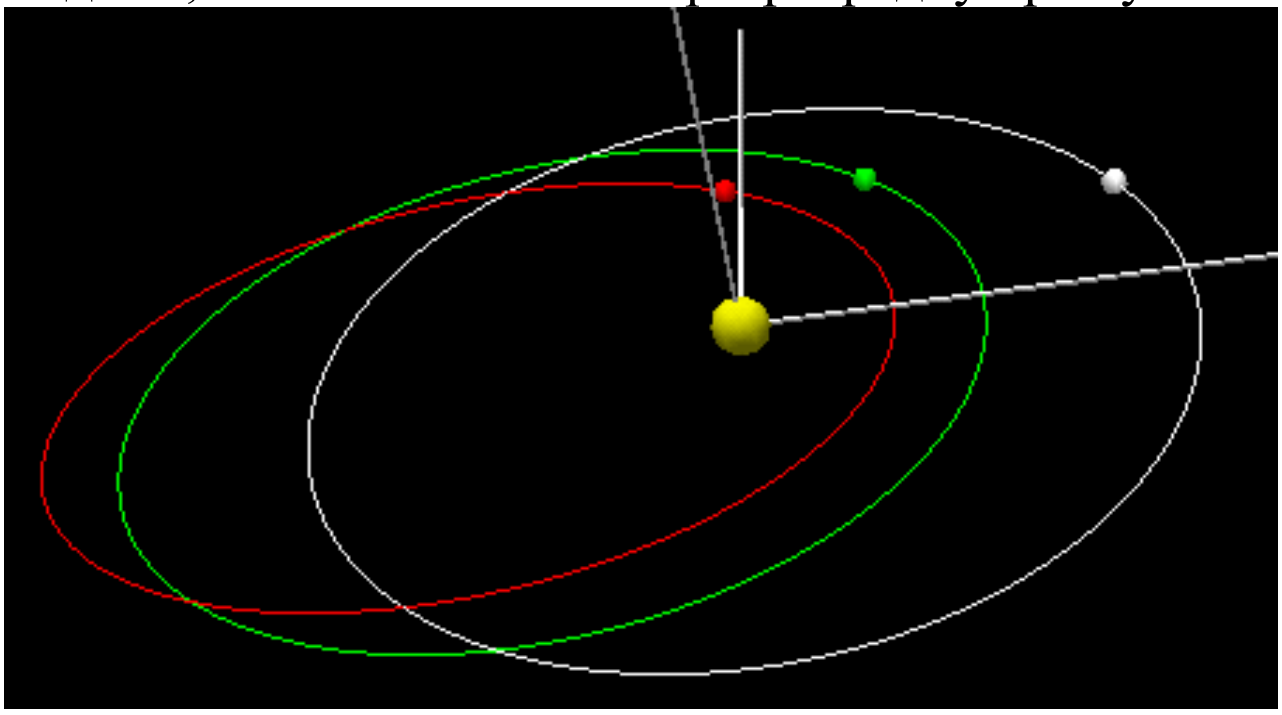

Рис. Б.10

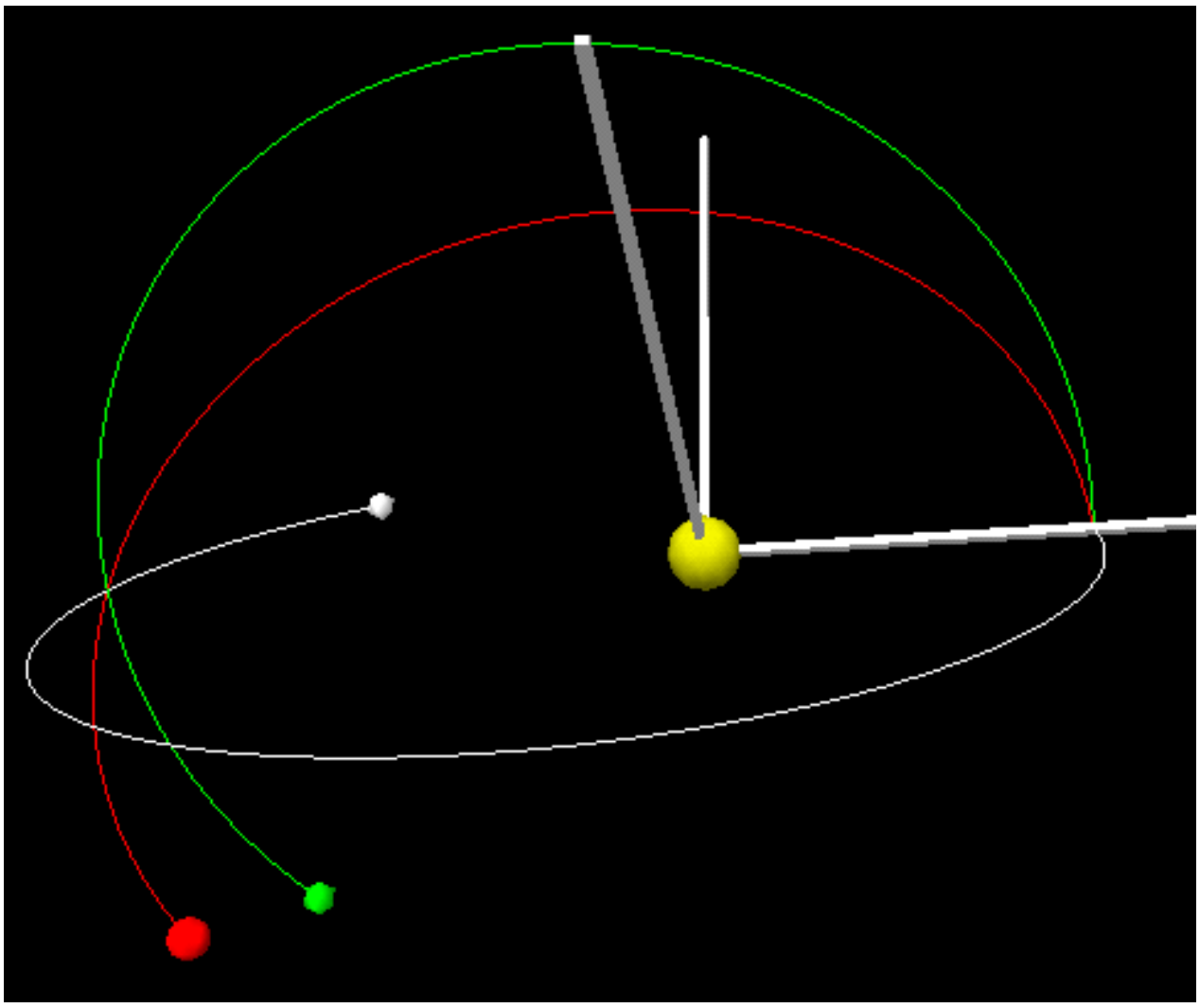

Рис. Б.11

**compare_lan.txt** – включає зірку і три планети з ідентичними орбітальними елементами, за винятком довготи точки сходження (рис. Б.12). Довготу точки сходження з усіх орбітальних елементів візуалізувати найважче, тому дана модель буде корисна для розуміння сутності цього поняття.

**compare_ma.txt** – включає зірку і три планети з ідентичними орбітальними елементами, за винятком початкових середніх аномалій – стартових позицій і швидкостей (рис. Б.13).

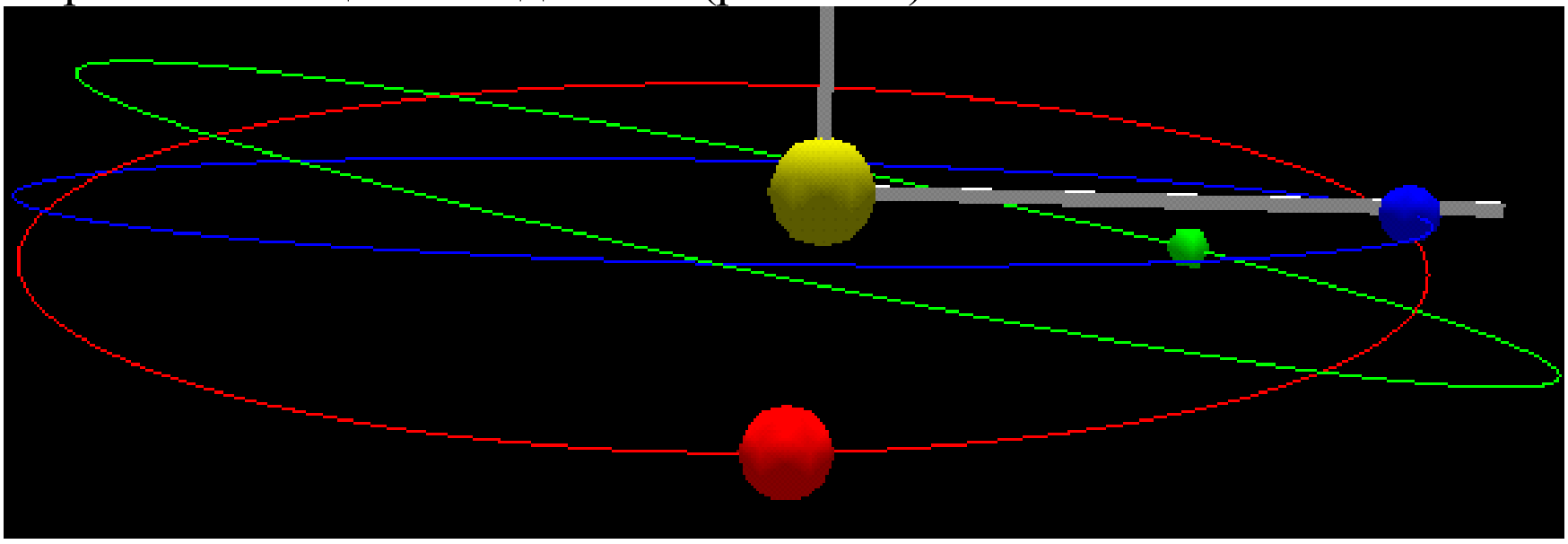

Рис. Б.12

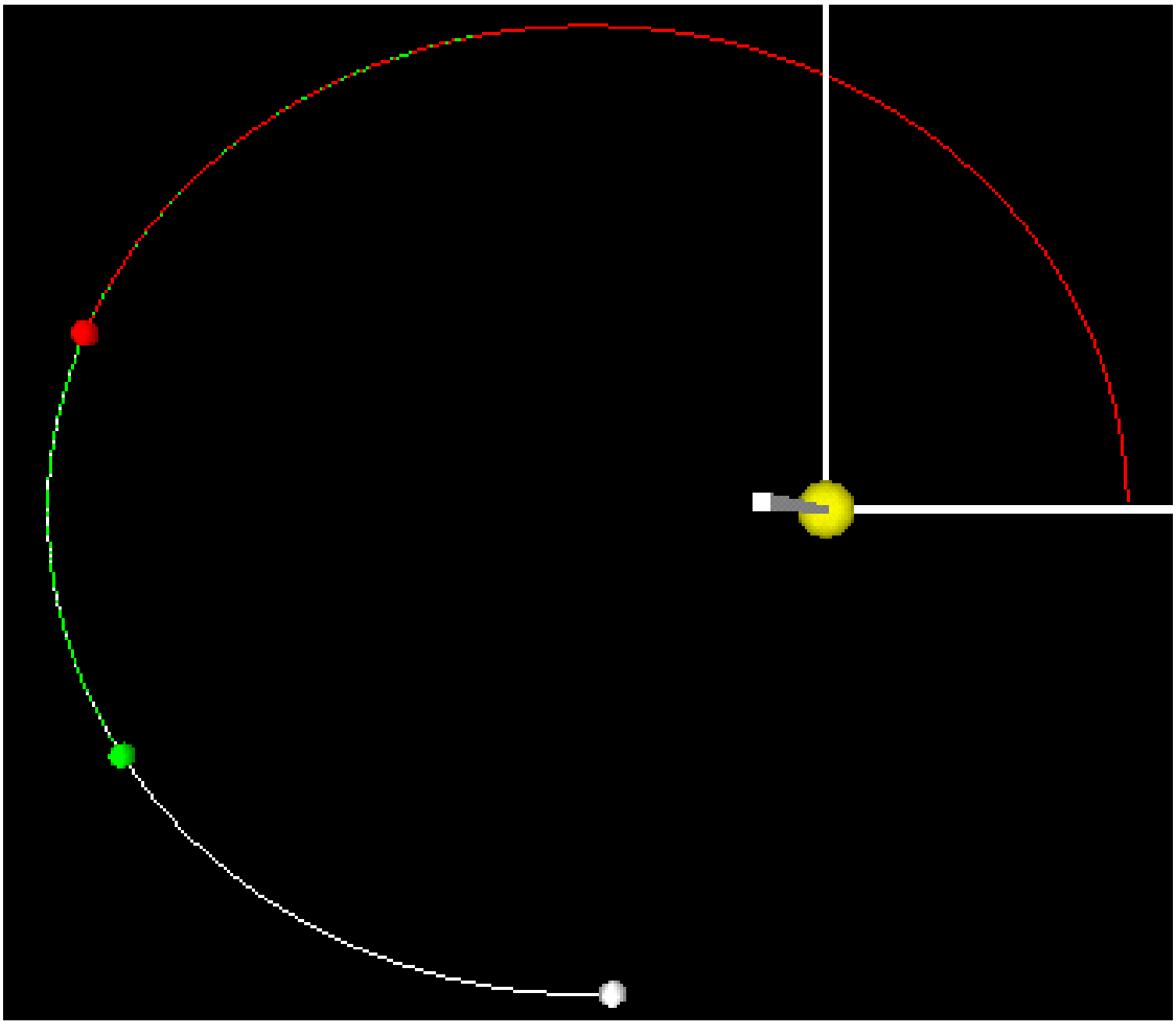

Рис. Б.13

**draw_full_solar_system.txt** – створює картину всіх дев'яти планет Сонячної системи. Радіус Сонця відображається в коректному масштабі (рис. Б.14).

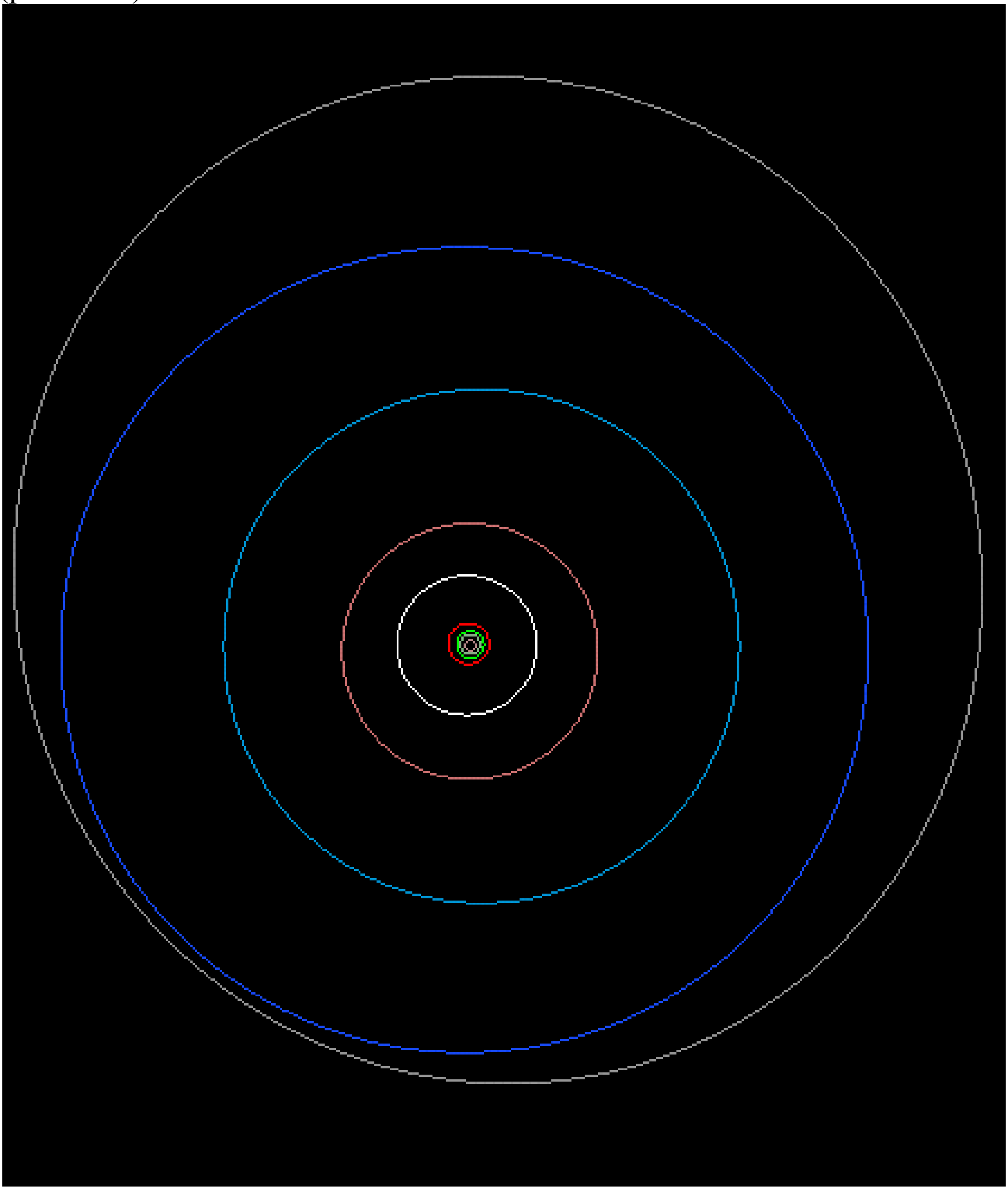

Рис. Б.14

**rogue_star_inner.txt** – у цій моделі невелика зірка (0,5 маси Сонця) блукає серед внутрішніх планет і спотворює їхньої орбіти (рис. Б.15).

**rogue_star_outer.txt** – у цій моделі зірка в 1,5 маси Сонця блукає серед зовнішніх планет і спотворює їхньої орбіти (рис. Б.16).

**sun_wobble.txt** – включає Сонце, Юпітер і Сатурн. При збільшенні зображення можна спостерігати рух Сонця під дією Юпітера і Сатурна (рис. Б.17).

Таким чином, при вивченні динамічних моделей механіки доцільно використати програмний комплекс VPNBody. Сферами застосування VPNBody є лекційні демонстрації планетних орбіт, орбітальних елементів, тривимірного характеру орбіт (орбітних нахилів), планетних конфігурацій (протистоянь, сполучень тощо), ретроградного (зворотного) руху, ідентифікація екзосистем. Моделі, створені за допомогою VPNBody, можуть використовуватися також для лабораторних робіт.

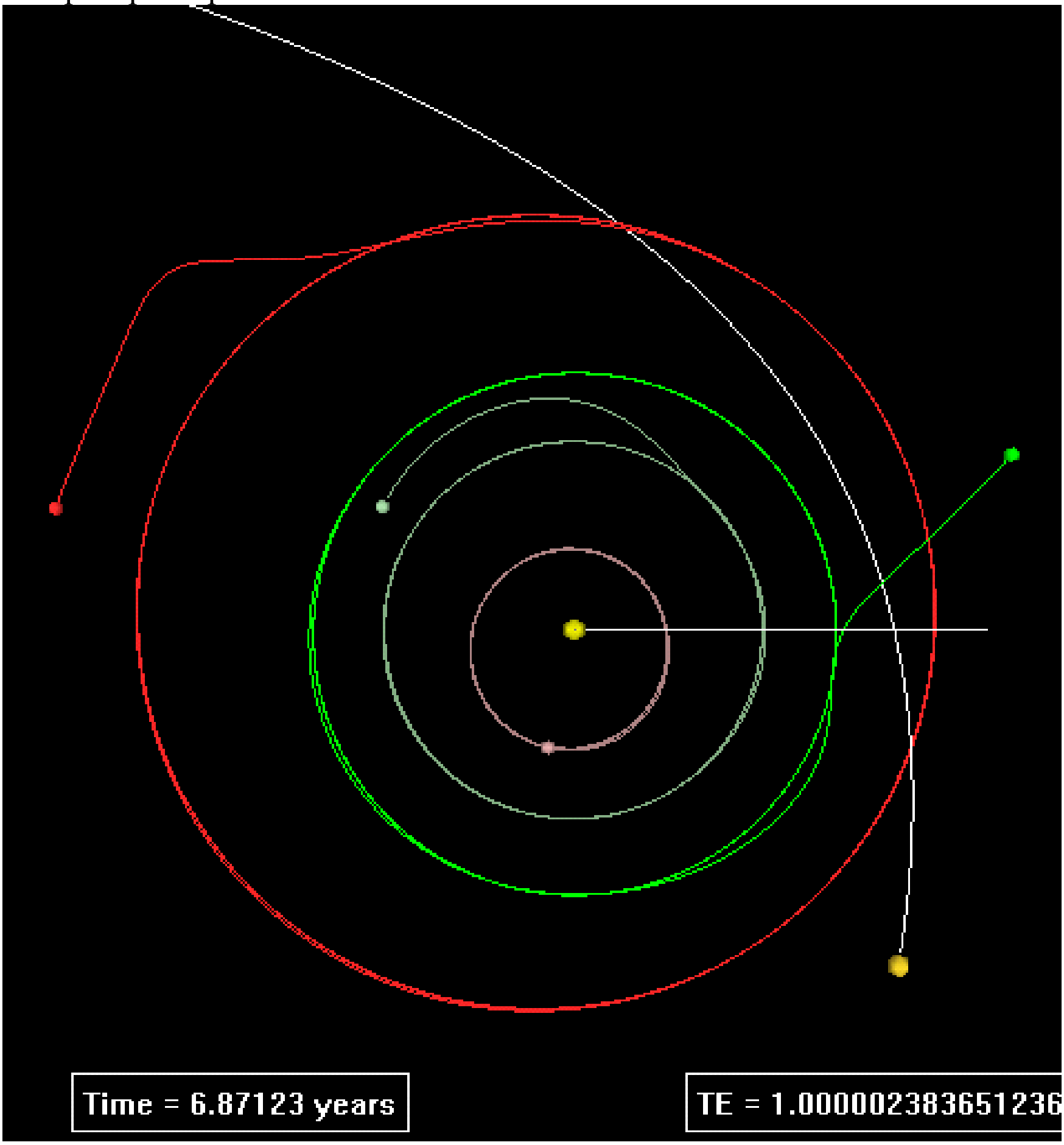

Рис. Б.15

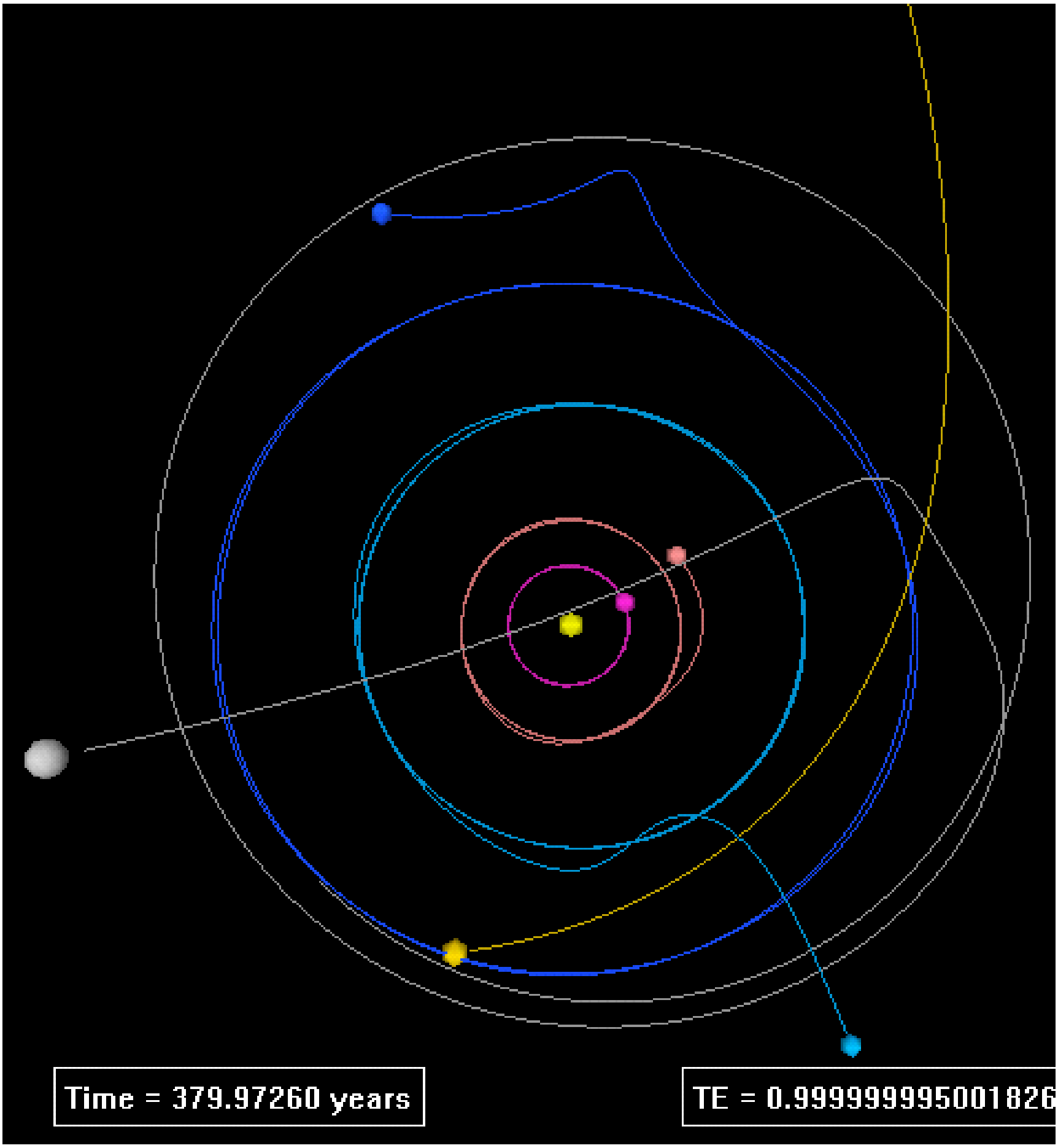

Рис. Б.16

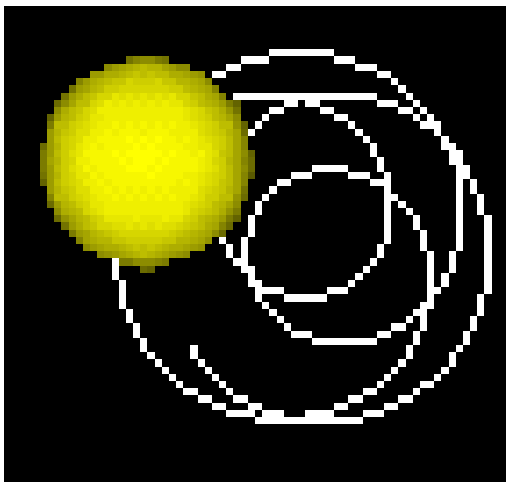

Рис. Б.17

## Додаток В
## Тематика проектів з курсу «Об’єктно-орієнтоване моделювання»

Дані проекти, запропоновані студентам спеціальності «Фізика», були виконані у двох середовищах – VPython (мовою Python) та Squeak (мовою Smalltalk). Наведемо інтерфейс моделей, виконаних у середовищі Squeak:

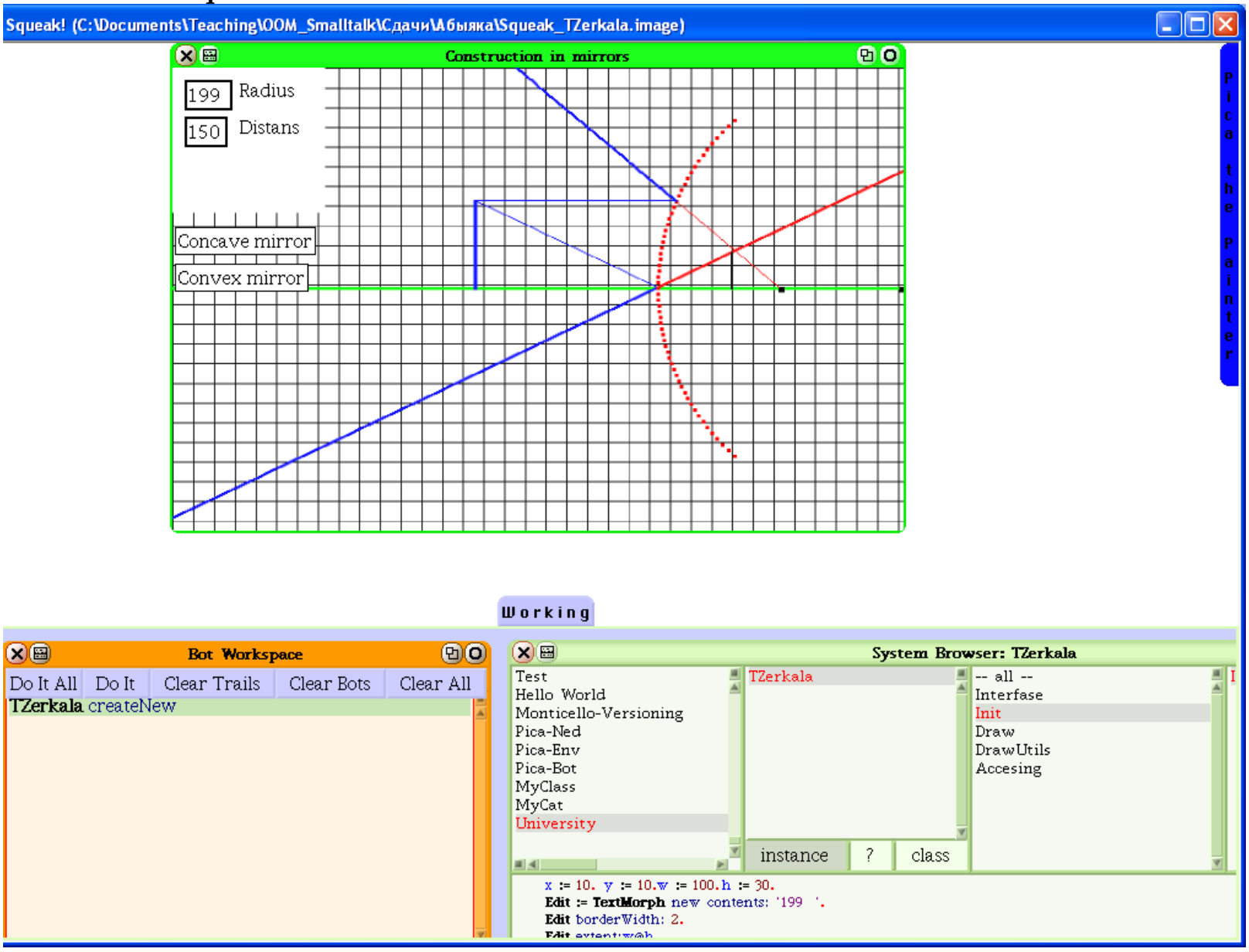

Рис. В.1. Моделювання ходу променів у дзеркалах

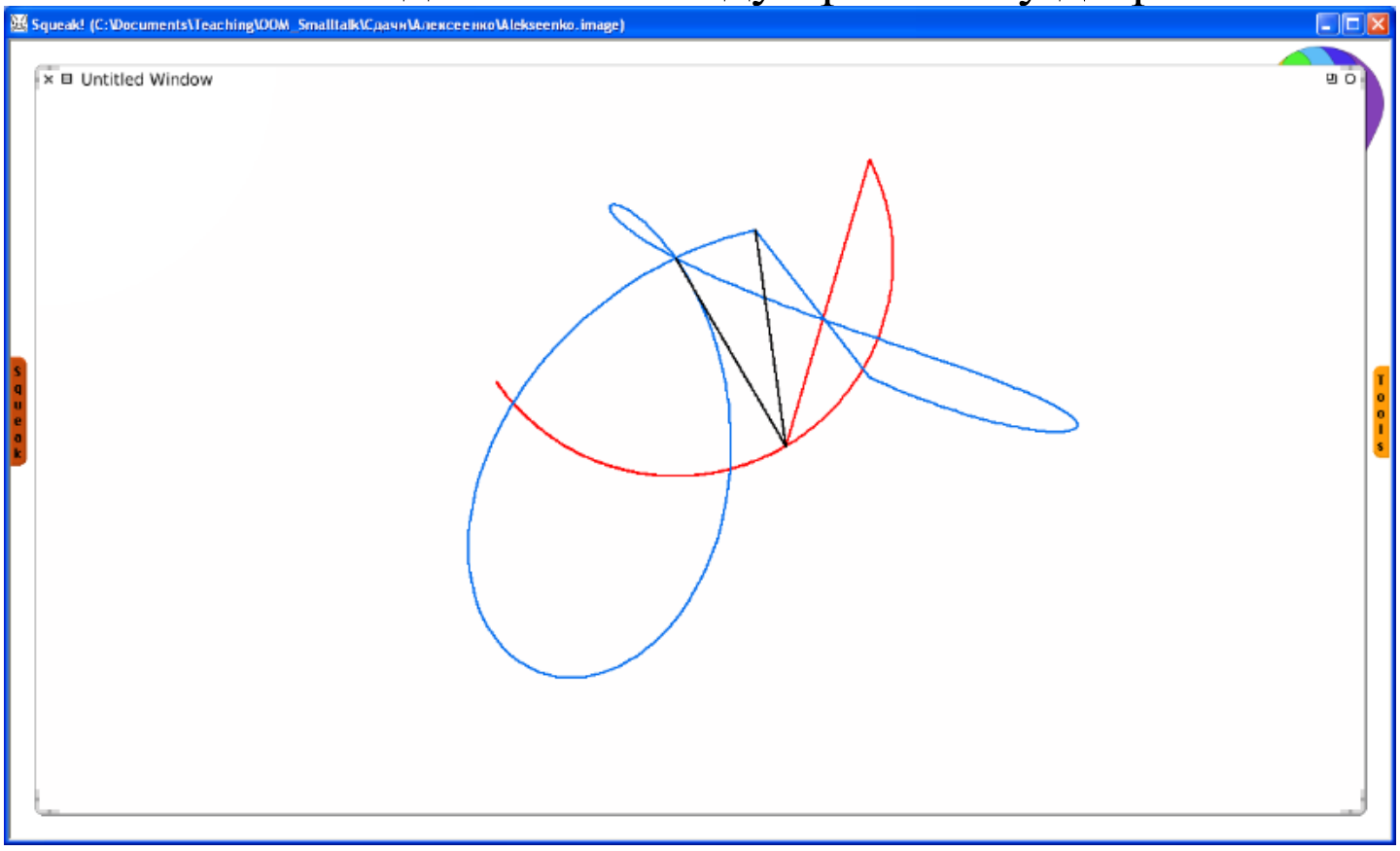

Рис. В.2. Моделювання коливання подвійного маятника

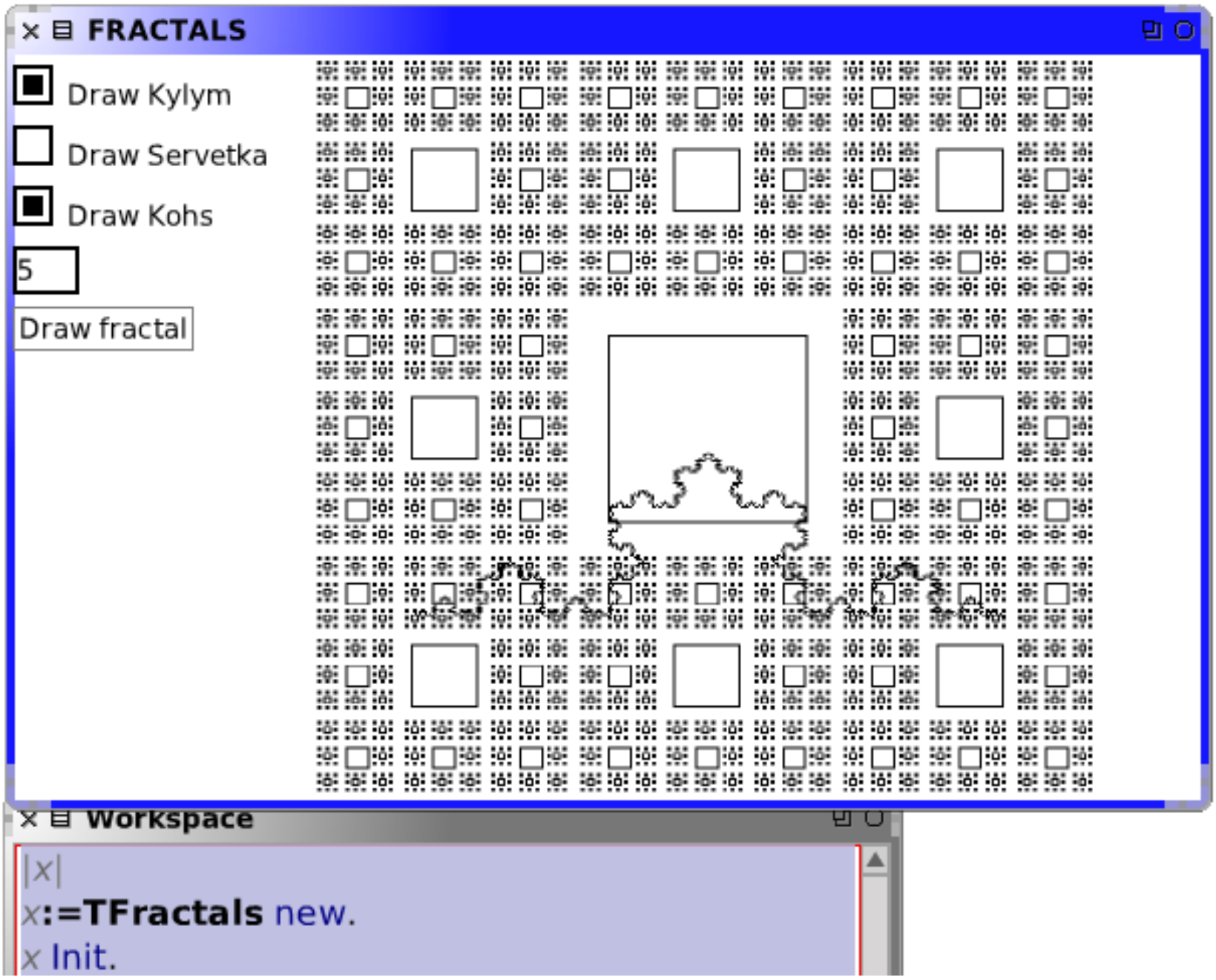

Рис. В.3. Моделювання регулярних фракталів

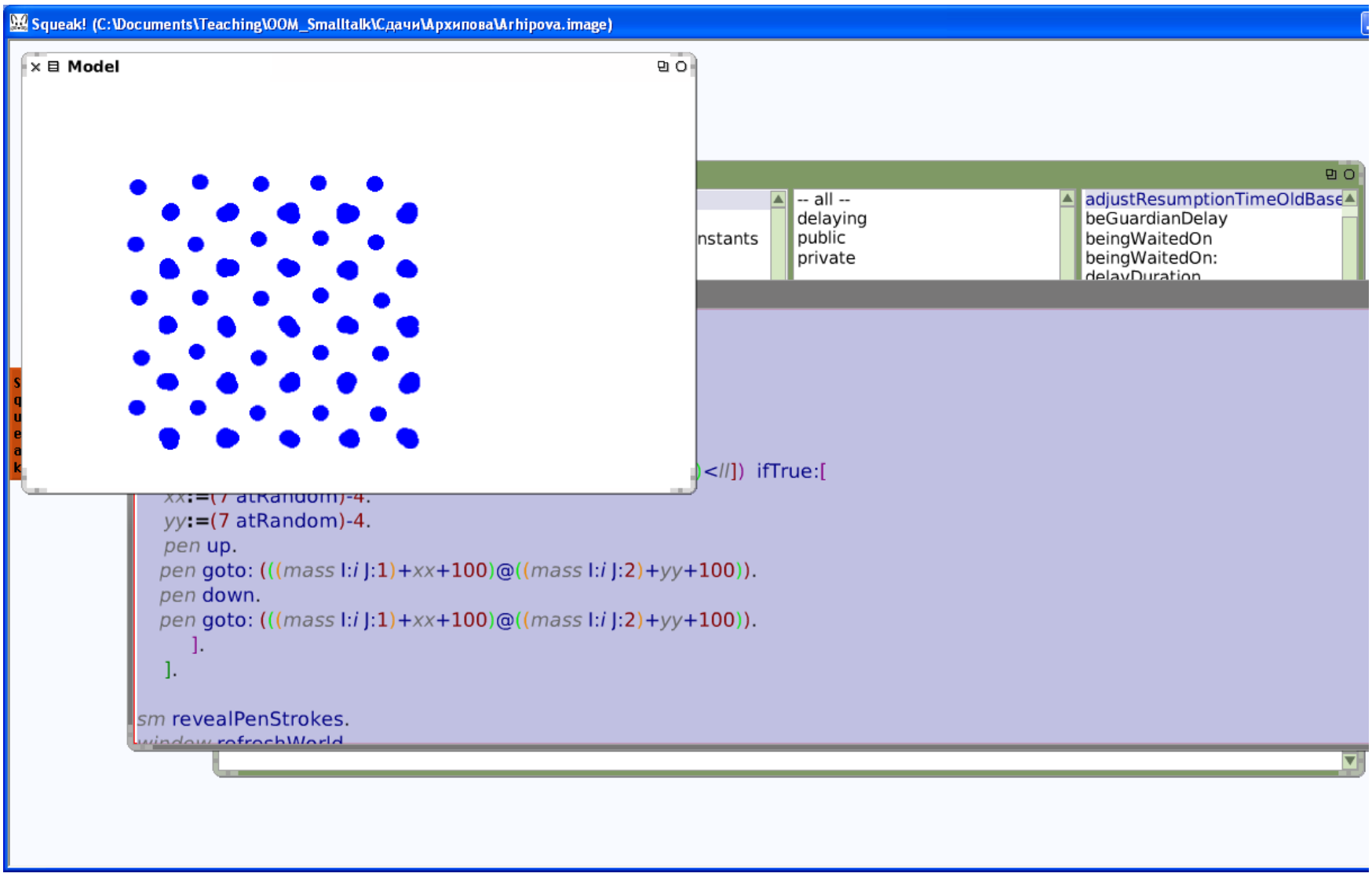

Рис. В.4. Моделювання коливань у кристалічній гратці алмазу

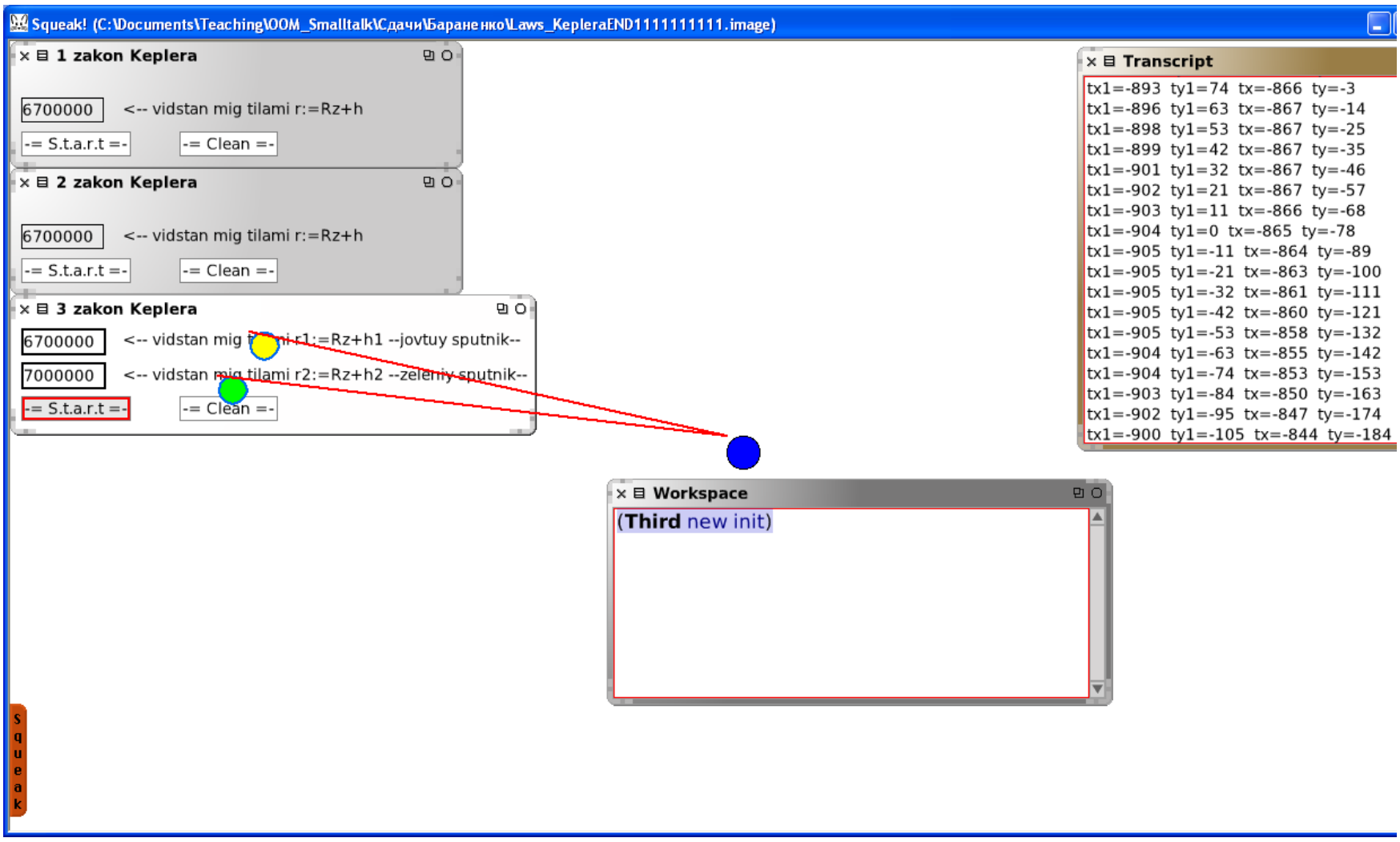

Рис. В.5. Моделювання законів Кеплера

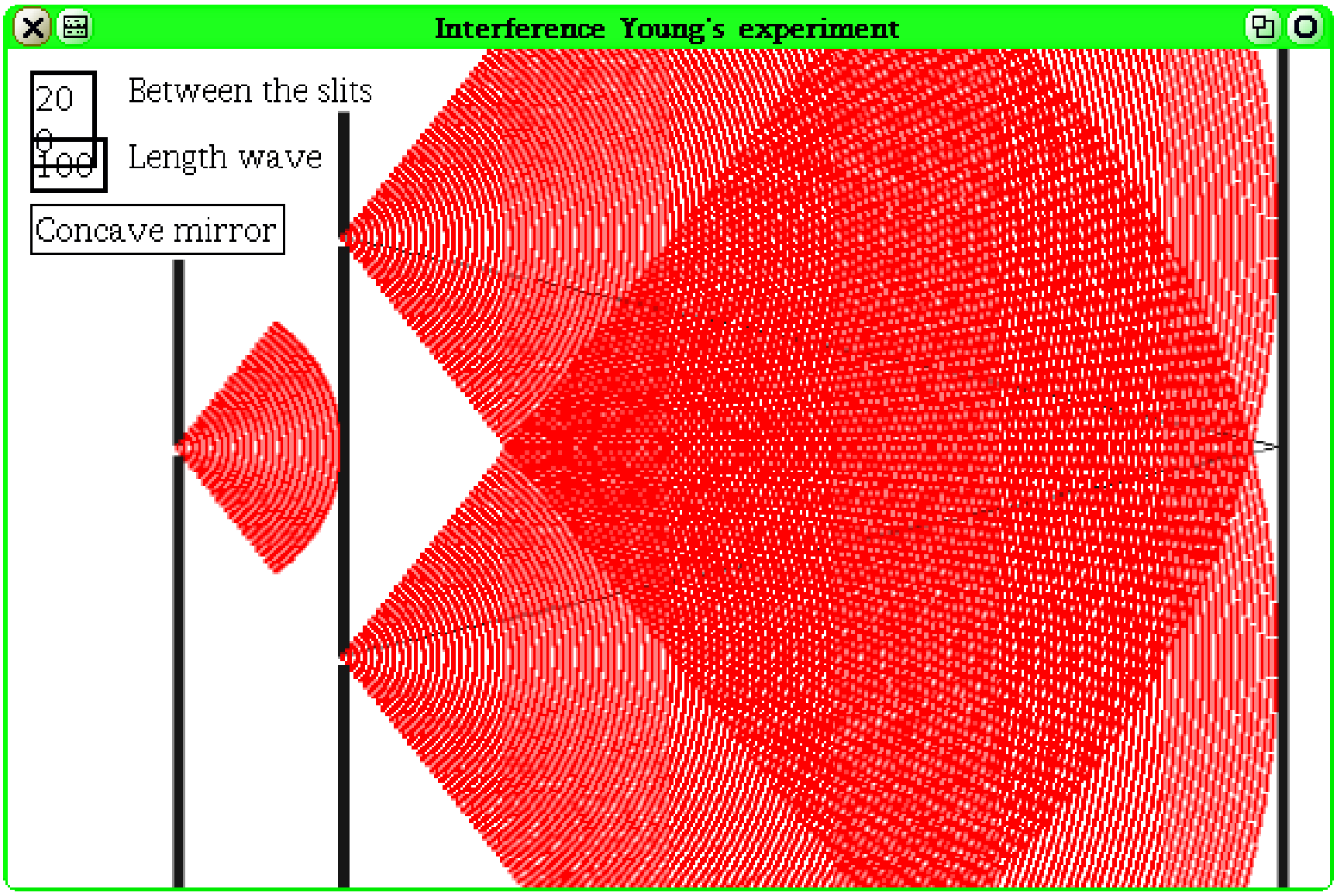

Рис. В.6. Моделювання експерименту Юнга

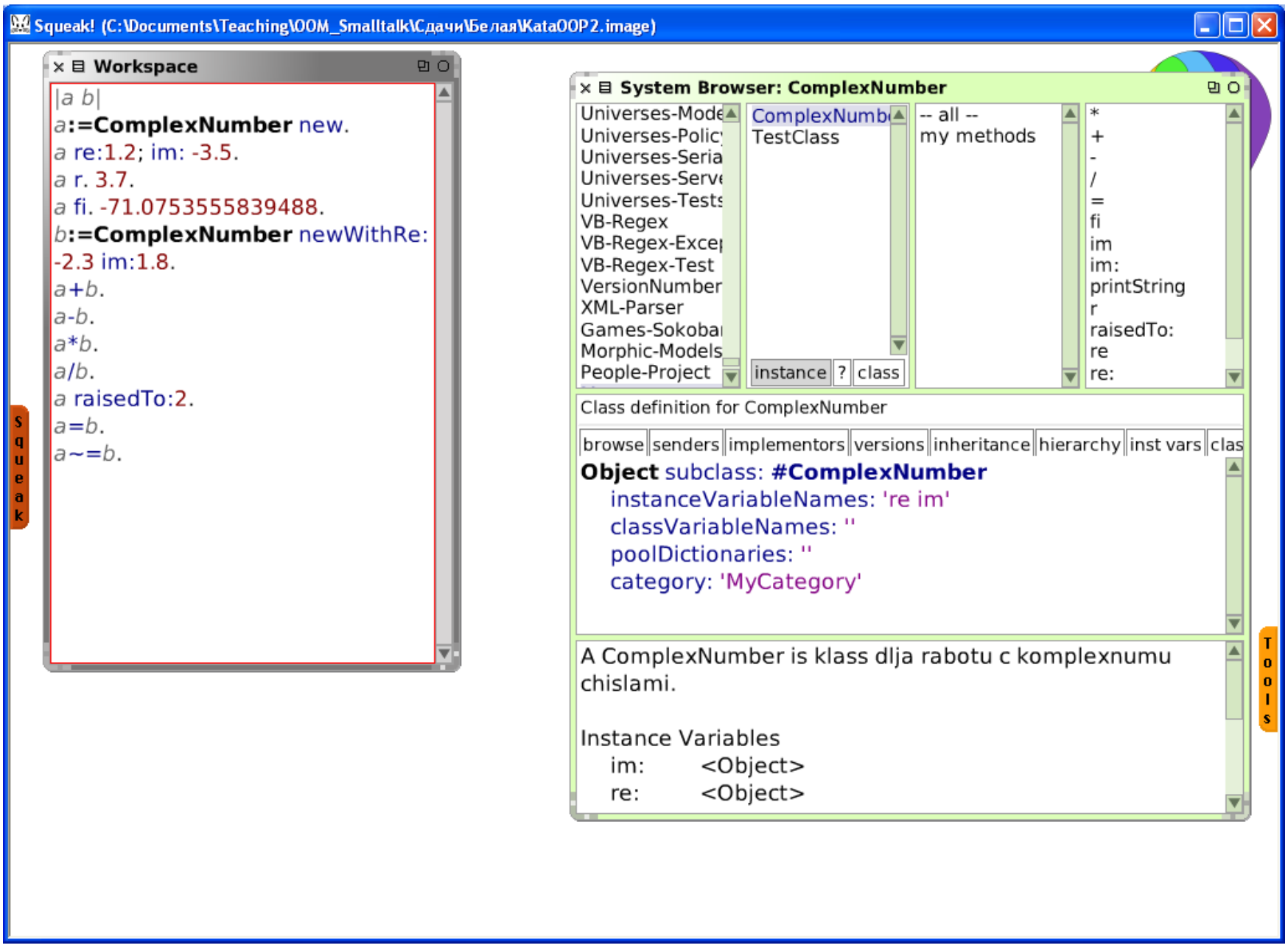

Рис. В.7. Моделювання комплексної арифметики

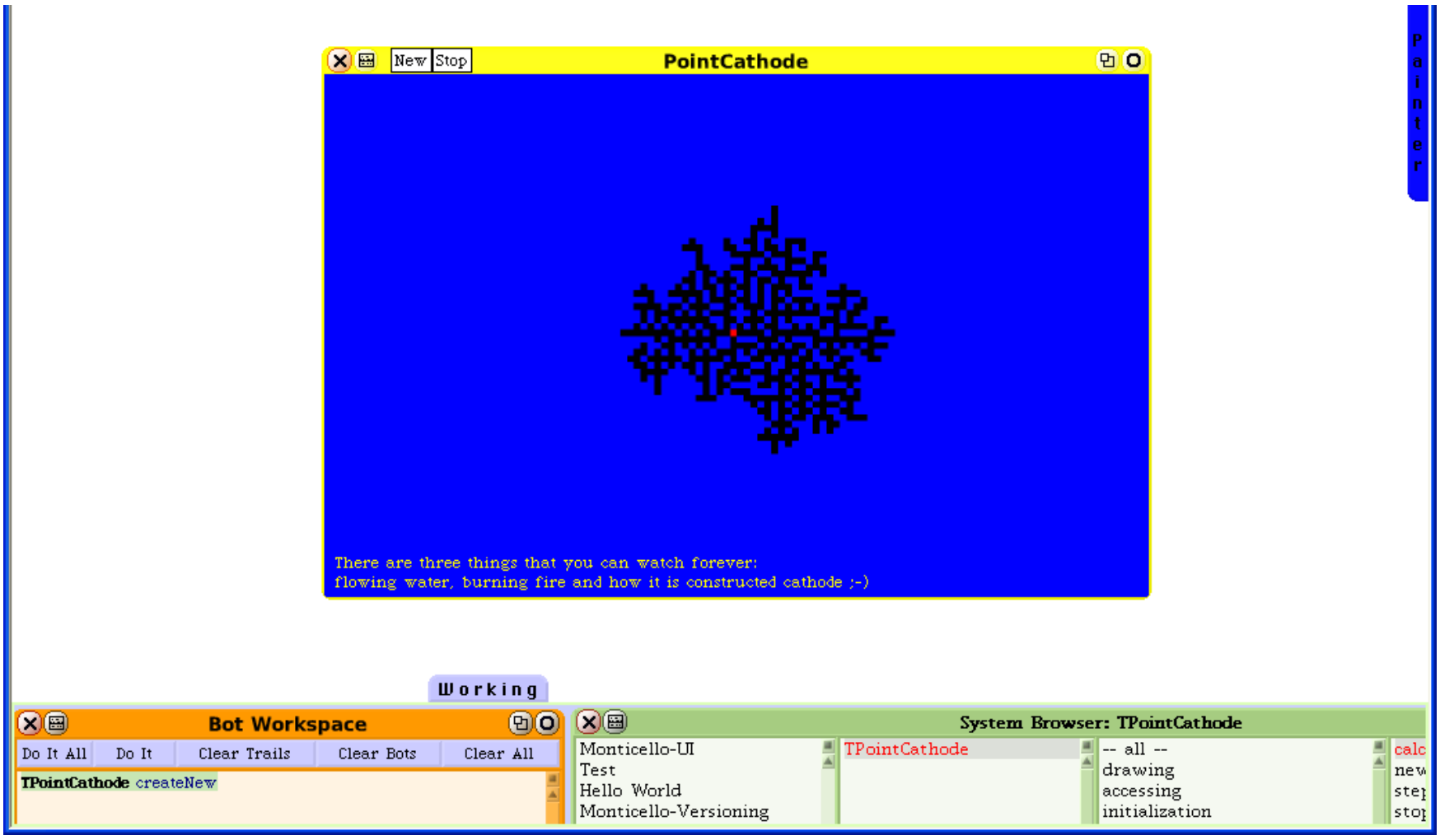

Рис. В.8. Модель електролізу на точковому катоді

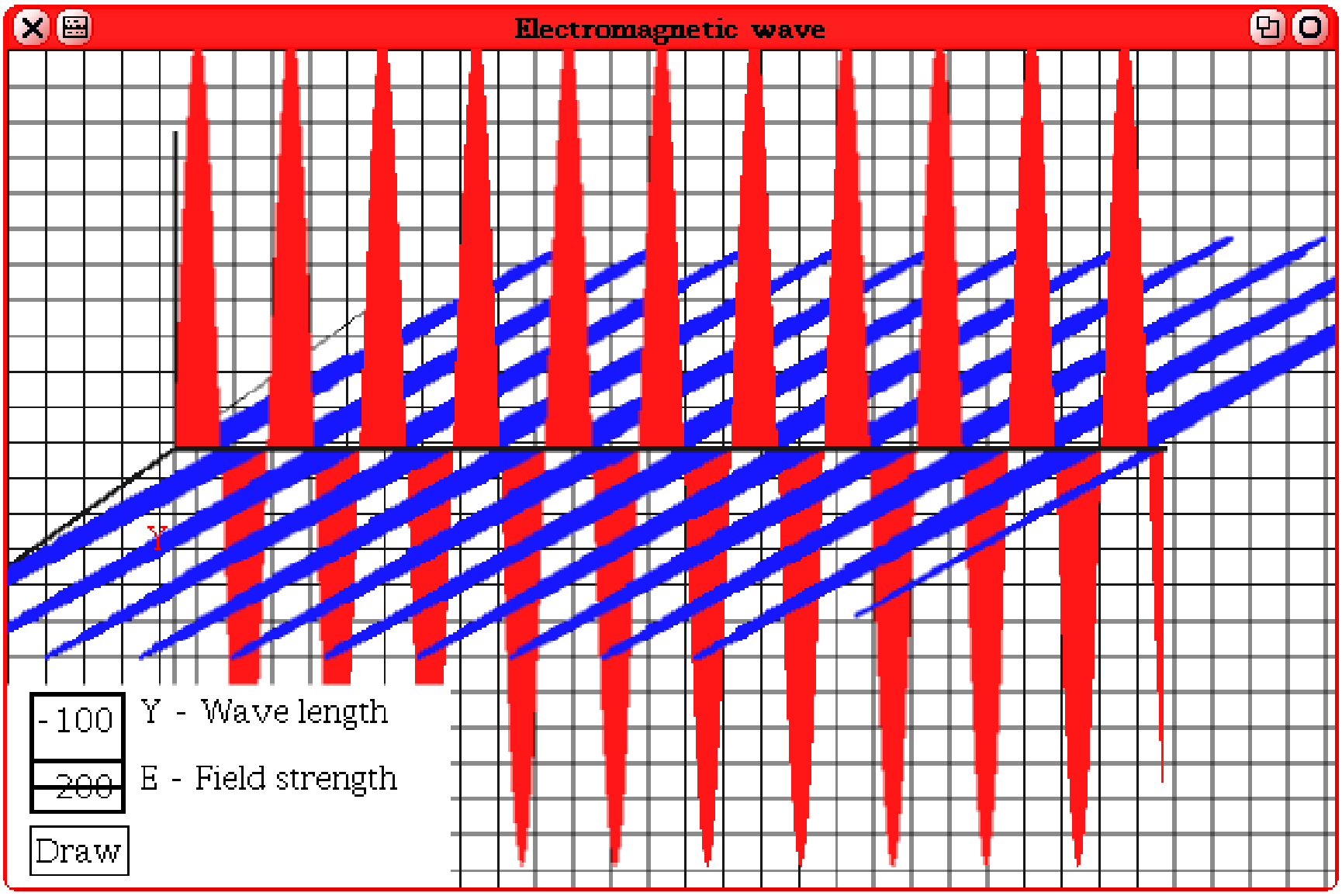

Рис. В.9. Моделювання електромагнітних хвиль

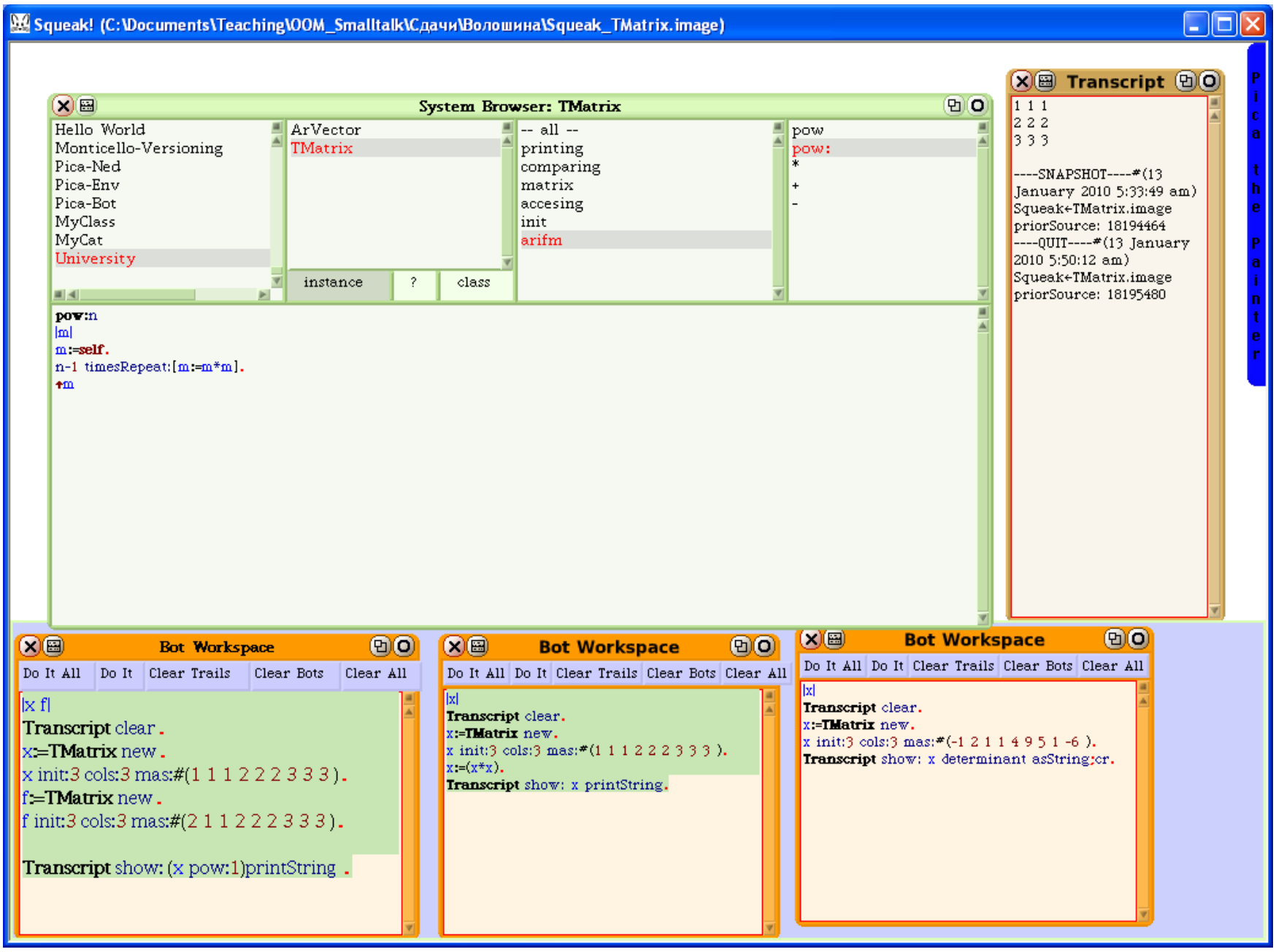

Рис. В.10. Клас для роботи з матрицями

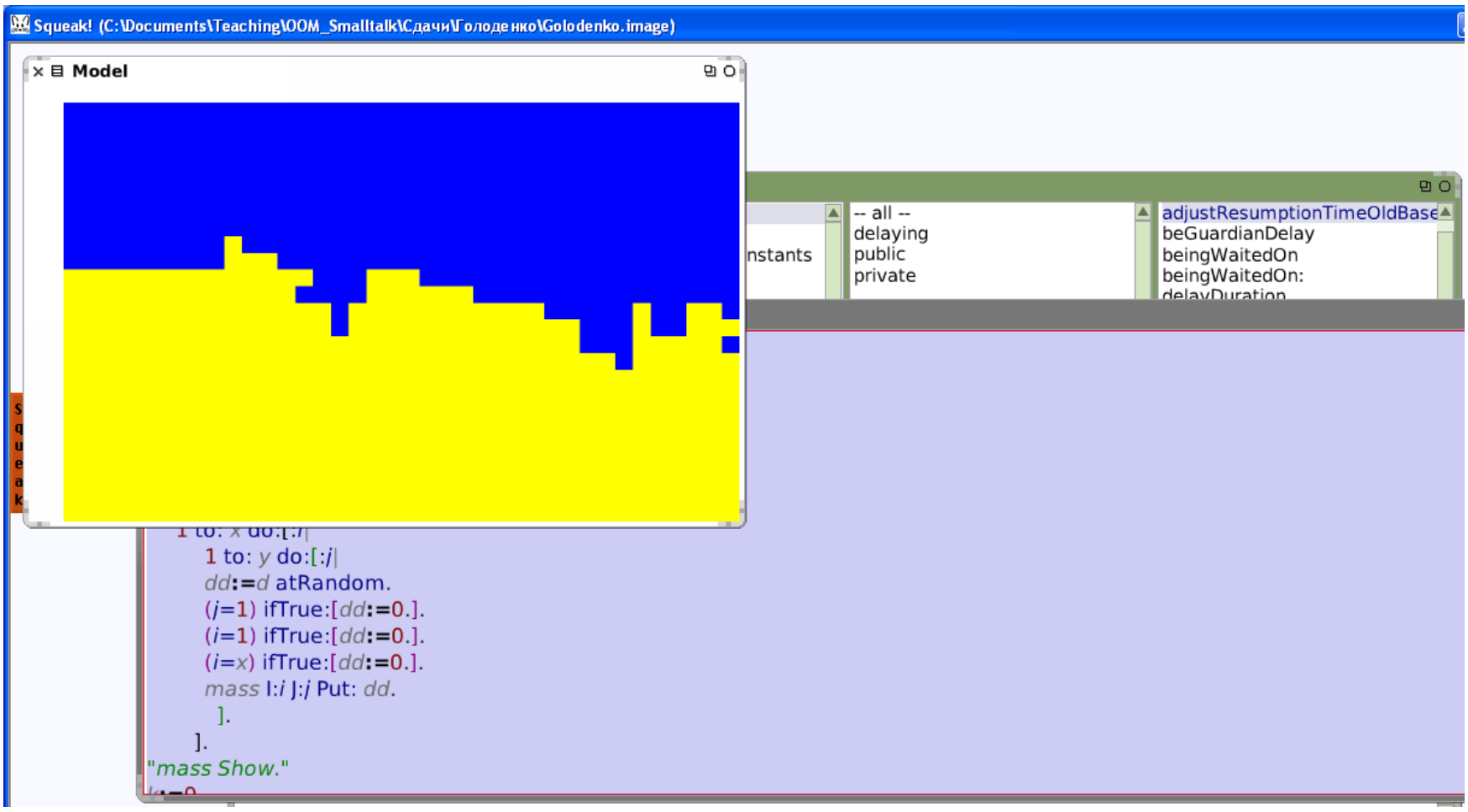

Рис. В.11. Моделювання утворення берегової лінії

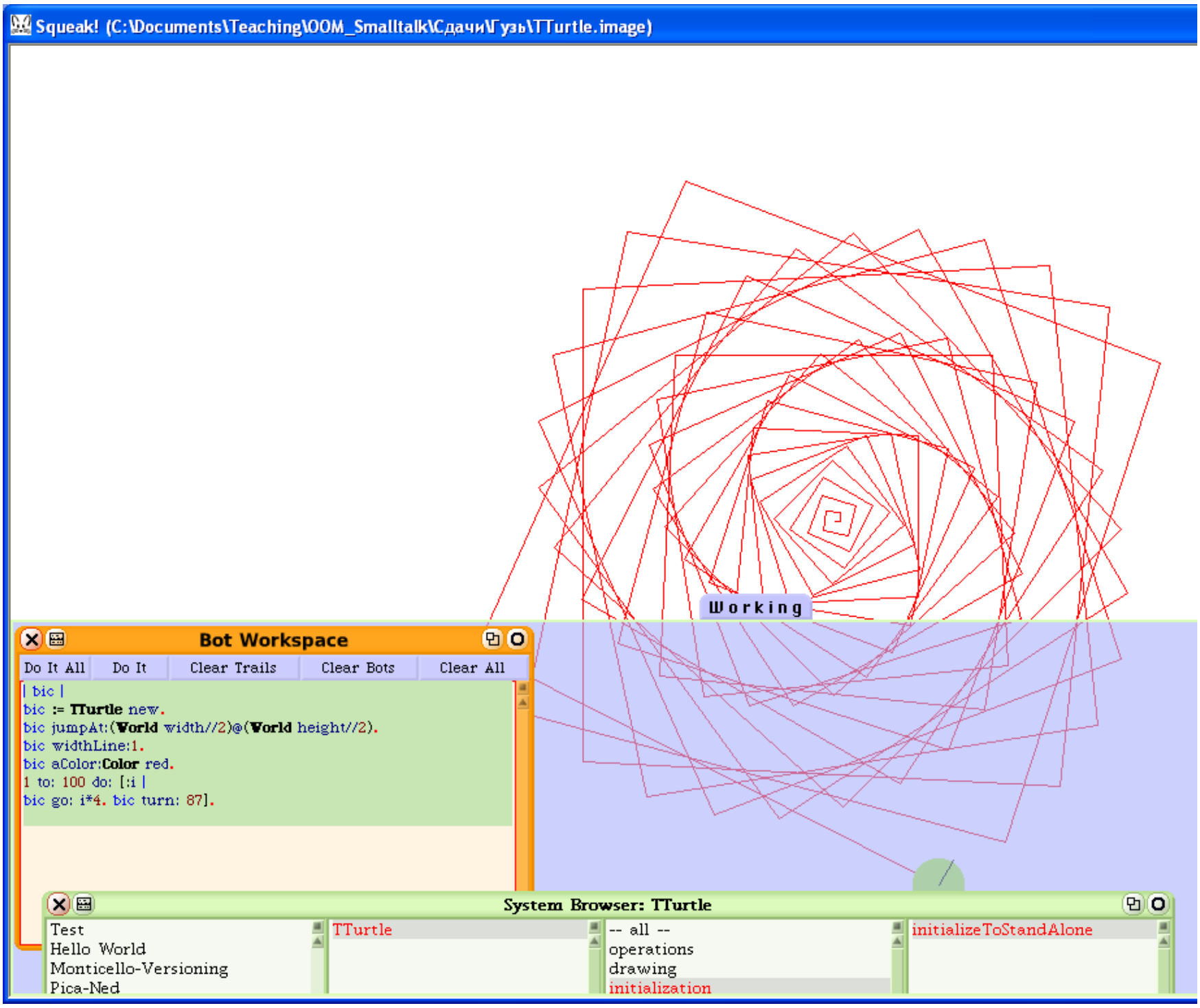

Рис. В.12. Моделювання Лого-«черепахи»

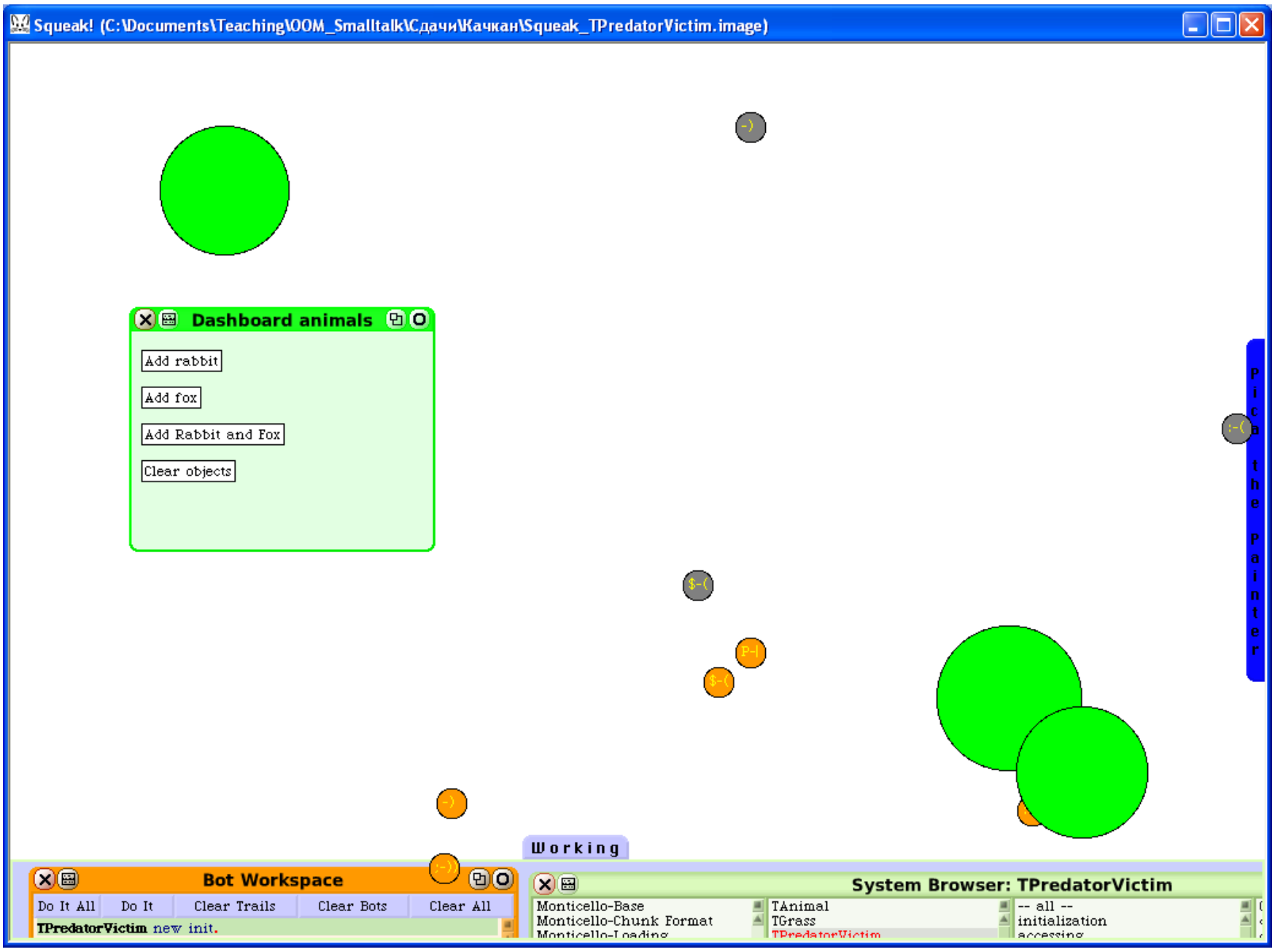

Рис. B.13. Клітковий автомат «Хижак-жертва»

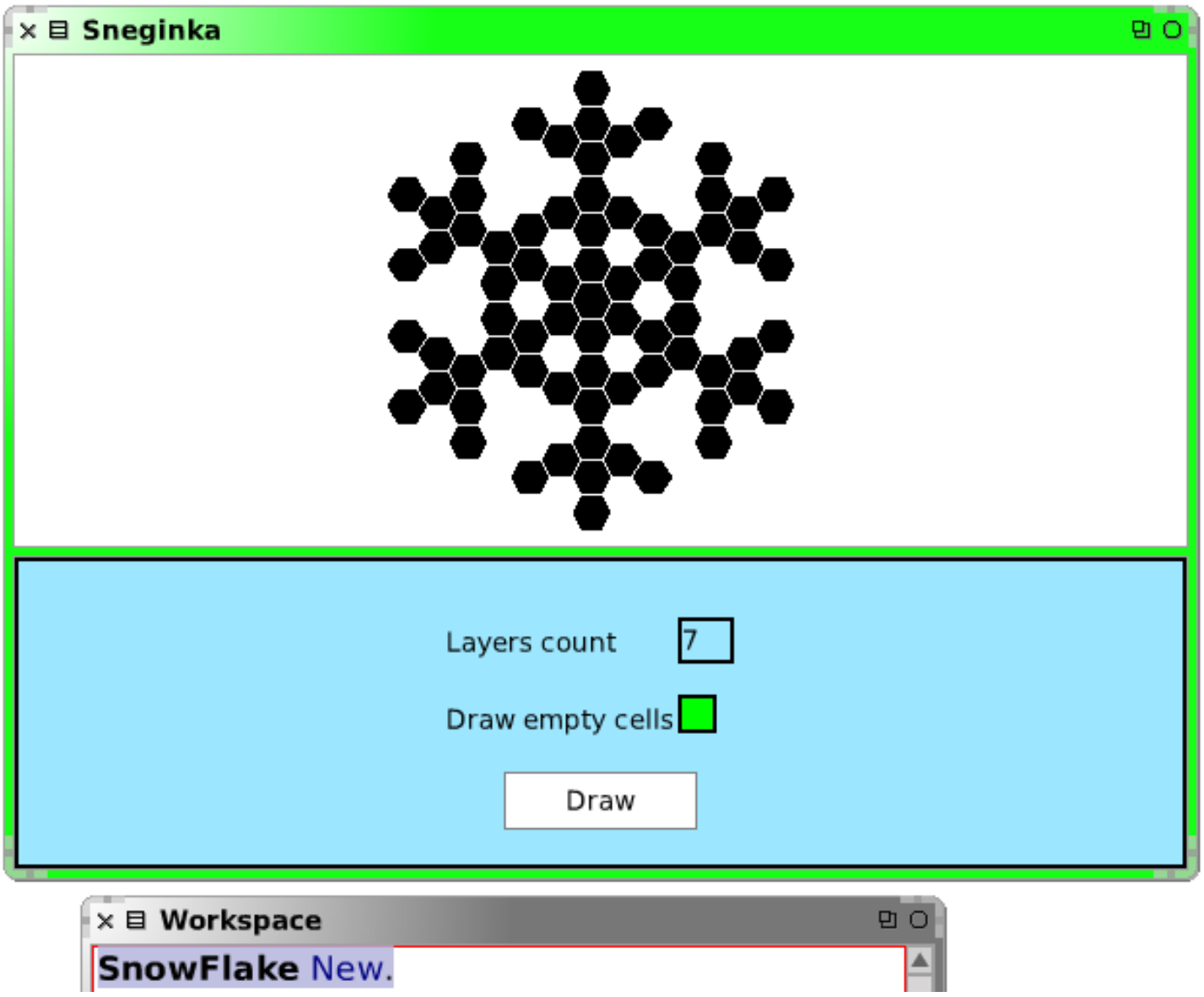

Рис. B.14. Клітковий автомат «Сніжинка»

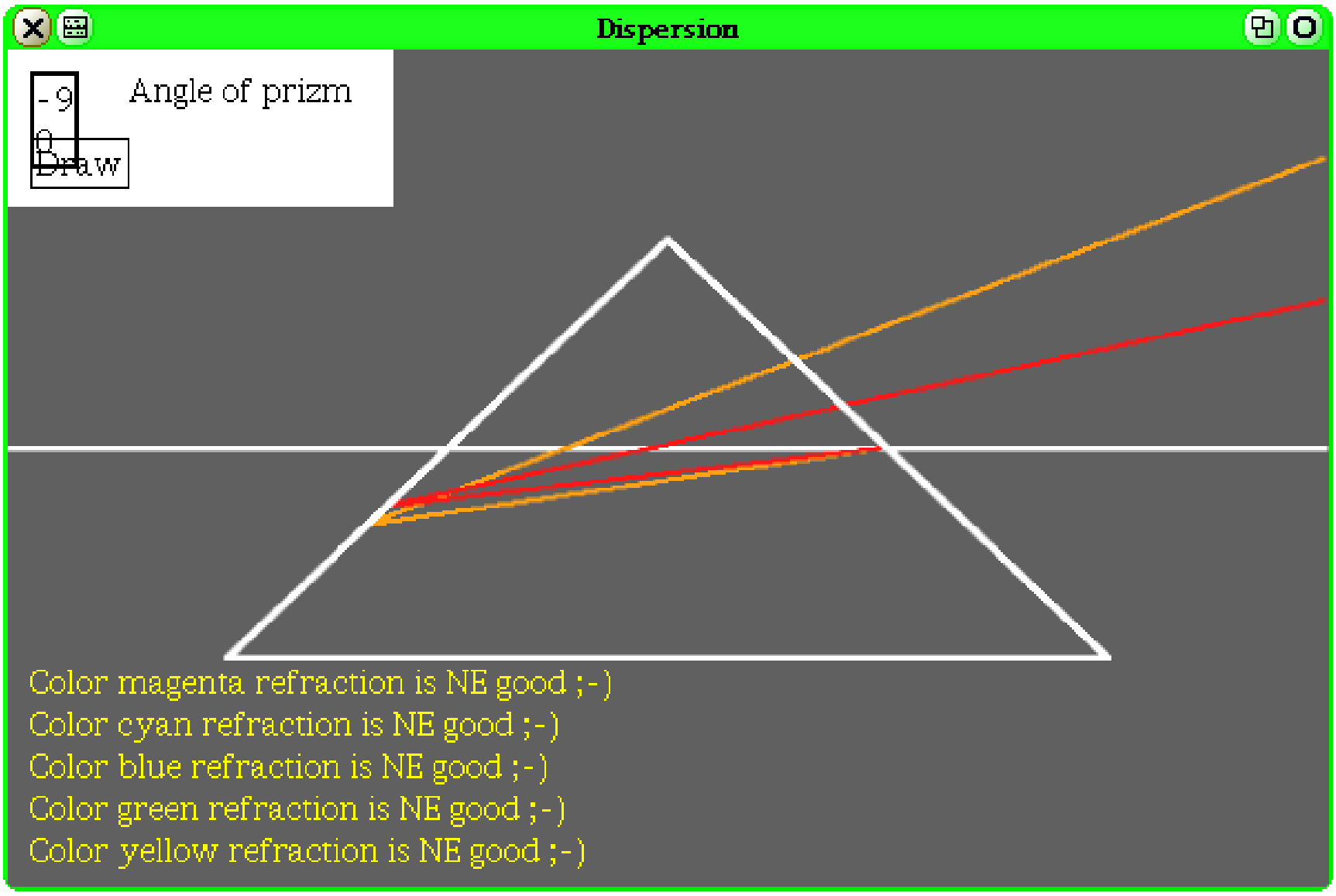

Рис. В.15. Моделювання дисперсії світла

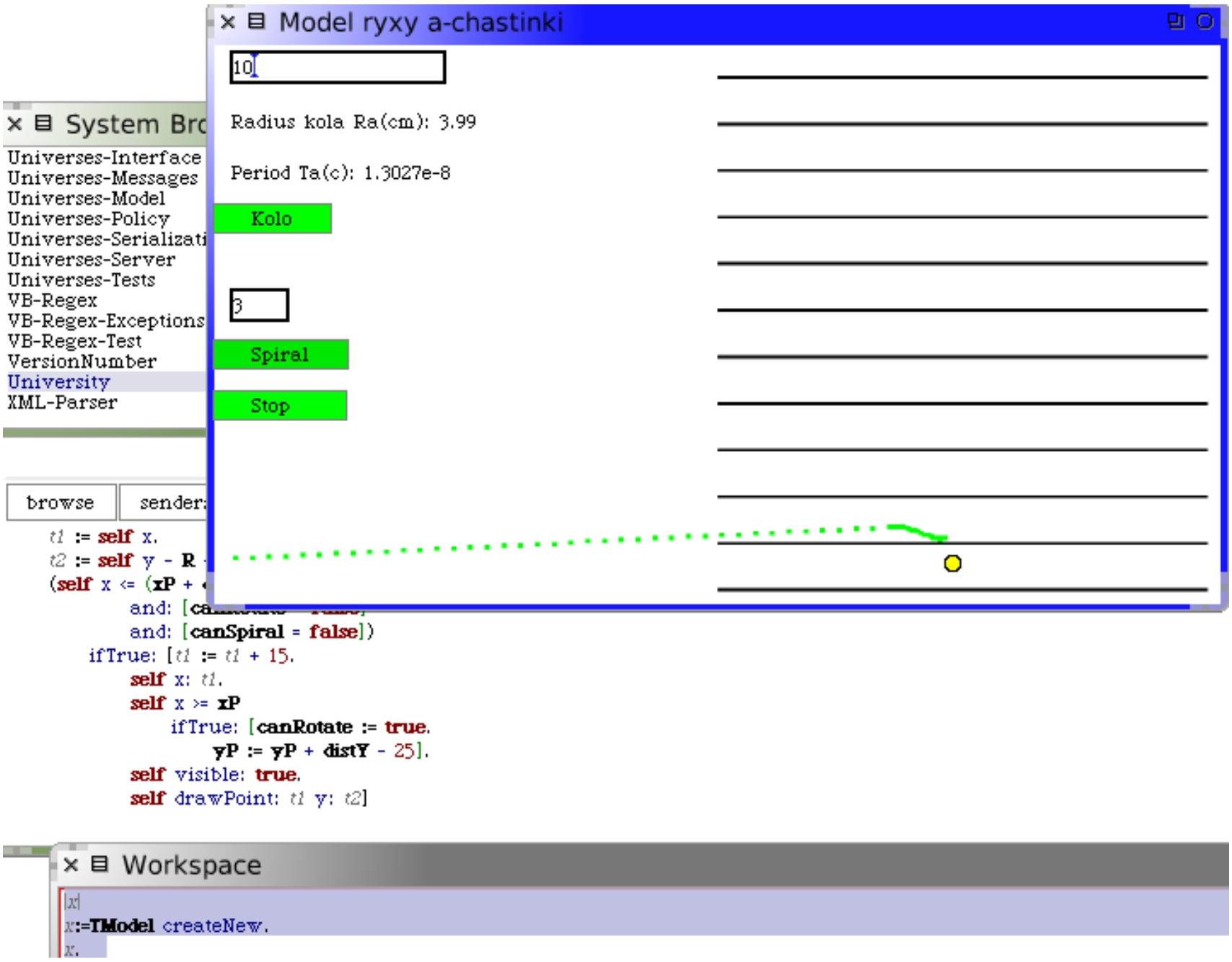

Рис. В.16. Модель руху зарядженої частинки у полі

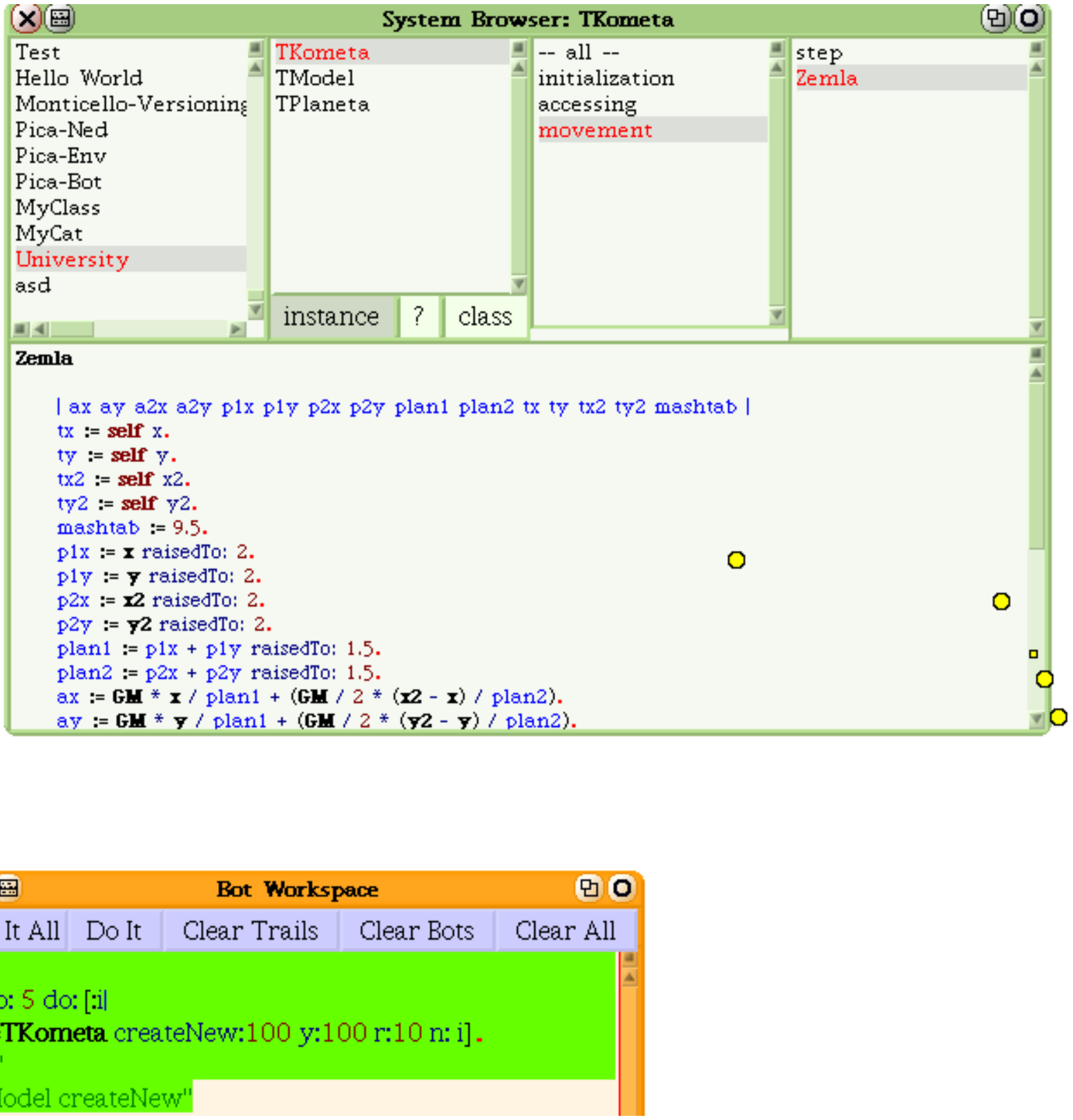

Рис. В.17. Модель руху планет Сонячної системи та комети

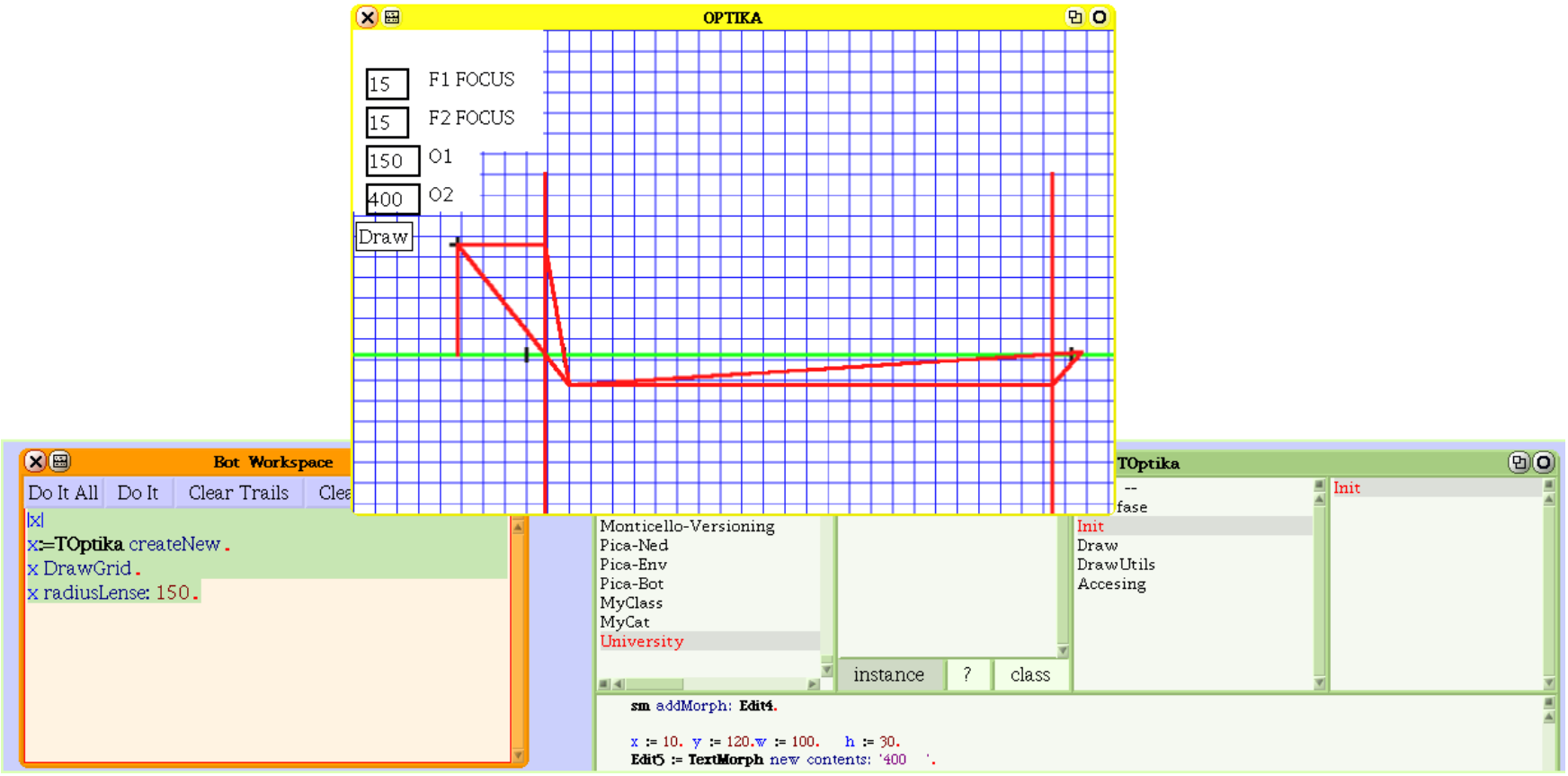

Рис. В.18. Моделювання ходу променів в тонких лінзах

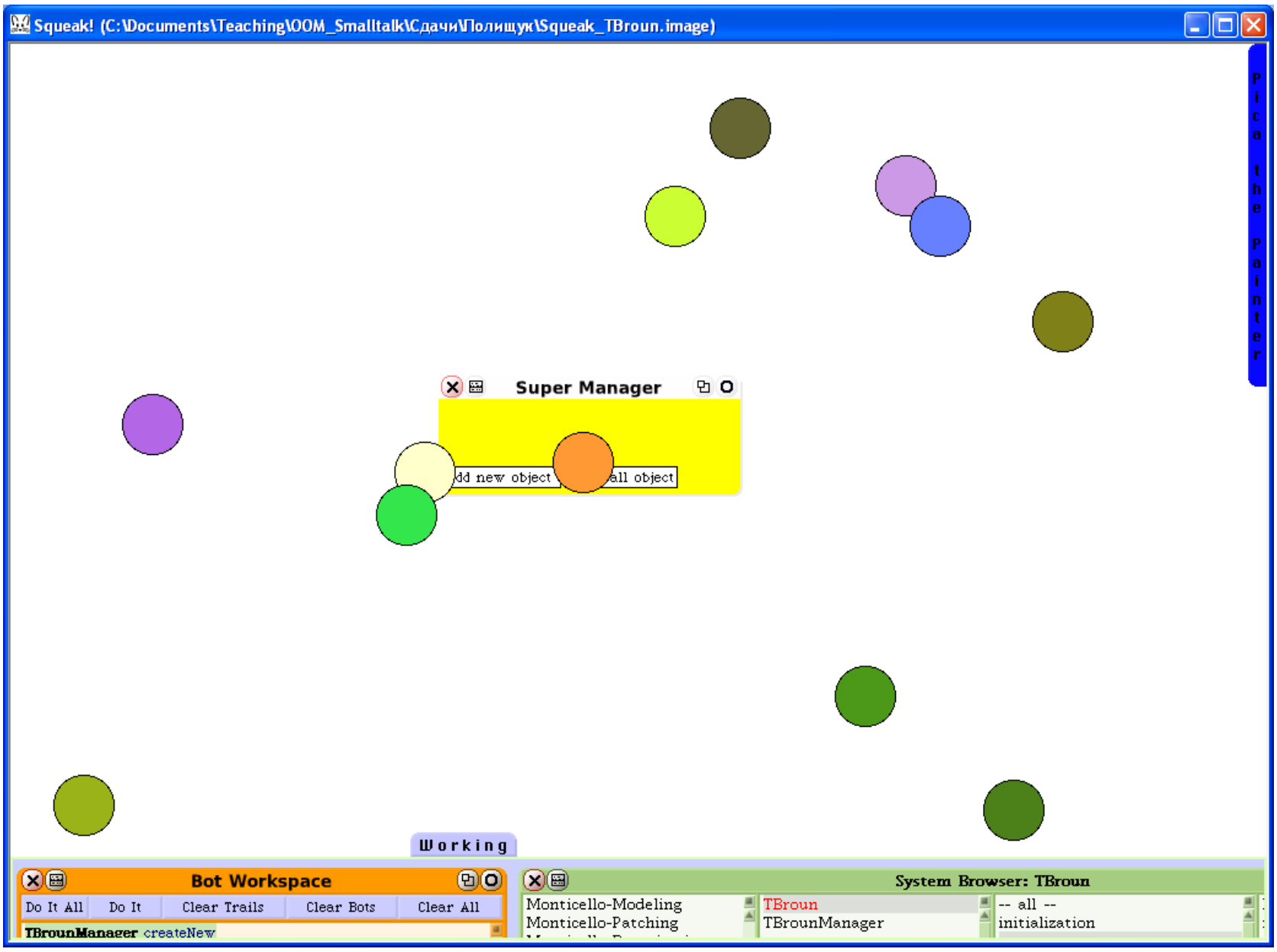

Рис. В.19. Модель броунівського руху

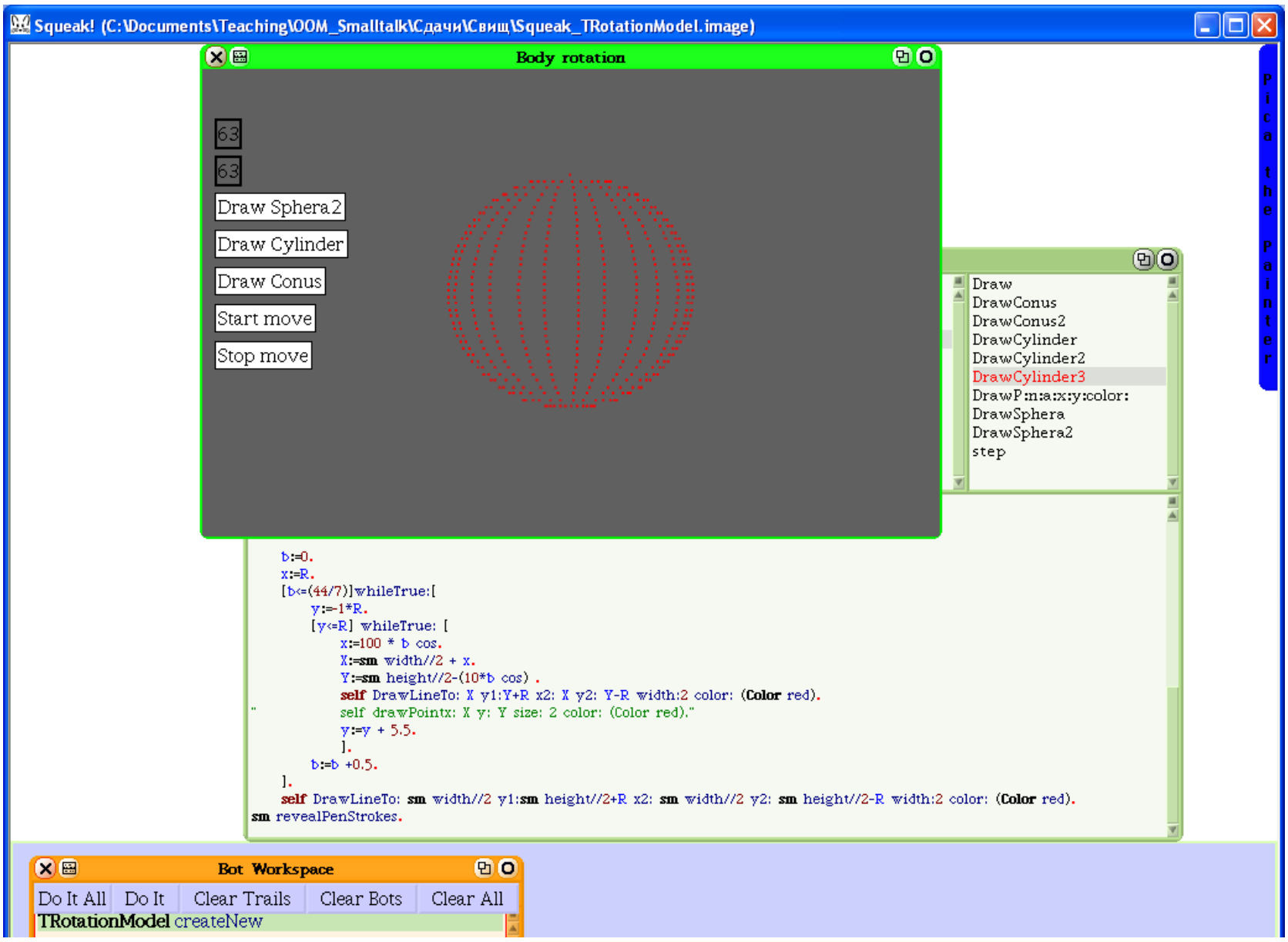

Рис. В.20. Модель обертання тіл

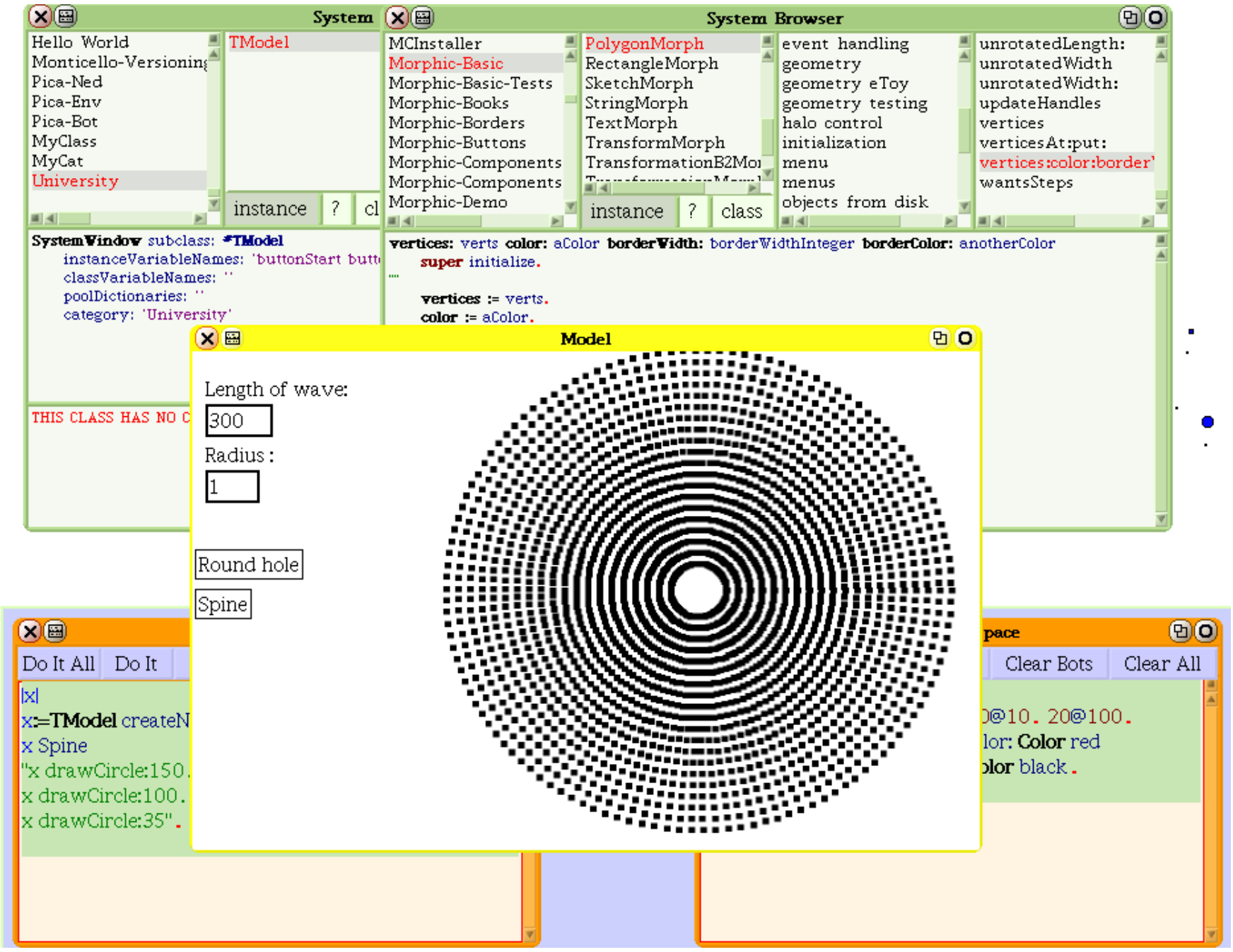

Рис. В.21. Моделювання явища дифракції

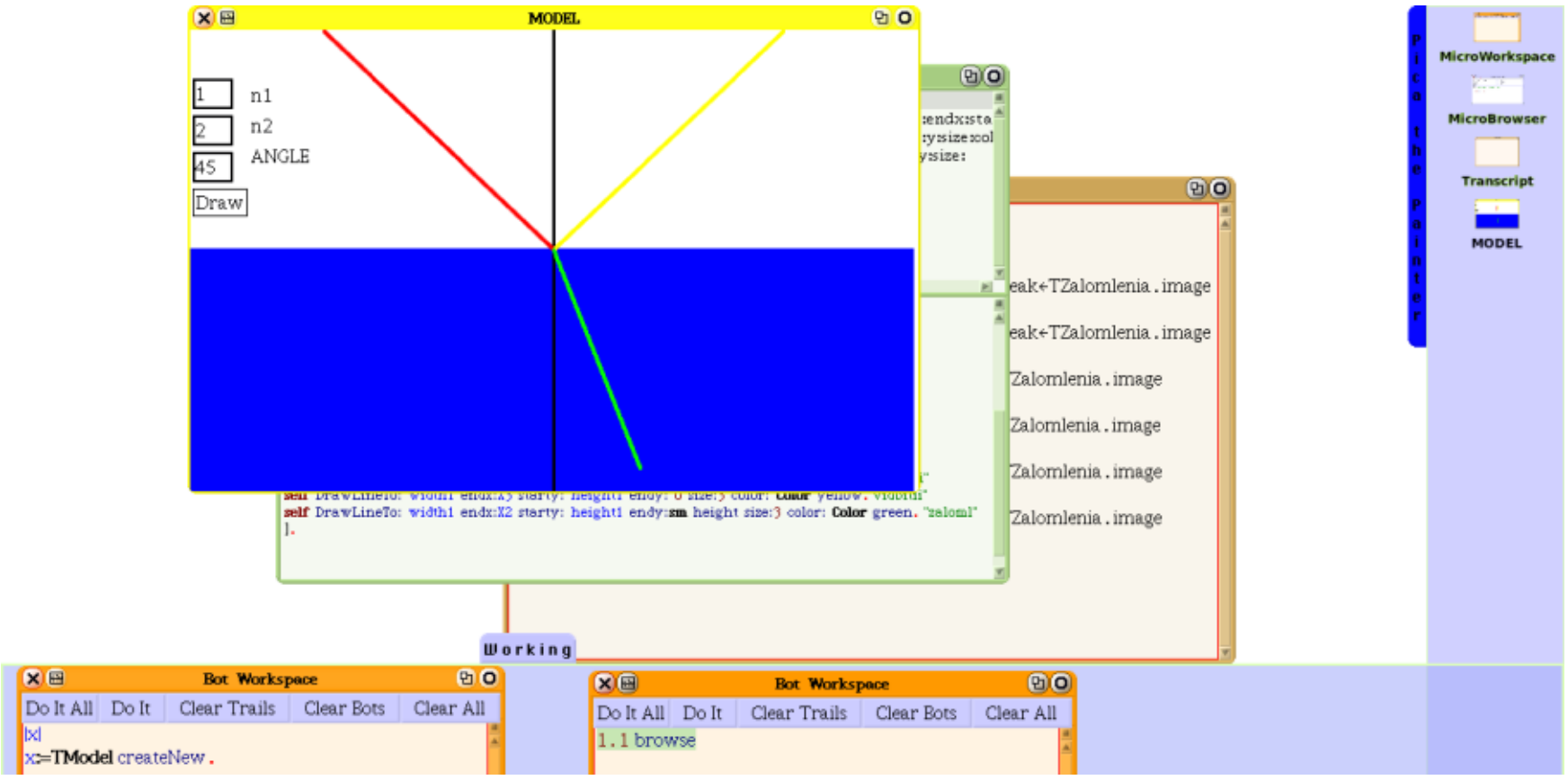

Рис. В.22. Моделювання явищ відбиття та заломлення світла у різних середовищах

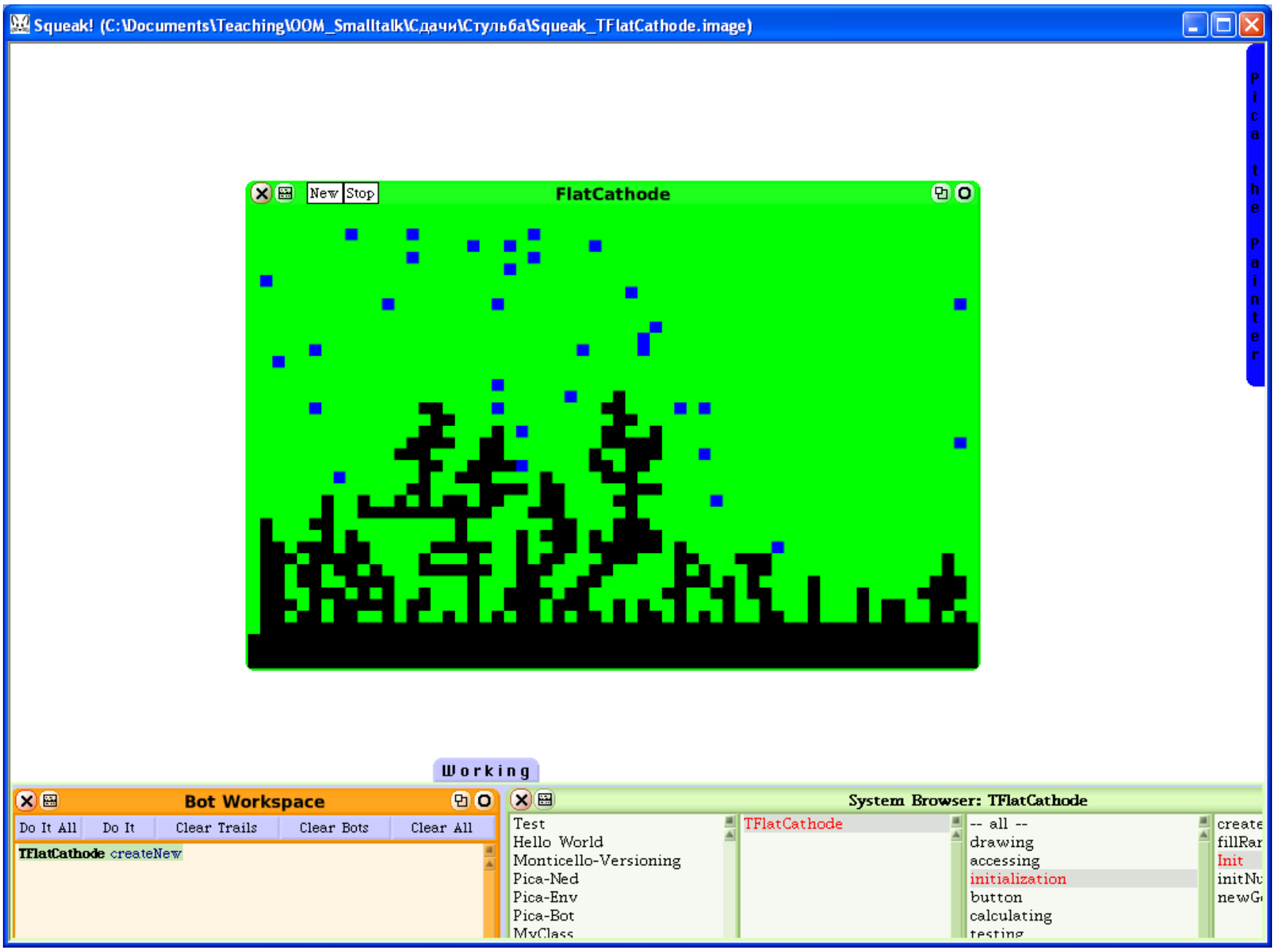

Рис. В.23. Модель електролізу на пласкому катоді

**Додаток Г**

**Робоча навчальна програма зі спецкурсу «Об'єктно-орієнтоване моделювання»**

**МІНІСТЕРСТВО ОСВІТИ І НАУКИ УКРАЇНИ**
**КРИВОРІЗЬКИЙ ДЕРЖАВНИЙ ПЕДАГОГІЧНИЙ УНІВЕРСИТЕТ**
**КАФЕДРА ІНФОРМАТИКИ ТА ПРИКЛАДНОЇ МАТЕМАТИКИ**

**ЗАТВЕРДЖУЮ**
Проректор з навчально-педагогічної роботи
____________________ В. П. Лисечко
«17» травня 2007 р.

**РОБОЧА НАВЧАЛЬНА ПРОГРАМА**
з навчальної дисципліни
*Об'єктно-орієнтоване програмування*
для студентів спеціальності 6.070100 Фізика (спеціалізація «Інформатика»)

Дані робочого навчального плану

| Форма навчання | Курс | Розподіл за семестрами | | | загальний обсяг | | з них годин | | | | |
|---|---|---|---|---|---|---|---|---|---|---|---|
| | | Екзамени | Заліки | Курс. роб. | години | кредити | всього ауд. | лекції | лабораторні | практичні | самостійна робота |
| Денна | 3 | 5 | | | 90 | 2 | 54 | 18 | 36 | | 18 |
| Заочна | | | | | | | | | | | |

Кривий Ріг
2007 р.

Робоча навчальна програма складена к. пед. н., доц., доцентом кафедри інформатики та прикладної математики С. О. Семеріковим та асистентом кафедри інформатики та прикладної математики О. І. Теплицьким

Робоча навчальна програма затверджена на засіданні кафедри інформатики та прикладної математики, протокол № 10 від «11» травня 2007 р.

Зав. кафедри ІПМ ______________ к. ф.-м. н., доц. Н. В. Моісеєнко

# ТЕМАТИЧНИЙ ПЛАН

## Загальні відомості

«Об’єктно-орієнтоване моделювання» – навчальна дисципліна, в якій вивчаються способи конструювання та дослідження об’єктно-орієнтованих моделей. Курс інтегрує знання з процедурного, об’єктно-орієнтованого, подієво-орієнтованого та візуального програмування, загальних принципів дослідження складних систем.

В результаті вивчення предмета студент повинен:

**знати:**

– принципи об’єктно-орієнтованого програмування, аналізу та проектування;

– основні етапи комп’ютерного моделювання;

– методи конструювання об’єктно-орієнтованих моделей у середовищах об’єктно-орієнтованого моделювання.

**уміти:**

– виділяти суттєві властивості об’єктів та зв’язків між ними та реалізовувати об’єктно-орієнтовану взаємодію мовою ООП;

– виконувати індивідуальні та колективні навчальні проекти з конструювання та дослідження об’єктно-орієнтованих моделей;

– добирати середовище моделювання, адекватне розв’язуваній задачі;

використовувати стандартні бібліотеки моделей та створювати власні.

**Примітка:** рекомендована форма проведення екзамену – захист індивідуальних та колективних дослідницьких проектів.

## Тематика курсу лекцій

| Номер теми | Найменування теми (модуля). Основні питання лекції та її зміст. Завдання для самостійної роботи | Денна форма, годин | |
|---|---|---|---|
| | | Лекції | Самост. роб. |
| | **Модуль 1**<br>**Вступ до об’єктно-орієнтованого моделювання** | | |
| 1 | *Базові поняття об’єктно-орієнтованого моделювання:*<br>– поняття про моделювання, види моделей, об’єктно-орієнтоване моделювання; | 2 | 0 |

| Номер теми | Найменування теми (модуля). Основні питання лекції та її зміст. Завдання для самостійної роботи | Денна форма, годин | |
|---|---|---|---|
| | | Лекції | Самост. роб. |
| | – об'єктно-орієнтоване програмування та об'єктно-орієнтовані мови;<br>– абстракція, інкапсуляція, спадкування, поліморфізм – основи об'єктно-орієнтованої методології;<br>– етапи об'єктно-орієнтованого моделювання: об'єктно-орієнтований аналіз, проектування, обчислювальний експеримент та аналіз його результатів. | | |
| 2 | *Огляд середовищ об'єктно-орієнтованого моделювання:*<br>– критерії вибору середовища моделювання;<br>– універсальні середовища об'єктно-орієнтованого моделювання (самостійно);<br>– середовища для конструювання динамічних моделей;<br>– середовища для конструювання імітаційних моделей. | 2 | 1 |
| **Модуль 2. Об'єктно-орієнтовані динамічні моделі** | | | |
| 3 | *Моделі математичної екології:*<br>– динаміка одновидової популяції;<br>– модель «Хижак-жертва»;<br>– вікові моделі Леслі. | 2 | 0 |
| 4 | *Моделі класичної механіки:*<br>– динаміка коливних систем (самостійно);<br>– рух тіл в полі сили тяжіння;<br>– моделі аеродинаміки. | 2 | 1 |
| 5 | *Моделі молекулярної фізики та фізики твердого тіла:*<br>– динаміка одноатомного газу (самостійно);<br>– молекулярна динаміка. | 2 | 1 |
| 6 | *Моделювання електричного та магнітного полів:* | 2 | 1 |

| Номер теми | Найменування теми (модуля). Основні питання лекції та її зміст. Завдання для самостійної роботи | Денна форма, годин | |
|---|---|---|---|
| | | Лекції | Самост. роб. |
| | – рух заряду в електричному полі;<br>– рух заряду в магнітному полі (самостійно); | | |
| **Модуль 3. Об'єктно-орієнтовані імітаційні моделі** | | | |
| 7 | *Кліткові автомати:*<br>– модель поширення чуток;<br>– модель «Хижак-жертва»;<br>– модель поширення пожежі (самостійно);<br>– гра «Життя». | 2 | 1 |
| 8 | *Метод Монте-Карло:*<br>– модель броунівського руху;<br>– модель відмов обладнання;<br>– модель росту кристалу. | 2 | 0 |
| 9 | *Моделі фрактальних об'єктів та процесів:*<br>– моделі регулярних фракталів (самостійно);<br>– задача перколяції;<br>– моделі електролізу;<br>– модель утворення берегової лінії. | 2 | 1 |
| | Разом: | 18 | 6 |

**Перелік тем лабораторних занять**

| Номер теми | Найменування теми (модуля). Основні питання лабораторного заняття, її зміст. Завдання для самостійної роботи | Денна форма, годин | |
|---|---|---|---|
| | | Аудит. роб. | Самост. роб. |
| **Модуль 1<br>Вступ до об'єктно-орієнтованого моделювання** | | | |
| 1 | Огляд середовищ об'єктно-орієнтованого | 2 | 0 |

| Номер теми | Найменування теми (модуля). Основні питання лабораторного заняття, її зміст. Завдання для самостійної роботи | Денна форма, годин | |
|---|---|---|---|
| | | Аудит. роб. | Самост. роб. |
| | моделювання: Sage, VPython (PyGeo, VPNBody), NetLogo, Squeak, Alice | | |
| 2 | Основи програмування мовою Python в середовищі VPython (PyGeo, VPNBody) | 2 | 4 |
| 3 | Основи програмування мовою Smalltalk в середовищі Squeak | 2 | 4 |
| 4 | Основи програмування мовою Java в середовищі Alice | 2 | 4 |
| **Модуль 2. Об'єктно-орієнтовані динамічні моделі** | | | |
| 5 | Динаміка одновидової популяції. Модель «Хижак-жертва». | 2 | 2 |
| 6 | Вікова модель одновидової популяції | 2 | 1 |
| 7 | Рух тіла в полі сили тяжіння | 2 | 1 |
| 8 | Моделювання абсолютних та відносних рухів планет | 2 | 1 |
| 9 | Задача про політ паперового літачка | 2 | 2 |
| 10 | Динаміка одноатомного газу | 2 | 1 |
| 11 | Молекулярна динаміка | 2 | 2 |
| 12 | Моделювання електричного та магнітного полів | 2 | 1 |
| **Модуль 3. Об'єктно-орієнтовані імітаційні моделі** | | | |
| 13 | Клітковий автомат «Життя» | 2 | 1 |
| 14 | Модель поширення чуток. Модель поширення пожежі | 2 | 1 |
| 15 | Імітаційна модель «Хижак-жертва» | 2 | 2 |
| 16 | Стохастичні моделі | 2 | 1 |
| 17 | Моделі регулярних фракталів. Перколяційний кластер на порозі протікання | 2 | 1 |
| 18 | Електроліз на точковому катоді. Електроліз на пласкому катоді. Модель утворення берегової лінії | 2 | 1 |
| | Разом: | 36 | 30 |

## Зміст самостійної роботи

### а. Лекції

| Номер теми | Найменування теми (модуля). Основні питання лекції та її зміст. Завдання для самостійної роботи | Денна форма, годин |
|---|---|---|
| **Модуль 1. Вступ до об'єктно-орієнтованого моделювання** | | |
| 2 | *Огляд середовищ об'єктно-орієнтованого моделювання:*<br>– універсальні середовища об'єктно-орієнтованого моделювання. | 1 |
| **Модуль 2. Об'єктно-орієнтовані динамічні моделі** | | |
| 4 | *Моделі класичної механіки:*<br>– динаміка коливних систем. | 1 |
| 5 | *Моделі молекулярної фізики та фізики твердого тіла:*<br>– динаміка одноатомного газу. | 1 |
| 6 | *Моделювання електричного та магнітного полів:*<br>– рух заряду в електричному полі;<br>– рух заряду в магнітному полі (самостійно); | 1 |
| **Модуль 3. Об'єктно-орієнтовані імітаційні моделі** | | |
| 7 | *Кліткові автомати:*<br>– модель поширення пожежі. | 1 |
| 9 | *Моделі фрактальних об'єктів та процесів:*<br>– моделі регулярних фракталів. | 1 |
| | Разом: | 6 |

### б. Лабораторні заняття

| Номер теми | Найменування теми (модуля). Основні питання лабораторного заняття, її зміст. Завдання для самостійної роботи | Денна форма, годин |
|---|---|---|
| **Модуль 1. Вступ до об'єктно-орієнтованого моделювання** | | |

| Номер теми | Найменування теми (модуля). Основні питання лабораторного заняття, її зміст. Завдання для самостійної роботи | Денна форма, годин |
|---|---|---|
| 2 | Основи програмування мовою Python в середовищі VPython. Програмні комплекси PyGeo, VPNBody. | 4 |
| 3 | Основи програмування мовою Smalltalk в середовищі Squeak. Об'єкти Morphic. | 4 |
| 4 | Основи програмування мовою Java в середовищі Alice. Експорт проектів Alice у NetBeans. | 4 |
| **Модуль 2. Об'єктно-орієнтовані динамічні моделі** | | |
| 5 | Модель «Хижак-жертва»: графічний аналіз результатів обчислювального експерименту. | 2 |
| 6 | Вікова модель одновидової популяції: удосконалення моделі. | 1 |
| 7 | Рух тіла в полі сили тяжіння: моделювання у середовищі VPNBody. | 1 |
| 8 | Моделювання абсолютних та відносних рухів планет: моделювання у середовищі VPNBody. | 1 |
| 9 | Задача про політ паперового літачка: моделювання «мертвої петлі». | 2 |
| 10 | Динаміка одноатомного газу: урахування різних розподілів випадкових чисел. | 1 |
| 11 | Молекулярна динаміка: моделювання динаміки інертного газу. | 2 |
| 12 | Моделювання магнітного поля. | 1 |
| **Модуль 3. Об'єктно-орієнтовані імітаційні моделі** | | |
| 13 | Узагальнений клітковий автомат «Життя». | 1 |
| 14 | Модель поширення пожежі. | 1 |
| 15 | Імітаційна модель «Хижак-жертва»: удосконалення моделі. | 2 |
| 16 | Моделювання роботи транзистору | 1 |
| 17 | Моделі регулярних фракталів | 1 |
| 18 | Модель утворення берегової лінії | 1 |
| | Разом: | 30 |

## Орієнтовна тематика індивідуальних та колективних дослідницьких проектів

Гра «Життя»
Узагальнена гра «Життя»
Електроліз на пласкому катоді
Електроліз на точковому катоді
Катастрофічна модель руху планет Сонячної системи
Модель руху планет Сонячної системи
Клас «Колода гральних карт».
Моделювання поведінки Лого-«черепахи»
Геометро-графічне моделювання
Розробка класу для побудови регулярних фракталів
Моделювання арифметичних векторів
Моделювання комплексних чисел
Моделювання тіла кватерніонів
Клас для роботи з нечіткими числами
Моделювання матричної арифметики над довільним кільцем
Розробка класу для роботи з поліномами
Моделювання дій над «довгими» цілими числами
Моделювання дій над «довгими» раціональними числами
Клітковий автомат «Сніжинка»
Клітковий автомат «Хижак-жертва»
Моделювання коливань подвійного маятника
Модель математичного більярду
Моделювання відносних рухів планет Сонячної системи
Моделювання дисперсії світла
Моделювання явища дифракції
Моделювання електромагнітних хвиль
Моделювання законів Кеплера
Моделювання явища інтерференції
Моделювання коливань у кристалічній гратці алмазу
Моделювання поширення світла у волоконно-оптичному кабелі
Моделювання регулярних фракталів
Моделювання утворення берегової лінії
Моделювання розподілу Максвелла
Моделювання ходу променів в тонких лінзах
Моделювання ходу променів у дзеркалах
Моделювання явищ відбиття та заломлення світла у різних середовищах
Модель поширення пожежі
Моделювання руху зарядженої частинки у електричному полі
Моделювання червоної межі для фотоефекту

## Перелік
## навчально-методичної літератури

**Додаток Д**

## Підготовка учителів природничо-математичних дисциплін у ЗВО України

*Таблиця Д.1*

**Ліцензійний обсяг, державне замовлення, кількість зарахованих на 1 курс абітурієнтів за денною формою навчання за природничо-математичними напрямами підготовки (за даними [124])**

| № | ЗВО | Л | Д | З | Δ |
|---|---|---|---|---|---|
| 1 | Полтавський національний педагогічний університет імені В. Г. Короленка | 315 | 172 | 191 | 19 |
| 2 | Республіканський вищий навчальний заклад «Кримський гуманітарний університет» (м. Ялта) | 50 | 13 | 13 | 0 |
| 3 | Державний вищий навчальний заклад «Донбаський державний педагогічний університет» | 190 | 43 | 43 | 0 |
| 4 | Сумський державний педагогічний університет імені А. С. Макаренка | 330 | 124 | 137 | 13 |
| 5 | Рівненський державний гуманітарний університет | 310 | 170 | 178 | 8 |
| 6 | Волинський національний університет імені Лесі Українки | 445 | 242 | 315 | 73 |
| 7 | Херсонський державний університет | 380 | 168 | 212 | 44 |
| 8 | Кіровоградський державний педагогічний університет імені Володимира Винниченка | 330 | 148 | 168 | 20 |
| 9 | Київський університет імені Бориса Грінченка | 75 | 30 | 34 | 4 |
| 10 | Харківський національний педагогічний університет імені Г. С. Сковороди | 405 | 85 | 115 | 30 |
| 11 | Чернівецький національний університет імені Юрія Федьковича | 845 | 214 | 275 | 61 |
| 12 | Державний вищий навчальний заклад «Запорізький національний університет» | 315 | 145 | 184 | 39 |
| 13 | Мелітопольський державний педагогічний університет імені Богдана Хмельницького | 430 | 123 | 167 | 44 |
| 14 | Державний заклад «Луганський національний університет імені Тараса Шевченка» | 410 | 130 | 162 | 32 |
| 15 | Дрогобицький державний педагогічний університет імені Івана Франка | 275 | 73 | 86 | 13 |
| 16 | Державний заклад «Південноукраїнський національний педагогічний університет імені | 135 | 65 | 68 | 3 |

| № | ЗВО | Л | Д | З | Δ |
|---|---|---|---|---|---|
| | К. Д. Ушинського» | | | | |
| 17 | Уманський державний педагогічний університет імені Павла Тичини | 280 | 90 | 143 | 53 |
| 18 | Закарпатський угорський інститут імені Ференца Ракоці ІІ | 75 | 0 | 33 | 33 |
| 19 | Тернопільський національний педагогічний університет імені Володимира Гнатюка | 425 | 175 | 214 | 39 |
| 20 | Черкаський національний університет імені Богдана Хмельницького | 265 | 145 | 157 | 12 |
| 21 | Житомирський державний університет імені Івана Франка | 400 | 145 | 209 | 64 |
| 22 | Ніжинський державний університеті імені Миколи Гоголя | 310 | 125 | 136 | 11 |
| 23 | Чернігівський національний педагогічний університет імені Т. Г. Шевченка | 340 | 150 | 144 | -6 |
| 24 | Кам'янець-Подільський національний університет імені Івана Огієнка | 290 | 167 | 173 | 6 |
| 25 | Бердянський державний педагогічний університет | 100 | 33 | 34 | 1 |
| 26 | Національний педагогічний університет імені М. П. Драгоманова | 595 | 236 | 280 | 44 |
| 27 | Державний вищий навчальний заклад «Ужгородський національний університет» | 430 | 221 | 259 | 38 |
| 28 | Донецький національний університет | 701 | 282 | 322 | 40 |
| 29 | Вінницький державний педагогічний університет імені Михайла Коцюбинського | 425 | 125 | 213 | 88 |
| 30 | Миколаївський національний університет імені В. О. Сухомлинського | 265 | 115 | 139 | 24 |
| 31 | Державний вищий навчальний заклад «Переяслав-Хмельницький державний педагогічний університет імені Григорія Сковороди» | 130 | 67 | 82 | 15 |
| 32 | Державний вищий навчальний заклад «Прикарпатський національний університет імені Василя Стефаника» | 505 | 175 | 281 | 106 |
| 33 | Таврійський національний університет імені В. І. Вернадського | 575 | 280 | 384 | 104 |
| 34 | Державний вищий навчальний заклад «Криворізький національний університет» | 484 | 156 | 222 | 66 |

| № | ЗВО | Л | Д | З | Δ |
|---|---|---|---|---|---|
| 35 | Глухівський національний педагогічний університет імені Олександра Довженка | 145 | 62 | 62 | 0 |
| | **Разом:** | **11980** | **4694** | **5835** | **1141** |

Умовні позначення: Л – ліцензійний обсяг, Д – обсяг державного замовлення, З – кількість абітурієнтів, зарахованих на перший курс, Δ = З–Д – перевищення кількості абітурієнтів, зарахованих на перший курс, над державним замовленням.

**Додаток Е**

## Методика навчання об’єктно-орієнтованого моделювання у соціально-конструктивістському середовищі на прикладі окремих тем

### *Е.1 Вступ до об’єктно-орієнтованого моделювання: початок роботи в середовищі Alice*

Запуск програми приводить до появи вікна відкриття проекту (рис. Е.1).

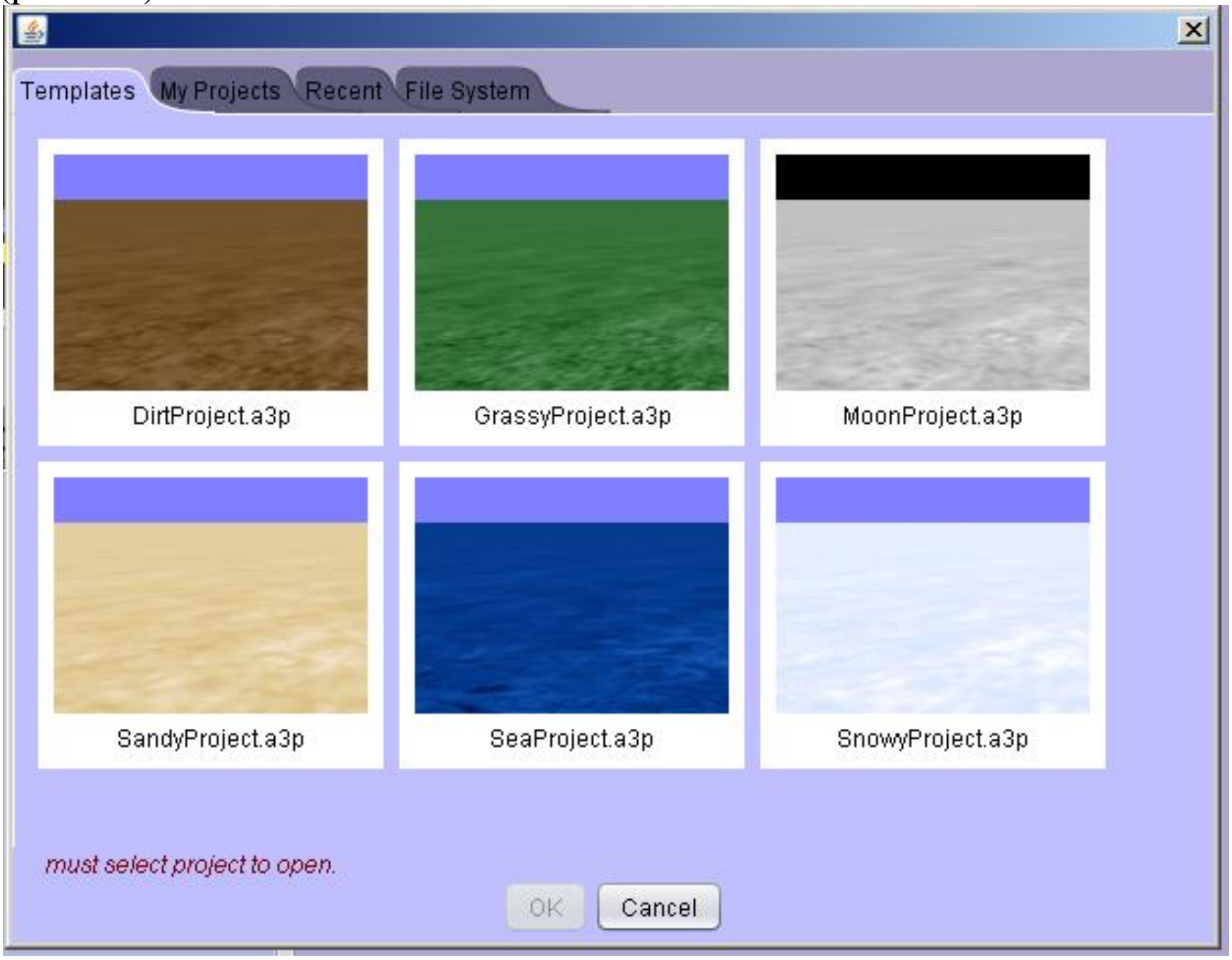

Рис. Е.1. Вікно відкриття проекту

Тут є декілька закладок: *My Projects* – містить посилання на проекти, збережені раніше; *Recent* – останні проекти, що відкривалися; *Templates* – стандартні світи Alice; *File System* – завантаження проектів з різних носіїв.

Обравши проект, ми потрапляємо на робочий екран Alice (рис. Е.2).

Під заголовком вікна ми бачимо рядок текстового меню: *Файл, Правка, Проект, Виконати, Вікно, Допомога*.

Меню *Файл* містить пункти *Новий* – створення нового проекту, *Відкрити* – відкриття існуючого, раніше збереженого проекту, *Останні*

*проекти* – відкриття останніх проектів, пункти *Зберегти, Зберегти як* – відрізняються лише варіантом збереження створеного проекту, *Повернутися до збереженого* – дозволяє повернутися до початкового виду проекту, *Export Video / Upload To YouTube™* – експортує готовий проект в Інтернет, *Вихід* – вихід із середовища.

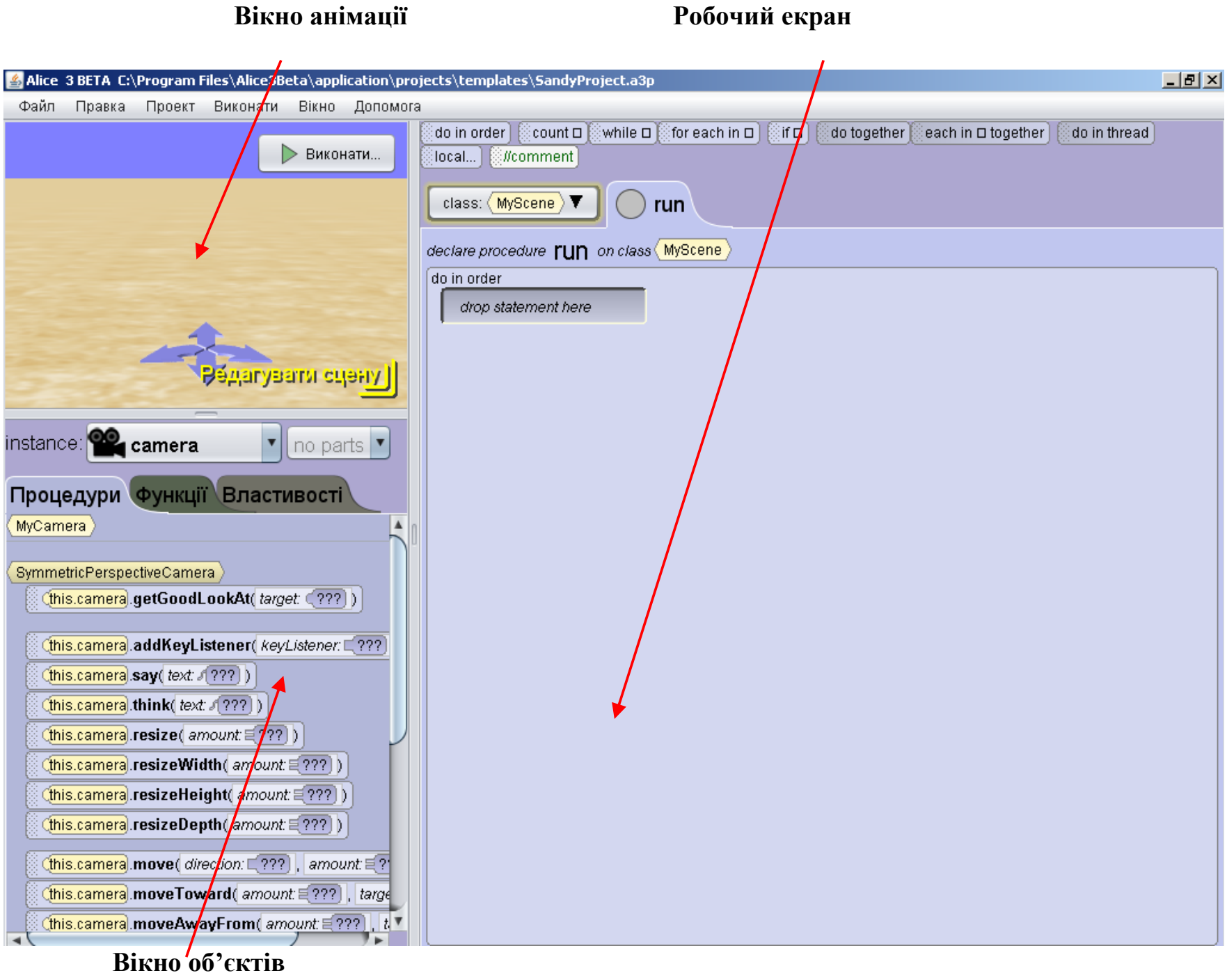

Рис. Е.2. Робочий екран Alice

Меню *Правка* містить пункти *Скасувати/Повторити* – відмовитись від/повернутися до змін, *Вирізати, Копіювати, Вставити*.

Меню *Проект* містить пункт *Управління ресурсами* – дозволяє добавити до/вилучити з існуючого проекту нові ресурси (аудіо або зображення), а також отримати статистику проекту, та пункт *Виконати* – запускає створений проект на виконання.

Окремо слід зупинитися на меню *Вікно*, в якому містяться пункти *Історія проекту* – відкриває вікно, в якому відображуються команди зміни програмного коду (натискаючи мишею на команди, можна вимикати/вмикати всі подальші дії з їх видаленням/відновленням), *Використання пам'яті* та *Установки* – налаштування зовнішнього вигляду середовища (рис. Е.3).

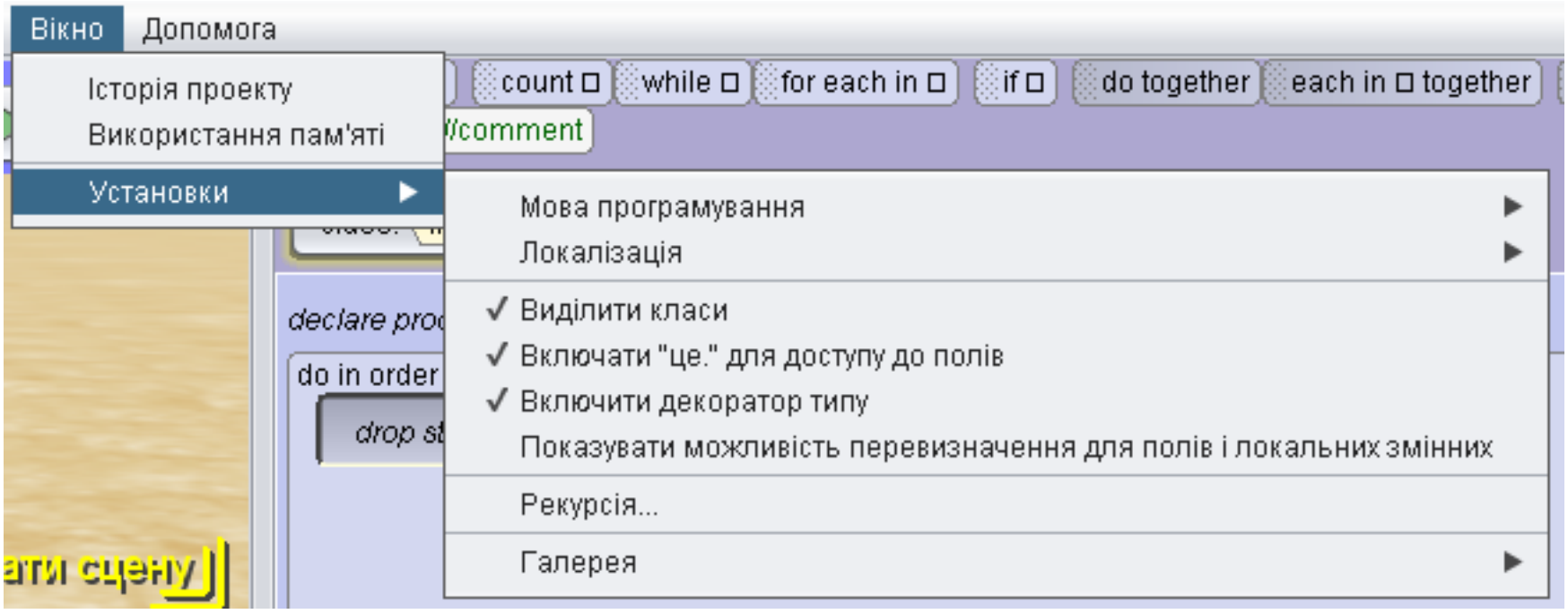

Рис. Е.3. Налаштування зовнішнього вигляду середовища

*Мова програмування* – вибір мови відображення програмного коду, *Локалізація* – мова інтерфейсу, *Виділити класи* – якщо не вибраний, то з'являється можливість редагувати процедури, функції і властивості всіх використовуваних класів, *Включати «це.» для доступу до полів* – відображувати це слово, воно позначає/або ні поточний об'єкт у програмному коді, *Включити декоратор типу* – чи показувати тип даних (клас) для виразів, *Рекурсія* – вимкнути/дозволити рекурсію.

Меню *Довідка* містить пункти *Довідка* – завантажує сторінку допомоги з мережі Інтернет, *Повідомити про помилку* – завантажує аналізатор вже описаних помилок у вигляді відповідей, *Запропонувати поліпшення* – дозволяє описати невідомі помилки, *Запит на додання нової функції* – дозволяє запитати опис невідомих помилок, *Показати попередження* – виводить попередження, що ми працюємо із бета-версією програми, *Показати системні властивості* – виводить відомості про використовувану операційну систему, *Проглянути примітки до випуску* – посилання на Інтернет-ресурс, що відображує зміни і оновлення поточної версії, *Про програму* – інформація про версію програми.

Для створення об'єкта та налаштування його початкового положення необхідно перейти в режим редагування сцени: 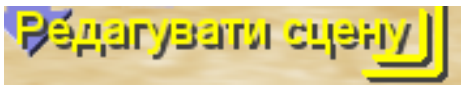

. Додавання об'єкту здійснюється у такий спосіб:

– вибираємо галерею, в якій міститься потрібний об'єкт (рис. Е.4);

– вибираємо об'єкт з галереї (рис. Е.5).

Додавши новий об'єкт, відразу дістаємо можливість переміщувати його по сцені і обертати відносно вертикальної осі. Більш точне налаштування положення об'єкту виконується за допомогою меню режимів налаштування положення об'єкту *Handle Style*: *Default* – налаштування за замовчанням, *Rotation* – обертання, *Move* – переміщення, *Resize* – зміна розміру. При виборі одного з пунктів (режимів налаштування) автоматично відбувається зміна маркерів біля об’єкту.

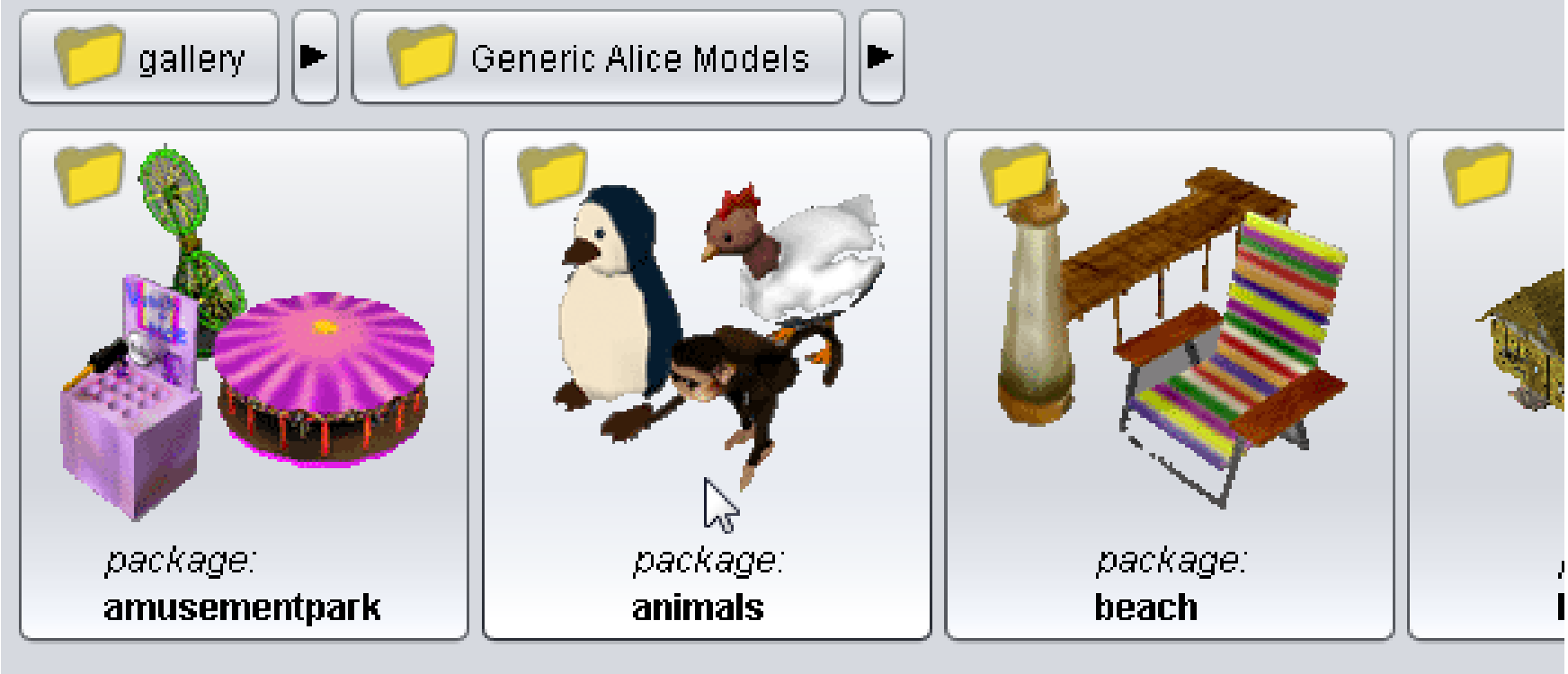

Рис. Е.4. Галерея об’єктів

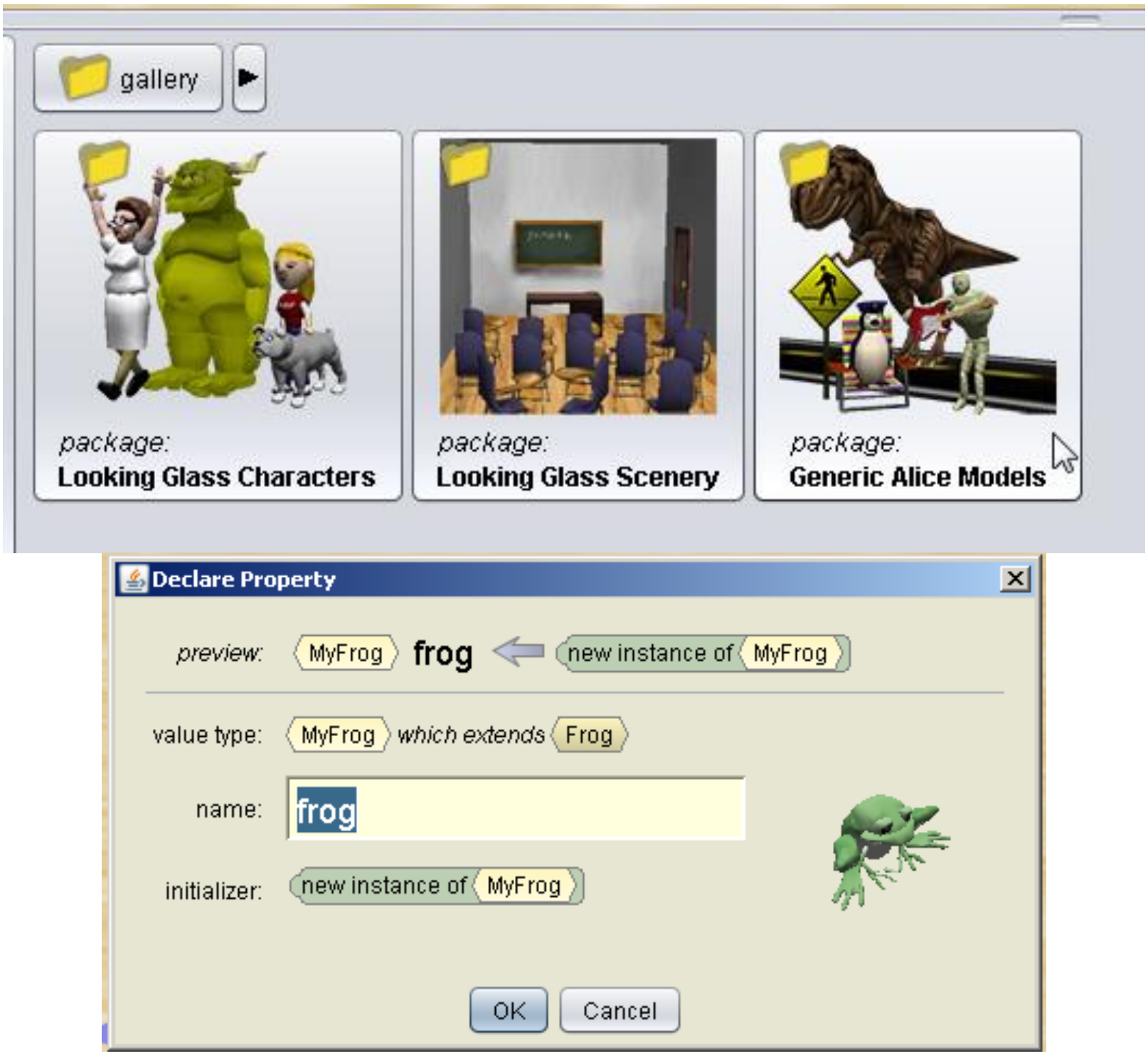

Рис. Е.5. Динамічне створення об’єкту *frog* класу *MyFrog*, що наслідує клас *Frog*

У режимі *Default*, натискуючи і утримуючи лівою кнопкою миші кільце навколо об’єкту, обертаємо об’єкт відносно вертикальної вісі або просто переміщуємо його мишею. Режим *Rotate* аналогічний попередньому, з тією лише різницею, що є можливість обертання в трьох площинах. У режимі *Move*, утримуючи відповідні стрілки мишею, можна переміщати його вертикально, горизонтально і фронтально. Режим *Resize* дозволяє змінювати розмір об’єкту в будь-якому з трьох вимірів, а також пропорційно.

Камеру перегляду можна переміщувати відносно об’єкту за допомогою відповідних стрілок-маркерів: 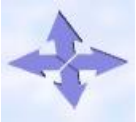 переміщує камеру по вертикалі і горизонталі, 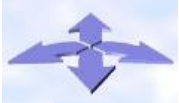 – фронтально і обертаючи відносно вертикальної осі камери, 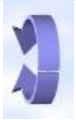 обертає відносно горизонтальної вісі камери. Слід відзначити, що за необхідності об’єкти можна позиціонувати за допомогою клавіш управління курсором («стрілок ←↑↓→»), але при цьому всі переміщення здійснюються дискретно, кратне певному кроку.

Як вже указувалося вище, при запуску середовища об’єктно-орієнтованого моделювання Alice з’являється діалогове вікно *Open Project*, проте в ході роботи над проектом інший проект можна відкрити за допомогою пункту меню *Файл*. Наприклад, послідовність дій *Файл→Відкрити→File System*→D:\Alice→ rabbit.a3p приведе до відкриття відповідного файлу, що входить до створеного нами дистрибутиву локалізованої версії Alice 3 Beta. Виконавши ці дії, отримаємо у вікні редагування програму (рис. Е.6):

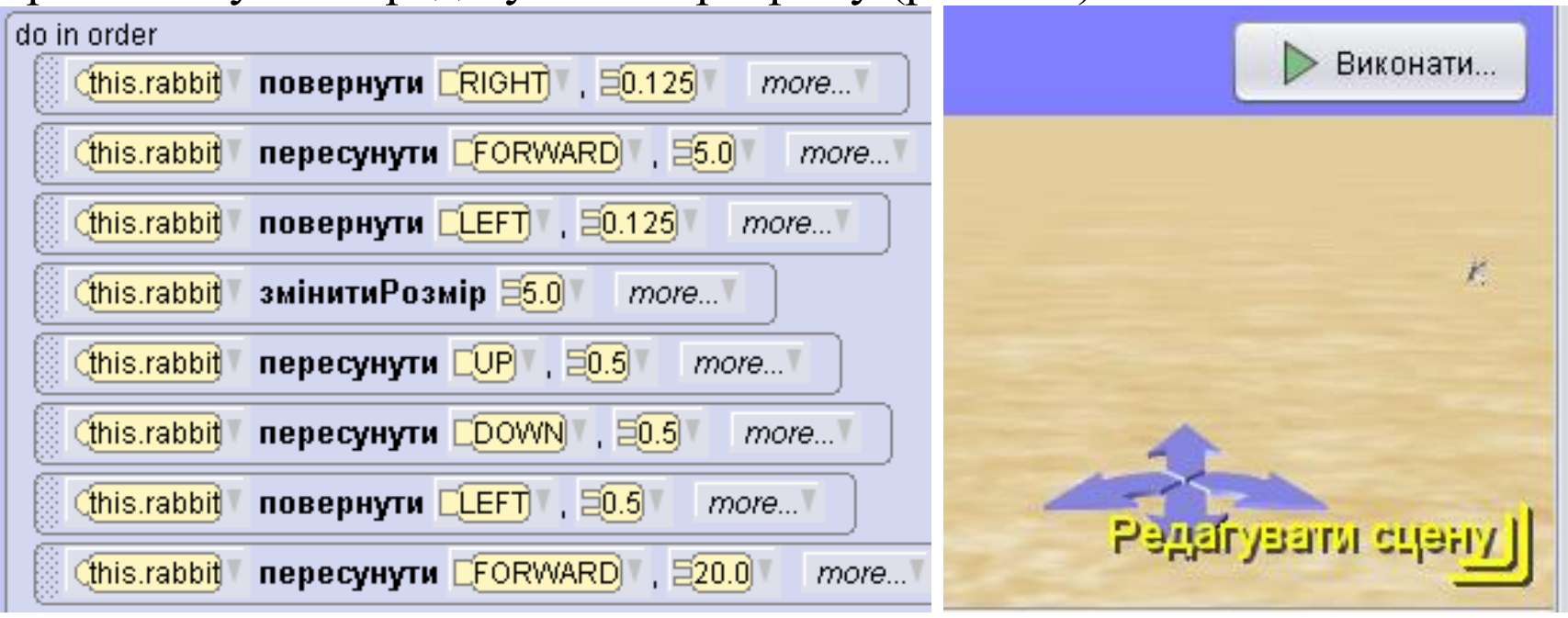

*а)* *б)*

Рис. Е.6. Подання програми (а) та сцени (б) у середовищі Alice

Натискання кнопки *Виконати* надає можливість запустити програму і подивимось на дії кролика.

Alice надає можливість створювати інтерактивні програми – програми, що складаються з об’єктів, які очікують на дії користувача (повідомлення), для виконання яких-небудь дій (реакції на отримані повідомлення).

Зупинимося на конкретному прикладі: створимо проект «Заєць і бджола».

1. Вибираємо шаблон *GrassyProject* (рис. Е.1).

2. Перейшовши до вікна редагування сцени, додаємо об’єкти *hare* (заєць) та *bee* (бджола) (зверніть увагу на шлях до потрібних об’єктів, рис. Е.7).

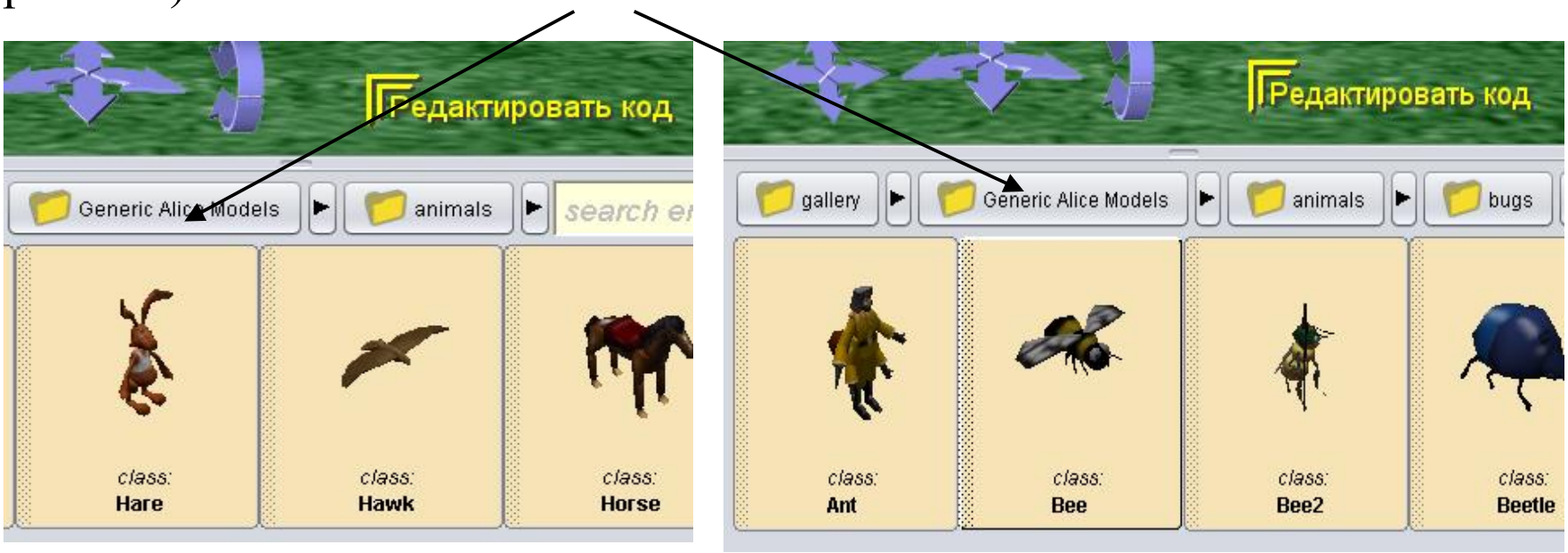

Рис. Е.7. Вибір об’єктів класів, похідних від *Hare* та *Bee*

3. Позиціонуємо об’єкти та за потреби змінюємо їхні розміри.

Об’єкти реагують тільки на дії користувача: досить натискання на будь-яку клавішу на клавіатурі, щоб відбулося виконання програми (заєць стрибнув, а бджола виконала пірует). Ці дії об’єктів досягаються за допомогою створення процедур *hop* і *piruet* для об’єктів *hare* і *bee* відповідно (рис. Е.8).

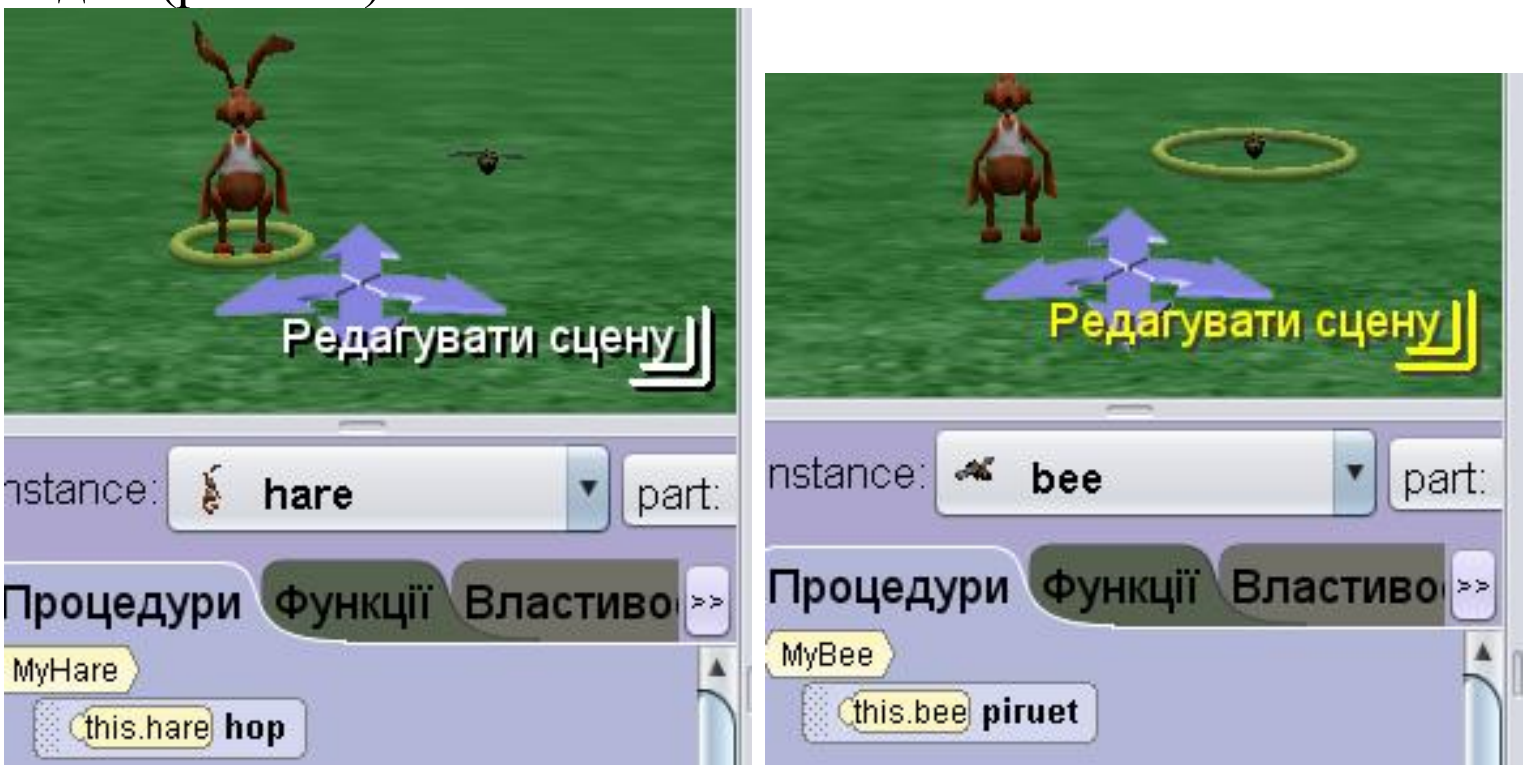

Рис. Е.8. Оголошення процедур *hop* у класі *MyHare* та *piruet* у класі *MyBee*

На рис. Е.9 показано код, що додається до процедури *run* класу *MyScene* (виконується після ініціалізації всіх об'єктів, розміщених на сцені). Слід звернути увагу на те, що був створений блок *do together* (для паралельного виконання дій), у якому викликаються процедури *piruet* об'єкту *bee* та *hop* об'єкту *hare*. Метод *додатиОбробникКлавіатури* визначає реакцію об'єкту *hare* на натискання будь-якої клавіші клавіатури (*keyPressed*).

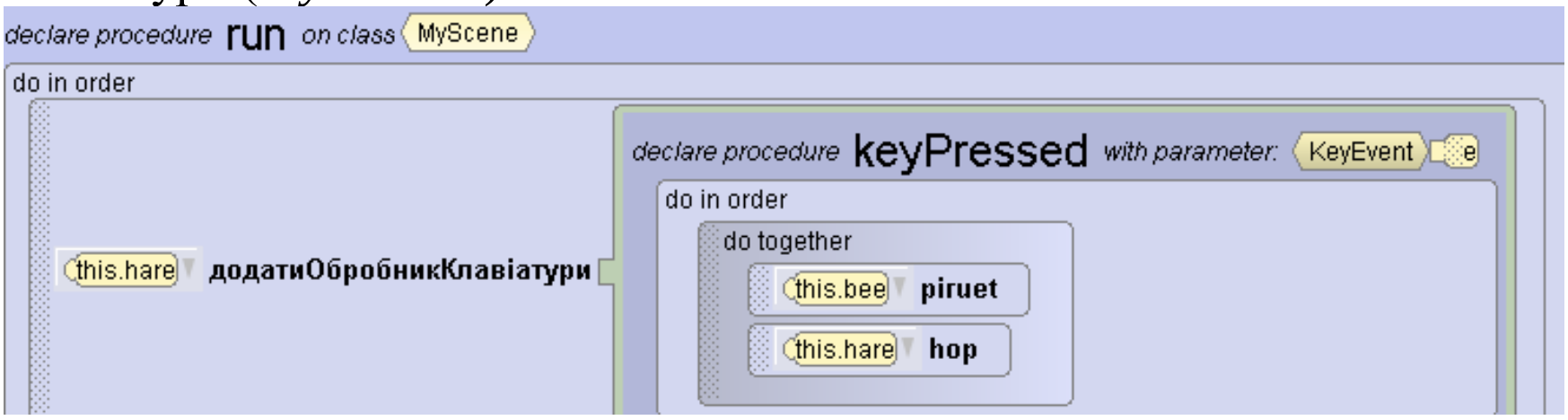

Рис. Е.9. Додавання обробника клавіатури для об'єкту *hare*

Для оголошення процедури необхідно вибрати пункт *Declare Procedure* для відповідного класу (рис. Е.10).

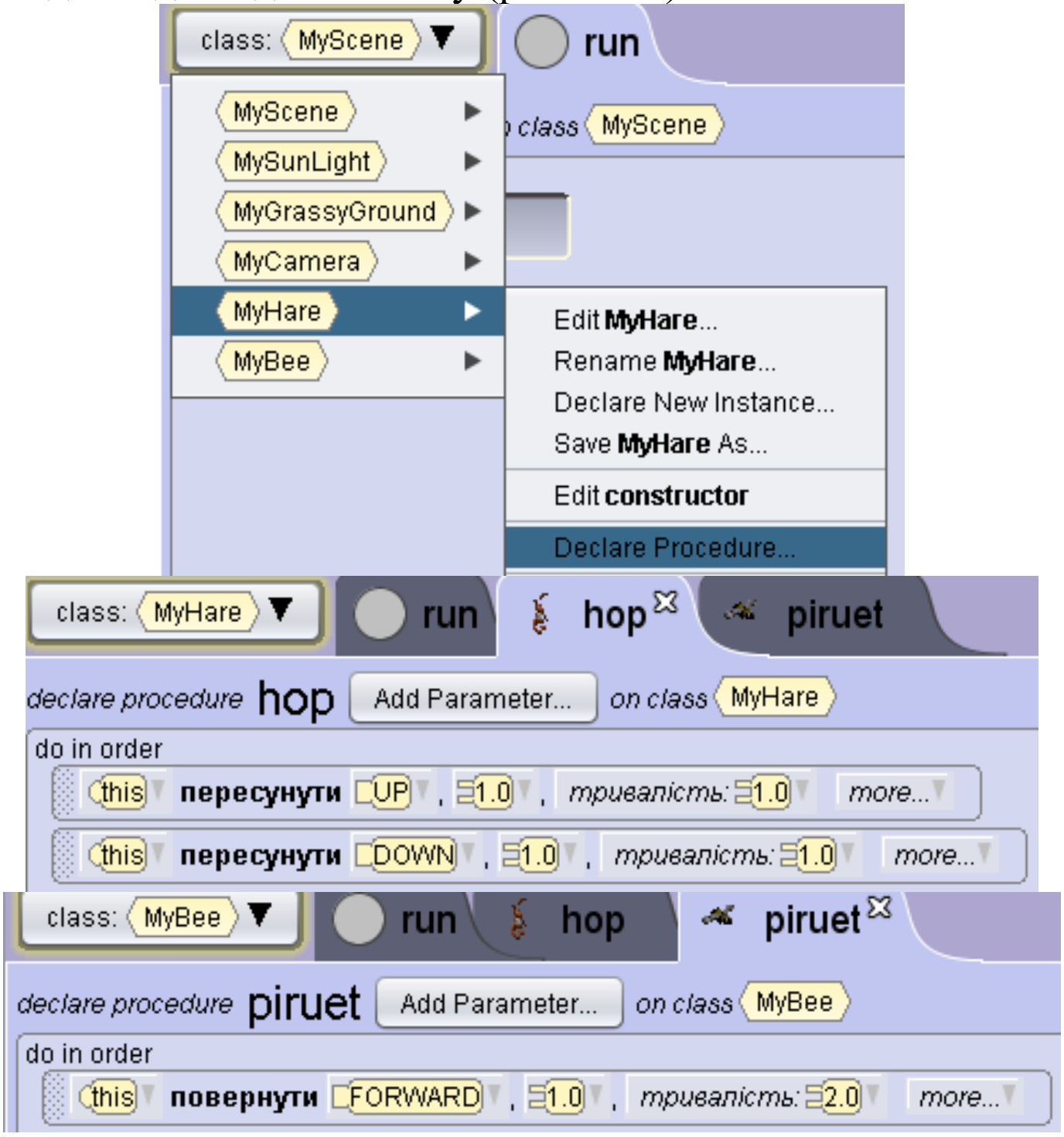

Рис. Е.10. Оголошення процедур *hop* та *piruet*

Слід звернути увагу на параметр *тривалість* процедур *повернути* та *пересунути*. Тут він використовується для того, щоб дії збігалися за часом виконання, тобто виконувалися *одночасно* (2 дії – по 1 секунді і одна – за 2 секунди).

Таким чином, створений проект складається з об’єктів (сцена, освітлення, камера, поверхня – створені при виборі шаблону проекту; заєць, бджола – додані користувачем), процедур (методів) та функцій (убудованих та створених користувачем), що описують реакції об’єктів на певні події (повідомлення), які змінюють властивості об’єктів.

Метод в Alice використовується як інструмент для відправлення об’єктові повідомлення із вказівкою виконати деякі дії. Важливою відзнакою є те, що метод не повертає значення. Функції в Alice також використовуються для відправки об’єкту повідомлення, в якому зазвичай об’єкт повинен повернути деяку інформацію. В деяких випадках функція може також наказати об’єктові виконати деякі дії.

Функція в Alice завжди повертає значення. Описати механізм повернення значення функції можна, сказавши, що при запуску функції вона звертається до якогось об’єкту і перевіряє значення деякого параметра цього об’єкту. Після цього повертається назад на те місце головної програми, звідки була запущена, тобто в програмі замість функції з’являється значення відповідного параметра об’єкту.

Всі об’єкти в Alice – сцена, камера, освітлення, поверхня тощо, а також об’єкти, створені з класів галереї Alice включають, окрім методів, також і функції. Ці функції називають примітивними. Окрім цього, користувач може створювати нові функції і закріплювати їх за потрібними об’єктами. Такі функції називають функціями, визначеними користувачем.

Як і в інших об’єктно-орієнтованих мовах програмування, в Alice властивості об’єктів (закладка *Властивості*) здатні повертати поточні значення характеристик об’єктів і призначати нові, тобто можуть виступати в якості функцій.

З рис. Е.11 видно, що функції-властивості, які повертають значення (жовтого кольору), відрізняються від функцій, що передають значення (блакитного кольору і наявністю поля для значення, що передається).

На першому занятті з початків об’єктно-орієнтованого моделювання у середовищі Alice слід звернути увагу ще на одну його особливість: при підведенні миші до функції одразу з’являться рамки навколо тих параметрів, куди цю функцію (а точніше, значення, яке вона повертає) можна вставити.

Подальшу роботу можна побудувати за таким планом:

1. Типи даних (числовий, логічний, текстовий, об’єктний) та

оголошення змінних. Час життя локальних змінних та змінних об’єкту (властивостей). Створення змінних.

2. Послідовні, циклічні структури і структури вибору.

3. Масиви об’єктів.

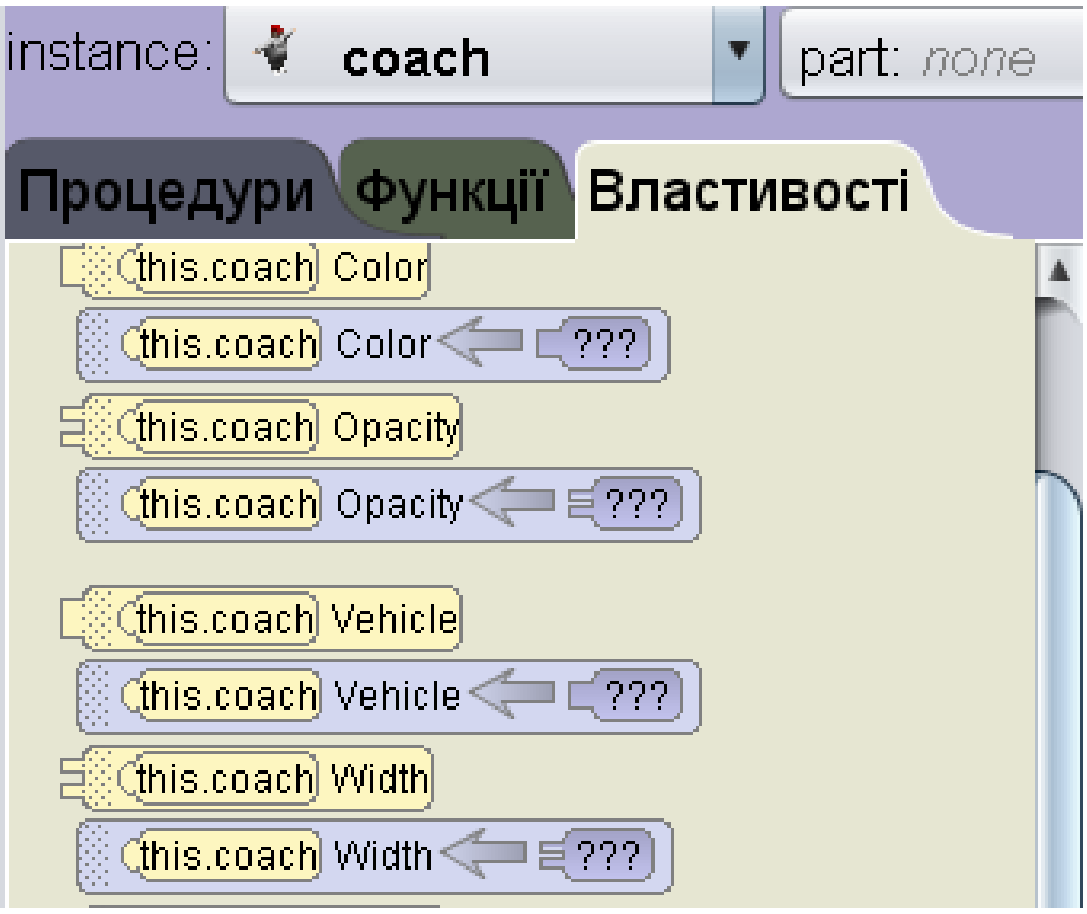

Рис. Е.11. Закладка *Властивості* об’єкту *coach*

Після завершення опрацювання першої частини посібника [348] можна перейти до другого та третього модулів курсу.

*Е.2 Об’єктно-орієнтовані імітаційні моделі: конструювання «Життя»*

У третьому модулі курсу розглядається ряд моделей, що мають загальну назву кліткові автомати. Кліткові автомати були вперше розглянуті Дж. фон Нейманом (John Von Neumann) та С. Уламом (Stanislaw Ulam) в 1948 р. як можлива ідеалізація біологічного самовідтворення [49; 50]. Зацікавленість ними обумовлена, головним чином, тим, що більшість з них є прикладами простих динамічних систем (розглянутих у другому модулі курсу), які дають впорядковані візерунки, що виникають із випадкових початкових умов. Тривимірний характер середовища Alice надає можливість впевнитися в тому, що просторові візерунки багатьох кліткових автоматів нагадують візерунки, які можна спостерігати в природних явищах (поширений приклад – ріст кристалів). Саме тому кліткові автомати розглядають як корисні й цікаві комп’ютерні моделі, які зберігають привабливі можливості для опису складних фізичних систем.

Кліткові автомати являють собою моделі фізичних систем, в яких простір і час дискретні, самі ж фізичні величини (у разі потреби) набувають скінченної множини дискретних значень. Для прикладу уявимо регулярну решітку клітин (комірок), кожна з яких може знаходитися у скінченому числі можливих станів, наприклад, 0 або 1. Стан системи повністю визначається значеннями змінних в кожній клітині.

Важливими особливостями кліткових автоматів є такі:

1. Стан кожної комірки поновлюється за скінчену послідовність кроків (зокрема, на кожному кроці).

2. Ці поновлення значень змінних в кожній комірці відбуваються одночасно (паралельно), виходячи із значень змінних на попередньому кроці.

3. Новий стан комірки залежить лише від локальних значень у сусідніх комірках.

Для характеристики кліткового автомата зазначимо такі його особливості:

1. Розташування клітин утворює деяку геометричну фігуру. Для моделі росту сніжинок досить двомірної шестикутної системи, але у більшості інших випадків обирають прямокутну решітку, що складається з квадратів. Існують тривимірні схеми (і навіть з більшою кількістю вимірів, але уявити їх складно).

2. За заданою схемою необхідно визначити те оточення, яке дана клітина «вивчає» при обчисленні свого наступного стану.

3. Кількість станів, які може приймати клітина, буває різною. Дж. фон Нейман у 1952 р. побудував систему, здатну до самовідтворення [50], у якій клітини мали 29 можливих станів, однак більшість автоматів значно простіші.

4. Головне джерело змін у світі кліткових автоматів – це величезна кількість можливих правил для визначення наступного стану клітини, виходячи із станів її сусідів у даний момент.

Всі досліджені правила зміни стану комірки можна поділити на чотири класи:

1) правила, за яких еволюція приводить систему до стійкого та однорідного стану;

2) правила, що ведуть до появи простих структур (стійких або періодичних), які в будь-якому випадку залишаються ізольованими одна від одної;

3) правила, які ведуть до появи хаотичних візерунків (хоча й не обов'язково випадкових);

4) правила, які породжують структури істотної просторової та часової складності.

Мабуть, найвідомішим клітковим автоматом є гра «Життя», запропонована у 1970 р. відомим алгебраїстом Дж. Х. Конвеєм (John

Horton Conway). Та навряд чи вона отримала таке поширення, якби не статті відомого популяризатора науки М. Гарднера (Martin Gardner), у яких вперше вона була представлена широкому загалу [50].

Ситуації, що виникають у процесі гри, дуже нагадують реальні процеси, що відбуваються при зародженні, розвитку та загибелі колонії організмів. Умови народження та загибелі визначаються виключно взаємним розташуванням учасників, а правила гри жорстко визначають, де та коли відбуваються народження та смерть. Гра складається з «циклів життя», або з послідовності дискретних кроків, за допомогою яких імітується зміна поколінь.

Для реалізації гри «Життя» сам Дж. Х. Конвей спочатку використовував скінченну шахову дошку, у клітинках якої розташовувались шашки одного кольору. Проте на краях дошки правила гри не спрацьовували, тому йому довелося вдатися до концепції нескінченної (замкненої з усіх боків) дошки, що моделювала планету, повністю вкриту океаном. Острови в океані розташовані на рівних відстанях уздовж меридіанів і паралелей. На кожному острові може мешкати лише одна істота – конвік. Найближчими до кожного острова є завжди 8 сусідніх островів.

Для моделювання простору гри «Життя» оберемо морську сцену із переліку шаблонів, що їх надає Alice. Аналогом острову може бути будь-який кулястий об'єкт, наполовину занурений в море (наприклад бейсбольний м'яч, який є в галереї Alice).

Для моделювання оберемо фрагмент зі 100 островів (сітка розміром 10×10, рис. Е.12).

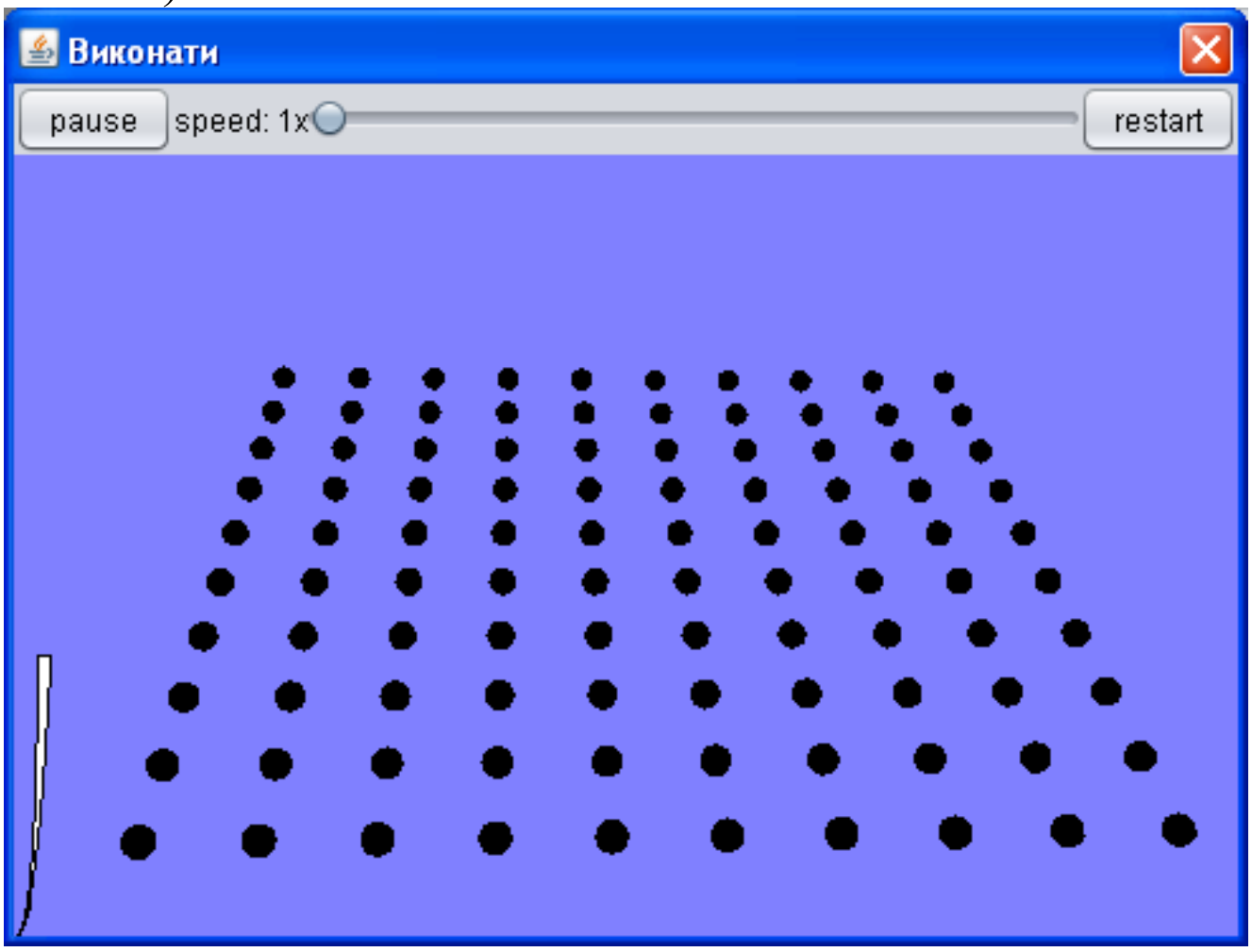

Рис. Е.12. Ігровий простір

Для забезпечення умови необмеженості обраного фрагменту вдамося до «склеювання» його границь, а саме: вважатимемо, що верхній і нижній краї фрагменту межують один з одним (те ж саме відбувається з лівим і правим краями). В результаті з'являється нова властивість: переміщення деякого об'єкта за правий край поля приводить до появи цього об'єкта на лівому краю і навпаки. Так само переміщення об'єкта під нижній край поля приводить до його появи на верхньому краю і навпаки. У такий спосіб виявляється можливим замість необмеженого поля обійтися полем скінченного і не дуже великого розміру.

Спочатку заповнимо морську сцену сотнею однотипних об'єктів. Для цього достатньо створити один об'єкт класу *Baseball*. За замовчанням ім'я створеного похідного класу буде *MyBaseball*. Назвемо його *convik00*. Змінимо ім'я класу *MyBaseball* на *Convik*, обравши пункт *Rename* (перейменування) у полі «*class:*», після чого оголошуємо створений об'єкт *convik00* вибором пункту «*Declare new instance…*», надаючи кожному новому об'єкту ім'я «*convikXY*», де *XY* – число від 01 до 99. Всі клоновані у зазначений спосіб об'єкти мають спільне розташування та інші властивості, розрізнюючись лише ім'ям.

Клас *Convik* зараз є ідентичним класу *Baseball* у всьому, за виключенням імені. Внесемо до нього наступні зміни:

1. Додамо дві властивості класу (рис. Е.13):

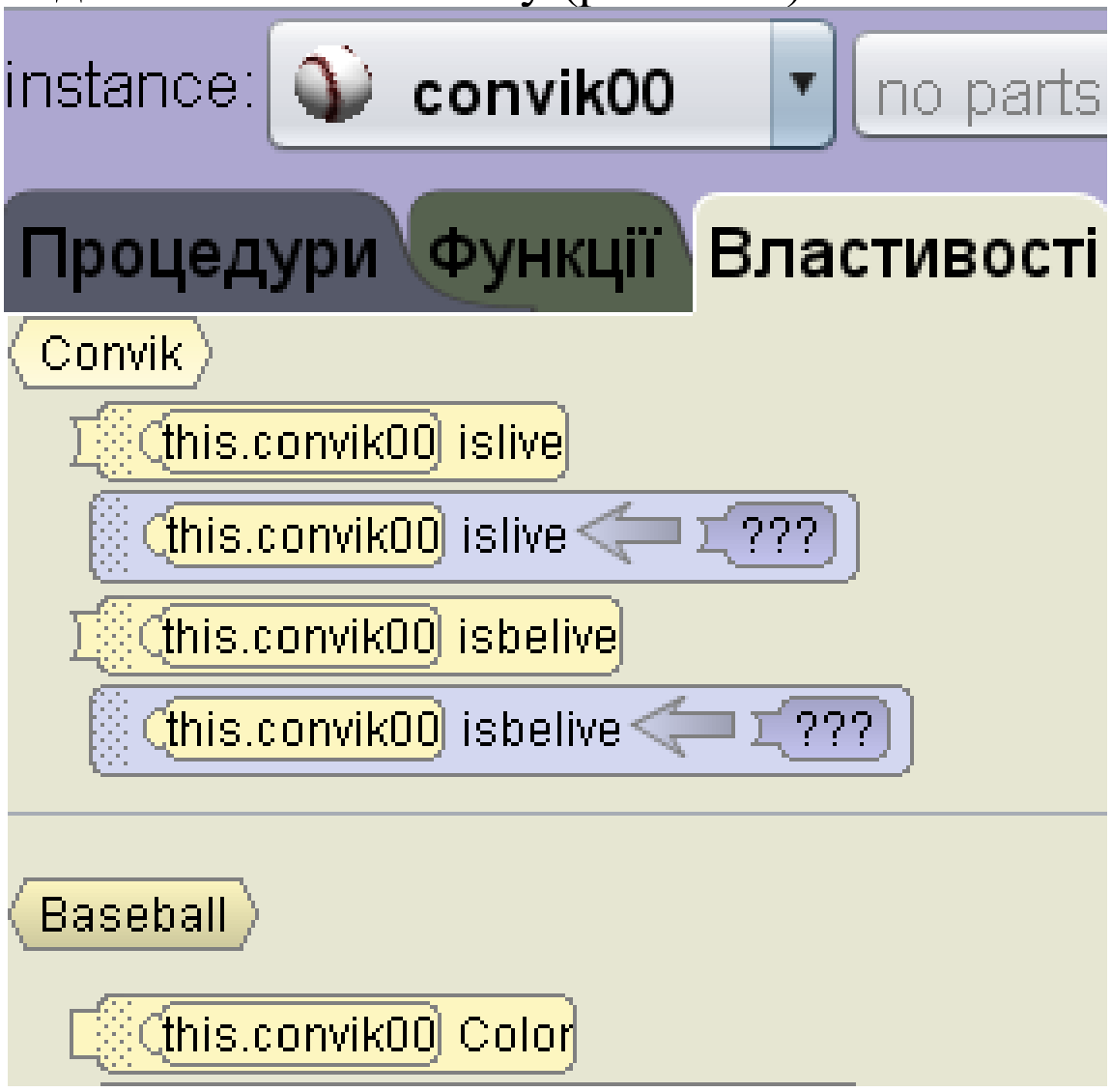

Рис. Е.13. Властивості класу *Convik* наслідують властивості батьківського класу *Baseball*

– *islive* – логічна змінна: значення *true* означає, що острів заселений, *false* – що не заселений;

– *isbelive* – логічна змінна: значення *true* для якої означає, що після застосування правил гри «Життя» острів має бути заселений, а *false* – звільнений.

2. Додамо дві процедури (рис. Е.14):

– *draw* – обирає колір острова в залежності від того, острів заселений (він білий), чи ні (чорний);

– *swap* – у залежності від того, чи змінює острів стан (із заселеного на незаселений чи навпаки), змінюватиме його колір.

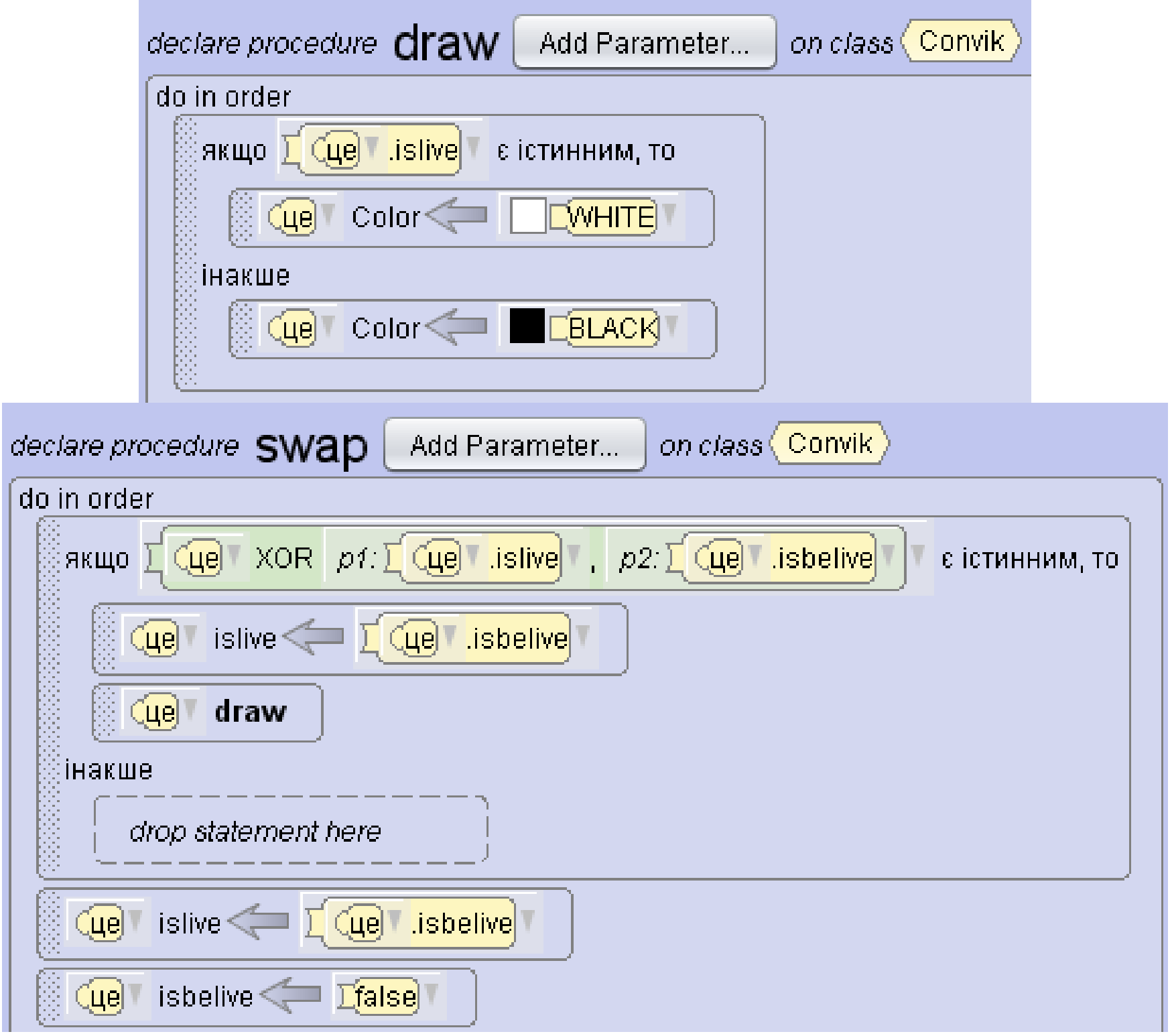

Рис. Е.14. Процедури *draw* та *swap* класу *Convik*, що використовують нові властивості цього класу

3. Додамо логічну функцію *XOR*, що повертатиме значення *true* у тому випадку, коли її параметри різняться і *false* навпаки (рис. Е.15). Ця функція є допоміжною для процедури *swap*.

Також змінимо конструктор класу *Convik*, задавши за замовчанням такі значення властивостей: *islive* – *false* (незаселений), *isbelive* – *false* (не претендує на заселення), тому колір острова встановимо чорний. Додатково в конструкторі визначимо обробник миші, який при виборі об'єкта мишею змінюватиме стан із заселеного на незаселений (і навпаки) одночасно, відповідно, змінюючи колір об'єкта (рис. Е.16).

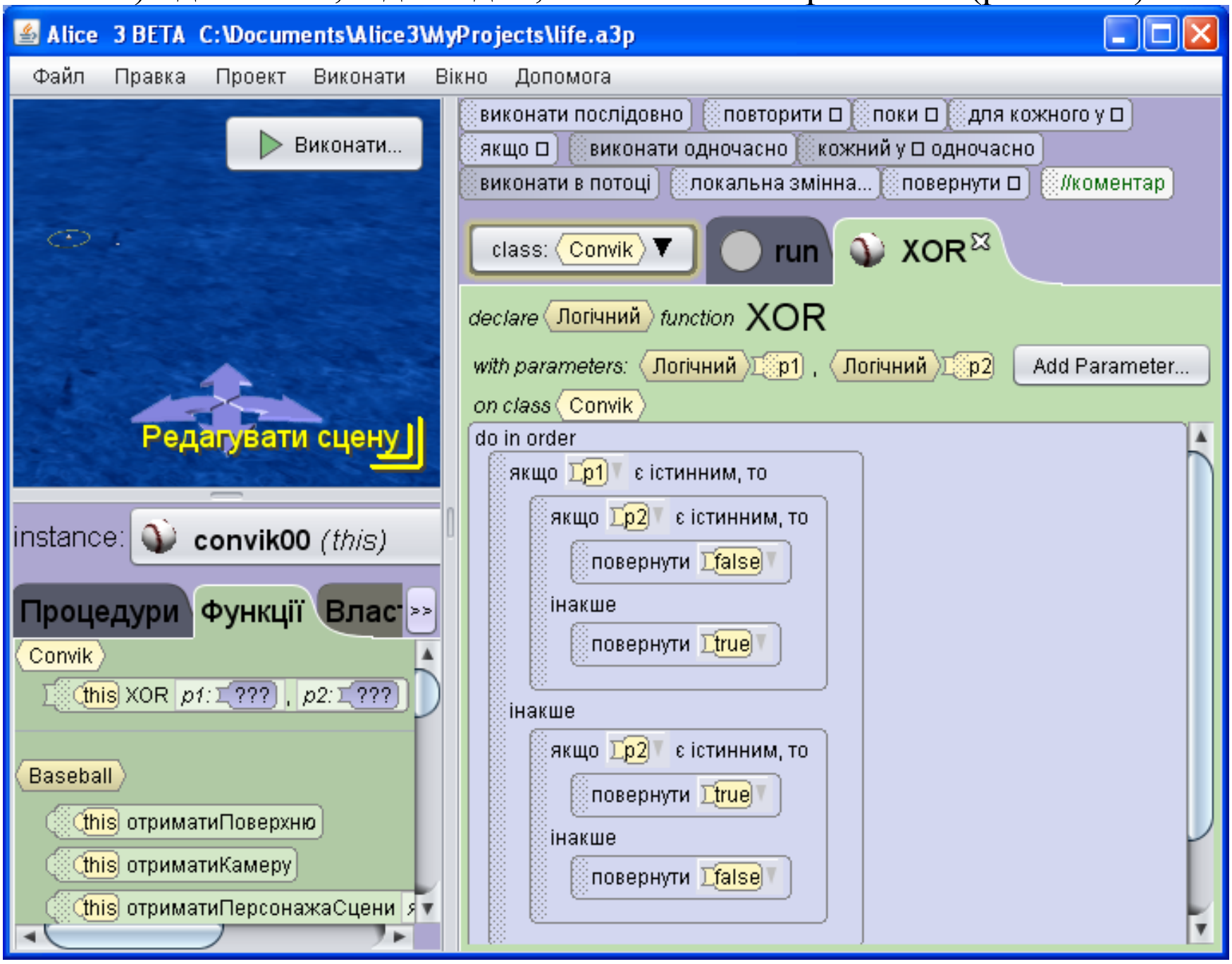

Рис. Е.15. Функція *XOR* – метод класу *Convik*, що доповнює методи батьківського класу *Baseball*

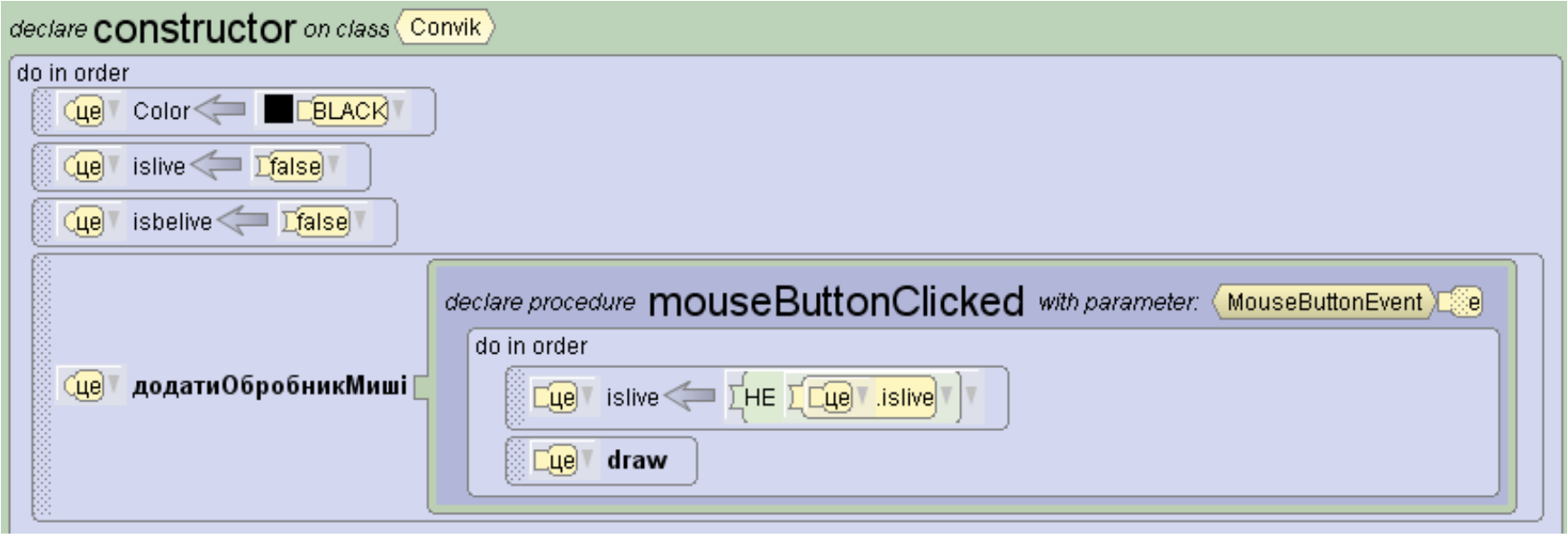

Рис. Е.16. Зміни, що вносяться до конструктора класу *Convik*

Створений клас *Convik* може бути збережений за допомогою панелі класів (*Save Convik As*…) у вигляді бінарного файлу класу Alice (Convik.a3c), який, за замовченням, зберігається в папці My Classes).

Зв’язування об’єктів класу *Convik* зі сценою виконується автоматично при додаванні об’єктів до сцени: всі додані об’єкти стають властивостями сцени – стандартні об’єкти сцени, такі як *sunLight* (освітлення), *seaSurface* («морська поверхня»), *camera* та додані нами 100 об’єктів класу *Convik* (рис. Е.17). Для роботи з однотипними об’єктами доцільно використати масив, який створимо як властивість об’єкта *scene* під ім’ям *convfield* (рис. Е.18). Це буде масив із 100 створених об’єктів (з індексами від 0 до 99).

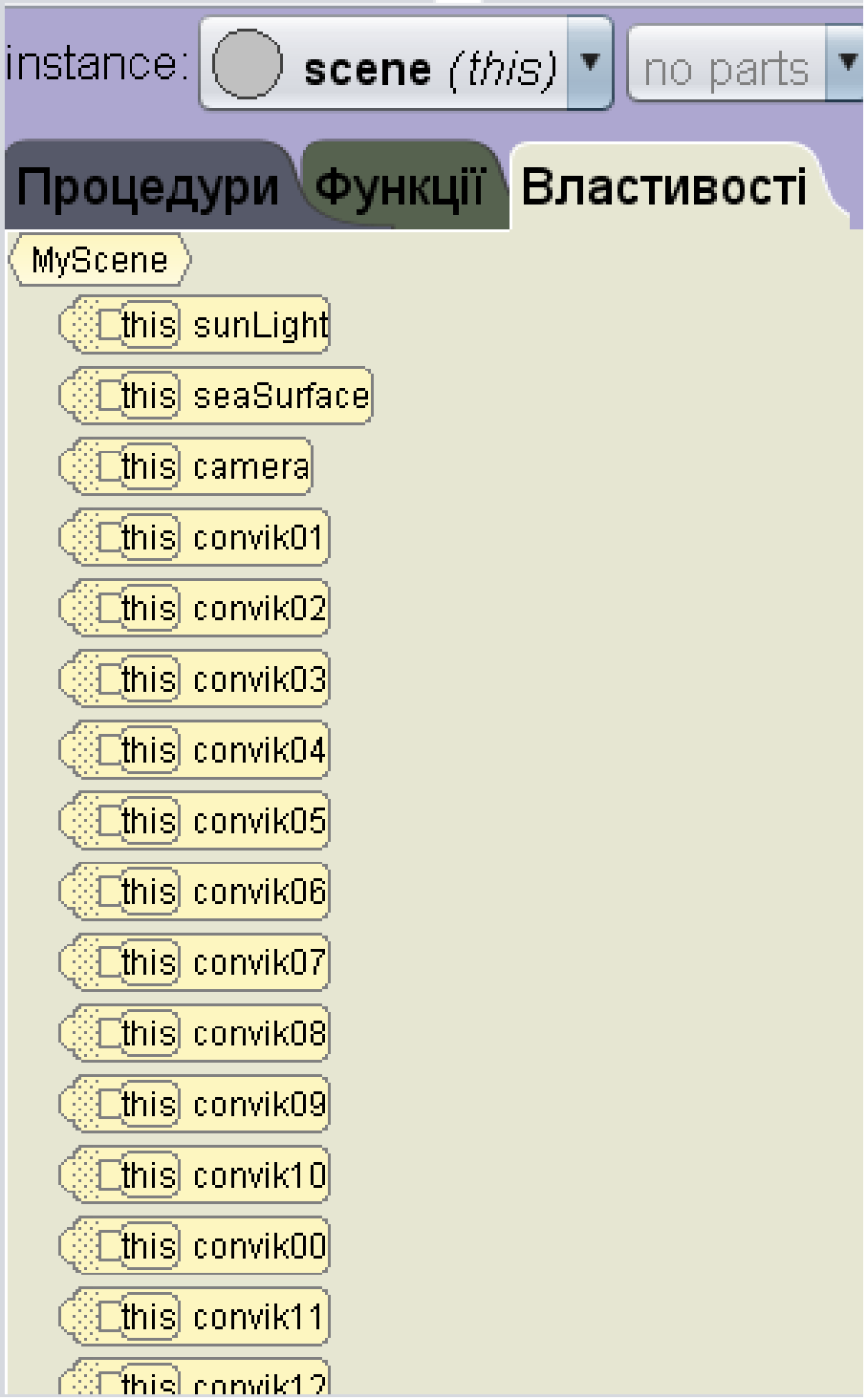

Рис. Е.17. Розміщення об’єктів на сцені як створення нових властивостей об’єкта *scene*

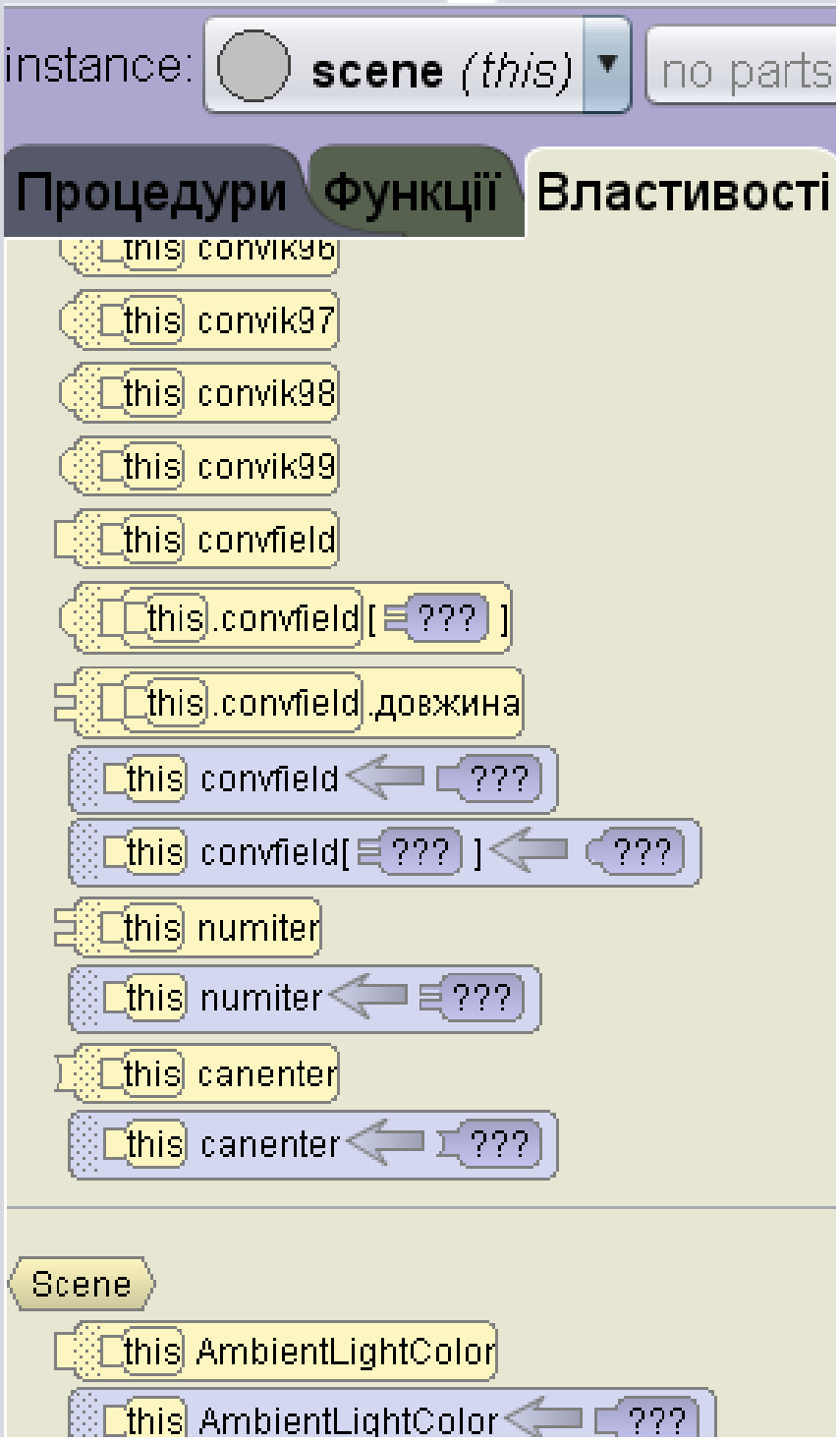

Рис. Е.18. Нові властивості класу *MyScene* та успадковані властивості класу *Scene*

Конструктор класу *MyScene* (успадкованого від *Scene*) викликає процедуру *performMySetUp*, яку використаємо для початкового налаштування створених об’єктів (рис. Е.19):

1. Якщо при створенні конвіків вони «розбіглися» по сцені (через невдале ручне розташування) – пересунемо їх до якогось одного за

допомогою стандартної процедури *пересунутиДо*.

2. Встановимо прозорість морської поверхні у 0.

3. Зорієнтуємо камеру так, щоб сцену було краще видно (методами *пересунутиУНапрямку* та *повернути*).

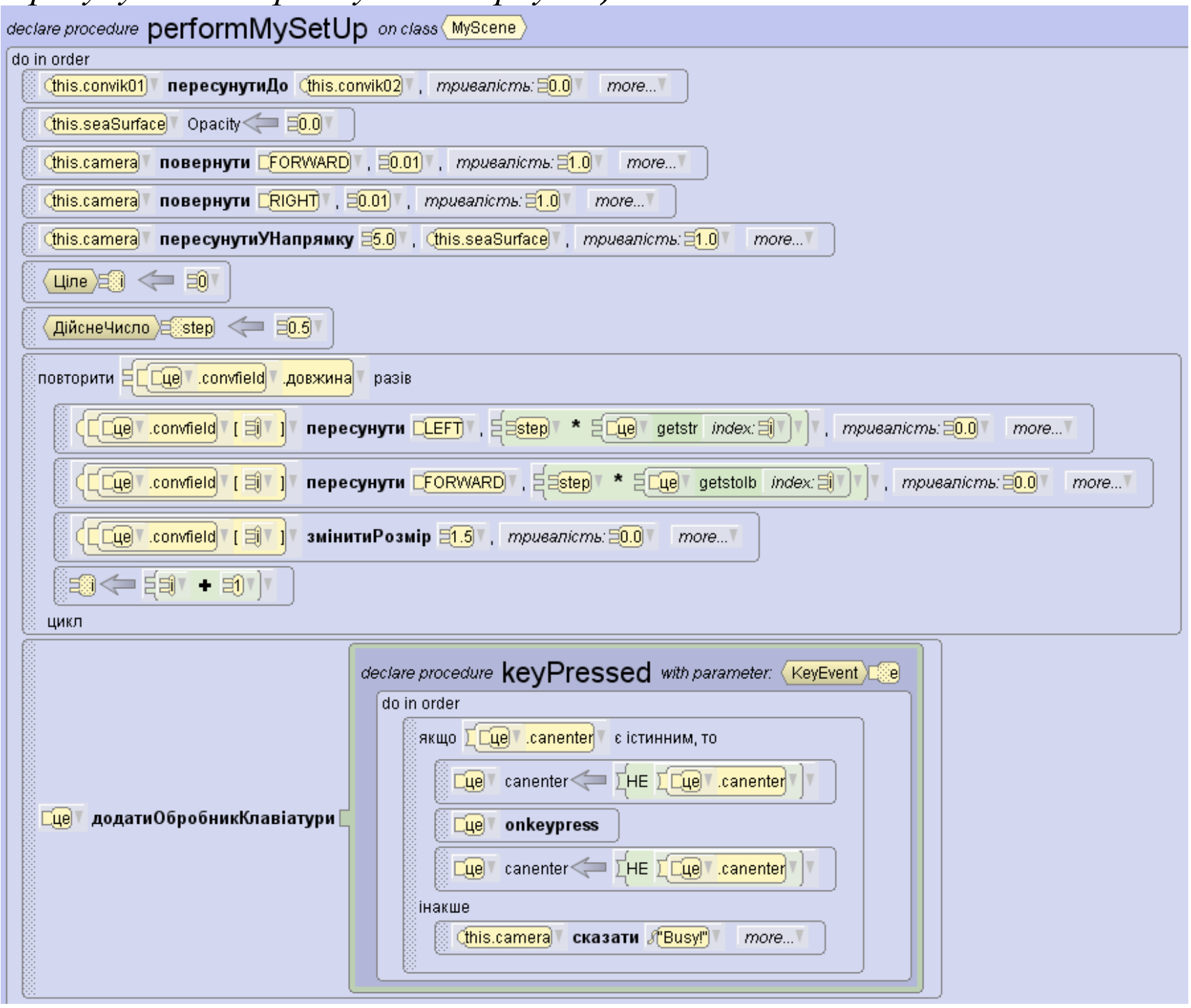

Рис. Е.19. Початкове налаштування сцени

Розташуємо острови на паралелях та меридіанах шляхом пересування вліво та вперед на відстань, що відповідає положенню острова в нашому фрагменті: якщо *step* – це відстань між островами на паралелі (меридіані), то уліво зсунемо на *step*номер рядка острову* у фрагменті архіпелагу, а вперед – на *step*номер стовпця*.

Остання дія у *performMySetUp* – це додавання обробника клавіатури: за натисканням на будь-яку клавішу викликатимемо процедуру *onkeypress*, що виконуватиме один крок еволюції (одну ітерацію) гри «Життя». Враховуючи, що під час виконання кроку еволюції втручання користувача у роботу моделі є недоцільним, виконаємо блокування обробки натискань клавіш за допомогою нової властивості класу *MyScene* – логічної змінної *canenter* із початковим значенням *true* (рис. Е.18). Якщо ця властивість дорівнює *true*, можна виконувати новий крок еволюції, інакше варто повідомити користувача, що модель зайнята роботою. Для цього перед та після виклику процедури виконання кроку еволюції змінимо значення властивості *canenter* на протилежне.

Для того, щоб сигналізувати про завершення кроку еволюції, створимо ще одну властивість: *numiter* – цілочисельну змінну із початковим значенням 1.

Для роботи процедури *onkeypress* створимо наступні додаткові функції (рис. Е.20):

– *getstr* та *getstolb* – повертають номер рядка та стовпця відповідно за одновимірним поданням індексу елементу;

– *getindex* – повертає одновимірний індекс за номером рядка та стовпця;

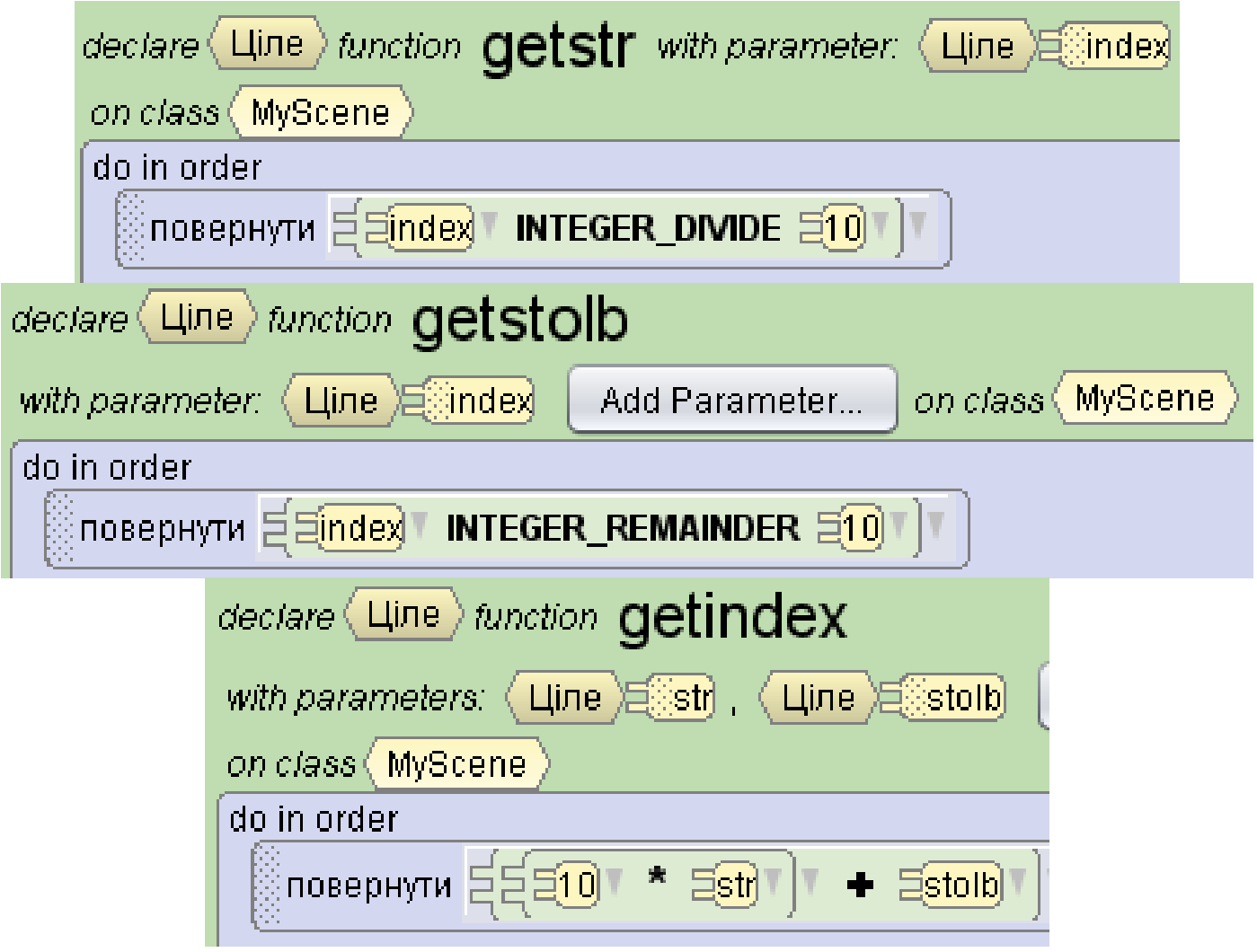

Рис. Е.20. Функції перетворення координат

Необхідність у цих функціях зумовлена відмінністю між внутрішнім поданням об’єктів масиву (одновимірним від 0 до 99) та зовнішнім їх поданням на екрані (двовимірна сітка 10×10 з індексами рядків та стовпців від 0 до 9).

– *left, right, up, down* – набір функцій, що реалізує склейку країв фрагменту (рис. Е.21);

– *getneigbcount* – повертає кількість заселених островів, що межують із даним: на сітці це такі острови, що розташовані зліва та вгорі, вгорі, справа та вгорі, зліва, справа, зліва та знизу, знизу, справа та знизу (рис. Е.22).

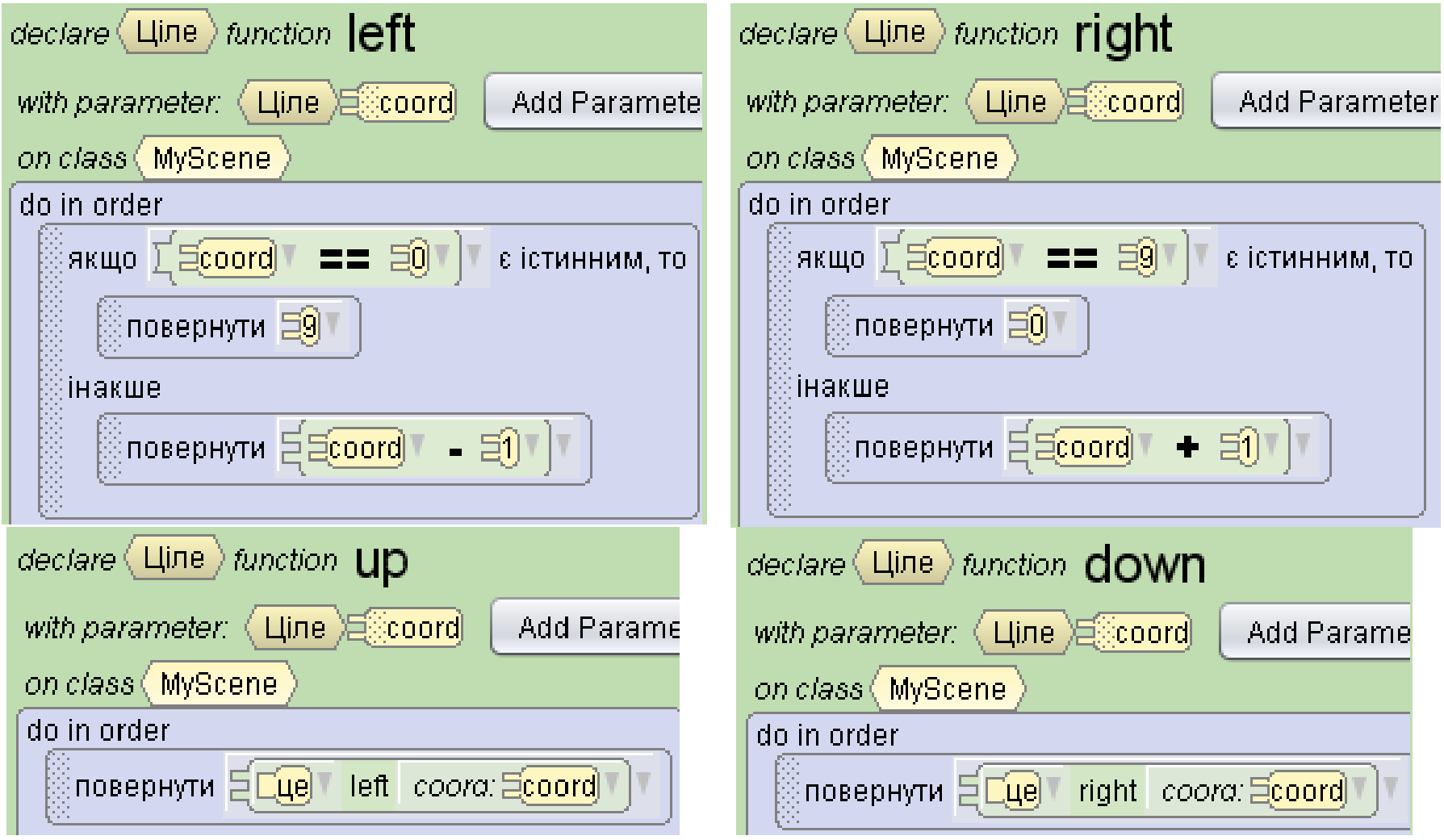

Рис. Е.21. Функції склеювання країв

При натисканні будь-якої клавіші процедура *onkeypress* виконує наступні дії (рис. Е.23):

1) встановимо номер кроку еволюції (номер ітерації) у 0;

2) для кожного острова:

а) підрахуємо кількість сусідніх заселених островів;

б) визначимо значення властивості *isbelive* у такий спосіб: якщо острів незаселений і кількість сусідніх заселених островів дорівнює трьом, то острів стає кандидатом на заселення; інакше, якщо острів заселений та кількість сусідніх заселених островів – 2 чи 3, то острів зберігає свій статус заселеного; інакше – наступний статус острова буде «незаселений».

3) для кожного острова викличемо процедуру *swap*.

4) виведемо номер кроку еволюції (номер ітерації) та збільшимо його на одиницю.

На прикладі цієї моделі можна проілюструвати основні принципи об’єктно-орієнтованого програмування:

1) *наслідування* – існуючих можливостей класів недостатньо для побудови моделі, тому довелося не просто створити класи-нащадки, й змінити їх функціональність у порівнянні з класами-батьками: додати нові дані (властивості) та методи (процедури та функції);

2) *інкапсуляція* – дії, що є специфічними для класу, реалізуються у його властивостях, доступ до яких з інших класів можливий лише через створення та виклик їх методів;

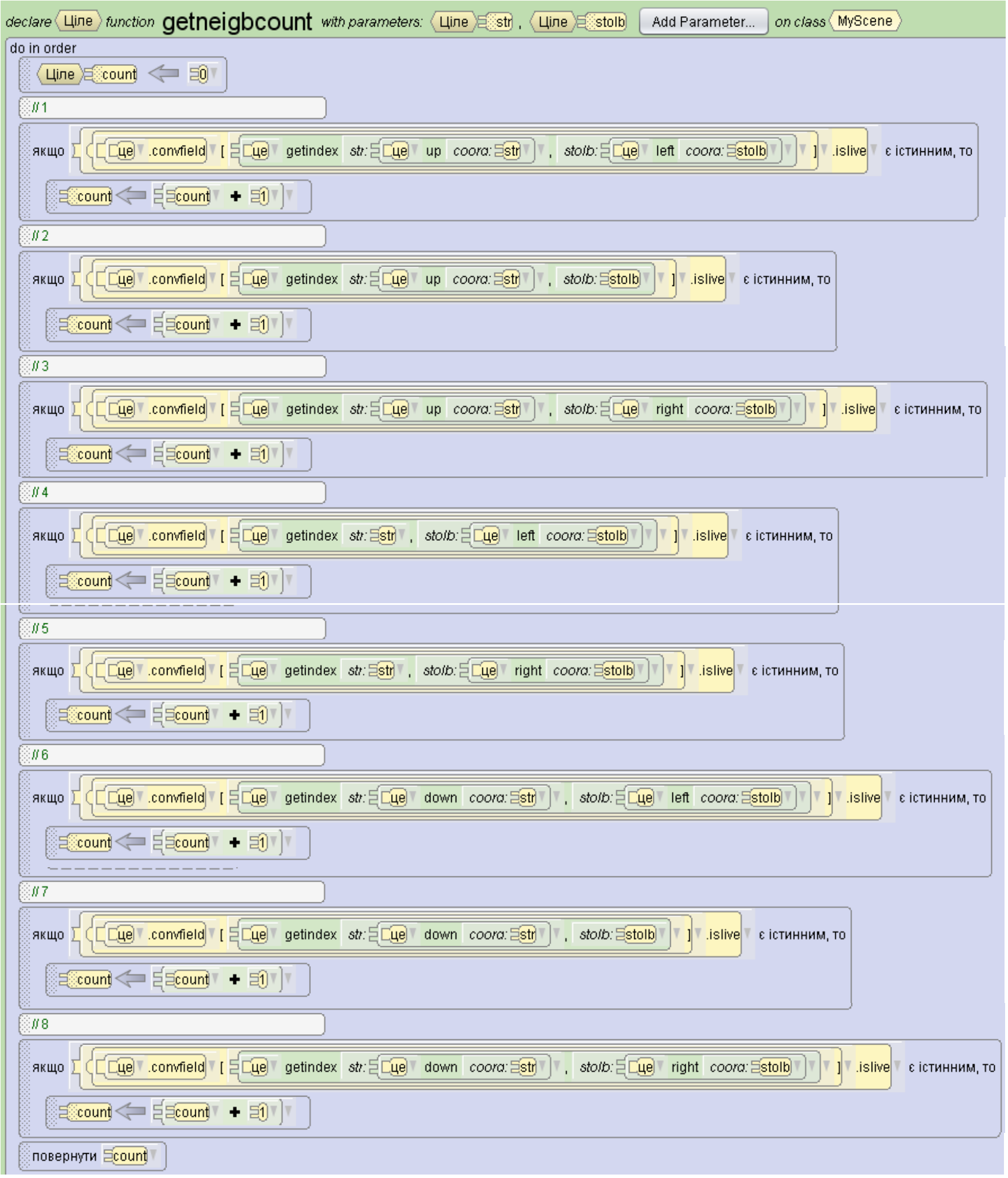

Рис. Е.22. Функція підрахунку кількості заселених островів-сусідів

3) *абстрагування* – реалізується функціонально-процедурною декомпозицією по методах класів у такий спосіб, щоб кожен метод був осяжним та зрозумілим, та створенням спеціалізованих класів (*Convik* та *MyScene*);

4) *модульність* – реалізується через можливість зберігання та повторного використання розроблених класів у нових проектах;

5) *ієрархія* – реалізується у системі класів Alice (як убудованих, так і створених користувачем), що походять від єдиного абстрактного класу

*Object*;

6) *типізація* – реалізується будовою мови Alice та Java. Прикладом узгодження типів є можливість поєднання різних об'єктів процедурами *сказати* та *подумати*;

7) *паралелізм* – може бути реалізований у процедурі *onkeypress*: визначення того, який буде наступний стан острова (заселений, чи ні), не залежить від порядку опрацювання елементів;

8) *стійкість* – властивість об'єктів існувати в часі та у просторі забезпечується можливостями середовища Alice (збереженням об'єктів у складі проекту) та його розміщенням у Internet.

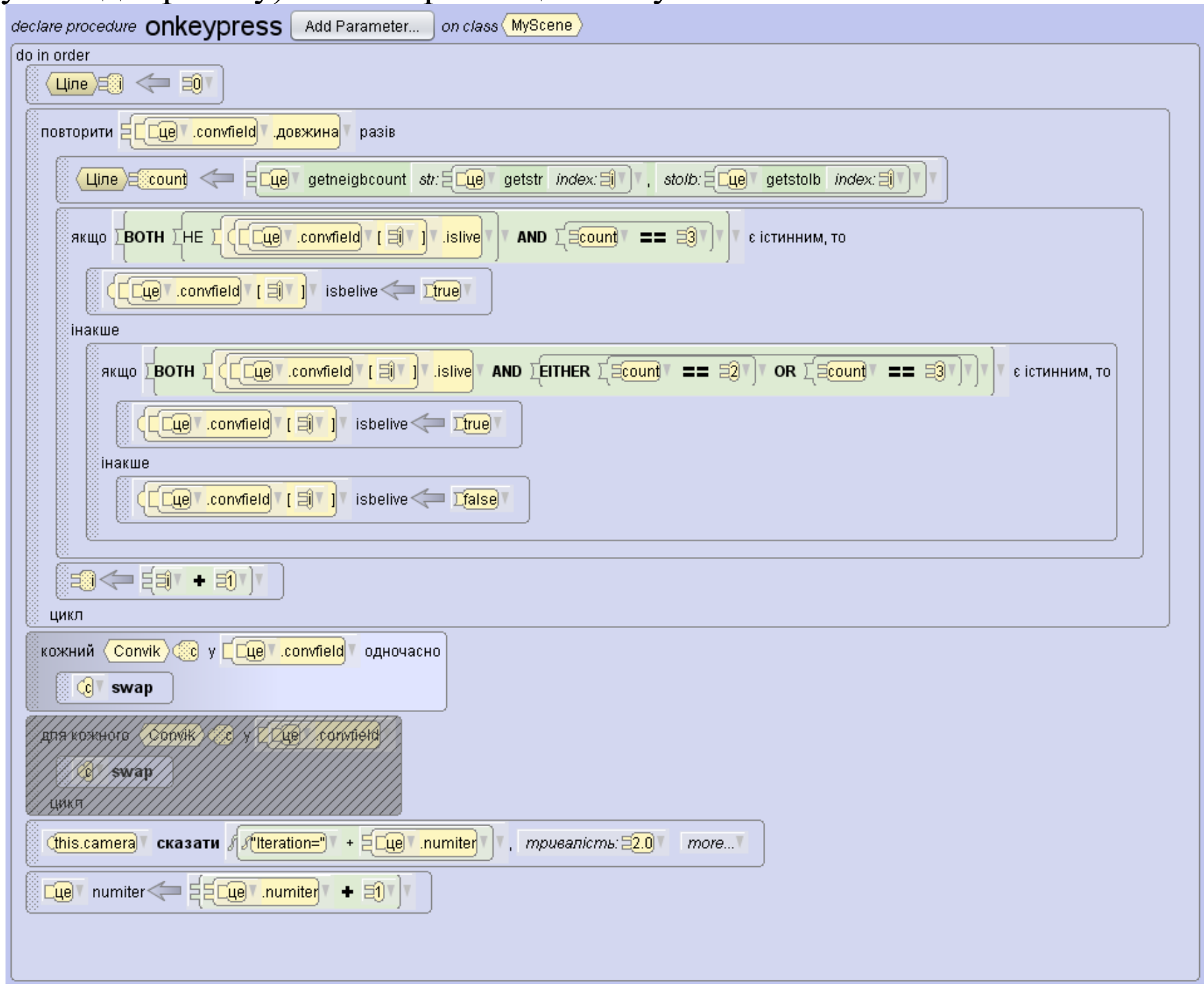

Рис. Е.23. Процедура *onkeypress* реалізує один крок еволюції

У даній моделі відсутні приклади *поліморфізму* – можливості визначення об'єкту у процесі звернення до нього, проте, враховуючи, що мова Alice є Java-подібною, а у Java всі методи є віртуальними, то й у Alice всі процедури та функції є поліморфними. Ознакою «чистоти» застосування концепцій об'єктно-орієнтованого програмування при розробці даної моделі є також те, що нам не довелося створити жодного рядка у процедурі *run* класу *MyScene*, як це робилось у п. Е.1. Чи не найкращою ілюстрацією корисності концепції паралелізму є багатократне (до 100 разів) прискорення швидкості зміни стану острову простою заміною циклу «*для кожного об'єкту у масиві*» (показаний на рис. Е.23 як закоментований) на «*кожний об'єкт у масиві одночасно*» у функції *onkeypress*, адже при виклику процедури *swap* тривалість одночасного стану оновлення островів практично не відрізняється від тривалості оновлення стану одного острова.

**О. І. Теплицький, І. О. Теплицький, С. О. Семеріков, В. М. Соловйов**

**Професійна підготовка учителів природничо-математичних дисциплін засобами комп'ютерного моделювання: соціально-конструктивістський підхід**

E-mail: semerikov@ccjournals.eu